\documentclass[twocolumn,epjc3]{svjour3}

\usepackage[a4paper, total={6in, 8in}, left=1in, right=1in, top=1.2in, bottom=1.2in]{geometry}

\usepackage[numbers,sort&compress]{natbib}
\RequirePackage[T1]{fontenc}
\smartqed
\RequirePackage{graphicx}
\RequirePackage[colorlinks,citecolor=blue,urlcolor=blue,linkcolor=blue]{hyperref}
\graphicspath{{figures/}} 

\usepackage{epstopdf}
\usepackage{amsmath}
\usepackage{amsfonts}
\usepackage{amssymb}
\usepackage{textcomp}
\usepackage[utf8]{inputenc}
\usepackage{hyperref}
\usepackage{slashed}
\usepackage{feynmp-auto}
\usepackage{bm}
\usepackage{subfigure}
\usepackage{orcidlink}
\usepackage{physics}
\usepackage{nicefrac}

\usepackage{ulem}
\usepackage{cancel}
\usepackage{cuted}
\usepackage{mathtools,empheq}

\usepackage{mathrsfs}

\usepackage{dsfont}

\usepackage{tikz}

\newcommand{\nn}{{\nonumber}}

\newcommand{\im}{{\rm Im}}

\newcommand{\be}{\begin{eqnarray}}
\newcommand{\ee}{\end{eqnarray}}
\newcommand{\ba}{\begin{array}}
\newcommand{\ea}{\end{array}}
\newcommand{\bi}{\begin{itemize}}
\newcommand{\ei}{\end{itemize}}
\newcommand{\amp}{\mathcal{A}} 

\newcommand{\pbbi}{\mathbb{P}_{\rm (i)}}
\newcommand{\pbbii}{\mathbb{P}_{\rm (ii)}}

\newcommand{\wpbbi}{\widetilde{\mathbb{P}}_{\rm (i)}}

\newcommand{\wpbbiii}{\widetilde{\mathbb{P}}_{\rm (iii)}}

\newcommand{\cffH}{\mathcal{H}}

\newcommand{\bp}{\bar{p}}

\newcommand{\hplus}{H^{(+)}}

\newcommand{\scale}{\mathbb{Q}}

\newcommand{\bu}{{\bar{u}}}

\newcommand{\dxi}{\partial_{\xi}}

\newcommand\Li[2]{{\textrm{Li}_{#1}\left(#2\right)}}  
\newcommand\Ln[1]{{\,\textrm{ln}\left(#1\right)}}  

\journalname{Eur. Phys. J. A}

\begin{document}

\title{Three-dimensional imaging of hadrons  with hard exclusive reactions: advances in experiment, theory, phenomenology, and lattice QCD}

\author{ 
 M.~Bo\"er\thanksref{addr-vt, e5} 
\and A.~Camsonne\thanksref{addr1,e5}
\and M.~Constantinou\orcidlink{0000-0002-6988-1745}\thanksref{addrTemple,e5}
\and H.S.~Jo\orcidlink{0009-0005-5672-6948}\thanksref{addr3,e5,e4}
\and K.~Joo\orcidlink{0000-0002-5265-6867}\thanksref{addr1a, e5}
\and K.~Semenov-Tian-Shansky\orcidlink{0000-0001-8159-0900}\thanksref{addr3,addr10a,e5,e4}  
\and H.-D.~Son\orcidlink{0000-0002-9601-3724}\thanksref{addr5, e5}
\and P.~Sznajder\orcidlink{0000-0002-2684-803X}\thanksref{addr6, e5}
\and C.~Van~Hulse\orcidlink{0000-0002-5397-6782}\thanksref{addr-Alcala,e5} 
\and {E.~Voutier}\orcidlink{0000-0003-4990-3369} ~\thanksref{addr12, e5}
    \and J.~Wagner\orcidlink{0000-0001-8335-7096}\thanksref{addr6, e5}
\and    
A.~Afanasev\orcidlink{0000-0003-0679-3307}\thanksref{addr-gwu}
\and  J.S.~Alvarado\orcidlink{0000-0001-8642-3036}\thanksref{addr12} 
\and  S.~Bhattacharya \orcidlink{0000-0001-8536-082X}\thanksref{addr-uconn} 
 \and D. Biswas\thanksref{addr-vt} 
    \and Xu Cao\orcidlink{0000-0003-3827-1781}\thanksref{addr-imp}  
    \and H.-M.~Choi\orcidlink{0000-0003-1604-7279}\thanksref{addr3}
    \and K.~Cichy\orcidlink{0000-0002-5705-3256}\thanksref{addr-amu}
    \and N.~Crnković\thanksref{IRB}
    \and W.~Hamdi\orcidlink{0009-0004-4045-723X}\thanksref{addr-Tunisia} 
    \and M.~Hoballah\thanksref{addr12}
    \and G.M.~Huber\orcidlink{0000-0002-5658-1065}\thanksref{addr-regina} 
    \and P.~T.~P.~Hutauruk\orcidlink{0000-0002-4225-7109}\thanksref{addr8}
    \and A.~Jentsch\thanksref{addr-BNL}
    \and C.-R.~Ji\orcidlink{0000-0002-3024-5186}\thanksref{addr2}       
    \and H.-Ch.~Kim\orcidlink{0000-0002-8718-8661}\thanksref{addr5} 
    \and B.~Kriesten\orcidlink{0000-0002-9587-8777}\thanksref{addr-argonne}
    \and Huey-Wen Lin\orcidlink{0000-0001-6281-944X}\thanksref{addr-msu}
    \and P.-J. Lin\orcidlink{0000-0001-7073-6839}\thanksref{addr-ncu}
    \and V.~Mart\'inez-Fern\'andez\orcidlink{0000-0002-0581-7154}\thanksref{addr-irfu, addr-cfns}
    \and M.~Mazouz\orcidlink{0000-0003-2375-4676}\thanksref{addr-Tunisia} 
    \and Z.-E.~Meziani\orcidlink{0000-0001-9450-2914}\thanksref{addr-argonne1} 
    \and M.~Nefedov\orcidlink{0000-0002-1046-9625}\thanksref{BenGurion}
    \and K.~Passek-K.\orcidlink{0000-0001-6520-3640}\thanksref{IRB}
    \and B.~Pire\orcidlink{0000-0003-4882-7800}\thanksref{addrX}
    \and P.~Rossi\orcidlink{0000-0002-2366-1085}\thanksref{addr4,addr9}  	
    \and O.~Teryaev\orcidlink{0000-0001-7002-9093}\thanksref{addr-Dubna}
    \and A.~W.~Thomas\orcidlink{0000-0003-0026-499X}\thanksref{addr7} 
    \and N.~Tomida\thanksref{addr-kyoto}    
    \and {Z.~W.~Zhao}\thanksref{addr-duke}
    }

\thankstext{e5}{Editors}
\thankstext{e4}{Corresponding authors, e-mail: ksemenov@knu.ac.kr, hyonsuk@knu.ac.kr}

\institute{
Virginia Tech, Physics department, Blacksburg, VA 24061, USA
\label{addr-vt}
\and 
Thomas Jefferson National Accelerator Facility, Newport News, VA 23606, USA 
\label{addr1}
\and
Temple University, Philadelphia, PA 19122-1801, USA 
\label{addrTemple}  
\and
Department of Physics, Kyungpook National University, Daegu 41566, Republic of Korea 
\label{addr3}
\and 
University of Connecticut, Storrs, CT 06269, USA 
\label{addr1a}
\and
NRC ``Kurchatov Institute'' - PNPI, Gatchina 188300, Russia
\label{addr10a}
\and
Department of Physics, Inha University, Incheon 22212, Republic of Korea 
\label{addr5}
\and 
National Centre for Nuclear Research (NCBJ), 02-093 Warsaw, Poland 
\label{addr6}
\and
University of Alcal\'{a}, Alcal\'{a} de Henares, Madrid, Spain
\label{addr-Alcala}
\and
Universit\'{e} Paris-Saclay, CNRS/IN2P3, IJCLab, 91405 Orsay, France
\label{addr12}
\and
Department of Physics, George Washington University, Washington, DC 20052, USA
\label{addr-gwu}
\and
Department of Physics, University of Connecticut, Storrs, CT 06269, USA
\label{addr-uconn}
\and
Institute of Modern Physics, Chinese Academy of Sciences, Lanzhou 730000, China
\label{addr-imp}
\and
Faculty of Physics and Astronomy, Adam Mickiewicz University,  61-614 Poznań, Poland 
\label{addr-amu}
\and
Division of Theoretical Physics, Rudjer Bošković Institute (RBI), HR-10002 Zagreb, Croatia
\label{IRB}
\and
Monastir University, Faculty of Sciences of Monastir, LPQS Laboratory, Monastir 5019, Tunisia
\label{addr-Tunisia}
\and
 University of Regina, Regina, SK S4S~0A2, Canada
 \label{addr-regina}          
\and
Department of Physics, Pukyong National University (PKNU), Busan 48513, Republic of Korea
\label{addr8}
\and 
Department of Physics, Brookhaven National Laboratory, Upton, New York 11973, U.S.A.\label{addr-BNL}
\and
Department of Physics and Astronomy, North Carolina State University, Raleigh, NC 27695-8202, USA 
\label{addr2}
\and 
High Energy Physics Division, Argonne National Laboratory, Lemont, IL 60439, USA 
\label{addr-argonne}
\and
Department of Physics and Astronomy, Michigan State University, East Lansing, Michigan 48824, USA
\label{addr-msu}
\and
Center for High Energy and High Field Physics and Department of Physics, National Central University, Zhongli 320317, Taiwan
\label{addr-ncu}
\and
IRFU, CEA, Universit\'e Paris-Saclay, F-91191 Gif-sur-Yvette, France
\label{addr-irfu}
\and
Center for Frontiers in Nuclear Science, Stony Brook University, Stony Brook, NY 11794, USA
\label{addr-cfns}
\and 
Physics Division, Argonne National Laboratory, Lemont, IL 60439, USA 
\label{addr-argonne1}           
\and
Physics Department, Ben-Gurion University of the Negev, 84105, Beer Sheva, Israel
\label{BenGurion}
\and CPHT, CNRS, \'Ecole polytechnique, I.P. Paris, 91128 Palaiseau, France
\label{addrX}
\and
Thomas Jefferson National Accelerator Facility, Newport News, VA 23606, USA
\label{addr4}
\and
INFN Laboratori Nazionali di Frascati, 00044, Italy 
\label{addr9}
\and   
Bogoliubov Laboratory of Theoretical Physics, Joint Institute for Nuclear Research, 141980 Dubna, Russia   
\label{addr-Dubna} 
\and
CSSM, Department of Physics, University of Adelaide, Adelaide SA 5005, Australia
\label{addr7}
\and
Department of Physics, Kyoto University, Kyoto, 606-8502, Japan
\label{addr-kyoto}
\and
Department of Physics, Duke University, Durham, NC 27708, USA
\label{addr-duke}
}

\date{\today}

\maketitle

\sloppy 


\begin{abstract}
Generalized Parton Distributions (GPDs) have emerged as a powerful framework for exploring the internal structure of hadrons in terms of their partonic constituents. Over the past three decades, the field has witnessed significant theoretical and experimental advancements. The interpretation of GPDs in impact parameter space offers a vivid three-dimensional visualization of hadron structure, correlating longitudinal momentum and transverse spatial distributions, thereby  enabling tomographic imaging of hadrons. 

Furthermore, the link between GPDs and the matrix elements of the QCD energy-momentum tensor provides access to fundamental properties of hadrons, including spin decomposition and internal pressure distributions. Notably, recent analyses of Deeply Virtual Compton Scattering (DVCS) data have enabled the empirical extraction of the quark pressure profile inside the proton.

{Motivated by the rapidly evolving experimental landscape, this white paper provides a timely and focused overview of recent developments in GPD theory, phenomenology, and lattice QCD studies. Its scope is shaped by the needs and opportunities of forthcoming experimental programs, and it highlights advances that are particularly relevant for the next generation of dedicated measurements, including the extended Jefferson Lab 12 GeV program and its potential 22 GeV upgrade, J-PARC, COMPASS/AMBER, LHC ultra-peripheral collisions, and the future electron-ion colliders EIC and EicC.}

\end{abstract}


\newpage
\setcounter{tocdepth}{2}
\tableofcontents

\section{Introduction}
\label{sec:introduction}

In the roughly 50 years since the discovery of Quantum Chromodynamics (QCD, see 
{\it e.g.},
\cite{Gross:2022hyw} for a review) we have learned a great deal about strongly interacting systems. Impressive though the achievements have been, there is even more to understand. These challenges range from the elastic form factors and parton distribution functions of the nucleon to the hadron spectrum and atomic nuclei.
The understanding of hadron structure in terms of  the fundamental degrees of freedom
of QCD -- quarks and gluons -- remains a profound and persistent problem.

The main development path for studying hadronic structure consists in using small-sized quark and gluon probes created within hard reactions. {The examples include} Deep-Inelastic Scattering (DIS), semi-inclusive DIS, Drell-Yan processes, and
hard exclusive reactions 
{such as}
Deeply Virtual Compton Scattering (DVCS)
and hard exclusive electroproduction of mesons, 
{commonly} abbreviated  as DVMP (Deeply Virtual Meson Production), or DEMP 
(Deep Exclusive Meson Production). The crucial feature of hard reactions is the ability to separate the
perturbative and non-perturbative stages of interaction, relying on the factorization theorems. The interaction of a hard probe is then described using perturbative QCD, while the response of the target hadron is specified in terms of non-perturbative objects that encode hadronic structural information.
These non-perturbative objects, defined through hadronic matrix elements of specific QCD operators, can be assigned a rigorous meaning in QCD and provide
access to various aspects of hadronic structure. Depending on the nature of the hard processes, these non-perturbative objects comprise Parton Distribution
Functions (PDFs) for inclusive reactions, Transverse-Momentum-Dependent parton distributions (TMDs) for semi-inclusive reactions, and Generalized Parton Distributions
(GPDs), Distribution Amplitudes (DAs), Generalized Distribution Amplitudes (GDAs), and Transition Distribution Amplitudes (TDAs) for exclusive reactions.
Alongside studies of more conventional quantities, such as hadronic form factors, this provides a broad program for investigating the 3D structure of hadrons, see  {\it e.g.},
\cite{Diehl:2023nmm, Lorce:2025aqp}.

In this white paper, we primarily focus on hadronic structure studies associated with GPDs (and much related matrix elements, DAs, GDAs, and TDAs) probed in hard exclusive reactions. By fully specifying the final state and maintaining complete control over the reaction mechanism, hard exclusive reactions allow for a clear interpretation of observables in terms of QCD factorization theorems and facilitate an accurate mastering of  QCD evolution effects.

Generalized parton distributions (GPDs) emerged in the mid-1990s through complementary developments. An operator-based foundation was established by D. Müller  {\it et al.} through studies of the evolution properties of hadronic matrix elements of QCD light-cone operators~\cite{Muller:1994ses}. Subsequently, X. D. Ji introduced GPDs as fundamental descriptors of nucleon structure, establishing their connection to the QCD energy–momentum tensor, deriving the nucleon spin sum rule, and formulating the description of deeply virtual Compton scattering and deeply virtual meson production in terms of GPDs~\cite{Ji:1996ek,Ji:1996nm}. Alternative theoretical descriptions of non-forward parton distributions were presented in Refs.~\cite{Radyushkin:1996nd,Radyushkin:1997ki}. {A historical overview is given in Ref.~\cite{Ji:2016djn}.}
Rigorous proofs of factorization theorems were provided for these processes at the leading twist-$2$ level, {\it i.e.},
for transversely polarized virtual photons in DVCS ~\cite{Collins:1998be} and longitudinally polarized virtual photons in DVMP~\cite{Collins:1996fb}. Different aspects of the physics associated with GPDs are reflected in the detailed review papers~\cite{Goeke:2001tz,Diehl:2003ny,Belitsky:2005qn,Boffi:2007yc,Guidal:2013rya,Mueller:2014hsa}.

{
From the experimental perspective, pioneering measurements of exclusive processes were performed by the HERA experiments (HERMES, H1 and ZEUS) at the Deutsches Elektronen Synchrotron (DESY), through the study of DVMP~\cite{H1:1996gwv,H1:1997gev,HERMES:2000jnb} and DVCS~\cite{HERMES:2001bob,H1:2001nez,ZEUS:2003pwh}, and by the CLAS experiment at the Continuous Electron Beam Accelerator Facility (CEBAF) at $6$~GeV at Jefferson Lab (JLab), through the study of DVCS~\cite{CLAS:2001wjj}.  The experimental program on exclusive processes was subsequently extended at the HERA experiments, the JLab CLAS and Hall A experiments, and at the COMPASS experiment at CERN. In addition, studies of hadronic structure using GPDs became a major focus of the CEBAF research program at $12$~GeV~\cite{Dudek:2012vr,Burkert:2018nvj}, which is currently under way and has started to provide a growing number of results.}

The potential of GPDs for studying hadronic structure became evident after  X.D.~Ji's paper~\cite{Ji:1996ek} established them as a tool to investigate the origin of the nucleon's spin,
with the help of the sum rule quantifying the total angular momentum carried by partons. This ultimately revealed an opportunity to access various mechanical characteristics of hadrons, such as the distribution of pressure and shear forces 
\cite{Polyakov:2002yz}, 
which are encoded in the hadronic matrix elements of the QCD Energy-Momentum Tensor (EMT) operator and expressed in terms of so-called Gravitational Form Factors (GFFs) of hadrons. The latter can be written as the first Mellin moment of GPDs.
These insights have inspired extensive experimental 
\cite{Burkert:2018bqq,Kumericki:2019ddg,Duran:2022xag}
and theoretical activity over the past decade, as reviewed in~\cite{Polyakov:2018zvc,Burkert:2023wzr,Lorce:2025oot}.

Another exciting application of GPDs, first revealed by M.~Burkardt
\cite{ Burkardt:2000za},
is the possibility of performing spatial imaging (femto-photography~\cite{Ralston:2001xs}) of hadrons in the transverse plane, providing a comprehensive description of hadron structure in both longitudinal momentum and transverse spatial dimensions.

Beyond the nucleon, nuclear GPDs offer novel means to study the properties of nuclei, bridging the gap between traditional nuclear structure models and the underlying quark-gluon dynamics governed by QCD.
In particular, GPDs of light nuclei 
\cite{Berger:2001zb, Scopetta:2004kj}, 
which are accessible in coherent exclusive electroproduction experiments 
\cite{Cano:2003ju, Kirchner:2003wt,Martinez-Fernandez:2026web,Martinez-Fernandez:2026zog}, 
offer unique access to the gluon content of nuclei, which is challenging to study using conventional methods. Understanding the gluon distribution is crucial for explaining phenomena like nuclear binding, shadowing effects, and modifications of parton distributions in nuclei.
The GPDs can be used to investigate how gluon distributions change in the nuclear medium compared to free nucleons. This has implications for understanding the EMC effect 
\cite{EuropeanMuon:1983wih,Geesaman:1995yd}, 
as well as attempts to study its origin~\cite{Weinstein:2010rt,CLAS:2019vsb,Hen:2016kwk,Wang:2020uhj,Cloet:2006bq,Thomas:2018kcx,Guichon:1995ue,Guichon:2018uew}
and other nuclear modifications observed in deep inelastic scattering experiments, see {\it e.g.}, Ref.~\cite{Klasen:2023uqj} and references therein.
Currently, there have been only a few preliminary studies on how the measurements of GPDs might contribute~\cite{Guzey:2008fe,CLAS:2018ddh,Fucini:2020lxi}, 
but this topic certainly deserves dedicated investigation.

As for other exclusive processes, the phenomenon of color transparency is a signature of the factorization property of amplitudes in processes  where an emerging meson is created in a short distance process~\cite{Jain:1995dd,Szumila-Vance:2025xvz}. The topic, addressed in the early days of GPD physics~\cite{Liuti:2004hd}, has been discussed recently both experimentally~\cite{CLAS:2012tlh} and theoretically~\cite{ Jain:2022xzo, Goharipour:2025kif, Huber:2022wns, Kong:2025del}. More experimental data on nuclear transparency are clearly needed, in particular for a variety of nuclei.

Another area where the study of GPDs may provide new insights is hadron spectroscopy and the exploration of properties of hadronic resonances~\cite{Diehl:2024bmd}. Non-diagonal exclusive electroproduction processes offer a means to access transition GPDs, which provide information on the quark-gluon structure of resonances. Specifically, one can study the mechanical properties of excited states  encoded in transitional GFFs. Moreover, this approach can introduce new tools to investigate the controversial nature of
 baryon excitations, such as the Roper resonance, $N^*(1440)$~\cite{Wu:2017qve,Burkert:2017djo,Burkert:2025coj}, and the first excited state of the $\Lambda$-hyperon, $\Lambda(1405)$~\cite{Veit:1984jr,Hall:2014uca,Mai:2020ltx}.
 Additionally, hard electroproduction processes provide an opportunity to address various problems in meson spectroscopy, {\it e.g.}, the search for  exotic $J^{P C}=1^{-+}$ hybrid mesons 
~\cite{Anikin:2004vc,Anikin:2004ja}.

The last decade has also seen considerable progress on the theory side and in the development of analysis tools that are gradually transforming GPD studies from model-driven exploration into a precision science. Below, we list, in an unavoidably incomplete manner, some of these developments:
\begin{itemize}
\item  
Sophisticated global analyses aimed at extracting information on GPDs from multiple-channel observables, as exemplified by Refs.~\cite{Kumericki:2015lhb,Moutarde:2018kwr,Guo:2023ahv,Cuic:2023mki,Kriesten:2020apm,Guo:2025muf,Panjsheeri:2025vpa,Goharipour:2025lep}, along with developments of the open-source frameworks Gepard~\cite{gepard} and PARTONS~\cite{Berthou:2015oaw}, and the creation of an open database for GPD analysis~\cite{Burkert:2025gzu}.

\item The first extraction of the proton $D$-term form factor from JLab  DVCS data revealed the quark component of the pressure distribution in the proton~\cite{Burkert:2018bqq}. As pointed out in~\cite{Kumericki:2019ddg}, further analyses with current and future experimental data are expected to enhance our understanding of the $D$-term and its relation to the mechanical properties of hadrons, as discussed in
Refs.~\cite{Polyakov:2018rew,Lorce:2025oot,Ji:2025qax}.
Moreover, gluonic GFFs of the proton have recently 
been extracted from threshold $J/\psi$ photoproduction 
data~\cite{Duran:2022xag} within the threshold approximation, 
{albeit in a model-dependent manner.}
This 
may lead to
a better understanding of the role of the gluonic component in shaping the properties of nucleons. 

\item 
Significant progress has been achieved in next-to-leading order (NLO) and next-to-next-to-leading order (NNLO)~\cite{Braun:2020yib,Braun:2021grd}, higher-twist~\cite{Braun:2025xlp}, and kinematic~\cite{Braun:2014sta,Braun:2022qly} corrections, providing the theoretical precision necessary for comparison with modern experimental data.   

\item The present day high precision DVCS data are shown to be sensitive to twist-3 and/or higher-order contributions. In particular, the analysis in Ref.~\cite{Defurne:2017paw} highlighted the crucial role of gluons in describing the DVCS process, either through quark-gluon correlations occurring in higher twist diagrams or through NLO effects involving gluon {tensor} GPDs.

\item Considerable development  has occurred in the application of Lattice QCD methods. The exploration of parton densities on the Euclidean lattice has enabled direct access to the $x$-dependence of parton distributions through the use of parton quasi-distributions and large momentum effective theory (LaMET)~\cite{Ji:2013dva}. This is further complemented by approaches utilizing pseudo-distributions~\cite{Radyushkin:2017cyf,Orginos:2017kos}. Numerous studies have been conducted from both {the theoretical and the lattice} QCD perspectives; see Refs.~\cite{Ji:2020ect,Cichy:2021lih,Lin:2025hka} for a review. 

\item 
Light-front (LF) quantization offers a frame-independent Hamiltonian formulation in which hadron structure is encoded 
in light-front wave functions~\cite{Brodsky:1997de}. 
In this basis, GPDs admit an intuitive overlap representation that connects partonic Fock components to observables and implements positivity constraints transparently~\cite{Diehl:2000xz,Brodsky:2001xy}. 
There  has also been growing interest in exploring 
GPDs within the framework of light-front holographic QCD (LFHQCD), an approach to hadron structure that is based on the holographic embedding of the light-front dynamics in a higher-dimensional gravity theory
\cite{Nishio:2014rya, Nishio:2014eua,deTeramond:2018ecg,Mamo:2024jwp, Mamo:2024vjh}.

\item The calculation of GPDs in effective hadronic models such as the chiral quark–soliton model ($\chi$QSM), MIT bag model, Nambu–Jona-Lasinio (NJL) model, light-front constituent quark models, etc., play a crucial role in bridging QCD symmetries and observable hadron structure.
Recent progress includes accounting for the    
instanton effects in twist-$3$ GPDs in the $\chi$QSM~\cite{Kim:2023pll},    
and the calculation of the poorly constrained  
tensor GPDs in the large $N_c$ limit
\cite{Kim:2024ibz}, as well as in the MIT bag model~\cite{Tezgin:2024tfh}.

\item Unraveling new connections between GPDs and quantum anomalies~\cite{Bhattacharya:2022xxw,Bhattacharya:2023wvy} has provided  strong motivation to further advance GPD studies. It was discovered that the trace anomaly and chiral anomaly lead to relationships for the form factors of local QCD operators, which can be expressed in terms of the Mellin moments of GPDs.

\item  
There are also attempts to compute GPDs with the help of the functional approach based on the Dyson-Schwinger (DS) and Bethe-Salpeter (BS) equations~\cite{Mezrag:2023nkp}.
The basic calculation of pion GPD within the   rainbow-ladder truncation scheme has been performed in~\cite{Mezrag:2014jka}. Some useful steps toward generalizing of the approach for a nucleon are presented in Ref.~\cite{Bednar:2018htv}.

\item  
A major coordinated effort 
{in GPD physics}
is the Quark–Gluon Tomography (QGT) Topical Collaboration, supported by the U.S. Department of Energy, Office of Science, Office of Nuclear Physics~\cite{QGTcollab}.
{It exemplifies the broader trend toward the formation of large-scale collaborations in this field.}

\end{itemize}

We are now at a decisive moment for the future of hadronic physics over the  upcoming decades, as several experimental facility projects are progressing through the design stage and developing their research programs. This includes the project of the Electron Ion Collider (EIC) 
\cite{AbdulKhalek:2021gbh,Burkert:2022hjz}
at the Brookhaven National Laboratory (BNL), and the complementary Electron ion collider in China (EicC)
\cite{CAO:2020gbr,CAO:2024fdz,Anderle:2021wcy}, proposed as an extension of the Heavy Ion Accelerator Facility (HIAF)~\cite{Yang:2013yeb}.
Additionally, there is a proposal to add a positron ring to the JLab CEBAF installation 
\cite{Afanasev:2019xmr},
and, potentially, to implement the 22~GeV upgrade for JLab 
\cite{Accardi:2023chb}.

Complementary options to study GPDs in experiments with secondary beams are  being considered for J-PARC~\cite{DY_PRD2016, Tomida_MENU2023, Takahashi:2019xcq}.
There are also prospects to access GPDs through the cross channel counterpart of the DVCS and DVMP reactions, such as exclusive proton-antiproton annihilation into two photons at large $s$ and $t$, with
the \={P}ANDA experiment at GSI-FAIR~\cite{PANDA:2009yku}.
Moreover, while not a dedicated facility for GPD studies, the NICA heavy-ion collider at JINR Dubna 
\cite{Kekelidze:2016hhw}
supports an experimental program through the Spin Physics Detector (SPD) that includes investigations of spin-dependent GPDs, gluon distributions, and the spatial structure of partons in the nucleon.

\begin{figure}[hbt]
\includegraphics[width=0.98\columnwidth]{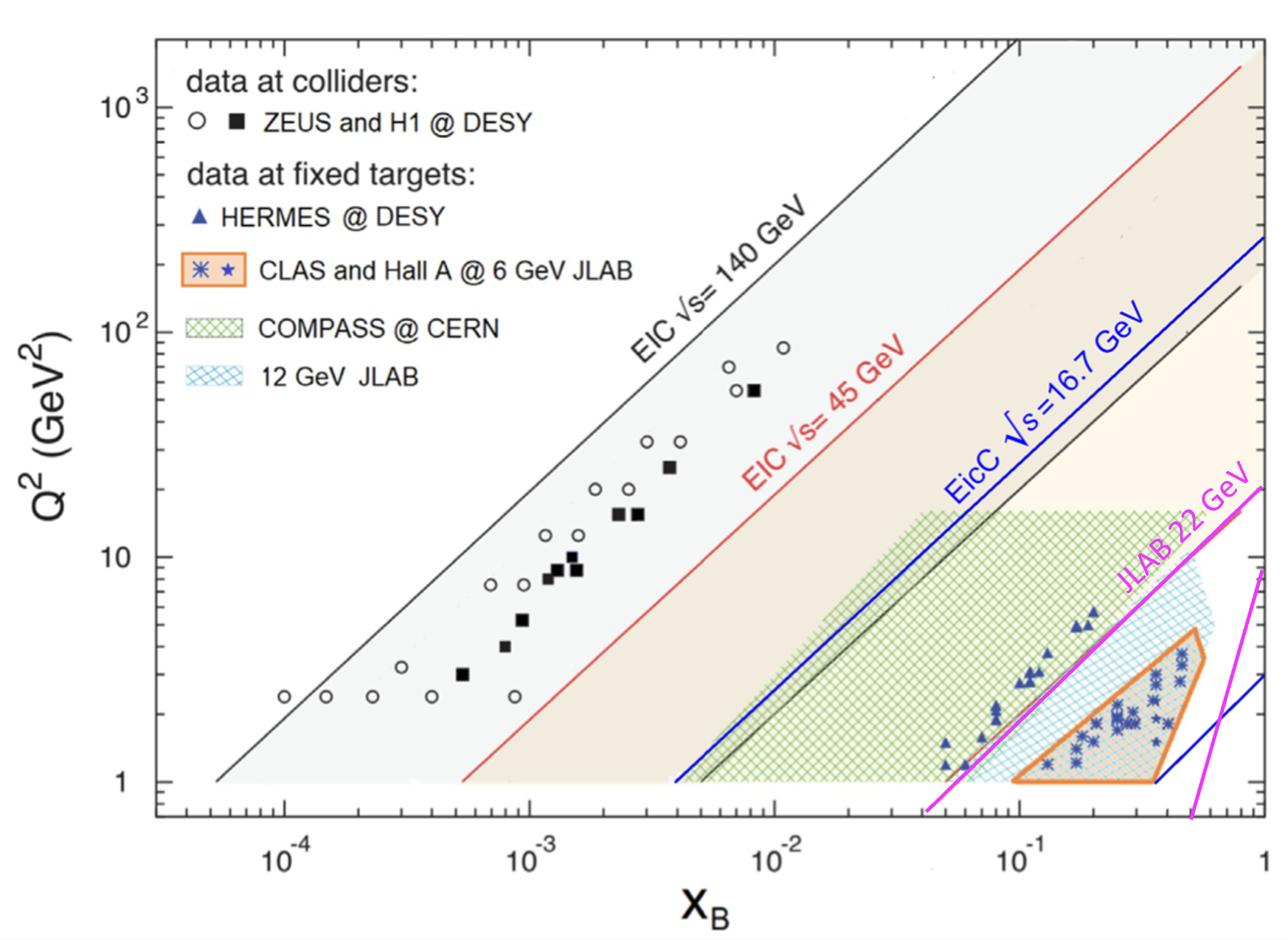}
\caption{Overview on the kinematic coverage of previous, current, and future DIS experiments. As an example, the available kinematic points from published data of the DVCS
process are given for the different experiments. This figure is taken from 
Ref~\cite{Diehl:2023nmm}. 
We additionally included the kinematical range of the planned EicC experimental facility taken from Fig.~2 of Ref.~\cite{Cao:2023wyz} 
and JLab 22~GeV, Fig.~17 of 
Ref.~\cite{Accardi:2023chb}.}
\label{fig:kinematic_coverage}
\end{figure}

The workshop ``3D Structure of the Nucleon via Generalized Parton Distributions'' 
[\href{https://indico.knu.ac.kr/event/706/}{https://indico.knu.ac.kr/event/706/}]
was held in Incheon, Republic of Korea, on June  25-28, 2024, and was supported 
by the \href{http://apctp.org/}{Asia Pacific Center for Theoretical Physics} (APCTP),  \href{https://chep.knu.ac.kr/}{Center for High Energy Physics} (CHEP), and
\href{https://rsri.knu.ac.kr/}{Radiation Science Research Institute} of \href{https://en.knu.ac.kr/main/main.htm}{Kyungpook National University} (KNU), Daegu, Republic of Korea.  
During the four days of the workshop, nineteen presentations were given and followed by extensive discussions. The aim of the workshop  was to review the current status of studies of the 3D structure of the nucleon via GPDs and to formulate a plan for future development of the field.
Particular emphasis was placed on formulating a road map for the Korean hadronic physics community to engage in developing  research programs for potential future experiments and to enhance scientific collaboration with foreign research institutions.

The workshop ``Towards improved hadron tomography with hard exclusive reactions'' was held in Trento, Italy, on August 5-9, 2024, by ECT*-Trento [\href{https://indico.ectstar.eu/event/211/}{https://indico.ectstar.eu/event/211/}]  and 
supported by the
\href{https://web.infn.it/EURO-LABS/}{EUROpean Laboratories for Accelerator Based Sciences} (EURO-LABS),
\href{https://www.infn.it/en/}{Istituto Nazionale di Fisica Nucleare} (INFN), and the
\href{https://jsallc.org/}{Jefferson Science Associates} (JSA).
The workshop included 34 presentations followed by a round-table discussion.
A key focus of this workshop was on ``novel'' reactions that provide access to GPDs, as well as on innovative techniques for modeling and projecting observables relevant to upcoming measurements.
Special attention was given to developments in lattice QCD and the integration of emerging computational approaches--such as machine learning and quantum computing--into theoretical frameworks and data analysis strategies. Physics topics covered included hard exclusive production of mesons, lepton pairs, and meson pairs, investigated across both fixed-target and collider experiments. Ultra-peripheral collisions and their role in accessing GPDs, Generalized Transverse Momentum Distributions (GTMDs), {GFFs}, and spin-related observables were also a central focus of the workshop.

This paper offers a selective overview focused on the topics covered  at the Incheon and Trento workshops and highlights the research perspectives that emerged from them.  
Although being selective,  it provides a broad and timely overview of the current status of hadron structure studies through hard exclusive reactions, while identifying some of the most promising directions for future theoretical and experimental developments. In particular, it reflects the community’s efforts to establish guidelines for upcoming experimental programs,  deepen international collaborations, and integrate novel approaches -- from lattice QCD to machine learning -- into the broader framework of hadron structure studies.


\section{A brief overview of GPDs}
\label{sec:gpd_overview}

\begin{figure*}[h]
 \centering
 \includegraphics[width=0.31\linewidth]{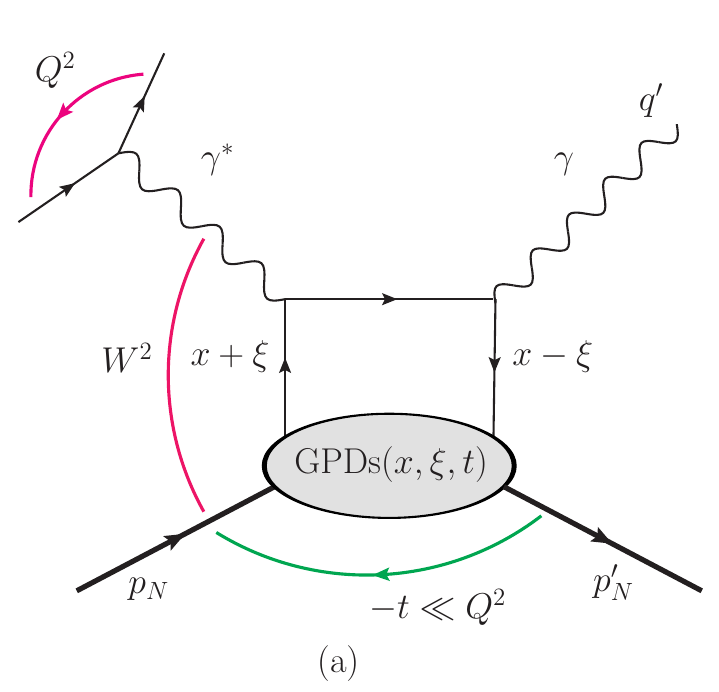}
 \includegraphics[width=0.31\linewidth]{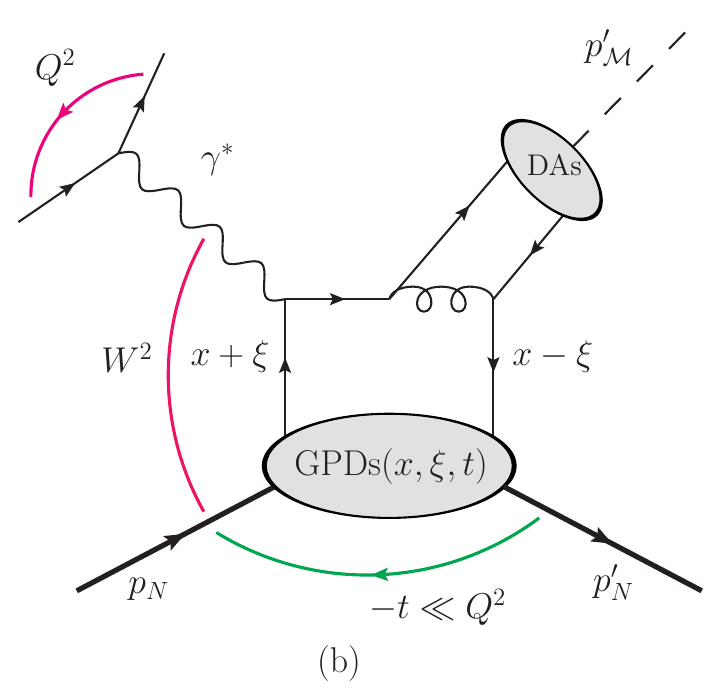}
 \includegraphics[width=0.31\linewidth]{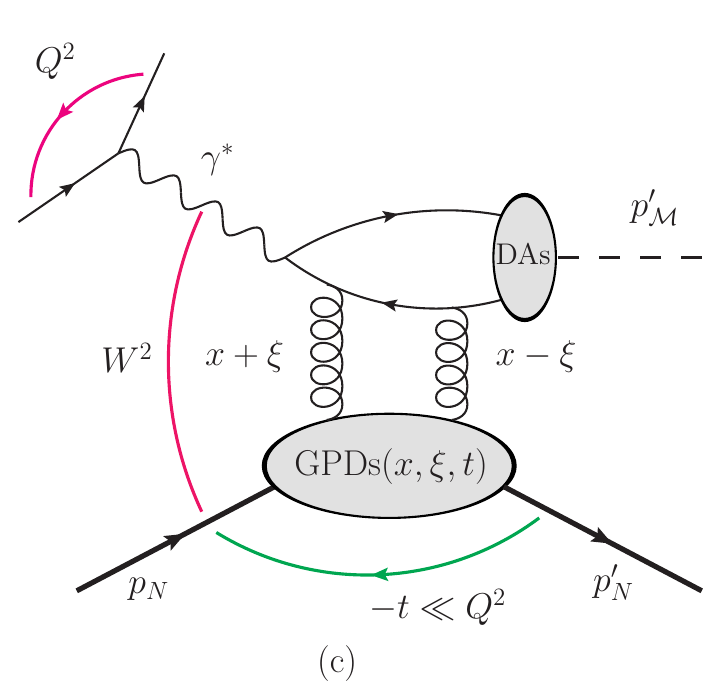}
 \caption{ 
{Dominant mechanisms} for exclusive processes involving GPDs: deeply virtual Compton scattering (a);
hard exclusive electroproduction of {flavor non-singlet} mesons (b); and {flavor singlet mesons} (c). 
$Q^2=-q^2$
is the virtuality of the virtual photon; 
$x$ 
is the average fractional longitudinal momentum of the active parton, and  
$\xi$  
is the skewness variable that characterizes the longitudinal momentum transfer between 
the initial and the final nucleon states. The Mandelstam variable 
$t$ 
represents the invariant momentum transfer between the initial and final nucleons.  
The DAs refer to the distribution amplitudes of the produced mesons. 
 }
 \label{Fig_DVCS_DVMP}
\end{figure*}

The concept of generalized parton distributions emerged in the 1990s as a natural extension of the partonic model framework. In early studies by the Leipzig group
\cite{Muller:1994ses},
GPDs arose as non-forward matrix elements of non-local QCD operators on the light cone, exhibiting a generic QCD scale evolution that ``interpolates'' between the familiar Dokshitzer–Gribov–Lipatov–Altarelli–Parisi (DGLAP) \cite{Gribov:1972ri,Altarelli:1977zs,Dokshitzer:1977sg} evolution of usual parton distribution functions and the Efremov–Radyushkin–Brodsky–Lepage~(ERBL) \cite{Efremov:1979qk,Lepage:1979zb} evolution of meson distribution amplitudes. The field gained significant attention in 1996, 
when the role of generalized parton distributions was established in the theoretical description of deeply virtual Compton scattering 
~\cite{Ji:1996ek,Ji:1996nm}.
J.~Collins, L.~Frankfurt, and M.~Strikman provided the QCD factorization theorems for
hard exclusive meson production
\cite{Collins:1996fb}
and for DVCS
\cite{Collins:1998be}, see Fig.~\ref{Fig_DVCS_DVMP}
for the corresponding reaction mechanisms.

Moreover, X.-D.~Ji demonstrated 
that GPDs obey the sum rule \cite{Ji:1996nm,Ji:1996ek}, which provides access to the total angular momentum of quarks and gluons, thus offering a new tool to address the nucleon spin problem~\cite{Thomas:2008ga}. These developments also revealed the role of GPDs in connecting inclusive observables (PDFs) with exclusive ones (elastic form factors), establishing GPDs as a fundamental framework for describing hadron structure in QCD.

The primary purpose of this section is to introduce a consistent system of notations used throughout this white paper. For a more detailed discussion, we direct the reader to a collection of excellent review papers 
\cite{Goeke:2001tz,Diehl:2003ny,Belitsky:2005qn,Boffi:2007yc} on the topic.

\subsection{Definition and basic properties of leading twist nucleon GPDs}

GPDs are defined as hadronic matrix elements of renormalized 
light-cone QCD operators carrying specific quantum numbers. These operators are classified by their (geometrical) twist, defined as the difference between their canonical dimension and spin. The leading-twist (twist-$2$) quark and gluon GPDs involve two parton fields separated by a distance $\kappa$ along a light ray, defined by a light-like vector $n$ ($n^2=0$).  

Throughout this paper, we adopt the standard notations for the light-cone expansion of momenta. We introduce a 
second independent light-cone vector 
$\tilde n$ 
with 
$\tilde n^2 = 0$ 
and 
$n \cdot\tilde n = 1$. 
The  light-cone components of an arbitrary $4$-vector $v$ are defined as
\begin{align}
v^+ \equiv n\cdot v, \hspace{1em} v^- \equiv \tilde{n} \cdot v,
\hspace{1em}
v = v^+ \tilde n + v^- n + v_\perp.
\label{lightcone}
\end{align}

In the quark sector, the leading twist-$2$ GPDs are define from hadronic matrix elements of the light-cone operator
\be
\bar{\psi}^q(-\kappa n / 2)[-\kappa n / 2, \kappa n / 2] \Gamma \psi^q(\kappa n / 2),
\label{Def_operator}
\ee
where $\psi^q$ ($\bar{\psi}^q$) stand for quark (antiquark) fields of flavor $q=u,\,d,\, \ldots$; 
$
[-\kappa n/2, \kappa n/2]$
denotes the gauge link (Wilson line)
ensuring the gauge invariance of the non-local operator (\ref{Def_operator}).  
The spinor matrix $\Gamma$  specifies
spin projection of the quark fields. For the leading twist-2, there are three different spin projections
\be
\Gamma = \{ \gamma^+,\,\gamma^+\gamma^5,\,\sigma^{+\perp} \},
\ee
corresponding to vector (unpolarized), axial vector (helicity-polarized), and
tensor (transversity-polarized) operators (\ref{Def_operator}) 
with 
$\sigma^{\mu \nu} \equiv \tfrac{i}{2} (\gamma^\mu \gamma^\nu - \gamma^\nu \gamma^\mu)$%
\footnote{We use the common shortened notations: for arbitrary 4-vectors $v$, $w$,
$\sigma^{\mu v} \equiv \sigma^{\mu \alpha} v_\alpha$ and  
$\sigma^{w v} \equiv \sigma^{\alpha \beta}  w_\alpha v_\beta$.}.

In the gluon sector, the leading twist-$2$ light-cone operators read
\be
 G_a^{\alpha \beta}\left( \kappa n/2 \right) [-\kappa n / 2, \kappa n / 2]_{ab} \Gamma_{\alpha \beta \gamma \delta} G_b^{\gamma \delta}\left(  \kappa n/2\right), 
 \label{Def_operator_G}
\ee
where $G_a^{\alpha \beta}$ ($ G_b^{\gamma \delta}$) stand for the gluon field strength tensor operator with $[-\kappa n / 2, \kappa n / 2]_{ab}$ denoting the Wilson line in the adjoint representation of the gauge group. The projectors 
$\Gamma_{\alpha \beta \gamma \delta}$ for the unpolarized, polarized and the transversity-like (gluon helicity flip) operators, respectively, read 
\be
&&
\Gamma_{\alpha \beta \gamma \delta}= \left\{ 
n_\alpha n_\gamma g_{\beta \delta}, \,
\frac{1}{2} n_\alpha n^\rho \varepsilon_{\rho \beta \gamma \delta}, \right. \nn \\  && \left.
n_\alpha n_\gamma \left( \frac{1}{2} g_{ i \beta}^ \bot g_{ j \delta}^\bot+ \frac{1}{2} g^\bot_{j \beta} g^\bot_{i \delta} -\frac{1}{2} g^\bot_{ij}  g^\bot_{\beta \delta} \right)
\right\},
\ee
where 
$i,\,j=1,2$, 
and 
$g^{\perp}_{\mu \nu} \equiv g_{\mu \nu}-n_\mu \tilde{n}_\nu-\tilde{n}_\mu n_\nu$
is the transverse metric.

For simplicity we set the gauge links to unity in 
(\ref{Def_operator}) 
and 
(\ref{Def_operator_G}) 
by choosing the appropriate light-like gauge. Therefore the gauge links will be omitted it in the following.

For the case of a nucleon target, there turn out to be a total of $8$ leading-twist-$2$ {GPDs} in the quark sector ~\cite{Diehl:2003ny}. 
These include 2 {vector} nucleon {GPDs} 
({unpolarized GPDs}):
\be
&& \int \frac{d\kappa}{4 \pi} e^{i \kappa(n \cdot P) x}
\nn \\
&& 
\! \! \! \left\langle N\left(p^{\prime}_N, \lambda_N^{\prime}\right)\right| \bar{\psi}^q(-\kappa n / 2) \gamma^{+} \psi^q(\kappa n / 2)\left|N\left(p_N, \lambda_N\right)\right\rangle
\nn \\
&& =\frac{1}{2 P^+} \bar{u}\left(p^{\prime}_N, \lambda_N^{\prime}\right)\Big[H^q(x, \xi, t) \gamma^{+}
\nn \\ && 
+E^q(x, \xi, t) \frac{i \sigma^{+\alpha} \Delta_\alpha}{2 M_N}\Big] u\left(p_N, \lambda_N\right) ;
\label{Def_vector_GPD}
\ee
2 {axial-vector} GPDs 
({helicity-polarized GPDs})
\be
&&
\! \! \! \! \! \!\int  \frac{d \kappa}{4 \pi} e^{i \kappa(n \cdot P) x}
\nn \\ && 
\! \! \! \! \! \! \left\langle N\left(p^{\prime}_N, \lambda_N^{\prime}\right)\right| \bar{\psi}^q(-\kappa n / 2) \gamma^{+} \gamma_5 \psi^q(\kappa n / 2)\left|N\left(p_N, \lambda_N\right)\right\rangle 
\nn \\ &&
\! \! \! \! \! \! =  \frac{1}{2P^+} \bar{u}\left(p^{\prime}_N, \lambda_N^{\prime}\right)\Big[\tilde{H}^q(x, \xi, t) \gamma^{+} \gamma_5  
\nn \\ && 
\! \! \! \! \! \! +\tilde{E}^q(x, \xi, t) \frac{\gamma_5 \Delta^{+}}{2 M_N}\Big] u\left(p_N, \lambda_N\right);
\label{Def_axial_vector_GPD}
\ee
and $4$ {tensor} GPDs
 (transversity-polarized GPDs), 
$H_T^q$, $\tilde{H}_T^q$, $E_T^q$, $\tilde{E}_T^q$,
\cite{Diehl:2001pm}.
In 
(\ref{Def_vector_GPD}), 
(\ref{Def_axial_vector_GPD}) 
we employ the standard kinematics notations 
$P = \frac{1}{2}(p'_N+p_N)$ and  
$\Delta=p'_N-p_N$; 
$u$ ($\bar{u}$) 
stands for the Dirac spinor of {the} incoming (outgoing) nucleon with 
$\lambda_N$ ($\lambda_N^{\prime}$) 
referring to {the} nucleon's polarization variable. Similarly, in the gluon sector, there are $8$ nucleon GPDs with a parametrization analogous to that of the quark case; see Ref.~\cite{Diehl:2003ny} for 
detailed definitions. 

Invariant GPDs depend on the momentum fraction variable 
$x$, 
skewness variable 
$\xi$, 
defined with respect to the longitudinal momentum transfer
\be
\xi=-\frac{p^{\prime+}_N-p^{+}_N}{p^{+}_N+p^{\prime+}_N}=-\frac{\Delta^{+}}{2 P^{+}};
\label{Def_xi}
\ee
and invariant momentum transfer 
$t \equiv \Delta^2$, 
as well as on the factorization scale 
$\mu$ 
(not shown explicitly). 
For the description of the DVCS and DVMP processes shown in Fig.~\ref{Fig_DVCS_DVMP},
the skewness variable 
(\ref{Def_xi})
can be approximately expressed in terms of the Bjorken variable 
$x_B \equiv  {Q^2}/{(2 p_N \cdot q)}$. 
This relationship is given by 
$\xi \simeq x_B/(2-x_B)$.

\begin{figure*}[t]
\centering
\includegraphics[width=0.93\linewidth]{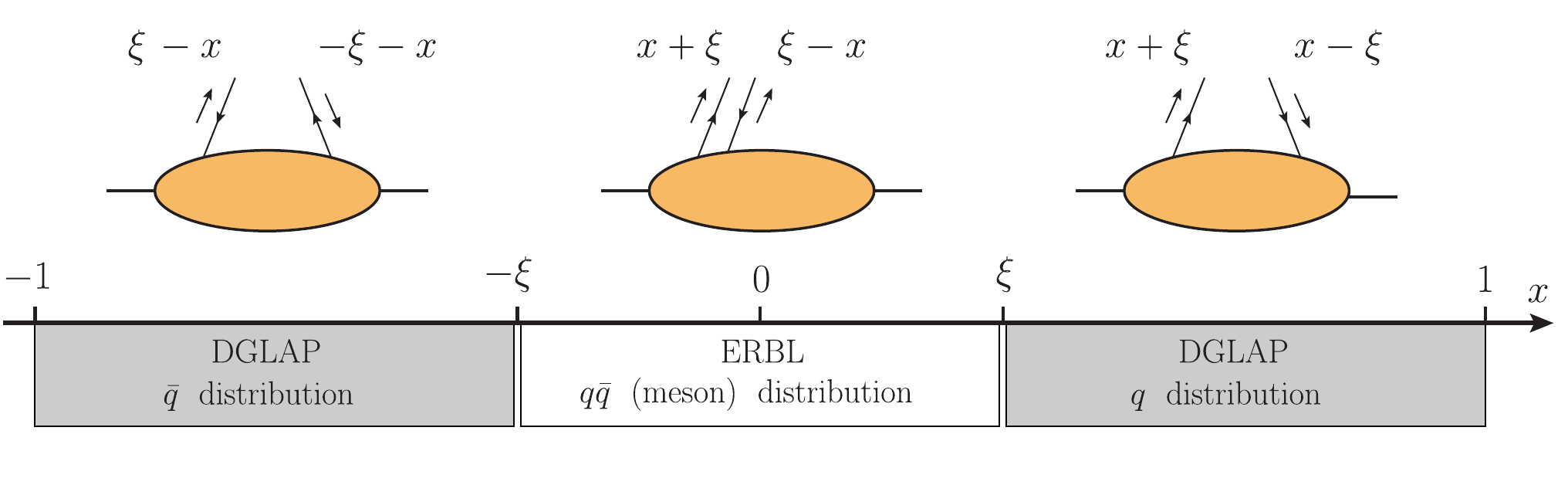}
\caption{ERBL and DGLAP regions of the GPDs and corresponding partonic interpretation in the three regions separated by the cross-over lines $x = \pm \xi$.
}
\label{fig:ERBL-DGLAP}
\end{figure*}

Time reversal symmetry results in all nucleon GPDs being real functions. Their support in 
$x$ 
and 
$\xi$ 
is restricted to the interval 
$x, \, \xi \in [-1;\,1]$. 
The support region in 
$x$ 
is separated into the inner, so-called
ERBL interval, 
$|x| \le \xi$, 
and two outer DGLAP intervals, 
$-1 \le x < -\xi$ 
and
$\xi < x \le 1$, 
in which GPDs admit different interpretation according to Fig.~\ref{fig:ERBL-DGLAP} 
and possess different types of renormalization scale dependence.

In the forward limit 
$\Delta=0$ 
both 
$\xi$ 
and 
$t$ 
are set to 
$0$, 
and GPDs reduce to appropriate PDFs. In particular, 
\be
H^q\left(x , 0,0\right)=\left\{\begin{aligned}
q\left(x\right), \ \ \ & x>0; \\
-\bar{q}\left(-x \right), \ \ \ & x<0,
\end{aligned}\right.
\ee
where 
$q(x)$ 
and 
$\bar{q}(x)$ 
receptively denote the unpolarized quark and antiquark densities of flavor $q$.

As a consequence of the Lorentz invariance, the $x$ Mellin moments of GPDs
map to towers of the local symmetric–traceless twist-$2$ operators, producing 
polynomials in the skewness variable $\xi$. Furthermore, due to
hermiticity, parity and time-reversal invariance, these polynomials are even in $\xi$,
except for the tensor GPD $\tilde{E}_T$. 

For example, for the case of the quark vector GPDs 
$H^q$ 
and 
$E^q$ 
(\ref{Def_vector_GPD}) 
the $N$-th Mellin moments read
\be
&& \int_{-1}^1 d x x^{N-1} H^q(x, \xi, t)=h_0^{(N)}(t)+h_2^{(N)}(t) \xi^2 \nn \\ && +\ldots+ \begin{cases}h_N^{(N)}(t) \xi^N & \text { for } N \text { even } \\
h_{N-1}^{(N)}(t) \xi^{N-1} & \text { for } N \text { odd, }\end{cases} \nn \\ &&
\int_{-1}^1 d x x^{N-1} E^q(x, \xi, t)=e_0^{(N)}(t)+e_2^{(N)}(t) \xi^2 \nn \\ &&+\ldots+ \begin{cases}e_N^{(N)}(t) \xi^N & \text { for } N\text { even } \\
e_{N-1}^{(N)}(t) \xi^{N-1} & \text { for } N \text { odd, }\end{cases}
\label{Mellin_Nth_HE}
\ee
where, for brevity, we do not show explicitly the flavor indices.
The coefficients 
$h_{2k}^{(N)}(t)$, $e_{2k}^{(N)}(t)$ 
at powers of 
$\xi$ 
can be expressed through form factors occurring in the parametrization 
of nucleon matrix elements local totally symmetric traceless twist-$2$
operators 
\be
&&
\mathcal{O}^{\mu \mu_1 \cdots \mu_{N-1}}(0) \nn \\ && =\bar{\psi}^q(0) \gamma^{(\mu} i \overleftrightarrow{ {\cal D}}^{\mu_1} \cdots i \overleftrightarrow{ {\cal D}}^{\left.\mu_{N-1}\right)} \psi^q(0),
\label{Def_tw2_local_Op}
\ee
where 
$(..)$ 
denotes symmetrization in all {uncontracted} Lorentz
indices and subtraction of trace; and 
$\overleftrightarrow{ {\cal D}}^{\mu} \equiv 
(\overrightarrow{ {\cal D}}^{\mu}-\overleftarrow{ {\cal D}}^{\mu})$ 
is the bi-directional covariant derivative. {For low values of $N$}, the form factors of local twist-$2$  operators,
such as
(\ref{Def_tw2_local_Op}), 
can be computed using lattice QCD methods; see {\it e.g.}, Ref.~\cite{Hagler:2003jd}.

The coefficients at the highest power of 
$\xi$ 
for even 
$N$ 
in 
(\ref{Mellin_Nth_HE})  
arise from the $D$-term, 
$D^q(z, t)$, 
with 
$z=x/\xi$, 
\cite{Polyakov:1999gs}. 
$D$-term has the pure ERBL support  
$|x| \le \xi$, 
and
\be
h_N^{(N)}(t)=-e_N^{(N)}(t)=\int_{-1}^1 d z z^{N-1} D^q(z, t).
\ee
It is convenient to expand the $D$-term in the Gegenbauer polynomials (eigenfunctions \cite{Lepage:1980fj} of the LO ERBL evolution equation )
\cite{Polyakov:2002wz}: 
\be 
D^q(z,t)= (1-z^2) \sum_{n \, {\rm odd}} d^q_n(t)  C_n^{3/2}(z).
\label{D_Gegenbauer}
\ee

The first Mellin moment of nucleon quark GPDs provides a useful constraint connecting vector GPD to nucleon electromagnetic form factors. Particularly,
\be
&&
\int_{-1}^1 d x H^q(x, \xi, t)=F_1^q(t), \nn \\ && \int_{-1}^1 d x E^q(x, \xi, t)=F_2^q(t),
\label{FF_1stmoment}
\ee
where 
$F_{1,2}^q(t)$ 
stand for the Dirac and Pauli nucleon form factors of quark flavor 
$q$. 
They are related to the physical nucleon form factors as 
$F_{1,2}^u=2 F_{1,2}^p+F_{1,2}^n$,  
$F_{1,2}^d=F_{1,2}^p+2 F_{1,2}^n$.
Similarly, the first Mellin moment of axial-vector GPDs is expressed in terms of axial nucleon form factors. 

\subsection{DVCS cross section and Compton Form Factors}
\label{sec:DVCS_and_CFFs}

{The simplest processes for acquiring experimental information on GPDs are 
DVCS and the hard exclusive electroproduction of mesons (DVMP).} The corresponding leading twist-$2$, leading order reaction mechanism involving quark GPDs, and also gluon GPDs for the case of production of vector mesons 
$\rho$, $\omega$, $\phi$, $J / \psi$, $\Upsilon$, 
is schematically presented in 
Fig.~\ref{Fig_DVCS_DVMP}. 

{ For definiteness, we will briefly discuss the relation between GPDs 
and the DVCS observables. DVCS is usually accessed through hard exclusive leptoproduction of a real photon off a hadron target:
$\ell N \to \ell' N' \gamma$. 
DVCS naturally interferes with the Bethe-Heitler (BH) reaction, in which the final-state photon is emitted from either the incoming or outgoing lepton. The corresponding five-fold  differential cross section section reads}
\be
\frac{d^5 \sigma}{d x_B d Q^2 dt d \phi d \phi_s}=\frac{\alpha^3_{\rm em} x_B}{16 \pi^2 Q^4 \sqrt{1+\varepsilon^2}}|\mathcal{T}|^2,
\label{CS_electroprod_gamma}
\ee
{ where $\phi$ is the 
angle between the leptonic and the photon-target scattering planes and the angle $\phi_s$ is the azimuthal angle of the transversal component of the target polarization vector. We rely on the Trento
convention for the definition of the relevant momenta and angles, see Ref.~\cite{Bacchetta:2004jz} for a detailed discussion.
In the right hand side of Eq.~(\ref{CS_electroprod_gamma}) $\alpha_{\rm em}$ is the electromagnetic fine structure constant; 
$\varepsilon= 2 x_B M_N/Q$ is the kinematical parameter.
$\mathcal{T}$ is a superposition of the DVCS and the Bethe-Heitler amplitudes:}
\be
|\mathcal{T}|^2=
\left|\mathcal{T}_{\mathrm{BH}}\right|^2+\left|\mathcal{T}_{\mathrm{DVCS}}\right|^2+\mathcal{I},
\ee
{where $\mathcal{I}$ stands for the interference term between 
the BH and DVCS. }


{For the 
case 
of the nucleon target, the leading twist-$2$, LO DVCS amplitude can be expressed in terms of the 
{complex-valued}
generalized Compton form factors (CFFs) defined by the convolutions of vector 
(\ref{Def_vector_GPD})
$F^q=\
\{H^q,\,E^q\}$; and axial-vector (\ref{Def_axial_vector_GPD}) $\tilde{F}^q=\
\{\tilde{H}^q,\,\tilde{E}^q\}$ quark GPDs with singular hard scattering kernels derived from the perturbative calculation of hard subprocess amplitudes:}     
\be
\left\{\begin{array}{c}
\mathcal{F}(\xi, t) \\
\widetilde{\mathcal{F}}(\xi, t)
\end{array}\right\}= && \sum_q e_q^2 \int_{-1}^1 \mathrm{~d} x\Big[\frac{1}{\xi-x-i \epsilon} \nn \\ && \mp \frac{1}{\xi+x-i \epsilon}\Big]\left\{\begin{array}{l}
F^q(x, \xi, t) \\
\widetilde{F}^q(x, \xi, t)
\end{array}\right\}. 
\label{Def_CFF}
\ee
 The sum in the r.h.s. of 
(\ref{Def_CFF}) 
stands over the flavor contents of the nucleon target ($uud$ for the proton and $udd$ for the neutron) weighted by the squares of the quark charges 
$e_q$.
{A major challenge in GPD phenomenology is the recovery of GPDs from experimentally determined CFFs. This requires addressing the formidable ``deconvolution problem'', i.e., the inversion of the convolution integrals relating GPDs to observables \cite{Bertone:2021yyz,Dutrieux:2021wll,Moffat:2023svr,Riberdy:2023awf,Cichy:2024afd}.}


The imaginary part of the CFFs 
(\ref{Def_CFF}) 
is given by the values of GPDs on the so-called cross-over lines 
$x= \pm \xi$. 
It can be related to the real part of the CFFs with the help of the fixed-$t$ dispersion relation 
\cite{Anikin:2007yh,Diehl:2007jb}. 
The case of vector quark GPDs requires a single subtraction:
\be
&& 
{\rm Re}  \mathcal{H}^q\left(\xi, t\right)=\frac{1}{\pi} {\cal P} \int_0^1 d \xi^{\prime}  \left(\frac{1}{\xi-\xi^{\prime}}-\frac{1}{\xi+\xi^{\prime}}\right) 
\nn \\ && \times
{\rm Im} \mathcal{H}^q\left(\xi^{\prime}, t\right)
+4 {\cal D}^q\left(t\right),
\label{Disp_rel_H}
\ee
where $\cal P$ denotes the principal value integral prescription. 
The subtraction constant is expressed as
\be
&&
4 {\cal D}^q(t)= \sum_q e_q^2 \int_{-1}^1 dz \frac{2z D^q(z,t)}{1-z} \nn \\ && = 4 \sum_q e_q^2
\sum_{\substack{n=1 \\ \text { odd }}}^\infty d_n^q(t),
\label{Subtr_C}
\ee 
where $d_n^q(t)$ are the coefficients of the Gegenbauer expansion of the $D$-term
(\ref{D_Gegenbauer}).

Due to symmetry constraints, DVCS provides access only the charge conjugation even ($C=+1$) combination of GPDs
\be
&&
F^{(+)}=F(x)-F(-x), \ \ F= \{H, \,E\}; \nn \\ && 
\tilde{F}^{(+)}=\tilde{F}(x)+\tilde{F}(-x), \ \ \tilde{F}= \{\tilde{H}, \,\tilde{E}\}.
\label{C_even_GPDs}
\ee
In this regard, DVMP processes offer enhanced flexibility by allowing for the selection of meson quantum numbers. This selection not only facilitates access to charge conjugation odd ($C=-1$) combinations of GPDs 
\be
&&
F^{(-)}=F(x)+F(-x), \ \ F= \{H, \,E\}; \nn \\ && 
\tilde{F}^{(-)}=\tilde{F}(x)-\tilde{F}(-x), \ \ \tilde{F}= \{\tilde{H}, \,\tilde{E}\},
\label{C_odd_GPDs}
\ee
but also enables the flavor decomposition of GPDs.

One of the central predictions of the GPD framework is universality: once defined at a 
given factorization scale, the same set of GPDs should describe a whole class of cross-channel counterpart hard exclusive reactions, 
with all process dependence encoded in perturbatively calculable coefficient functions.
Beyond the standard spacelike DVCS and DVMP channels, 
timelike Compton scattering (TCS) \cite{Berger:2001xd,Boer:2015fwa,Boer:2015cwa}, 
double deeply virtual Compton scattering (DDVCS) \cite{Guidal:2002kt,Deja:2023ahc}, and meson-beam induced hard exclusive reactions 
\cite{Berger:2001zn}
provide complementary tests of universality of GPDs. 
This forms a stringent, global consistency check: a universal set of GPDs must describe all these channels simultaneously. Any persistent mismatch would signal the need for improved GPD modeling or indicate limitations of the factorization framework, thereby motivating the systematic inclusion of appropriate corrections.

\begin{figure*}[h]
 \centering
 \includegraphics[width=0.31\linewidth]{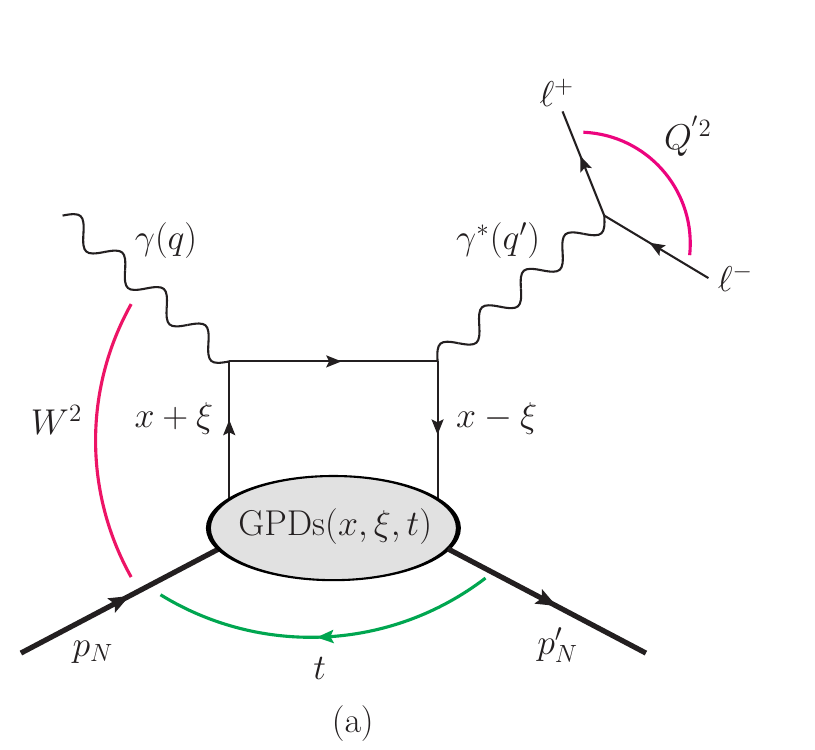}
 \includegraphics[width=0.31\linewidth]{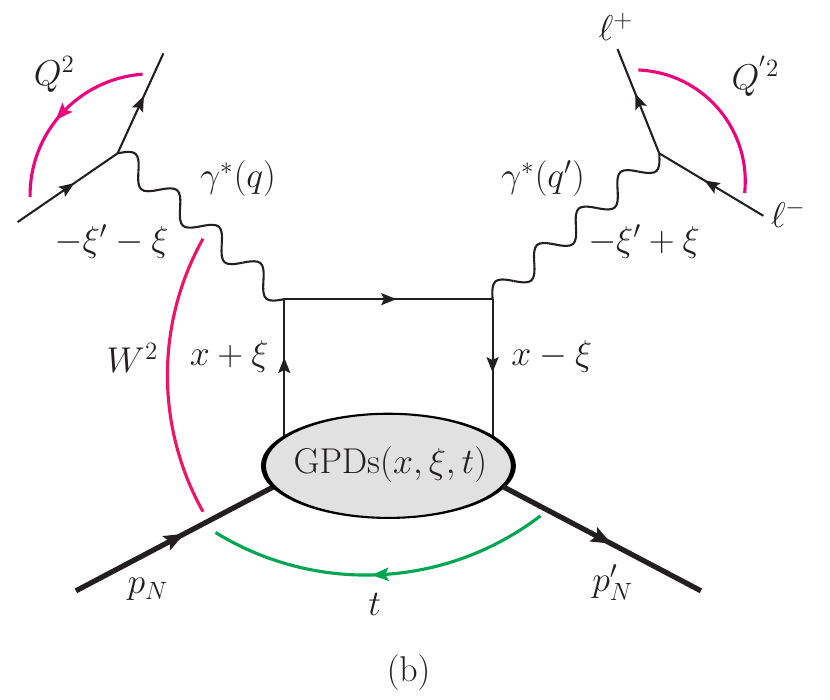}
 \includegraphics[width=0.34\linewidth]{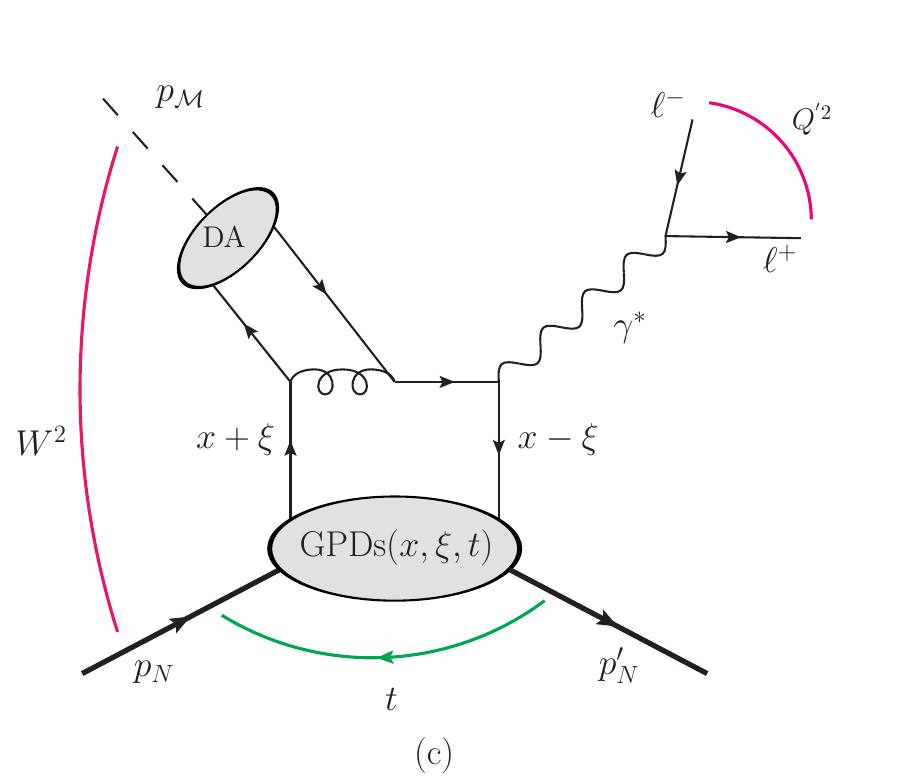}
 \caption{ 
Complimentary exclusive processes involving GPDs.
(a): timelike Compton scattering (TCS). (b): 
double deeply virtual Compton scattering (DDVCS); 
$Q^2$=$-q^2$ is the virtuality of the space-like initial photon, and $Q'^2$=$q'^2$ is the virtuality of the final time-like photon.
(c): Meson-beam induced production of a lepton pair.
}
 \label{Fig_TCS_DDVCS_MBIHER}
\end{figure*}

\subsection{Physical content of GPDs: 
3D imaging of hadrons and gravitational form factors}
\label{Sec_GFF_intro}

\begin{figure}[h]
 \centering
 \includegraphics[width=0.99\columnwidth]{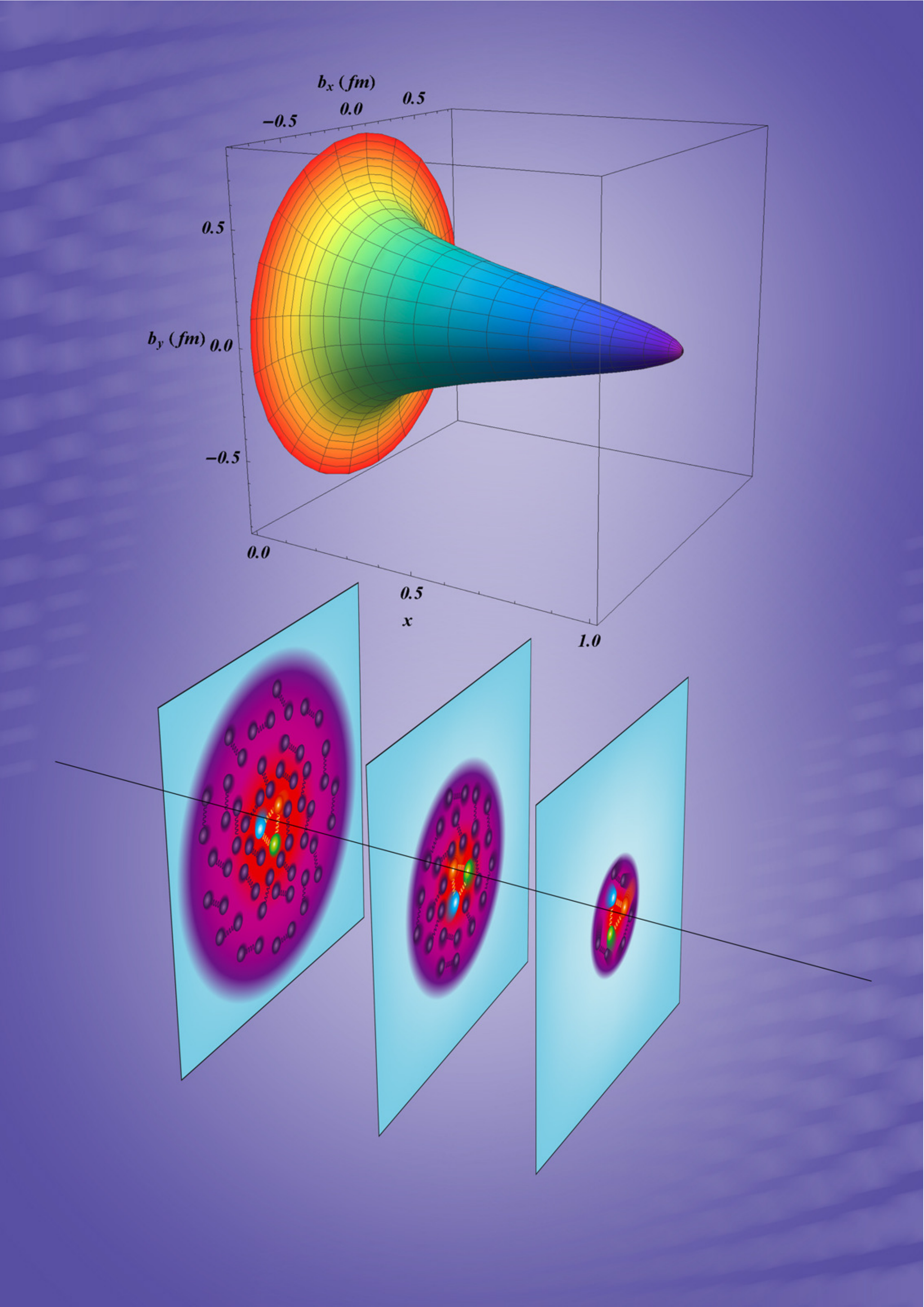}
 \caption{ Top panel: the $x$-dependence of the proton’s transverse charge radius. Bottom panel: artistic illustration of the increasing quark density and transverse extent as a function of $x$. The figure is adapted from Ref.~\cite{Dupre:2017hfs}.}
 \label{Fig_impact_parameter_space}
\end{figure}

A particularly illuminating perspective on GPDs arises from their representation in impact-parameter space. In the special case of zero skewness, 
$\xi=0$, 
the Fourier transform of GPDs with respect to the transverse momentum transfer $\Delta_\perp$ yields spatial densities of partons in the transverse plane at a fixed longitudinal momentum fraction 
$x$, 
as illustrated in Fig.~\ref{Fig_impact_parameter_space}. 
This interpretation, introduced by M.~Burkardt  
\cite{Burkardt:2000za,Burkardt:2002hr}, 
and further developed in subsequent works, connects GPDs to the concept of partonic tomography. It provides 3-dimensional ``images'' of  hadrons, with two dimensions representing the transverse coordinate space and the third corresponding to the longitudinal momentum distribution. This approach complements the one-dimensional momentum distributions accessible in inclusive deep inelastic scattering and the charge distributions encoded in form factors probed in elastic reactions.

The resulting distributions, 
$q(x,  {b}_\perp)$, 
describe the probability of finding a parton that carries a fraction 
$x$ of the longitudinal momentum at a transverse distance 
$b_\perp$ 
from the hadron's center of momentum. Unlike ordinary PDFs, which integrate out all spatial information, impact-parameter–dependent GPDs reveal the correlations between momentum and position degrees of freedom. They provide a natural bridge between the momentum-space picture familiar from high-energy scattering and the spatial densities discussed in hadron structure phenomenology.

For nonzero skewness, 
$\xi \neq 0$, 
the interpretation becomes more subtle~\cite{Diehl:2002he}, as the Fourier conjugate of GPD in 
${\Delta}_\perp$  
no longer corresponds directly to a probability density. Instead, one encounters off-diagonal distributions that encode quantum correlations between partons in different light-cone momentum states. However, even in this case, the impact-parameter representation remains a powerful tool: it clarifies how GPDs interpolate between form factors (fully integrated over $x$) and PDFs (zero momentum transfer), and illustrates how the transverse structure of the nucleon evolves with changing of longitudinal momentum fraction.

This framework has profound implications for ongoing experimental programs. The extraction of GPDs from deeply virtual Compton scattering (DVCS) and meson production processes, followed by their mapping into impact-parameter space, offers the prospect of femtometer-scale imaging of the nucleon. It directly connects to the broader goal of ``hadron tomography,'' revealing how the spatial distribution of quarks and gluons depends on their momentum fraction and polarization. In this sense, the impact-parameter interpretation transforms the study of GPDs into a form of QCD microscopy, providing an intuitive geometric picture of the inner structure of hadrons.

Another central motivation for studying GPDs is their unique ability to access the gravitational form factors  of hadrons ~\cite{Kobzarev:1962wt,Pagels:1966zza, Ji:1996ek}, 
see {\it e.g.},
Ref.~\cite{Burkert:2023wzr} 
for a review. These form factors are defined through matrix elements of the QCD energy–momentum tensor (EMT). Unlike electromagnetic form factors that probe charge and current distributions, {GFFs} may provide insights into the partonic buildup of hadron mass, angular momentum, as well as the mechanical properties of the hadronic medium, which turn to be key quantities for understanding the internal dynamics and stability of strongly interacting systems.

 While hadrons cannot be probed by real gravitational interactions, the EMT form factors provide the same information that, in principle, a graviton probe would measure. In this sense, GPDs offer means to study the ``gravitational structure'' of matter in the laboratory.

The generation of mass in QCD is qualitatively different from the Higgs mechanism in the electroweak sector~\cite{Higgs:1964pj,Englert:1964et,Guralnik:1964eu}. While the Higgs field gives masses to elementary fermions through electroweak symmetry breaking, most of the visible mass of the nucleon arises from strong-interaction dynamics, in particular conformal (QCD trace anomaly) and chiral symmetry breaking (quark masses)~\cite{Adler:1976zt,Collins:1976vm}. This distinction nevertheless suggests that scalar operators associated with mass generation provide especially important probes of hadron structure. In QCD, a natural scalar quantity is the trace of the energy-momentum tensor, whose nucleon matrix element is directly related to the mass decomposition of the nucleon. This motivates the study of a scalar density distribution and an associated scalar radius as spatial characterizations of the QCD dynamics responsible for the emergence of hadron mass.



The symmetric EMT current in QCD is derived using the Poincar{\'e}
transformation of the QCD action, with symmetrization imposed for 
particles with nonzero
spin~\cite{BELINFANTE1939887,Pauli:1940dq,Callan:1970ze}. 
One can also derive the symmetric EMT current by taking a functional derivative of the QCD action~\cite{Kobzarev:1962wt,Parker:2009uva} with respect to the metric tensor of a curved background field. Following to Ji's decomposition~\cite{Ji:1996nm} 
(see also Refs.~\cite{Leader:2013jra,Lorce:2017wkb}), 
we express the quark ($q$) and gluon parts ($g$) of the Belinfante-Rosenfeld-type QCD EMT currents as 
\be
&&
T_{\mu \nu}^{q}  
 = \frac{i}{4} 
    \bar{\psi}^{q} 
    \left( 
    \gamma_{ \{\mu} \overleftrightarrow{\mathcal{D}}_{\nu \}}
    \right) 
    \psi^{q}, 
\nn \\ &&   
    T_{\mu \nu}^{g}  
  = -G_{ \mu \rho}^b G_{\nu  }^{\ \rho,b} 
  + \frac{1}{4} g_{\mu \nu} G_{ \lambda \rho}^b G^{\lambda \rho,b},
\label{Def_EMT_current}  
\ee
where 
$\overleftrightarrow{\mathcal{D}}_{\mu} =
\overleftrightarrow{\partial}_{\mu} - 2 i g A_{\mu}$ 
denotes the bi-directional covariant derivative with $\overleftrightarrow{\partial}_{\mu} =  
\overrightarrow{\partial}_{\mu}- \overleftarrow{\partial}_{\mu}$, 
and the symmetrization notation is defined as
$v_{ \{ \mu }w_{ \nu \} } = v_{\mu} w_{\nu} + v_{\nu}
w_{\mu}$. 
$G^{ \mu \nu, b}$ 
represents the gluon field strength tensor, where
the superscript $b$ indicates the color index. 
The complete EMT current includes both the quark 
$(q)$ 
and gluon 
$(g)$
parts, and satisfies the equation of continuity (current conservation):  
\begin{align}
    T_{\mu \nu} 
  = \sum_{q} T_{\mu \nu}^{q} 
  + T_{\mu
    \nu}^{g},  \quad \partial^{\mu}T_{\mu\nu} = 0.
\end{align} 
The individual quark and gluon contributions of the EMT current are non-conserved.

For a spin-$1/2$ hadron, such as {the} nucleon, 
{the} 
hadronic matrix elements of both 
$a=q,\,g$ 
EMT can be parametrized in terms of 4 GFFs as follows 
\cite{Ji:1996ek},
\be 
&& \bra{p'_N,\lambda'_N} 
T^{a}_{\mu\nu}(0) \ket{p_N,\lambda_N}  \nn \\ && = 
    \bar u(p'_N,\lambda'_N) \bigg[ 
     A^a(t) \,\frac{P_\mu P_\nu}{M_N}
     +J^a(t) \,\frac{i P_{\{\mu}\sigma_{\nu\}\rho}\Delta^\rho}{2M_N} \nn \\ &&
     +D^a(t) \,\frac{\Delta_\mu\Delta_\nu - g_{\mu\nu}\Delta^2}{4 M_N} 
     \nn \\ &&
     +  \bar C^a(t) \, M_N g_{\mu\nu}
    \bigg] u(p_N,\lambda_N). 
    \label{EMT_FF_decomposition}
\ee
Here the normalization of the one-particle state for the nucleon is
defined as
$\langle N (p'_N, \lambda'_{N})| N (p_N, \lambda_{N}) \rangle = 2p^{0}_N (2\pi)^{3}
\delta_{\lambda'_N \lambda_N} \delta^{(3)}(\vec{p}'_N-\vec{p}_N)$, 
where 
$\lambda_N$ 
and 
$\lambda'_N$ 
represent the spin polarizations of the initial and final states, respectively. 

{An}
alternative 
parametrization of the nucleon matrix element of EMT is also 
employed
in the literature~\cite{Polyakov:2018zvc}, 
\be 
&& \bra{p'_N,\lambda'_N} 
T^{a}_{\mu\nu}(0) \ket{p_N,\lambda_N}  \nn \\ && = 
    \bar u(p'_N,\lambda'_N) \bigg[ 
     A^a(t) \,\frac{\gamma_{\{\mu} P_{\nu\}}}{2}
     +B^a(t) \,\frac{i P_{\{\mu}\sigma_{\nu\}\rho}\Delta^\rho}{4M_N}
      \nn \\ &&
     +C^a(t) \,\frac{\Delta_\mu\Delta_\nu - g_{\mu\nu}\Delta^2}{ M_N} 
     \nn \\ &&  +  \bar C^a(t) \, M_N g_{\mu\nu}
    \bigg] u(p_N,\lambda_N).  
    \label{EMT_FF_decomposition_2}
\ee
{The} two parametrizations are equivalent and can be transformed into one another using the Gordon identity for the corresponding spinor structures. {The form factors introduced in (\ref{EMT_FF_decomposition}) and (\ref{EMT_FF_decomposition_2})  are related as}
{\be
&&
J^a(t) = \frac{1}{2}[A^a(t) + B^a(t)], \nn \\ && 
D^a(t) = 4 C^a(t),
\ee
while $A^a(t)$ and $\bar C^a(t)$ are identical in both 
parametrizations. }
{The form factors $A^a$, $J^a$, and $D^a$ are respectively referred to as the mass-, spin-, and $D$-term form factors of the nucleon.}

{
Since the decomposed quark (gluon) EMT currents are not conserved separately, the form factors of individual quarks and gluons are both scale and scheme dependent.  The form factors $\bar{C}^a$ (or the ``cosmological-constant'' terms) represent the internal balance between the quarks and gluons in the quantities related to the diagonal components of the matrix elements. For instance, the $00$-th component 
reflects 
the role of $\bar C^a$ in the hadron mass decomposition~\cite{Ji:1994av,Lorce:2017xzd}, and the spatial isotropic part $(ii)$ 
characterizes 
the forces between the quark and gluon subsystems~\cite{Polyakov:2018exb,Lorce:2018egm}. Note that the total EMT current is conserved: the scale and renormalization-scheme dependence of the quark and gluon parts cancel each other out, and the non-conserved Lorentz structure vanishes, $\sum_q \bar{C}^q + \bar{C}^g = 0$.
}


Due to the polynomiality, 
$N=2$ 
Mellin moment of the quark GPDs 
$H^q$ 
and 
$E^q$ 
are written in terms of the {GFFs} as follows:
\begin{align}
\int_{-1}^{1}dx \, xH^{q}(x,\xi,t) &= A^q(t)
+ \xi^2 D^q(t), \nn  \\
\int_{-1}^{1}dx \, xE^{q}(x,\xi,t) &=- A^q(t)
+2 J^q(t) -\xi^2 D^q(t).
\label{N2-moments}
\end{align}
The form factor 
$A^q(t)$  
occurs in the momentum sum rule, which relates the form factor to the fraction of the nucleon momentum carried by quarks, and  
$B^q(t) \equiv - A^q(t)+2 J^q(t)$ 
is the quark contribution to the nucleon's anomalous gravitomagnetic moment 
\cite{Kobzarev:1962wt}. 
The form factor 
$D^q(t) \equiv \frac{4}{5} d_1^q(t)$ 
can be related to the subtraction constant 
(\ref{Subtr_C}) 
of the fixed-$t$ dispersion relation for the corresponding CFF.
Moreover, 
(\ref{N2-moments}) 
results in Ji's sum rule~\cite{Ji:1996nm,Ji:1996ek}
\begin{equation}
\label{ji_sum_rule}
\int_{-1}^{1}dx \, x[H^{q}(x,\xi,t=0)+E^{q}(x,\xi,t=0)]=2\, J^{q},
\end{equation}
which provides access to the total angular momentum 
$J^q$ 
carried by each quark flavor 
$q$.
Analogous relations can also be written for the gluon GPDs $H^{g}$ 
and 
$E^{g}$ linking them to the form factors of the nucleon matrix element of the gluonic part of the QCD EMT.
These relations already give GFFs of hadrons a solid physical meaning.

\begin{figure}[hbt]
\begin{center}
\includegraphics[width=0.99\columnwidth]{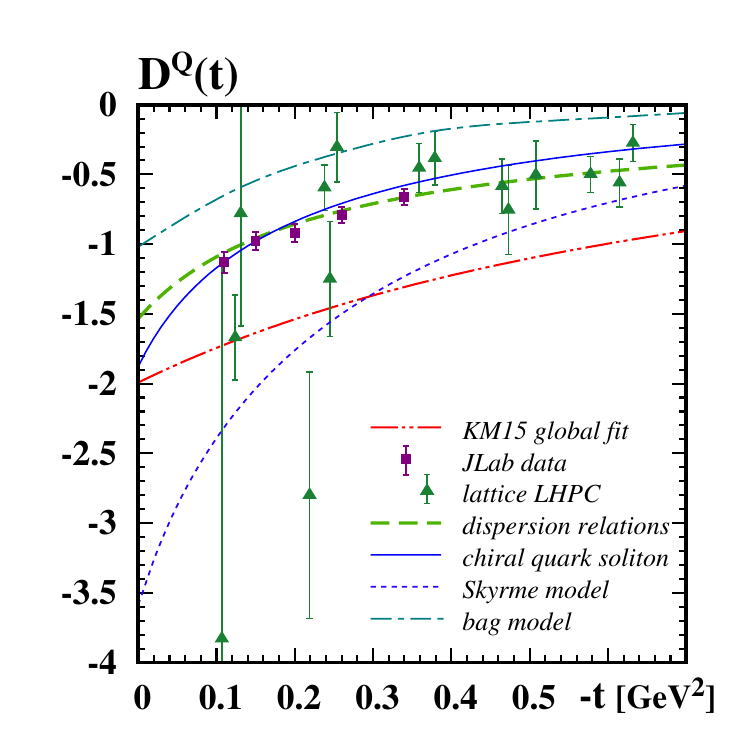}
\caption{
The $D^Q(t) \equiv \sum_q D^q(t)$ form factor obtained from the 
Kumeri\v{c}ki-M\"{u}ller
(KM15) fit~\cite{Kumericki:2015lhb} in comparison to $D^Q(t)$ from the JLab analysis~\cite{Burkert:2018bqq}, calculations from dispersion relations \cite{Pasquini:2014vua}, lattice QCD \cite{LHPC:2007blg}, and results from the bag \cite{Ji:1997gm}, chiral quark soliton \cite{Ossmann:2004bp}
and Skyrme \cite{Cebulla:2007ei} model. The JLab result \cite{Burkert:2018bqq} refers to a normalization point around $\mu^2 = 1.5$~GeV$^2$. The KM15 fit, dispersion
relations and lattice results show the contribution of quarks to the $D$-term at the QCD scale of 4 GeV$^2$. The bag, Skyrme, and
chiral quark soliton 
models show the total $D$-term form factor which is renormalization scale independent.
Figure is taken from Ref.~\cite{Polyakov:2018zvc}.}
\label{fig:Dterm}
\end{center}
\end{figure}

The third form factor occurring in the parametrization
(\ref{EMT_FF_decomposition}), 
$D^a(t)$, 
admits a mechanical interpretation in terms of pressure and shear force distributions within hadrons, as proposed by M.~Polyakov~\cite{Polyakov:2002yz,Polyakov:2018zvc} {by analogy with a perfect fluid EMT decomposition.}
In the Breit frame ($\Delta^0=0$), the spatial components of the nucleon matrix element of the 
total EMT operator can be interpreted as a stress tensor, analogous to the stress tensor of a continuous medium~\cite{Landau:1986aog},
\be
\int \frac{d^3 \Delta}{(2 \pi)^3} && e^{-i \vec{\Delta} \cdot \vec{r}} \nn \\ && \times \frac{\left\langle P+\frac{\Delta}{2}, \lambda^{\prime}_N \right| \hat{T}_{i j}(0)\left|P-\frac{\Delta}{2}, \lambda_N \right\rangle}{2 p_N^0} \nn \\ && =\delta_{i j} p(r)+\left(\frac{r_i r_j}{r^2}-\frac{1}{3} \delta_{i j}\right) s(r),
\ee
where 
$i,\,j$ 
stand for the spatial components and 
$r= |\vec{r}|$.
Within this framework, the isotropic component defines a 
pressure distribution 
$p(r)$, 
while the quadrupole component corresponds to shear forces 
$s(r)$. 
The $D$-term form factor 
\be
&&
D=M_N \int d^3 r \, r^2 p(r)=-\frac{4}{15} M_N \int d^3 r \, r^2 s(r), \nn \\ &&
\ee
appears as a global measure of these distributions, directly encoding the balance of repulsive and attractive forces inside hadron. The pressure interpretation implies that the spatial integral of the isotropic pressure must vanish, the so-called von Laue stability condition
\be
\int d^3 r \, p(r)=0.
\label{von_Laue_stability}
\ee
Therefore, the pressure distribution
$p(r)$ must change sign. 
Indeed, calculations in the chiral soliton models, bag models, lattice QCD, and dispersive analyses,   {\it cf}. Fig.~\ref{fig:Dterm} for an overview, consistently find that the pressure is positive in the core, reflecting repulsive forces from quark kinetic motion, and negative in the periphery, indicating the confining stresses that bind the system, see discussion in 
\cite{Lorce:2025oot}. 
This pattern naturally leads to a negative value of the $D$-term form factor, which has been conjectured to be a universal stability criterion for bound systems governed by short-range QCD forces.

The interpretation has raised important criticisms, see {\it e.g.}, Ref.~\cite{Ji:2025qax}. 
Unlike momentum and angular momentum, the pressure interpretation relies on treating the nucleon as a continuous medium, which is a questionable analogy given the quantum nature of QCD interactions. Additionally, the classical stability condition
(\ref{von_Laue_stability}) 
used to justify negative $D$-terms is not a genuine quantum requirement, and ambiguities in defining the EMT  further raise doubt about the uniqueness of such mechanical densities.
Moreover, some authors have questioned the use of the Breit frame, emphasizing substantial relativistic recoil corrections; others have objected that negative pressure is unphysical in standard thermodynamics. A further puzzle arises from the fact that while hadrons show negative $D$-terms, atomic systems, such as the hydrogen atom, yield positive $D$-terms, suggesting limitations of universality. 

These concerns were recently addressed in some detail in Ref.~\cite{Lorce:2025oot}, 
which argues that none of the objections raised truly invalidate the mechanical picture. It is pointed out that the EMT is a fundamental local field-theoretic operator, not requiring a kinetic-theory justification, and that negative pressures are not exotic but necessary for stable, self-bound systems. The case of atomic systems is understood to be qualitatively different, since long-range electromagnetic interactions fundamentally alter the asymptotics of the EMT distributions and lead to different signs of the $D$-term form factor.

Recent years have seen significant experimental advances in mapping the proton’s internal pressure distribution. The pioneering CLAS measurements at JLab 
\cite{Burkert:2018bqq} 
provided the first extraction of the $D$-term form factor, suggesting a pattern of strong repulsive pressure near the proton core balanced by confining negative pressure at larger radii. Subsequent re-analyses~\cite{Kumericki:2019ddg} 
highlighted the possible model dependence of this extraction, underscoring the need for higher precision and extended kinematic coverage. 
{
Extending the lever arms in $Q^2$ and $t$ is particularly crucial, allowing for a more refined interpretation of the subtraction constant in terms of evolution, higher-order and higher-twist effects~\cite{Dutrieux:2021nlz, Dutrieux:2024bgc,Martinez-Fernandez:2025rcg,Martinez-Fernandez:2025jvk}, while probing the $D$-term on other targets and with different processes will help with the flavor decomposition.
Consequently, 
} 
complementary measurements, along with forthcoming programs at the future EIC and EicC, are expected to provide {stronger} constraints on the $D$-term, enabling {a more reliable} empirical determination of the quark and gluon pressure distributions within the proton.

GFFs of the pion are of particular interest, as the pion is the (pseudo-)Goldstone boson associated with spontaneously broken chiral symmetry of the strong interaction. While the pion GFFs are theoretically well constrained within chiral perturbation theory~\cite{Donoghue:1991qv}, their experimental determination remains a challenging task. To access the pion GPDs, 
in Ref.~\cite{Amrath:2008vx} it was suggested to study the exclusive process 
$\gamma^* p \to \pi^+ n$, 
in the kinematic region dominated by the off-shell pion pole (so-called Sullivan process). Although the total cross section at JLab 12~GeV was estimated to be too small~\cite{Amrath:2008vx}, more recent studies have revisited this reaction for future EIC and EicC facilities, emphasizing the impact of the NLO contributions~\cite{Chavez:2021koz,Bhattacharya:2017bvs}. Complimentarily, 
in Ref.~\cite{Kumano:2017lhr} the pion GFFs were 
extracted by employing the crossing relation between GPDs and GDAs, using the Belle data $\gamma^*\gamma\to\pi^0\pi^0$~\cite{Belle:2015oin}. 
Very recently, 
Ref.~\cite{Hatta:2025ryj} proposed accessing the pion’s gluon GFFs through the Sullivan process in near-threshold quarkonium production and discussed its feasibility at JLab and future EIC experiments.

\section{Recent status of experiments}
\label{sec:experiments_and_data_analysis}

Measurements of hard exclusive reactions described in terms of GPDs constitute the physics programs of many past, present, and near-future experiments aimed at studying hadronic structure. Among them, HERMES, ZEUS, and H1 at HERA, COMPASS at CERN, and the Hall A/C and CLAS experiments at JLab have already provided data for DVCS and DVMP. The considerable experimental effort over the last two decades has resulted in a collection of data for many observables. This includes a variety of asymmetries combining data collected for unpolarized, longitudinally polarized, and transversely polarized targets with those collected for unpolarized and longitudinally polarized beams. Data for differential cross sections, their differences, and observables such as the slopes of distributions of DVCS events as a function of $t$ are also available. An overview of existing data on DVCS can be found, for instance, in Ref.~\cite{Kumericki:2016ehc}, while references for data on DVMP can be found through Refs.~\cite{COMPASS:2022xig,COMPASS:2016ium,Boer:2024ylx}.

First measurements of TCS at JLab \cite{CLAS:2021lky} highlight the role of TCS as a universality benchmark. 
A failure to describe TCS using GPDs tuned to DVCS would directly indicate the presence of missing 
higher-twist or soft mechanisms, or a breakdown of the factorization regime.

The experimental data for DVCS are not limited to proton targets but also include measurements on neutron~\cite{Benali:2020vma,CLAS:2024qhy} and helium~\cite{CLAS:2017udk}. The combination of proton and neutron data allows for a much-needed separation of flavors, which is essential since DVCS only probes a specific combination of quarks weighted by the square of their charges, see Eq.~\eqref{Def_CFF}. A recent example of such an analysis is presented in Ref.~\cite{Cuic:2020iwt}. The helium target, on the other hand, is interesting because it provides access to nuclear GPDs. As it is a spin-$0$ particle, the number of GPDs that need to be considered in the extraction is significantly reduced, making it an ideal case for studying nuclear effects~\cite{Dupre:2015jha,Fucini:2018gso}.

Exclusive pseudoscalar meson 
data has also been recorded along with DVCS experiments or in dedicated experiments 
\cite{CLAS:2019uzc,JeffersonLabHallA:2020dhq,JeffersonLabHallA:2017hky,JeffersonLabHallA:2016wye,CLAS:2022iqy, CLAS:2023wda}.
In the JLab kinematic regime, accessing GPDs through exclusive pion electroproduction remains challenging. The transverse cross section continues to dominate over the longitudinal one, whereas only the longitudinal component receives the leading–twist-2 contribution within the GPD factorization framework. This persistent dominance of transverse cross sections suggests that the onset of the QCD scaling regime, where collinear factorization is expected to apply, may be delayed at current energies.
On the other hand the transverse contribution can give access to higher twist contributions, for example, through the tensor GPDs~\cite{Goloskokov:2011rd}. 

Measurements of hard exclusive vector–meson electroproduction  with JLab 6 GeV CLAS \cite{CLAS:2008rpm} have made it very clear how hard it
is to reach in practice the ``clean'' GPD regime for vector meson DVMP. Within the JLab 6 GeV kinematics, in the valence region 
the transverse cross section 
$\sigma_T$ 
of $ep \to e'p'\rho^0$
remains comparable to, or larger than, the longitudinal piece 
$\sigma_L$.
At higher energies, the measurements  of spin–density matrix elements in exclusive $\rho^0$ muoproduction  was performed by the COMPASS \cite{COMPASS:2022xig}.
More recently measurement of exclusive electroproduction of  $\phi$-mesons near threshold in JLab Hall C has been proposed 
to probe the strangeness gravitational form factor \cite{Hatta:2021can,Klest:2025rek}.


\subsection{Neutron DVCS with CLAS12}

The study of GPDs at JLab is one of the central pillars of the CLAS12 physics program. Exclusive reactions, which provide the cleanest access to GPDs, are experimentally very challenging: their cross sections are typically small, and the complexity of the final states requires high statistics and excellent background suppression. Together with its luminosity ($10^{35}$~cm$^{-2}$s$^{-1}$) capability and large acceptance, the CLAS12 spectrometer~\cite{Burkert:2020akg} provides a unique instrument to study GPDs. 

Measuring DVCS on both protons and neutrons is essential to carry out the quark-flavor separation of GPDs. A particularly convenient observable in DVCS studies is the beam-spin asymmetry (BSA, hereafter also denoted by $A_{LU}$, where $L$ refers to the longitudinally polarized beam and $U$ to the unpolarized target), since many experimental systematic uncertainties and normalization factors largely cancel in the asymmetry ratio. It is extracted as
\begin{equation}\label{def_bsa}
A_{LU} = \dfrac{1}{P} \dfrac{N^+ - N^-}{N^+ + N^-},
\end{equation}
where $P$ is the average beam polarization, and
$N^{\pm}$ denote the yields of DVCS events for positive and negative beam helicities, respectively. For DVCS on the neutron, the BSA is particularly sensitive to the GPD $E$, which remains poorly constrained experimentally and phenomenologically. This lack of information contrasts with the importance of the GPD $E$ for the determination of the angular orbital momentum of partons via Ji's sum rule (\ref{ji_sum_rule}).

In a first approximation, the BSA (\ref{def_bsa}) relates to the CFFs (\ref{Def_CFF}) as~\cite{Guidal_2013}
\begin{equation}
\label{eq:1}
A_{LU} \propto \sin\phi  \Im
[F_1\mathcal{H}+\xi(F_1+F_2){\tilde{\mathcal{H}}}+kF_2\mathcal{E}],
\end{equation}
where $\phi$ is the angle between the lepton scattering and photon production planes, $F_1$ 
and $F_2$ are the Dirac and Pauli form factors, and 
$k=-t/4M^2_N$. Due to the different values of $F_1$ 
and $F_2$ for the proton and neutron, and to the small value of $\xi$, the BSA turns to be mainly sensitive to $\Im \mathcal{H}$ of the proton, if the target is a proton, and to $\Im \mathcal{E}$ of the neutron, if the target is a neutron. 

{For the recent results on DVCS off proton targets from Hall A and CLAS see Refs.~\cite{CLAS:2015bqi,JeffersonLabHallA:2022pnx,CLAS:2022syx}.} The importance of neutron targets in the DVCS phenomenology was established by a pioneering Hall A experiments~\cite{JeffersonLabHallA:2007jdm,Benali:2020vma} measuring polarized-beam cross section differences for DVCS off a neutron from a deuterium target by detecting the scattered electron and the DVCS/BH photon. The recent CLAS12 n-DVCS experiment measured the BSA off a neutron from a deuterium target detecting additionally the recoil neutron.

\begin{figure}[hbt]
\begin{center}
\includegraphics[width=0.99\columnwidth]{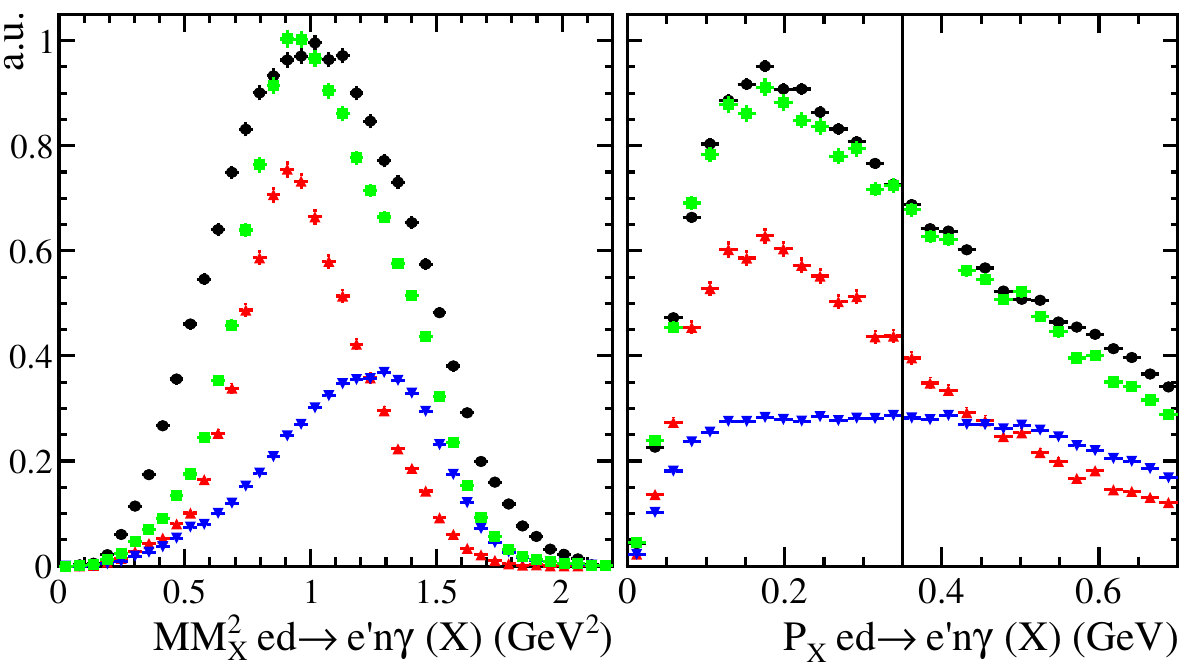}
\caption{Squared missing mass (left) and missing momentum (right) from $ed\to e'n \gamma X$. The line defines the applied cut on $P_X$. The data (black circles) are compared with simulations of neutron DVCS (red triangles) and of partially reconstructed $\pi^0$ background (blue upside-down triangles). The simulations are rescaled to match, approximately, the relative weights of each contribution to the data. The green squares are the sums of the two simulated contributions. The plots are taken from Ref.~\cite{CLAS:2024qhy}.}\label{fig:exclusivityndvcs}
\end{center}
\end{figure}

The experiment was conducted at JLab in Hall B, utilizing the CLAS12 and the longitudinally polarized electron beam produced by the CEBAF with an average 
$\sim$85\% beam polarization. The experiment ran  between February 2019 and January 2020, collecting  an integrated luminosity of roughly 285~fb$^{-1}$. Events with at least one electron, one photon, and one neutron were selected for the DVCS analysis. Several cuts were applied in order to ensure proper particle identification and select the relevant kinematic region for the DVCS reaction. Imposing $Q^2>1$~GeV$^{2}$ 
and $W>2$~GeV cuts places the measurement in a kinematic region where the GPD framework is expected to provide a useful description a reaction, while suppressing contributions from nucleon resonances. Nevertheless, non-negligible higher-twist and finite-$Q^2$ corrections may still be present, particularly at the lower end of the $Q^2$ range. Exclusivity cuts minimize the background coming from partially reconstructed $\pi^0$ decays from the $ed\to e'n\pi^0(p)$ reaction, where only one of the two photons from the $\pi^0$ decay is reconstructed. 
Fig.~\ref{fig:exclusivityndvcs} 
shows the squared missing mass of $X$ in $ed \to e'n\gamma X$ and the missing momentum $P_X$ 
for the data and the simulations for DVCS and for 
$\pi^0$, after applying exclusivity cuts. The data still contain some background from partially reconstructed $\pi^0$ decays. 

The $\pi^0$ contamination to the DVCS sample was evaluated from simulated ratio of partially reconstructed $e'n\pi^0$(1$\gamma$) 
events passing the DVCS selection criteria to the fully reconstructed $e'n\pi^0$ events and the detected number of reconstructed $e'n\pi^0$ events. This number was then subtracted from the yield of DVCS event candidates for each kinematic bin and helicity state. The 
$\pi^0$ contamination ranges from 10\% to 45\% depending on the kinematics. Further contamination of the DVCS samples arises from the Central Tracker of CLAS12~\cite{Burkert:2020akg}, leading to misidentification of protons. This background was reduced using a multivariate analysis technique relying on low-level features from the Central Neutron Detector (CND) and the Central Time-of-Flight (CTOF) device. 
This remaining contamination was estimated to be $\sim 5\%$ and was subtracted in the determination of the final BSA. Radiative corrections were evaluated according to Ref.~\cite{Akushevich:2017kct} and found to be negligible. Various sources of systematic uncertainty were studied leading to a total average of 
$\sim 0.01$, largely dominated by the systematics attached to exclusivity cuts. 

\begin{figure}[hbt]
\begin{center}
\includegraphics[width=0.99\columnwidth]{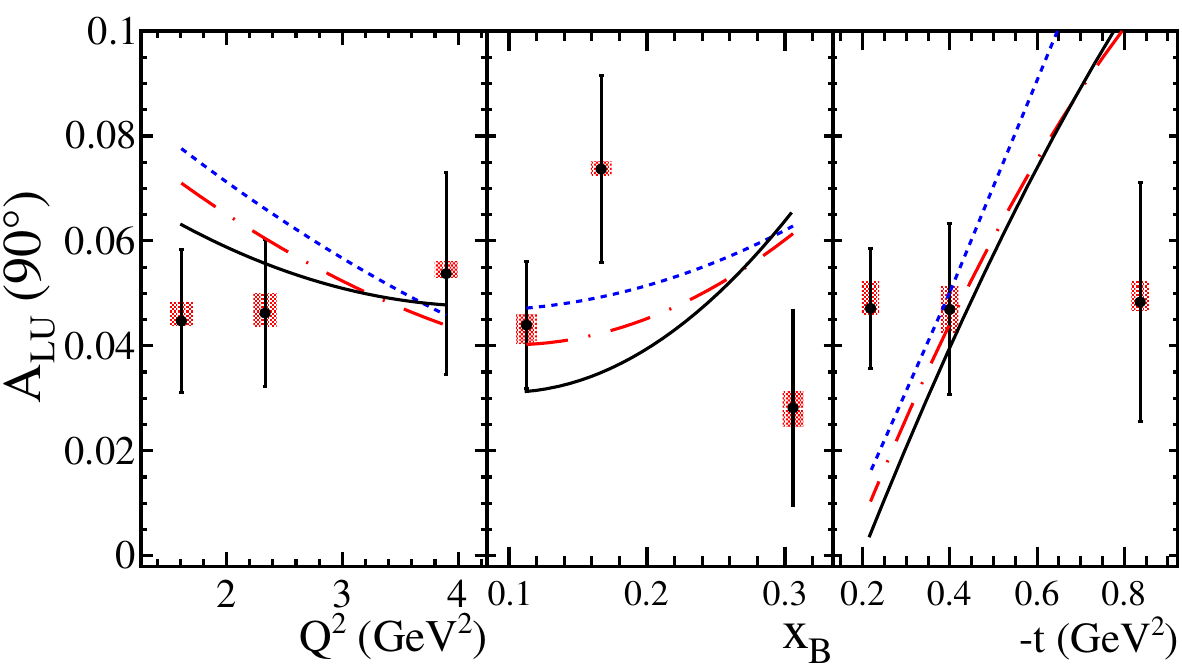}
\caption{\label{fig:sinphiq2} The $\sin\phi$ amplitude of $A_{LU}$ for nDVCS as a function of $Q^2$ (left), $x_B$ (middle), and $-t$ (right). The (red online) bands represent the systematic uncertainties. The VGG model~\cite{vgg} predictions for three of the combinations of $J^u$ and $J^d$ yielding the best $\chi^2$ are compared to the data: solid (black online) line for $J^u=0.35$, $J^d=0.05$, dashed-dotted (red online) line for $J^u=-0.2$, $J^d=0.15$, and (blue online) dotted line for $J^u=-0.45$, $J^d=0.2$.
This figure  is taken from Ref.~\cite{CLAS:2024qhy}.}\label{fig_alu}
\end{center}
\end{figure}
The BSAs were extracted in bins of either $Q^2$, $x_{B}$, or $t$. They exhibited the expected sinusoidal shape arising from the DVCS-BH interference, and have been fitted by the function $A_{LU}(90^{\circ})\sin\phi$. Its amplitude is about a few percent, that is a 4 times smaller than the amplitude measured at the same kinematics for the proton \cite{CLAS:2022syx}.  Fig.~\ref{fig_alu} shows the amplitude $A_{LU}(90^{\circ})$ as a function of 
$Q^2$ (left), $x_B$ (middle), and $-t$ (right). The data are compared to predictions for DVCS on a free neutron within the Vanderhaeghen-Guichon-Guidal (VGG) model 
\cite{vgg,vgg1} for different values of the quark total angular momenta $J^u$ and $J^d$. Three of the curves yielding the best 
$\chi^2$-agreement with experimental data are retained for Fig.~\ref{fig_alu}. Considering $\chi^2$ values within $3\sigma$ from the minimum, the data favor $d$ quark angular momenta $0<J^d<0.2$ while no constraints can be imposed on $J^u$. 

\begin{figure}[hbt]
\begin{center}
\includegraphics[width=0.99\columnwidth]{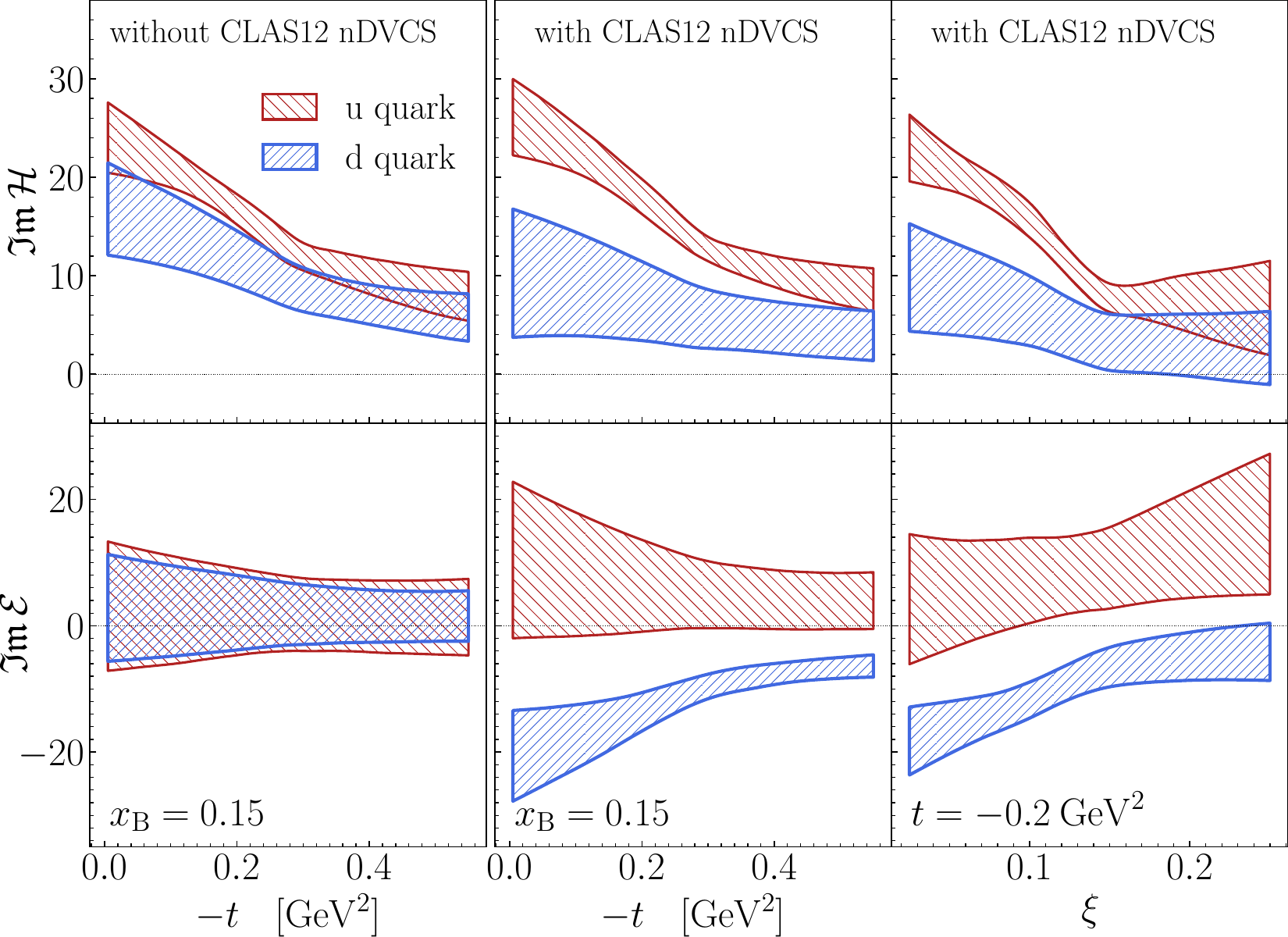}
\caption{\label{fig:flavsep} Extraction of up ($u$, coarser shading, red online) and down ($d$, finer shading, blue online) quark contributions to $\Im \mathcal{H}$ (top) and $\Im \mathcal{E}$ (bottom) as a function of $-t$ (left and middle) and $\xi$ (right). The leftmost column shows the extraction of the two CFFs without the CLAS12 nDVCS data, which are instead included in the other two columns. This figure is taken from Ref.~\cite{CLAS:2024qhy}.}
\end{center}
\end{figure}

The sensitivity of the CLAS12 n-DVCS BSA to CFFs, in particular to 
$\Im \mathcal{E}$, was tested by including these data in a global neural network fit of CFFs~\cite{CLAS:2024qhy}. The networks were trained on 200 Monte Carlo replicas of the experimental data, which ensured that uncertainty of the resulting CFFs is faithfully propagated from the experimental measurements. Fig.~\ref{fig:flavsep} shows the up and down quark $\Im \mathcal{H}$ and $\Im \mathcal{E}$
CFFs, extracted by fits to old CLAS \cite{CLAS:2015bqi, CLAS:2015uuo}
and HERMES \cite{HERMES:2010dsx,HERMES:2012gbh} proton data, to recent CLAS12 proton data \cite{CLAS:2022syx}, and to the new CLAS12 neutron data. The inclusion of the CLAS12 n-DVCS data clearly allows for the flavor separation of $\Im\mathcal{E}$. 


\subsection{NPS}

Exclusive neutron DVCS relies critically on the precise reconstruction of the real photon. The Neutral Particle Spectrometer (NPS) delivers this capability through its high segmentation and calibrated electromagnetic response, allowing accurate determination of the photon energy. The experimental program for DVCS off neutrons employs the  NPS installed in Hall~C at JLab. The NPS is a high-resolution electromagnetic calorimeter designed to detect photons with excellent spatial and energy resolution, enabling the exclusive reconstruction of DVCS final states. Its combination with the High Momentum Spectrometer (HMS) for electron detection allows for precision measurements of exclusive processes at a luminosity of $5\times 10^{38}$~cm$^{-2}$s$^{-1}$.

{The basic topology of the reaction} involves the coincident detection of the scattered electron in the HMS and the DVCS photon in the NPS. The recoiling nucleon, either a proton or a neutron, is not directly detected but identified via the missing-mass technique. {A cut around the nucleon mass, $M_X^2\simeq M_N^2$, is therefore an important exclusivity requirement, but it is not by itself sufficient to isolate neutron DVCS in deuterium. The LD$_2$ sample contains both neutron-DVCS and proton-DVCS events, as well as contributions from accidental coincidences, misidentified 
$\pi^0\to\gamma\gamma$ events, and coherent DVCS on the deuteron. For this reason, LH$_2$ data taken under matched kinematic conditions are used to subtract the proton-DVCS component from the LD$_2$ spectrum.}

Photons are measured in the NPS calorimeter composed of 1080 lead-tungstate
(PbWO$_4$) crystals. Its high density, short radiation length, and excellent radiation hardness, make it ideal for operation at the high luminosities available in Hall~C. The calorimeter's compactness and fine granularity enable sub-centimeter position resolution and an energy resolution better than {2\%/${\sqrt{E}}$} for photon energies in the multi-GeV range. The NPS was designed to allow for the simultaneous mapping of DVCS cross sections from 2 to 5~GeV$^2$, and Bjorken-$x_B$ from 0.25 to
0.6, thus probing the transition region between valence and sea quark dynamics. 

\begin{table}[h!]
	\centering
	\begin{tabular}{ccc}
		\hline
		$x_B$ & $E_{\mathrm{beam}}$ (GeV) & $Q^2$ (GeV$^2$) \\
		\hline
		0.25 & 6.6, 8.8, 11 & 2.1--2.4 \\
		0.36 & 6.6, 8.8, 11 & 3.0--3.4 \\
		0.50 & 6.6, 8.8, 11 & 3.4--4.4 \\
		0.60 & 6.6, 8.8, 11 & 5.1--5.6 \\
		\hline 
	\end{tabular}
	\caption{Representative neutron DVCS kinematics in the NPS program.}
	\label{tab:kinematics}
\end{table}
Data have been collected during the 2023--2024 run period with beam 
energies of 6.6, 8.8, and 11~GeV. These different beam energies provide 
lever arms in $Q^2$ and $x_B$ for disentangling the real and imaginary parts of the CFFs and for testing the scaling behavior predicted by QCD factorization. Table~\ref{tab:kinematics} summarizes the representative kinematic points covered during the campaign. The simultaneous coverage of low and high $x_B$ values corresponds to an overall momentum-transfer range of $|t| \in [0.06,\,1.30]~\mathrm{GeV}^2$, offering direct sensitivity to the transverse localization of quarks in the nucleon. 

\subsubsection{Detector performance}

\begin{figure}[t!]
	\centering
	\includegraphics[width=0.47\textwidth]{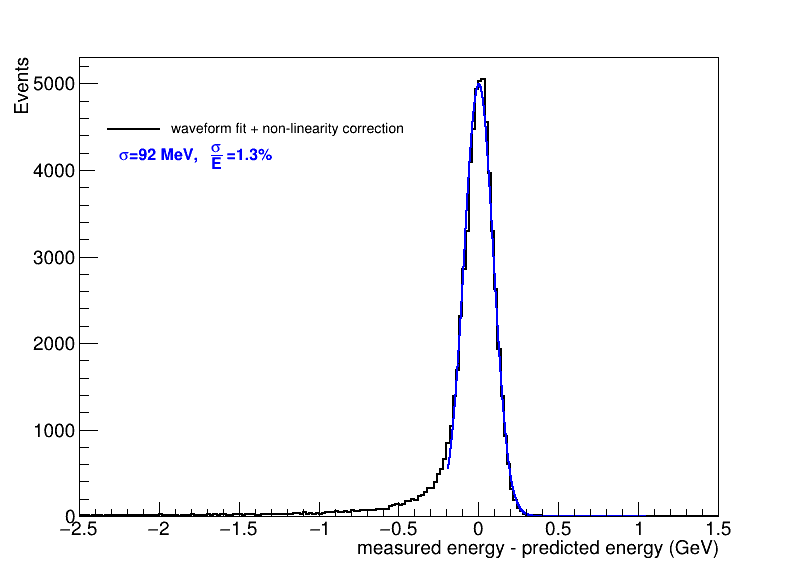}
	\includegraphics[width=0.43\textwidth]{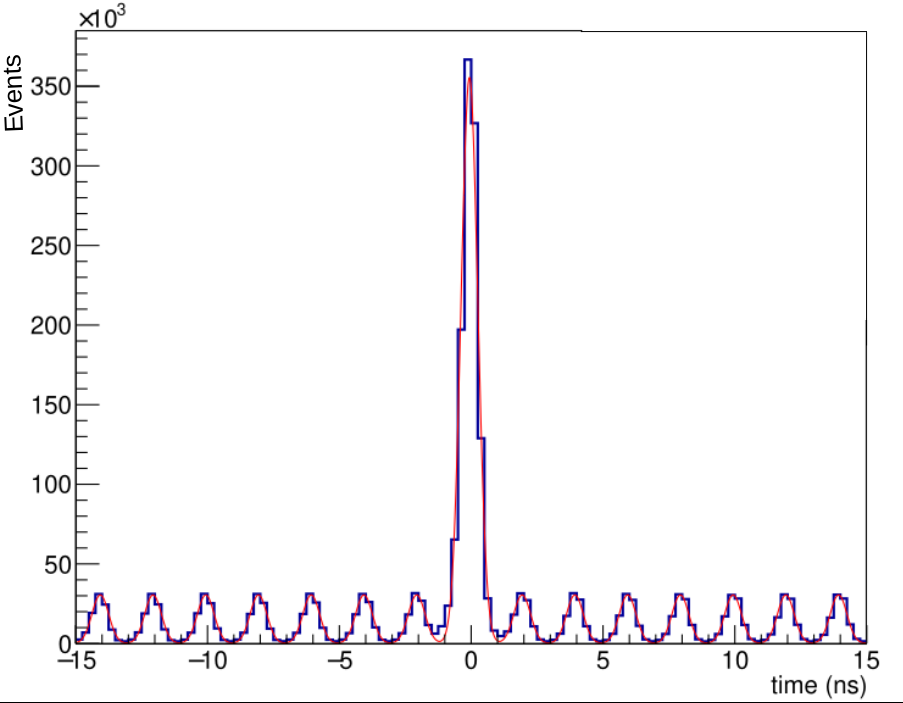}
	\caption{ {NPS calorimeter performance from an elastic $H(e,e'p)$ calibration sample. The scattered electron is detected in the NPS and the proton is detected in the HMS; the proton is reconstructed from the elastic kinematics and is not part of the nominal neutron-DVCS detection topology. Top: distribution of measured minus predicted electron energy at $E_\gamma=7.3\,\mathrm{GeV}$, yielding a resolution $\sigma_E/E=1.3\%$. Bottom: HMS--NPS time-difference spectrum. The central peak is the prompt electron--proton coincidence, while the smaller structures displaced by about the CEBAF RF period correspond to neighboring beam bunch/accidental coincidences. The Gaussian width of the prompt peak is $\sigma_t=0.5\,\mathrm{ns}$.} }
	\label{fig:nps_performance}
\end{figure}

The understanding of the detector performance is a prerequisite for the reliable extraction of DVCS observables. In particular, the energy resolution and timing resolution of the NPS calorimeter determine the accuracy of photon reconstruction and the rejection of accidental backgrounds. The NPS calibration was performed with elastic $H(e,e'p)$ events. In this calibration procedure, the scattered electron is detected in the NPS and the associated proton is detected in the HMS. The comparison between the measured and predicted electron energy gives the NPS energy response after waveform fitting and non-linearity correction. The resulting distribution has a width $\sigma_E=92\,\mathrm{MeV}$ at $E=7.3\,\mathrm{GeV}$, corresponding to $\sigma_E/E=1.3\%$. This level of precision for a high-luminosity environment confirms 
the good uniformity and optical quality of the PbWO$_4$ crystals used in 
the array. Equally important, the time resolution of the calorimeter  impacts the ability to distinguish true photon-electron coincidences from random accidentals. The timing performance of the NPS was evaluated using the same elastic calibration sample. A Gaussian fit to the time difference 
distribution between the electron and the reference HMS trigger 
yields a timing resolution of $\sigma_t = 0.5$~ns 
(Fig.~\ref{fig:nps_performance}). This corresponds to 
a coincidence accuracy well within the 2~ns beam structure of CEBAF, 
allowing the clear identification of events belonging to the same beam bunch. 

The combination of high energy and timing resolution provides the foundation for precise DVCS measurements. Accurate photon energy determination minimizes smearing in the reconstructed $t$ and missing-mass distributions, while excellent timing resolution improves 
signal purity. Together, these capabilities enable the NPS to operate at 
high luminosities with low systematic uncertainty, extending the reach of 
exclusive photon measurements in Hall~C to an unprecedented level of precision.

\subsubsection{Neutral pion reconstruction and exclusivity}
 
\begin{figure}[h!]
	\centering
	\includegraphics[width=0.48\textwidth]{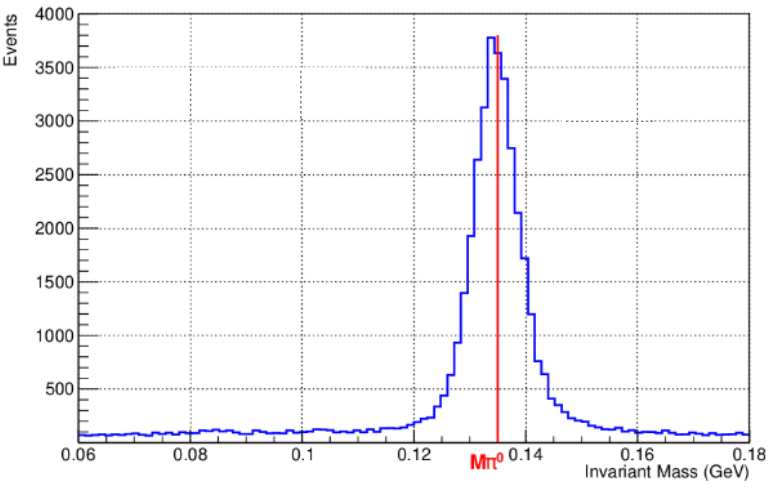}
	\caption{Reconstruction of the $\pi^0 \to \gamma\gamma$ decay in LD$_2$ data.
		 Invariant-mass spectrum of the two-photon system showing a Gaussian 
		width $\sigma = 4.1$~MeV around $M_{\pi^0} = 135$~MeV, obtained after 
		energy calibration and waveform correction.}
	\label{fig:pi0_reco}
\end{figure}

Neutral pion production represents the dominant background channel in DVCS analysis, and its accurate identification is essential for the subtraction of $\pi^0$ contamination in the D$(e,e’\gamma)$X missing-mass. 
Fig.~\ref{fig:pi0_reco} illustrates the two-photon invariant-mass spectrum reconstructed in LD$_2$ data, which provides a direct validation of the NPS energy and position reconstruction for $\pi^0\to\gamma\gamma$ events. The peak at $M_{\pi^0}\simeq135$~MeV, with a Gaussian width of $\sigma=4.1\,\mathrm{MeV}$, demonstrates that two-photon decays are well reconstructed. In symmetric decays, both photons carry approximately equal energy and are both detected in the NPS. In contrast, asymmetric decays produce one high-energy photon within the NPS acceptance and one low-energy photon that may escape detection, leading to the partial reconstruction responsible for the $\pi^0$ contamination seen in the $D(e,e’\gamma)X$ missing-mass analysis. This {$\pi^0$} contamination is evaluated from the number of {$\pi^0$} symmetric decays detected in the NPS.

\subsubsection{Preliminary results}

\begin{figure}[t!]
	\centering
	\includegraphics[width=0.48\textwidth]{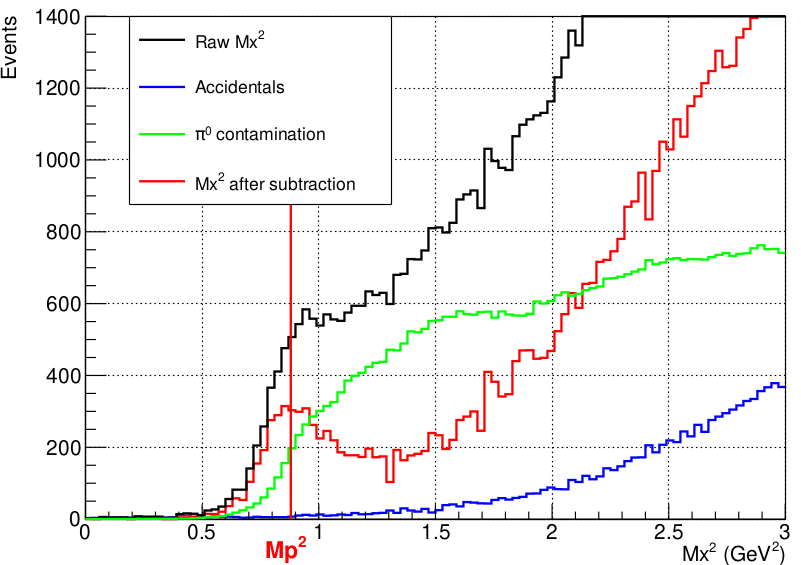}
	\includegraphics[width=0.48\textwidth]{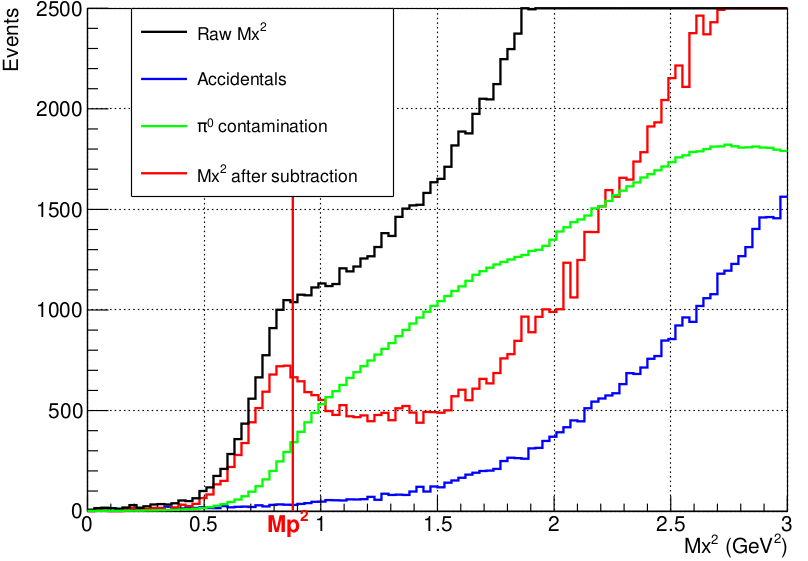}
	\caption{Preliminary missing mass squared distributions for DVCS analysis.
		Top: Hydrogen data ($eH \to e'\gamma X$). {Bottom: deuterium data, $eN\to e'\gamma X$ with $N=p$ or $n$ bound in deuterium. The HMS detects the scattered electron and the NPS detects the real photon; the recoil nucleon is reconstructed through $M_X^2$.} }
	\label{fig:mx2plots}
\end{figure}
The extraction of the exclusive $eN \rightarrow e'N'\gamma$ signal requires precise reconstruction of the  $eN \rightarrow e'\gamma X$ missing mass squared ($M_X^2$) distribution from the detected electron and photon. True DVCS events correspond to $M_X^2 = M_N^2$ within resolution effect, where 
$M_N$ is the nucleon mass. Fig.~\ref{fig:mx2plots} illustrates the subtraction of the accidental and $\pi^0$ contamination contributions from the raw hydrogen (LH$_2$) and deuterium (LD$_2$) data. The resulting distributions (red histograms) show a pronounced and narrow peak at 
$M_X^2 = M_N^2$, consistent with exclusive photon production. The top panel (LH$_2$) demonstrates the validation of the method on hydrogen, 
while the bottom panel (LD$_2$) presents the corresponding deuterium data. The excellent agreement between the peak positions and the expected nucleon mass squared indicates the robustness of the subtraction procedure and the exclusivity of the selected DVCS events. {For the acquired LD$_2$ data, the neutron-DVCS yield is obtained only after subtracting the proton-DVCS contribution constrained by the matched LH$_2$ data. A possible coherent-DVCS contribution from the deuteron, together with any residual background under the nucleon peak, will be evaluated as part of a more sophisticated analysis.}


\subsection{COMPASS}

One of the core missions of COMPASS, a fixed target experiment located at CERN, is to unveil the multi-dimensional nucleon structure through various reaction processes. To access GPDs, COMPASS conducted exclusive measurements of DVCS and the production of a variety of mesons. For exclusive measurements, the 160~GeV muon beams provided by the M2 beamline at the SPS are utilized. Originating dominantly from the decay of mesons produced by the primary proton beam, the muon beams are naturally polarized at about 80\%, and the muons of different signs carry opposite polarization. For exclusive productions, transversely polarized $\mathrm{NH}_3$ and unpolarized liquid hydrogen targets were employed. The final-state particles generated in exclusive reactions are measured by the COMPASS forward spectrometer~\cite{Mallot:2004gk}, which is about 50~m long and can be subdivided into two stages focusing on particle detection at large and small angular acceptance. Two dipole magnets and various detector stations for precise tracking or particle identification are installed. Three electromagnetic calorimeter (ECAL) stations for the detection of photons produced in DVCS or exclusive $\pi^0$ decay, named as ECAL0, ECAL1, and ECAL2, constitute a total of more than six thousand calorimeter cells in them. To ensure the exclusivity of events, a recoil-proton detector, CAMERA, was installed. The CAMERA is a barrel-shaped detector that consists of two rings of 24 slabs scintillators, with the liquid hydrogen target sitting at the ring center. The  time-of-flight information of the recoil protons from exclusive reactions are registered, improving the $t$ resolution, particularly for small $t$.

Over the span of twenty years of the COMPASS data taking, early exclusive vector meson production data were collected in 2007 and 2010, a 4 weeks pilot run for GPD studies was carried out in 2012, and dedicated runs were conducted in 2016 and 2017. The COMPASS results of the analyses on different exclusive channels from these data sets are discussed in the following subsections. 

\subsubsection{DVCS at COMPASS}

The differential cross section of the exclusive muoproduction of photons consists of contributions from the DVCS, BH, and the interference between them, denoted as $d\sigma^{DVCS}$, $d\sigma^{BH}$, and $d\sigma^{I}$, 
correspondingly. Using the oppositely polarized 
$\mu^+$ and $\mu^-$ beams and the unpolarized LH$_2$ target, COMPASS measured the beam charge-spin sum 
\be 
\mathcal{S}_{CS,U}\equiv{}d\sigma(\mu^{+\leftarrow})+d\sigma(\mu^{-\rightarrow}),
\ee
and the difference,
\be
\mathcal{D}_{CS,U}\equiv{}d\sigma(\mu^{+\leftarrow})-d\sigma(\mu^{-\rightarrow}),
\ee
where the arrows represent the helicities of the muons. The leading-order and leading twist contributions in $d\sigma^{DVCS}$ and $d\sigma^{I}$ are proportional to certain combinations of the CFFs. In the COMPASS kinematic domain and a LH$_2$ target, these BH removed observables are essentially the imaginary and real parts of the CFF $\mathcal{H}$, Im$\mathcal{H}$ and Re$\mathcal{H}$.

\begin{figure}[t!]
 \begin{center}
  \includegraphics[width=0.43\textwidth]{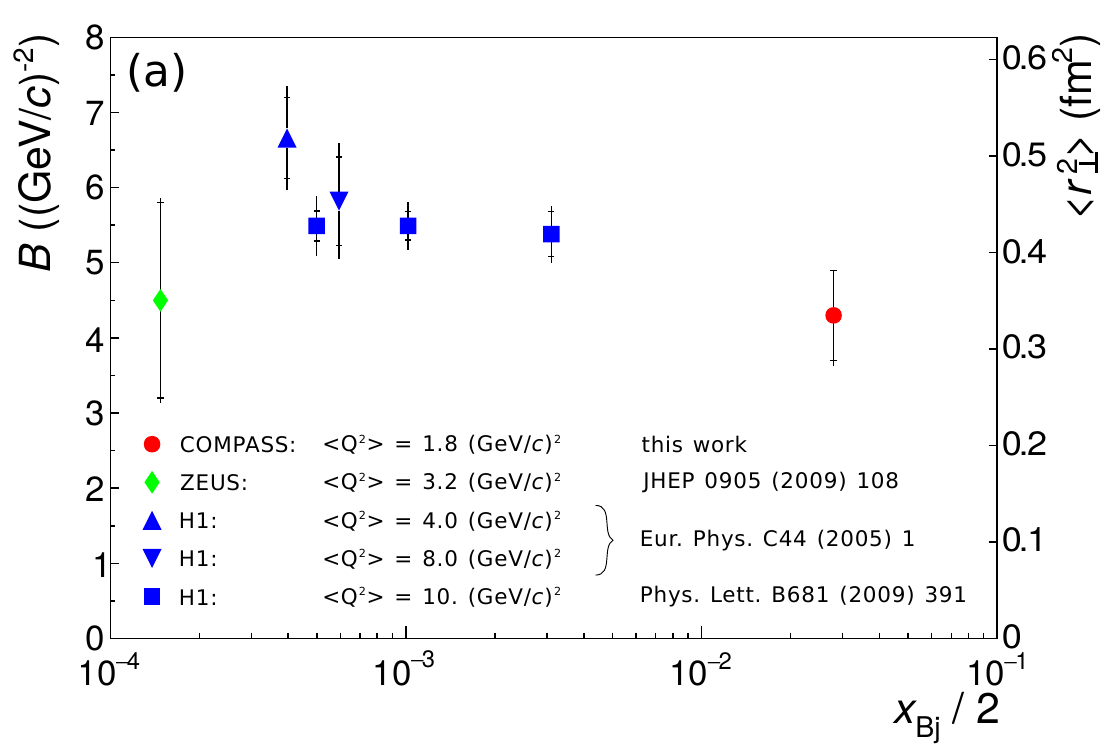} \ \ \ \
  \includegraphics[width=0.43\textwidth]{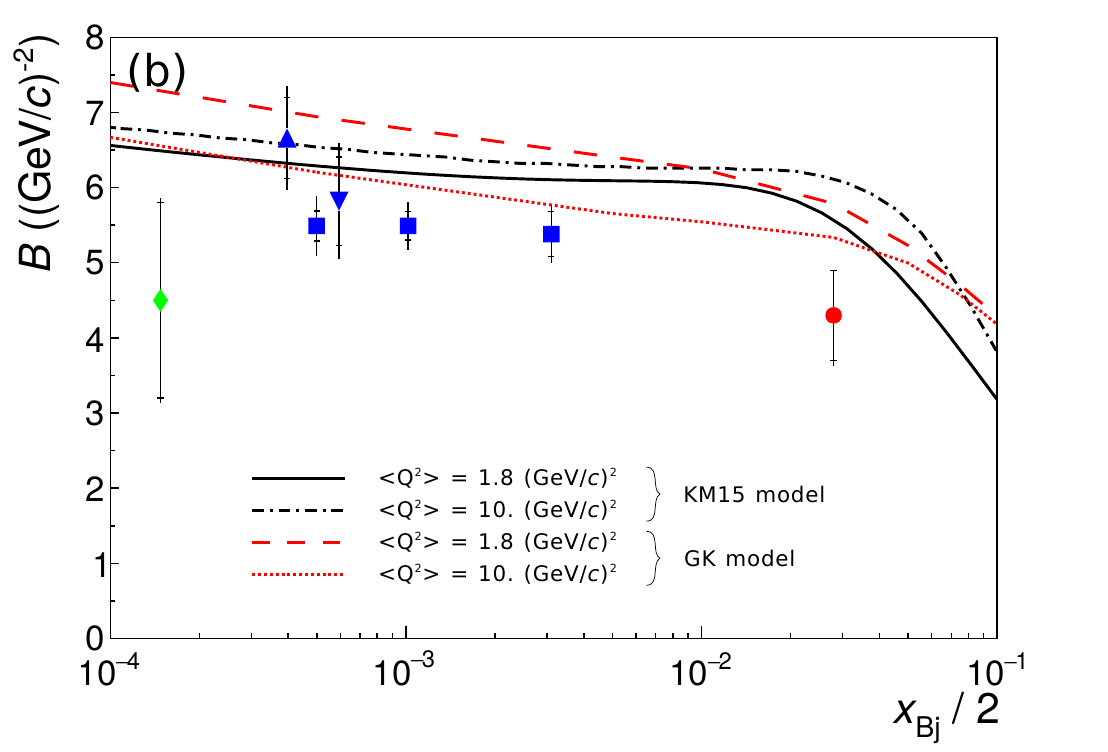}
   \caption{Upper panel: COMPASS result on the $t$-slope parameter $B$ and the corresponding average of squared transverse extension of partons inside the proton $\langle r_{\perp}^2 \rangle$, together with the measurements of H1 and ZEUS. Lower panel: Same results compared with the predictions of the GK and KM15 models. The plots are taken from~\cite{COMPASS:2018pup}.}
  \label{fig:compass_dvcs}
 \end{center}
\end{figure}

\begin{figure*}[t!]
 \begin{center}
  \includegraphics[width=0.92\textwidth]{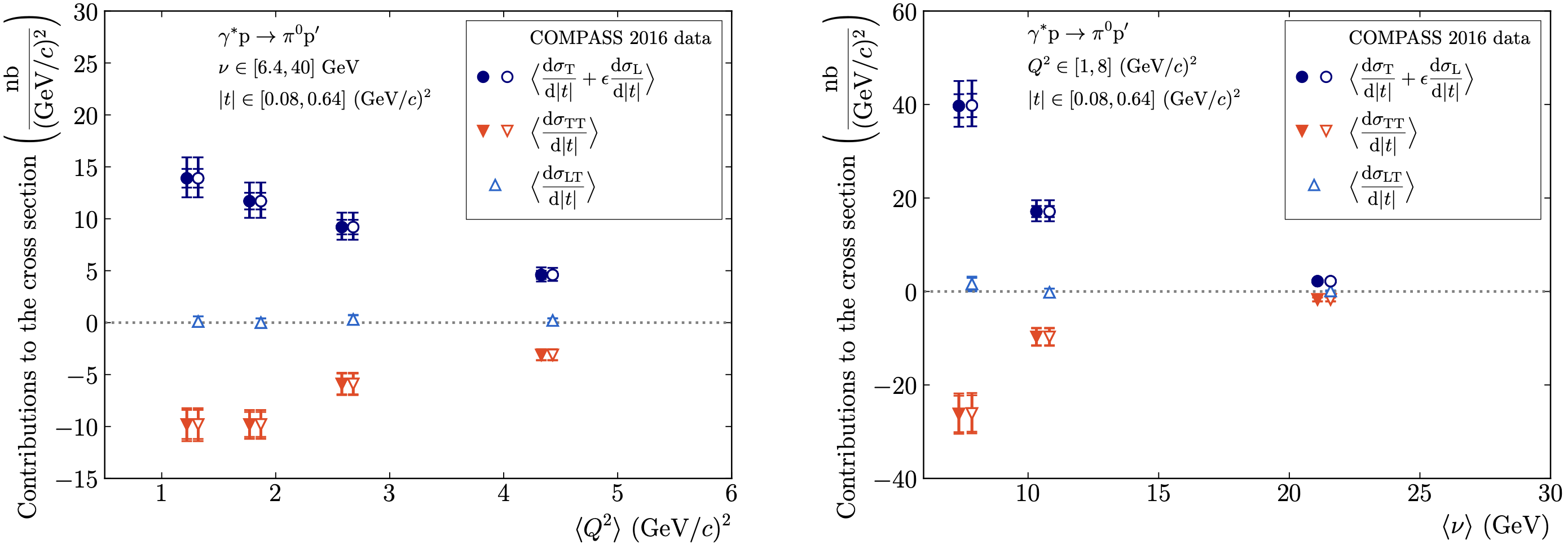} \ \    \caption{Cross-section terms extracted with the 2016 data as a function of $Q^2$ (left) and $\nu$ (right). While the $\langle{}Q^2\rangle$ ($\langle\nu\rangle$) in the left (right) plot changes, the corresponding $\langle\nu\rangle$ ($\langle{}Q^2\rangle$) is held almost constant, with varying $x_B$. The open markers represent the results by fitting {the} azimuthal-angle $\phi$ distribution with all three terms, while the closed markers show the results from the fit with only two terms, assuming $d\sigma_{LT}/d|t|=0$. The plots are taken from~\cite{COMPASS:2024hvm}.}
  \label{fig:compass_pi02024}
 \end{center}
\end{figure*}

In the first DVCS results of the COMPASS~\cite{COMPASS:2018pup}, the beam charge-spin sum was measured using the 2012 data. In the analysis, the single photon events selected are required to have reasonable values in the so-called exclusivity variables, which compare recoil-proton related quantities measured by the CAMERA with those determined by the rest of the spectrometer. The main background for the exclusive single photon events comes from photons resulting from $\pi^0$ decay. In the case where both photons from $\pi^0$ decay are detected, a DVCS candidate photon is removed if the invariant mass of its combination with another photon falls into the $\pi^0$ mass range. On the other hand, for the background where the other photon from $\pi^0$ decay is missing, their contribution is estimated by Monte Carlo. In the virtual photon energy ($\nu$) range of 10--32~GeV, with $\langle{}x_B\rangle\approx0.063$ and $\langle{}Q^2\rangle\approx2.1$~GeV$^2$, the DVCS events are expected to contribute significantly. In this 
$\nu$-range, the $t$-dependence of the cross section sum can be described satisfactorily by a single exponential function $e^{-B|t|}$. The slope of the exponential fitting function $B$ is $(4.3 \pm0.6_{\rm stat}{}^{+0.1}_{-0.3}|_{\rm sys})\,\mathrm{GeV}^{-2}$. It can be converted to the transverse extension of partons in the proton as $\sqrt{\langle r_{\perp}^2 \rangle}=(0.58 \pm 0.04_{\rm stat}{}^{+0.01}_{-0.02}|_{\rm sys}\pm{}0.04_{\rm model})$~fm. This result is shown in 
Fig.~\ref{fig:compass_dvcs} together with the measurements of H1 and ZEUS and the model predictions of Goloskokov-Kroll (GK) and Kumeri\v{c}ki-M\"{u}ller (KM15).

The DVCS analysis with the COMPASS 2016 dataset has been performed. The already released preliminary result from these new data shows a noticeably higher $B$ relative to the one extracted from the 2012 data. While investigations and cross-checks are still underway, the GPD working group of COMPASS is more confident in the new results due to the better beam quality and the more advanced analysis techniques employed. With the incorporation of the full 2016, and 2017 datasets, COMPASS expects to provide 
$\sqrt{\langle r_{\perp}^2 \rangle}$ 
values measured at three different 
$\langle x_B\rangle$ values. In addition, COMPASS will perform beam charge-spin asymmetry measurements using the full 2016 and 2017 datasets. With the full statistics of the data, it is expected that the corresponding uncertainties of the measured asymmetries can be reasonably controlled for the extraction of Re$\mathcal{H}$.

\subsubsection{Deeply virtual exclusive meson production at COMPASS}

The COMPASS experiment is capable of the exclusive muoproduction of a number of different mesons, such as 
$\pi^0$, $\rho$, $\omega$, $\phi$, and $J/\psi$. 
These measurements can provide information on GPDs that is not accessible through DVCS. It can be used for the GPD flavor decomposition and provides further insights into the validity of the QCD collinear factorization reaction mechanism.

\paragraph{Exclusive $\pi^0$ muoproduction} 
The differential cross section of exclusive 
$\pi^0$ muoproduction measured at COMPASS can be sensitive to not only the axial vector GPDs, 
$\tilde{H}$ and $\tilde{E}$, but also the tensor ones, 
$H_T$ and $\tilde{E}_T$. 
Similar to the DVCS case, the exclusive 
$\pi^0$ events are selected by applying cuts on the exclusivity variables. The contribution from non-exclusive $\pi^0$ background is estimated by Monte Carlo simulation. The non-exclusive DIS background and the exclusive $\pi^0$ events are generated by the LEPTO 6.5.1 and the HEPGEN++ generators. The exclusivity variables of experimental data are fitted by corresponding combinations of these two Monte Carlo data to determine their relative contribution, and thus the yield of non-exclusive background can be extracted.

The spin-independent cross section of exclusive 
$\pi^0$ production consists of contributions from the transversely and longitudinally polarized virtual photons, denoted as $\sigma_T$ and $\sigma_L$, 
and terms of their interference, $\sigma_{TT}$ 
and $\sigma_{LT}$. While different terms can be extracted via their corresponding modulations in the azimuthal angle $\phi$ between the scattering plane and the hadron production plane, $\sigma_T$ and $\sigma_L$ 
cannot be separated in the COMPASS measurements and are thus acquired as $\sigma_T+\epsilon\sigma_L$, 
where $\epsilon$ represents the virtual-photon polarization parameter. In the exclusive 
$\pi^0$ result using 2012 data~\cite{COMPASS:2019fea}, 
the cross-section terms are determined at the average kinematics of $\langle Q^2\rangle= 2.0$~$\mathrm{GeV}^2$, 
$\langle \nu\rangle=12.8$~$\mathrm{GeV}$,
$\langle x_B\rangle= 0.093$ 
and 
$\langle -t\rangle= 0.256$~$\mathrm{GeV}^2$. 
A large negative 
$\sigma_{TT}$ 
and a smaller positive 
$\sigma_{LT}$ 
are acquired, which suggests the significance of the contribution from the transversely polarized virtual photon. New results from the 2016 dataset were released late 2024~\cite{COMPASS:2024hvm}, extracting cross section in the kinematic domain of 
\begin{align*}
&1~\mathrm{GeV}^2 < \;Q^2 < 8~\mathrm{GeV}^2, \cr
&6.4~\mathrm{GeV} < \;\nu < 40~\mathrm{GeV}, \cr
&0.08~\mathrm{GeV}^2 < \;|t| < 0.64~\mathrm{GeV}^2, 
\end{align*}
with larger ranges in $Q^2$ and $\nu$ compared to 2012 data. A consistency check between the results of two different datasets in the same kinematic phase space shows reasonable agreement while 2016 data suggests a more gentle decrease of the cross section with increasing $|t|$. The higher statistics of the 2016 data allows to study the dependence of the differential cross section as a function of $Q^2$ and $\nu$. Cross sections were determined in four subdivisions in $Q^2$, and three in $\nu$; As can be seen in Fig.~\ref{fig:compass_pi02024},   while the $d\sigma_{LT}/d|t|$ values are consistent with zero, a noticeable evolution with respect to 
$Q^2$ and $\nu$ can be seen in both 
$d\sigma_T/d|t|+\epsilon{}d\sigma_L/d|t|$ 
and $d\sigma_{TT}/d|t|$. These new inputs provide valuable information for more stringent constraints on the GPDs involved.

\paragraph{Exclusive muoproduction of vector mesons} 
In the initial study of exclusive 
$\rho^0$ production~\cite{COMPASS:2012ngo, COMPASS:2013fsk} 
with COMPASS 2007 and 2010 data, transverse target spin asymmetries were measured, and the significance of the contribution from $H_T$ was suggested. On the other hand, the contribution of GPD $E$, which can also be identified in the non-zero asymmetry, is suppressed due to the cancellation between the opposite-sign behavior of $u$ and $d$ quarks. This cancellation is not expected in the case of exclusive production of $\omega$, and indeed a much larger asymmetry was observed~\cite{COMPASS:2016ium}. Nevertheless, the disentanglement of the $E$ contribution in the asymmetry is obscured due to the involvement of the pion pole.

\begin{figure}[htb]
 \begin{center}
  \includegraphics[width=0.99\columnwidth]{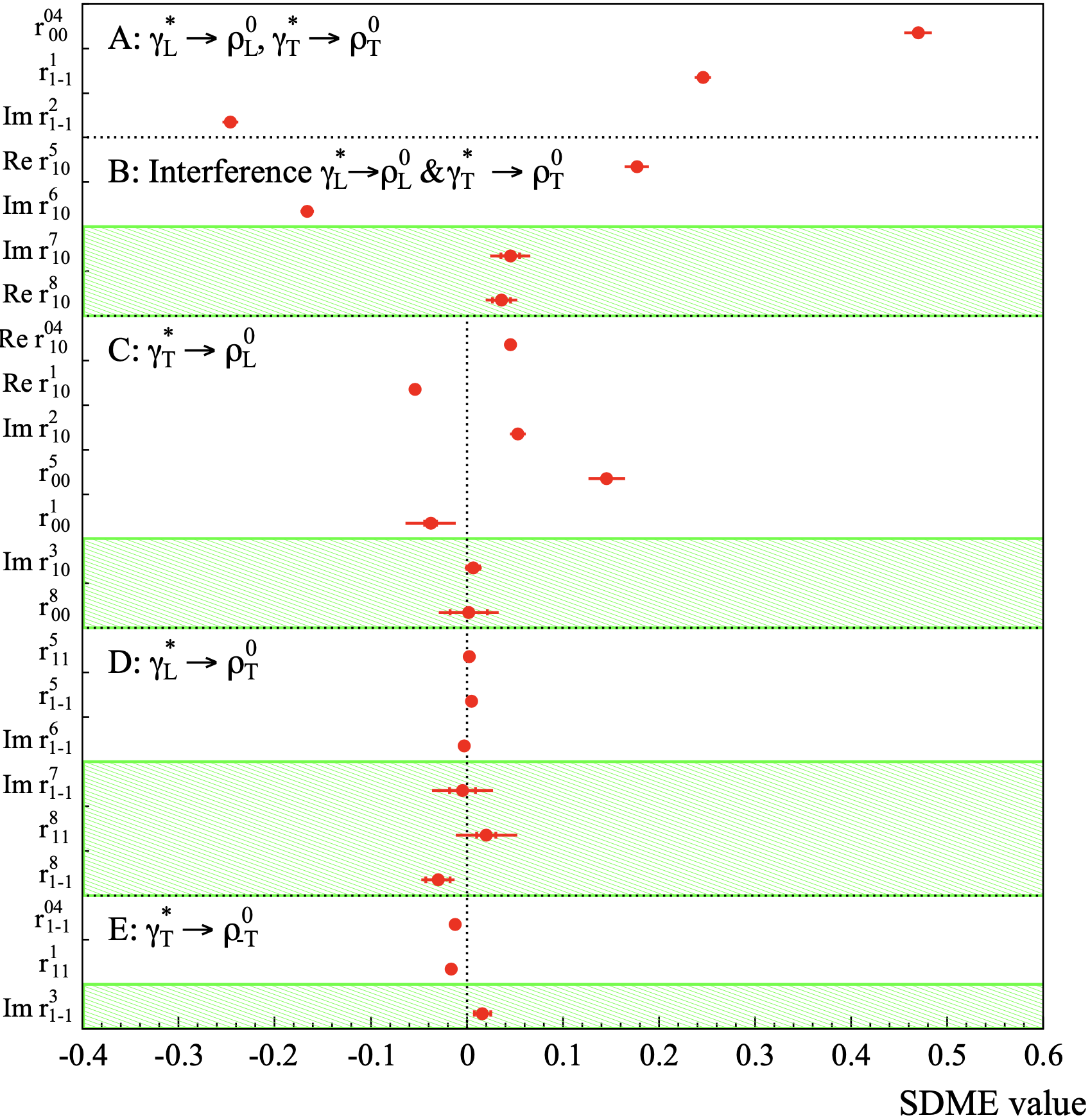} \ \    \caption{The 23 SDMEs extracted in the exclusive $\rho^0$ muoproduction. The unpolarized (polarized) SDMEs are shown in the unshaded (shaded) regions. This picture is taken from~\cite{COMPASS:2022xig}.}
  \label{fig:compass_rho_sdme}
 \end{center}
\end{figure}

With the COMPASS 2012 data, the spin density matrix elements (SDMEs) are determined from the angular distributions of particles produced with exclusive vector 
mesons~\cite{COMPASS:2020zre, COMPASS:2022xig}. 
A total of 15 unpolarized SDMEs and 8 polarized ones were extracted, which describe how the spin components of the virtual photons are transferred to the mesons created. These 23 SDMEs can be organized into five classes, depending on the helicity transition from 
$\gamma^*$ 
to vector meson, as shown in Fig.~\ref{fig:compass_rho_sdme} for the $\rho^0$ case as class A to class E. The GPDs that describe the helicity flip of the active quark can be related to the violation of $s$-channel helicity conservation (SCHC). Within SCHC, the SDMEs of classes C, D and E are all expected to be zero, and a clear violation of SCHC in class C is shown in both cases of exclusive $\omega$ and $\rho^0$. This behavior can be attributed to the presence of chiral-odd GPDs, for example, the $H_T$ and $\bar{E}_T$ as described in the GK model. In addition to SDME values, the asymmetry between the natural parity exchange (NPE) and the unnatural parity exchange (UPE) were studied. The NPE-to-UPE asymmetry for $\gamma^*_T\rightarrow V_T$ for a vector meson $V$ can be evaluated by the SDMEs, and while the dominance of NPE in the production of $\rho^0$ is suggested~\cite{COMPASS:2022xig}, NPE is comparable to UPE on average~\cite{COMPASS:2020zre} in the $\omega$ case. Furthermore, in the $\omega$ production, the dominance of UPE at small $W$ and $p_T^2$ is observed while the asymmetry shows no significant $Q^2$ dependence. These kinematic dependencies could serve as inputs for the modeling of chiral-even GPDs involved.
 
Beyond the progress mentioned above, there are ongoing studies on the production of other mesons, such as the SDMEs of exclusive $\phi$ and the cross section of exclusive $J/\psi$. Continuing efforts are being made to include all the COMPASS data available for all the exclusive channels accessible with COMPASS.


\subsection{Kaon-LT}

Pions and kaons are among the most prominent strongly interacting particles alongside the nucleon, as they are the Goldstone bosons of spontaneously broken chiral symmetry of QCD. Thus, it is important to study their internal structure and how this reflects their Goldstone boson
nature, a question particularly relevant for understanding the origin of mass generation in QCD.  

The hard contribution to the 
$\pi^+$ 
form factor can be calculated exactly within the framework of pQCD, and at asymptotically high 
$Q^2$ 
it takes a particularly simple form 
\cite{PETERLEPAGE1979359} 
\be
F_\pi(Q^2) \xrightarrow[Q^2 \rightarrow \infty]{ } 16 \pi \alpha_s (Q^2) f_\pi^2 / Q^2,
\ee 
where 
$f_{\pi}$ 
is the 
$\pi^+$ 
decay constant.  In general, pions also contain soft
contributions that are expected to dominate at lower 
$Q^2$. 
The actual behavior of 
$F_{\pi}$ 
as a function of 
$Q^2$, 
as QCD transitions smoothly from the non-perturbative (long-distance scale) confinement regime to the perturbative (short-distance scale) regime, is an important test of our
understanding of QCD. Since QCD calculations cannot 
yet be performed rigorously in the confinement regime, experimental data from JLab play a vital role in validating the theoretical approaches employed. In particular, due to the charged pion's relatively simple quark-antiquark ($q\bar{q}$) valence structure and its experimental accessibility, the pion elastic form factor ($F_{\pi}$) offers our best hope of directly observing QCD's transition from color-confinement at long distance scales to asymptotic freedom at short distances.  It is worth highlighting that in QCD the difference between the kaon and pion charge form factors is of the order of 20\% at 
$Q^2 \sim$~5~GeV$^2$~\cite{Gao:2017mmp} and disappears at asymptotic $Q^2$ as ($1/\ln Q^2$). Thus, the acquisition of experimental data for both form factors covering a wide 
$Q^2$ range is of particular interest.

Current experimental information on the pion and kaon form factors is limited, particularly at large $Q^2$ 
\cite{Horn:2016rip}. Measurement of the $\pi^+$ electromagnetic form factor for $Q^2>0.3$~GeV$^2$ can be accomplished by the detection of the exclusive reaction $p(e,e'\pi^+)n$ at low $-t$. This is best described as quasi-elastic ($t$-channel) scattering of the electron from the virtual $\pi^+$ cloud of the proton, where 
$t=(p_{p}-p_{n})^2$ is the invariant momentum transfer to the target nucleon. Scattering from the $\pi^+$ cloud dominates the longitudinal photon cross section ($d\sigma_L/dt$),  when $|t|\ll M_p^2$. To reduce background contributions, one preferably separates the components of the cross section due to longitudinal ($L$) and transverse ($T$) virtual photons (and the $LT$, $TT$ interference contributions), via the Rosenbluth separation. The Rosenbluth separation involves the absolute subtraction of two measurements determined at high and low virtual-photon polarization 
($\epsilon_{\rm Hi}$, $\epsilon_{\rm Lo}$), corresponding to high and low electron beam energies. The resulting errors on $\sigma_L$ 
and $\sigma_T$ are magnified by 
$1/\delta\epsilon=(\epsilon_{\rm Hi}-\epsilon_{\rm Lo})^{-1}$.  
To keep the uncertainties in $\sigma_L$ to an acceptable level, 
$\delta\epsilon >0.2$ is typically required,  {\it i.e.} an uncertainty magnification of no more than 5.  
The measurements require continuous, high intensity electron beams, and detectors with good particle identification and reproducible systematics. JLab Hall~C is the only facility worldwide capable of such studies.  

Hall C is equipped with two moderate momentum resolution 
($<0.1\%$) and small solid angle (4--6 msr) magnetic spectrometers that can be set to a wide variety of scattering angles and momenta to detect the scattered charged particles. In this measurement, the Super High Momentum Spectrometer (SHMS) detects the high momentum, forward going meson, while the High Momentum Spectrometer (HMS) is used to detect the scattered electron. The focal plane instrumentation of both spectrometers is designed to allow reliable tracking and particle identification (PID) over a wide range of rates and momenta. Hall~C can operate at a luminosity of $10^{39}$~cm$^{-2} \cdot$s$^{-1}$. This, in combination with well-understood spectrometer magnetic optics and focal plane detectors, is needed to reliably measure the smallest cross sections. The $F_{\pi}$ experiment is particularly demanding, as it requires very forward angle $\pi^+$ data to be acquired, with systematic and statistical uncertainties (in
$d\sigma/dt$) totaling under 4\% (in quadrature).

\begin{figure}[t!]
\centering\includegraphics*[width=0.99\linewidth]{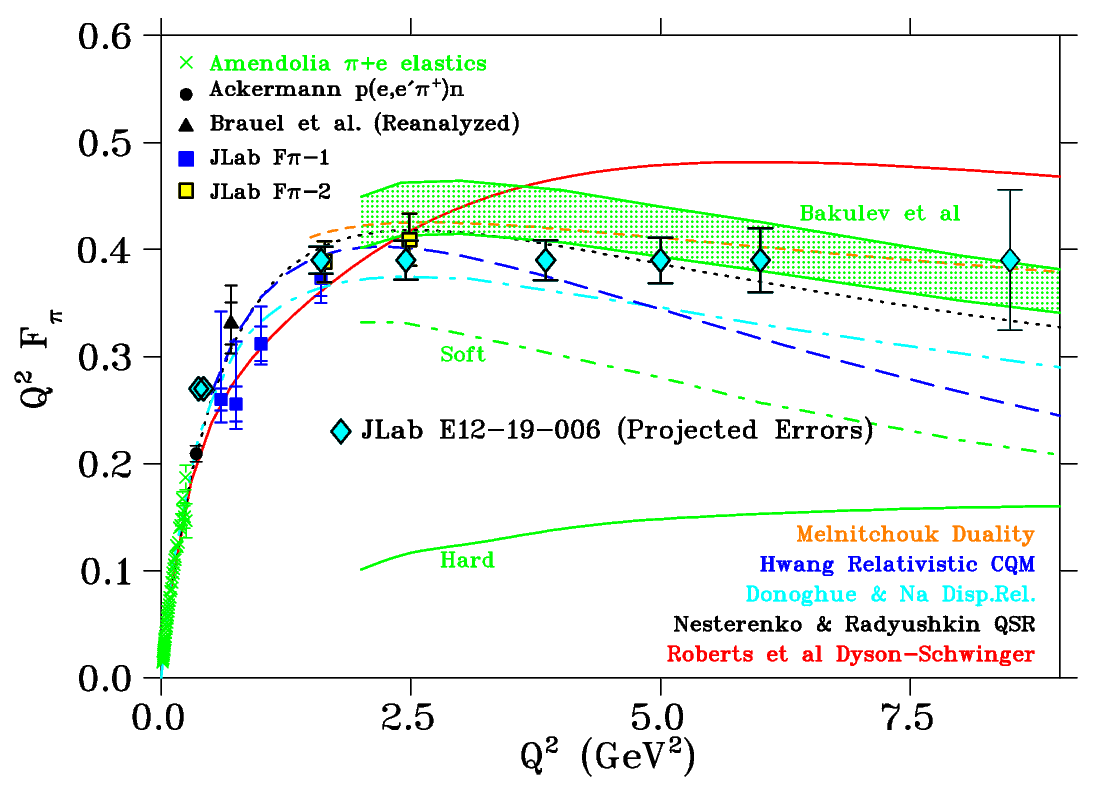}
\caption{
Existing data (green \cite{Amendolia:1984nz,Amendolia:1986wj};
black \cite{Ackermann:1977rp,Brauel:1979zk,Huber:2008id};
blue and yellow \cite{Horn:2006tm,Tadevosyan:2007yd,Huber:2008id})
and projected uncertainties for future data on the pion form factor from JLab (cyan~\cite{E12-19-006}), in comparison to a variety of hadronic structure models (orange dash \cite{Melnitchouk:2002gh};
blue long-dash \cite{Geng:2001de}; 
cyan dot-dash \cite{Donoghue:1996bt}; 
black dot \cite{Nesterenko:1982gc}; 
red solid \cite{Chang:2013nia}; 
and green \cite{Bakulev:2004cu}, where Hard is pQCD with analytic running coupling, and the band is Hard+Soft including non-perturbative uncertainties). The vertical axis position of the projected data is arbitrary and the error bars are based on the actual statistics obtained and projected systematic uncertainties. 
}
\label{fig:fpi_proj}
\end{figure}

The value of $F_{\pi}(Q^2)$ is determined by comparing the measured 
$\sigma_L$ values at small $-t$ to the best available electro-production model. The obtained $F_{\pi}$ values are, in principle, dependent upon the model used, but one anticipates this dependence to be reduced at sufficiently small $-t$. Measurements over a range of $-t$ are an essential part of the model validation process.
The JLab 6~GeV experiments were instrumental in establishing the reliability of this technique up to 
$Q^2$=2.45~GeV$^2$~\cite{Huber:2008id, Horn:2016rip, Horn:2007ug, Volmer:2000ek, Horn:2006tm, Tadevosyan:2007yd, Blok:2008jy, Huber:2014ius, Huber:2014kar}, and extensive further tests are planned as part of JLab experiment E12-19-006~\cite{E12-19-006} which will cover the region in $Q^2$ that transitions from a pion form factor description from large-distance to short-distance QCD phenomena, up to $Q^2$=8.5~GeV$^2$.  Data taking was completed in 2022, and first results are anticipated to be released in 2026.  Projected kinematic reach and experimental uncertainties are shown in Fig.~\ref{fig:fpi_proj}.

The reliability of the electro-production method to determine the 
$K^+$ form factor is not yet fully established.
$L/T$ separated kaon electro-production cross sections have been extracted at different values of $-t$ using JLab data, with the most recent results being for $Q^2$=1.00-2.35~GeV$^2$ 
in \cite{Carmignotto:2018uqj}, and older results in
~\cite{Mohring:2002tr,Coman:2009jk}. 
In \cite{Carmignotto:2018uqj}, the successful method from 
\cite{Horn:2006tm, Blok:2008jy} 
was applied to determine the kaon form factor from 
$p(e,e^\prime K^+)\Lambda$ 
data, but the uncertainties are very large, primarily due to inadequate statistics. JLab 12~GeV experiment
E12-09-011~\cite{E12-09-011} 
acquired higher statistics data for the 
$p(e,e^\prime K^+)\Lambda$, 
$p(e,e^\prime K^+)\Sigma^0$ 
reactions above the resonance region, $W=\sqrt{(p_{K}+p_{\Lambda,\Sigma})^2}>2.5$~GeV, 
to search for evidence of scattering from the proton's ``kaon cloud''. The data are still being analyzed, with $L/T$-separated cross sections up to $Q^2$=5.5~GeV$^2$ expected in the next year.
If the anticipated data confirm that the scattering from the virtual 
$K^+$ 
in the nucleon dominates at low four-momentum transfer to the target  
$|t|\ll M_p^2$, 
the experiment will yield the world's first quality 
$F_K$ 
data for 
$Q^2>0.2$~GeV$^2$.


\subsection{$J/\psi$-007}

The study of nucleon and nuclear structure has entered a transformative stage, driven by recent experimental and theoretical advances that enable direct access to GFFs. As already emphasized, these fundamental quantities encode the intrinsic mechanical properties of hadrons, including their mass and internal energy distributions. When generalized to nuclei, GFFs open new avenues for exploring the dynamics of nuclear matter particularly gluons which play an essential role as mediators of the strong force and the dominant contributors to the nucleon’s mass and scalar energy~\cite{Burkert:2023wzr}. 

Modern measurements of near-threshold $J/\psi$ photo-production off the proton~\cite{Duran:2022xag, GlueX:2019mkq, GlueX:2023pev} have provided a comprehensive experimental foundation for investigating the gluonic structure of the nucleon. The most recent differential cross-section data were primarily collected at JLab employing different detector systems following the 12 GeV upgrade: Hall C’s high-resolution spectrometers ($J/\psi$-007), Hall D’s large-acceptance GlueX detector, and Hall B’s CLAS12 detector. In Hall C, differential cross sections for $J/\psi$ photo-production off the proton were measured using the $e^+e^-$-decay channel, covering photon energies up to 10.6~GeV and momentum transfers up to 4~GeV$^2$~\cite{Duran:2022xag}. In Hall D, the GlueX collaboration has reported two independent sets of measurements based on separate running periods. The first dataset~\cite{GlueX:2019mkq} presents the $t$-distribution for an average photon energy of $E_{\gamma}$ = 10.72~GeV. The more recent analysis~\cite{GlueX:2023pev} extends this work by providing $t$-distributions in three distinct photon-energy bins. Another near-threshold photo-production experiment was performed using the CLAS12  detector which results has not yet been published~\cite{ref:CLAS12jpsi}.  Together, these datasets cover the entire near-threshold energy region, from $E_{\gamma}$ = 8.2~GeV up to 11.6~GeV, providing an ideal basis for a global fit of the differential cross sections and extraction of the gluonic GFFs $A^g(t)$ and $D^g(t)$. 

\begin{figure*}[t!]
 \begin{center}
  \includegraphics[width=0.79\textwidth]{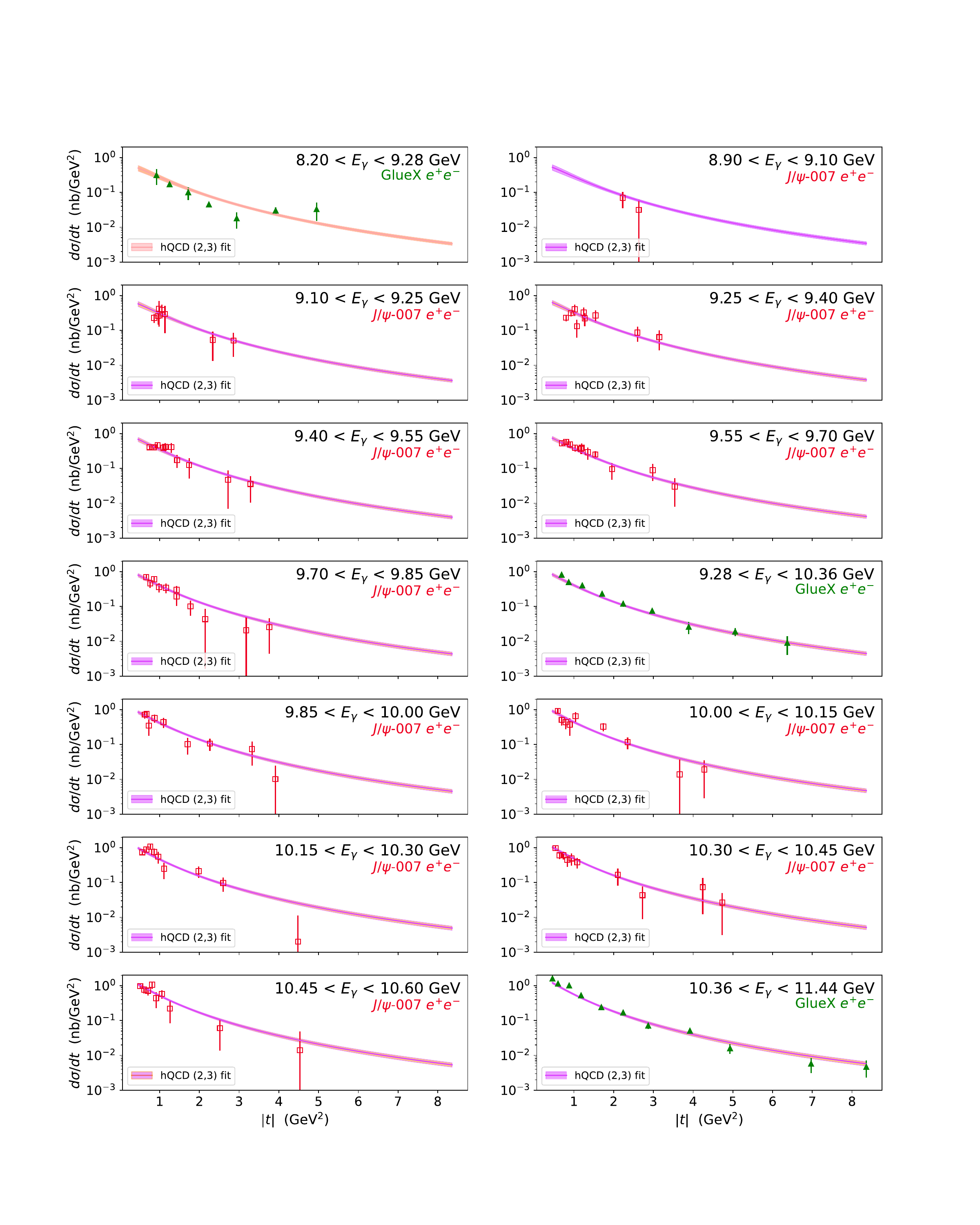}
   \caption{The $t$ dependence of the $J/\psi$ differential photo-production cross section on the proton near threshold using the $J/\psi$ $e^+e^-$-decay channel. The red points are those of the Hall C $J/\psi$-007 experiment~\cite{Duran:2022xag} and the green points are those of GlueX~\cite{GlueX:2023pev}. The purple band corresponds to a 2D fit of these data (except for the first GlueX photon energy data) using the holographic model~\cite{Mamo:2019mka,Mamo:2022eui}. The pink band in the first panel is a prediction of the fit for this photon energy bin.    
   }
  \label{fig:dxsec-jpsi007-glueX}
 \end{center}
\end{figure*}

{Using the $J/\psi$ $e^+e^-$-decay channel data from GlueX and $J/\psi$-007 we employed two different models for a two-dimensional fit in $E_{\gamma}$ and $t$ to extract the gluon EMT FFs $A^g(t)$ and $D^g(t)$: the QCD holographic model (hQCD) developed in Ref.~\cite{Mamo:2022eui} and the GPD model from Refs. \cite{Guo:2021ibg,Guo:2023pqw}.}

{It is worth noting for completeness that several models have been developed to evaluate the differential cross sections of $J/\psi$ in the near-threshold region with varying degrees of success \cite{Kim:2025oyo,Sakinah:2024cza,Pire:2022kwu, JointPhysicsAnalysisCenter:2023qgg,Tang:2025qqe}.}

%
%

First, we used the 
hQCD model
to perform the fit of the differential cross sections. This model effectively captures the non-perturbative interaction between the $J/\psi$ dipole and the nucleon, interpreting the exchange as a coherent sum of a graviton-like tensor glueball ($2^{++}$) and a dilaton-like scalar glueball ($0^{++}$). Within the hQCD model, the differential cross section is expressed as~\cite{Mamo:2022eui}: 
\be
\frac{d\sigma}{dt}
= && \mathcal{N}^2_e 
\frac{e^2}{64\pi (s - M^2_N)^2}
\frac{\bigl[A^g(t) + \eta^2 D^g(t)\bigr]^2}{[A^g(0)]^2} \nn \\ && \times
F(s) 
\frac{(2t + 8M^2_N)}{4M^2_N} ,
\label{eq:dxsec}
\ee
where $s$ is the square of the photon–nucleon center-of-mass energy, and $\mathcal{N}_e = 7.768~{\rm nb \cdot GeV^{-6}}$ is a normalization factor. The function $F(s)$, defined in Eq. (8.6) of Ref.~\cite{Mamo:2019mka}, encapsulates a nontrivial $s$-dependence of the cross section (see also Eq. (8.4) therein). Importantly, this expression avoids the high-energy approximation $s \gg -t$. The gluonic form factors $A^g(t)$ and $D^g(t)$ defined in Eqs.~(\ref{EMT_FF_decomposition}) and (\ref{EMT_FF_decomposition_2}) are modeled using a dipole-type parameterization consistent with lattice QCD results~\cite{Hackett:2023rif} and the expected perturbative QCD scaling behavior at large momentum transfer~\cite{Hoodbhoy:2003uu,Sun:2021gmi}:
\begin{equation}
A^g(t) = \left(\frac{1}{1 - t/m_A^2}\right)^2,
\ \ \
C^g(t) = \left(\frac{1}{1 - t/m_C^2}\right)^{n},
\label{eq:ff}
\end{equation}
with $n = 2$ for the dipole–dipole case and $n = 3$ for the dipole–tripole choice. The parameters $m_A$, $m_C$, and $C^g(0)$ are free fit parameters, while $A^g(0)$ is fixed to the experimental value $A^g(0) = 0.414 \pm 0.008$ from the CTEQ18 global analysis~\cite{Hou:2019efy}. 

Second, we used a GPD model to extract the same gluons GFFs as previously~\cite{Guo:2021ibg,Guo:2023pqw}. In this approach it is clear that higher order terms play an important role, and the condition on the skewness variable $\xi > 0.5$ needed to control the convergence of the expansion in $1/\xi$ filters out most of the world data leading to a less stable constraint on the form factors. A more recent analysis~\cite{Guo:2025jiz} at next-to-leading order was performed using DVCS and {DVMP}, as well as lattice QCD data, while still requiring $\xi > 0.5$. The cross section is expressed as follows ~\cite{Guo:2021ibg,Guo:2023pqw}:
\begin{equation}
\frac{d\sigma}{dt}= \frac{\alpha_{\rm em} m e^2_Q}{4(s-M^2_N)^2} \frac{(16\pi\alpha_s)^2}{3M^2_{J/\psi}}\vert \psi_{NR} \vert^2 
\vert G(t,\xi) \vert^2, 
\label{eqn:xsecgpd}
\end{equation}
where 
$M_{J/\psi}$ 
is the mass of the 
$J/\psi$. 
$\psi_{NR}$ is the value of the non-relativistic wave function of the heavy quarkonium at the origin. $G(t,\xi)$ contains the proton gluon GPDs, $\alpha_{\rm em}$ is the electromagnetic coupling constant, $\alpha_s$ is the strong running coupling constant and $e_Q$ is the charge of the heavy quark in the units of proton charge, and
\be 
&&
\vert G(t,\xi) \vert^2 = \frac{4}{\xi^4} \bigg\{ \left(1-\frac{t}{4M_N^2}\right) E_2^2 -2E_2(H_2+E_2) \nn \\ && 
+(1-\xi^2)(H_2+E_2)^2 \bigg\},
\label{eqn:amplisquared}
\ee
where
\be 
&&
H_2(t,\xi) \equiv \int_0^1dx H^g(x,\xi,t), 
\nn \\ &&  E_2(t,\xi) \equiv \int_0^1dx E^g(x,\xi,t). 
\ee
To perform our fits of the differential cross sections for this case we made the same assumptions regarding the form factors $t$ dependence and fixed $A(0)$ on the experimental data as previously stated. We also note that in the hQCD cross section {in} Eq.~(\ref{eq:dxsec}) the $t$-dependence of the skewness parameter 
\( \eta = M_{J/\psi}^2 / (4p_N \cdot q-M_{J/\psi}^2 + t) \)  
differs from that of 
\( \xi = (M_{J/\psi}^2 + t/2)/(4p_N \cdot q-M_{J/\psi}^2 + t) \)
employed in the standard GPD approach, Eq.~(\ref{eqn:xsecgpd}). Finally, since we are interested in the mass and scalar distributions we must define the total mass or scalar form factors \( G^{m,s}(t)= G^{m,s}_{g+q} (t) \), a sum of quarks and gluons contributions following a linear combination of GFFs according to~~\cite{Ji:2021mtz}:
\be 
G^s (t)  = &&  M_N \bigg[ A^{g+q}(t) + \frac{t}{4 M^2_N} B^{g+q}(t)  \nn \\ && - \frac{3t}{4 M^2_N} D^{g+q}(t) \bigg]; \\
G^m (t)  = && M_N \bigg[ A^{g+q}(t) + \frac{t}{4 M^2_N} B^{g+q}(t) \nn \\ && - \frac{t}{4 M^2_N} D^{g+q}(t) \bigg].
\label{scalar-mass}
\ee
To determine the total mass FFs, we adopt a hybrid methodology that combines our experimental extraction of the gluon GFFs with lattice-QCD results for the quark contributions. This approach enables a quantitative evaluation of the mass and scalar form factors of the nucleon. It is important to note that both \(
\bar{C}^{q+g}(t) \) and \( B^{q+g}(t) \) need not be considered explicitly: the former vanishes identically, while the latter is found to be small in lattice calculations and exactly zero in holographic models.

\begin{figure}[t!]
 \begin{center}
  \includegraphics[width=0.99\columnwidth]{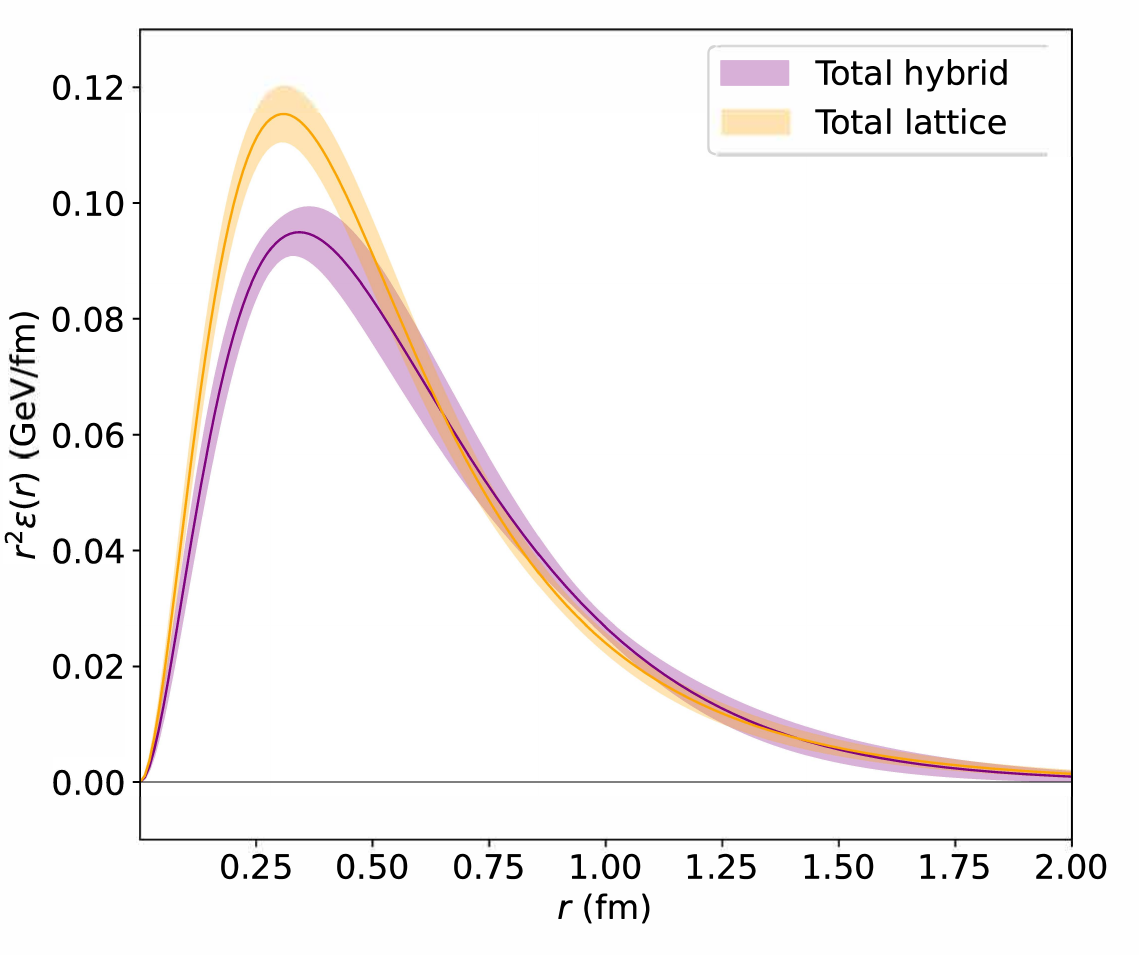}
   \caption{The total mass density where we combined the extracted gluon FFs from Fig.~\ref{fig:dxsec-jpsi007-glueX} and the quark FFs evaluated from the lattice QCD~\cite{Hackett:2023rif}. }
  \label{fig:mass-density}
 \end{center}
\end{figure}
With the mass form factor determined, we compute the corresponding mass-density profile as a function of the spatial coordinate relative to the nucleon center. The analysis is performed in the Breit frame, where the Fourier transforms of the form factors are evaluated~\cite{Lorce:2018egm}:
\begin{equation}
 \epsilon^{m,s}(r) = \int \frac{d^3 \vec{\Delta}}{(2\pi)^3} \, e^{-i \vec{\Delta}\cdot 
\vec{r}} \, G^{m,s}_{g+q} (t),
\label{Fourier_Epsilon_r}
\end{equation}
where $t= -\vec{\Delta}^2$. The Breit frame is selected not because it provides a purely probabilistic interpretation of the densities, but for practical consistency, as charge and magnetization distributions are conventionally presented in this frame. 

Shown in 
Fig.~\ref{fig:mass-density} 
is the mass density profile as a function of distance from the center of the nucleon in the Breit frame. It is important to note that 
\( A^q(0) = 0.51\pm 0.025 \) 
in the lattice calculation~\cite{Hackett:2023rif} 
at $\mu=2$~GeV in the $\overline{\rm MS}$ scheme,
while experimentally we know that 
\( A^q(0) \simeq 0.6 \). 
Here 
\( A^q \) 
was not renormalized to give 
\( A^q(0)[\mathrm{Lattice}] + A^g(0)[\mathrm{Experiment}] = 1.0 \). 
The agreement would improve if we set 
\( A^q(0) \simeq 0.6 \) 
for the experimental gluon-lattice quark hybrid case. For lattice 
\( A^{q+g} = 1.011\pm {0.037} \). It is clear from Fig.~\ref{fig:mass-density} 
that the proton mass root-square mean radius is smaller than the charge. We also find that the scalar energy density root-square mean radius is larger than the charge radius and seems to define the size of the proton.

\section{Advances in theoretical methods}

\subsection{Exclusive processes amplitudes beyond kinematic leading twist }
\label{sec:advences_in_theory_DVCS_ht}

As mentioned earlier, GPDs are functions of three variables (aside from scale dependence): the fraction of hadron's longitudinal momentum carried by the active parton ($x$), the skewness or the fraction of the longitudinal kick to the hadron ($\xi$), and the Mandelstam variable $t$. Disentangling their dependence on these variables is one of the major goals of the exclusive physics program on both the theoretical and experimental sides. To explore GPDs, various exclusive reactions are considered, as introduced in previous sections. Among these, DVCS stands out due to 
{its exceptional theoretical cleanliness and its comparatively direct interpretation in terms of CFFs.}
Consequently, DVCS has been extensively studied at different experimental facilities, earning the status of the {\it golden channel} for extracting GPDs. The lowest order approximation provides the $C$-even part of GPDs at 
$x=\xi$ ($F^{(+)}(\xi,\xi,t)$) 
through the imaginary part of 
the corresponding CFFs. This lowest order approximation typically involves a leading order analysis in the strong coupling (LO), while neglecting kinematic power corrections. The latter corresponds to disregarding the mass and momentum transfer to the hadron with respect to the hard scale, {\it i.e.}, the virtuality of the incoming photon, and is referred to as the (kinematic) leading twist approximation (LT).   

Other complementary processes exist, being TCS, a prominent reaction with a first measurement in 2021 at JLab's Hall {B}~\cite{CLAS:2021lky}. Both DVCS and TCS can be seen as limiting cases of the more general process, the DDVCS, which  at the lowest order (LO and LT), provides access to GPDs outside the line $x=\xi$ and thus over their full kinematic domain. Consequently, these three reactions can be studied within the DDVCS framework\footnote{For two-photon scattering, all-order factorization has recently been proven at next-to-leading power
in    the  $\Lambda_{\mathrm{QCD}} / Q$  and   $\sqrt{-t} / Q$  expansion~\cite{Schoenleber:2024ihr}.}. This will be the approach followed in this section.

For a proper handling of the scale ($\mu$) dependence of GPDs $F^{(+)}(x,\xi,t; \mu^2)$, next-to-leading order (NLO) corrections in $\alpha_s$ to the hard part of exclusive processes, as well as evolution equations for GPDs, are required.  Extensive literature exists on this topic; see  {\it e.g.},~Refs.~\cite{Belitsky:1997rh, Ji:1998xh,Ji:1997nk,Mankiewicz:1997bk,Pire:2011st,Ji:1996nm}.

\begin{figure}
    \centering
    \includegraphics[width=\columnwidth]{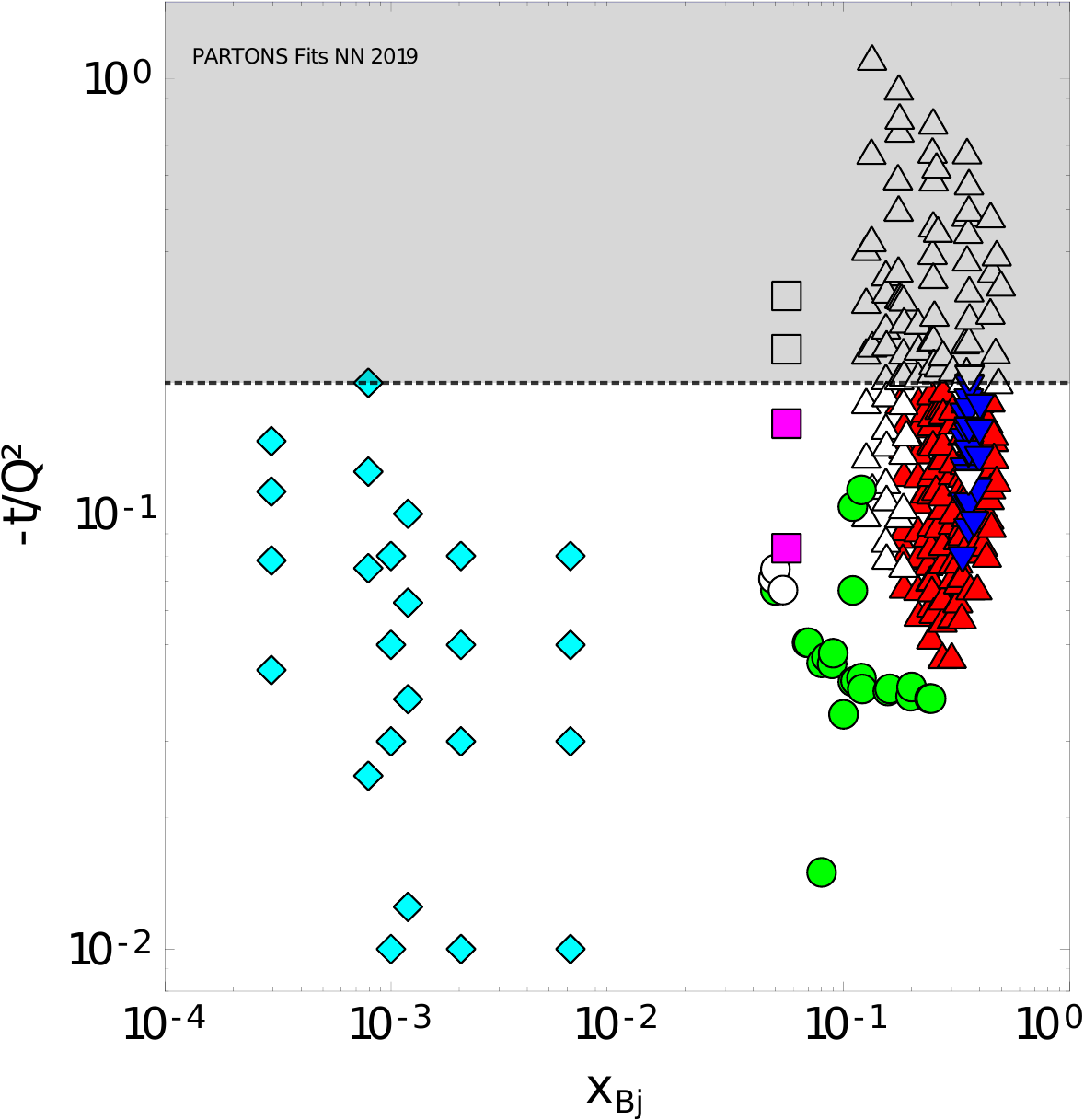} 
    \caption{{\small Plot from Ref.~\cite{Moutarde:2019tqa}. Data come from the Hall A (\textcolor{blue}{$\blacktriangledown$}, \textcolor{black}{$\triangledown$}), CLAS (\textcolor{red}{$\blacktriangle$}, \textcolor{black}{$\vartriangle$}), HERMES (\textcolor{green}{$\bullet$}, \textcolor{black}{$\circ$}), COMPASS (\textcolor{magenta}{$\blacksquare$}, \textcolor{black}{$\square$}) and HERA H1 and ZEUS (\textcolor{cyan}{$\blacklozenge$}, \textcolor{black}{$\lozenge$}) experiments. The gray bands (open markers) indicate phase-space areas (experimental points) being excluded in the analysis as they violate the author's chosen condition for LT: $|t|/Q^2 < 0.2$, $Q^2>1.5$~GeV$^2$.}}
    \label{fig::tOverQ2_vs_xB}
\end{figure}

The last piece of the 3D kinematic domain of GPDs is the dependence on $t$. At the leading twist, exclusive processes depend on it solely through the GPDs/CFFs. The validity of this approximation relies on the smallness of the ratio 
$|t|/\scale^2$, 
where 
$\scale^2$ 
represents the hard scale taking the form: 
$\scale^2=Q^2+t$ 
for DVCS, 
$\scale^2=Q'^2+t$ 
for TCS and 
$\scale^2=Q^2+Q'^2+t$ 
for DDVCS. Current and future experiments are far from achieving the LT limit, {\it i.e.}, 
$|t|/\scale^2 \to 0$. 
Therefore, powers of the form 
$(|t|/\scale^2)^{(\tau-2)/2}$ 
appear on the hard kernels, known as kinematic twist 
($\tau$) 
corrections. Accounting for these corrections improves accuracy and enables the incorporation of more data for extracting CFFs, avoiding arbitrary cuts that exclude points affected by twist powers, as shown by the gray band in~Fig.~\ref{fig::tOverQ2_vs_xB}. 
In fact, this artificial loss of data is partly responsible for the large uncertainty bands in certain kinematic ranges of the  neural-network CFF extractions, 
{\it cf}.~Ref.~\cite{Moutarde:2019tqa}. 
Furthermore, hadron tomography~\cite{Burkardt:2002hr} 
requires integration over all possible values of $t$, making the inclusion of these kinematic twist corrections essential. Consequently, recovering these kinematic factors must become an integral part of the standard for theoretical calculations, alongside with NLO corrections. 

For a specific class of processes, the works of V.~Braun, A.~Manashov and their collaborators
\cite{Braun:2012bg,Braun:2014sta,Braun:2016qlg,Braun:2020zjm,Braun:2022qly} have demonstrated that these corrections can be conveniently calculated using the conformal operator-product expansion (COPE). This approach  leverages  the conformal symmetry of the lowest order matrix elements of gauge-invariant QCD operators. 

In this section, we discuss recent results
\cite{Martinez-Fernandez:2025gub} 
of the application of the COPE for exclusive processes at LO including the kinematic twist-3 and twist-4 corrections for a spin-0 target. The advantages of considering spin-0 targets are twofold: 
{
\begin{itemize}
\item 
 Corresponding exclusive processes depend on a minimal set of CFFs, reducing both the complexity of the theoretical description and the number of observables required for their extraction compared to higher-spin hadrons.
\item  
 Spin-$0$ targets include nuclei such as helium-4, for which high-quality DVCS data is already available~\cite{CLAS:2021ovm}.
\end{itemize}}

The general framework of two-photon scattering allows, for the first time in Ref.~\cite{Martinez-Fernandez:2025gub}, the calculation of these corrections for TCS~\cite{Berger:2001xd}, and the comparison  with corresponding results previously obtained for DVCS~\cite{Braun:2012bg, Braun:2022qly}. The reaction reads
\begin{equation}
    \gamma^{(*)}(q)+N(p)\to \gamma^{(*)}(q')+N(p').
\end{equation}
For DVCS the incoming photon is virtual, while the outgoing one is real ($-q^2=Q^2>0$, $q'^2=Q'^2=0$); the other way around for TCS ($Q^2=0$, $Q'^2>0$); and for DDVCS both photons are virtual ($Q^2>0$,  $Q'^2>0$). We employ the standard notations
$\bar{p} =\frac{p+p'}{2}$,
$\Delta = p'-p$, $t = \Delta^2$.
The skewness variable takes the form:
\begin{align}\label{xiDDVCSphaseSpace}
    \xi  = -\frac{\Delta \cdot n}{2 \bar{p} \cdot  n}
     = \frac{\sqrt{ (Q^2+Q'^2)^2 + t^2 + 2t(Q^2-Q'^2) }}{2Q^2/x_B - Q^2 - Q'^2 + t}\,,
\end{align}
where $n$ is a longitudinal vector expressed through a combination of the photon momenta $q$ and $q'$.
This expression agrees with Eq.~(\ref{xi_sca}) in the LT limit, for more details, see Ref.~\cite{Martinez-Fernandez:2025gub}.

As follows from Ref.~\cite{Martinez-Fernandez:2025gub}, the Compton tensor can also be parameterized by means of helicity-dependent amplitudes~\cite{Belitsky:2012ch} that, for a spin-0 target, have the advantage of being fully equivalent to the corresponding CFFs, this is
\begin{equation}
    \amp^{AB}(\rho,\xi,t) = \cffH^{AB}(\rho,\xi,t)\,.
\end{equation}
Above, the indices $A$ and $B$ are the helicities of the incoming and outgoing photons, respectively; and $\rho$ stands for the generalized Bjorken variable:
\begin{equation}\label{rho}
     \rho = \xi \frac{q \cdot q'}{\Delta \cdot q'} = \xi\frac{Q^2 - Q'^2 + t}{Q^2+Q'^2+t}\,.
\end{equation}
This expression agrees with $\xi'$ of Eq.~(\ref{xip_sca}) at the LT limit. Here, we keep a different notation to stress out that Eq.~(\ref{rho}) is the correct one beyond LT. Due to parity conservation, the Compton tensor 
$\mathcal{T}^{\mu\nu}$
can be parametrized as
\begin{align}\label{Tvector-h-amplitudes2}
    \mathcal{T}^{\mu\nu} =  
    &  - \amp^{++}g_\perp^{\mu\nu} + \amp^{+-}\frac{1}{|\bp_\perp|^2}\left[ \bp^{\,\mu}_\perp\bp^{\,\nu}_\perp - \widetilde{\bp}^{\,\mu}_\perp\widetilde{\bp}^{\,\nu}_\perp \right] \nonumber\\
    &+\ \amp^{00} \frac{-i}{QQ'R^2}\left[ (q \cdot q')(Q'^2q^\mu q^\nu - Q^2 q'^\mu q'^\nu)\right. \nonumber\\
    &\left.
    + Q^2 Q'^2 q^\mu q'^\nu - (q \cdot q')^2 q'^\mu q^\nu \right] \nonumber\\
    & + \amp^{+0}\frac{i\sqrt{2}}{R|\bp_\perp|}\left[ Q'q^\mu - \frac{q \cdot q'}{Q'}q'^\mu \right]\bp^{\,\nu}_\perp \nonumber\\
    &-\amp^{0+}\frac{\sqrt{2}}{R|\bp_\perp|}\bp^{\,\mu}_\perp\left[ \frac{q \cdot q'}{Q}q^\nu + Q q'^\nu \right] 
    \,.
\end{align}
Here, $R = \sqrt{(q \cdot q')^2 + Q^2Q'^2}$, $\widetilde{\bp}^{\,\nu}_\perp = \epsilon_\perp^{\nu\mu}\bp_\mu$ and $|\bp_\perp|^2=-\bp_\perp^2 = -( M^2 - t\left( 1 - 1/\xi^2 \right)/4 ) > 0$. 

From Eq.~(\ref{Tvector-h-amplitudes2}), projectors~\cite{Martinez-Fernandez:2025gub} onto the different helicity-dependent amplitudes/CFFs can be read out and subsequently applied to the COPE~\cite{Braun:2020zjm,Braun:2022qly}. After twist expansion, the transverse-helicity conserving amplitude/CFF of DDVCS takes the form
\begin{align}
    &\amp^{++}(\rho,\xi,t) = \nonumber\\ 
    &\int_{-1}^1dx\ \Bigg\{ -\left(1-\frac{t}{2\scale^2}+\frac{t(\xi-\rho)}{\scale^2}\dxi\right)\frac{\hplus}{x-\rho+i0} \nonumber\\
    +& \frac{t}{\xi\scale^2}\Bigg[ \pbbi +  \pbbii - \frac{\xi L}{2}
    + \frac{\wpbbiii-\wpbbi}{2}  \nonumber\\
    &\ \qquad\qquad  - \frac{\xi}{x+\xi}\Bigg( \Ln{\frac{x-\rho+i0}{\xi-\rho+i0}} 
    \nonumber \\
    &\qquad\qquad- \frac{\xi+\rho}{2\xi}\Ln{\frac{-\xi-\rho+i0}{\xi-\rho+i0}} - \wpbbi \Bigg) \Bigg] \hplus \nonumber\\
     -&\frac{t}{\scale^2}\dxi\Bigg[ \Bigg( \pbbi+\pbbii - \frac{\xi L}{2}\nonumber\\
    &\ \qquad
     - \frac{\xi}{x+\xi}\Bigg( \Ln{\frac{x-\rho+i0}{\xi-\rho+i0}} 
     \nonumber \\ 
     &\ \qquad
     - \frac{\xi+\rho}{2\xi}\Ln{\frac{-\xi-\rho+i0}{\xi-\rho+i0}} - \wpbbi \Bigg) \Bigg)\hplus \Bigg] \nonumber\\
     +& \frac{\bp_\perp^2}{\scale^2}2\xi^3\dxi^2 \Bigg[ \Bigg( \pbbi+\pbbii- \frac{\xi L}{2}+\frac{\wpbbiii+\wpbbi}{2}  \Bigg)\hplus \Bigg] 
    \nonumber \\
  & \qquad\qquad \Bigg\}  + O(\textrm{tw-6})\,, \label{amp++_final}
\end{align}
where $H^{(+)}$ is the $C$-even component (\ref{C_even_GPDs}) of the GPD $H$. In (\ref{amp++_final}) we employ
the shortened notations for the functions:
\begin{align}\label{collectCoefficientsA++DDVCS}
    \pbbi(x/\xi,\rho/\xi) &=\frac{\xi-\rho}{x-\xi}\Li{2}{-\frac{x-\xi}{\xi-\rho+i0}}\,, \nn \\
    \wpbbi(x/\xi,\rho/\xi) &= -\frac{\xi-\rho}{x-\xi}\Ln{\frac{x-\rho+i0}{\xi-\rho+i0}}\,, \nn \\
    \pbbii(x/\xi,\rho/\xi) &= \frac{\xi-\rho}{x+\xi}\bigg[ \Li{2}{-\frac{x-\xi}{\xi-\rho+i0}} \nonumber \\
   & \qquad \qquad \qquad - (x\to -\xi) \bigg]\,, \nn \\
    \wpbbiii(x/\xi,\rho/\xi) &= -\frac{\xi+\rho}{x+\xi}\Ln{\frac{x-\rho+i0}{-\xi-\rho+i0}}\,,
\end{align}
and
\begin{align}\label{Ldef}
   & L(x,\xi,\rho) = \int_0^1dw\ \frac{-4}{x-\xi-w(x+\xi)}
   \nonumber \\
    &\times \int_0^1du\ \Ln{1 + \frac{\bu(x-\xi-w(x+\xi))}{\xi-\rho+i0}}\nonumber \\
    &\quad\quad \times \left[\Ln{\frac{\bu(1-w)}{1-\bu w}} + \frac{1}{1-\bu w}\right]
    \,,
\end{align}
where $\bu=1-u$.

This is the only amplitude/CFF carrying the LT on top of kinematic higher-twist corrections. The other amplitudes can be found, together with their respective DVCS and TCS limits, in Ref.~\cite{Martinez-Fernandez:2025gub} and they can be schematically sorted as
\begin{align}
    \amp^{+0},\,\amp^{0+} \sim O\left( \frac{|\xi \bp_\perp |}{\scale} \right) \to \textrm{twist-3} \,, \nonumber\\
    \amp^{+-},\,\amp^{00} \sim O\left( \frac{|t|}{\scale^2}, \frac{\xi^2 M^2}{\scale^2} \right) \to \textrm{twist-4} \,.
\end{align}

\begin{figure*}
    \centering
    \includegraphics[width=0.9\textwidth]{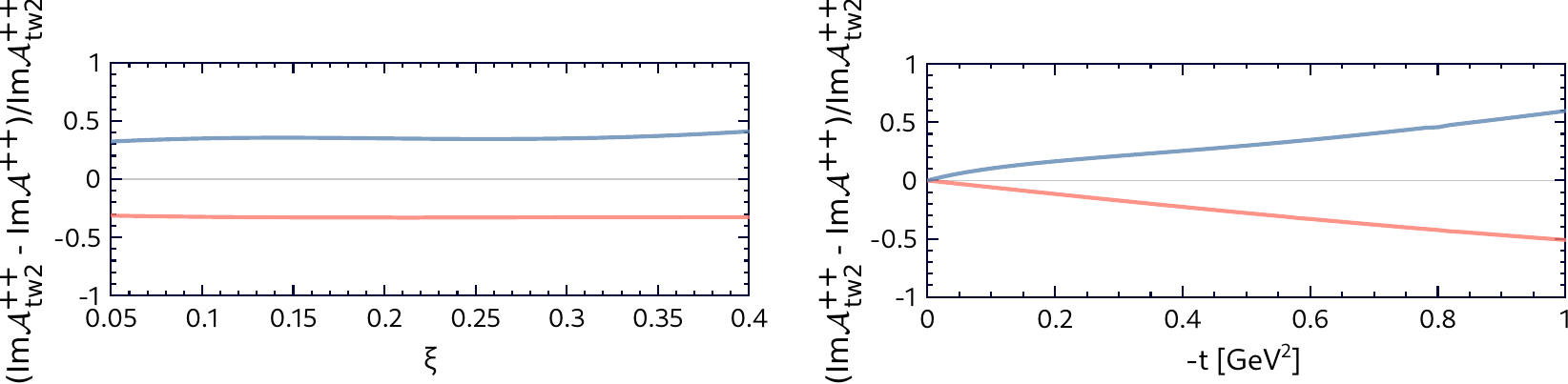}
    \caption{From~\cite{Martinez-Fernandez:2025gub}. The relative contribution of twist-4 corrections to twist-2 evaluation for the imaginary part of $\amp^{++}$ as a function of (left) $\xi$ at $t = -0.6\,\mathrm{GeV}^2$, and (right) $-t$ at $\xi=0.2$. Solid red (blue) curves correspond to DVCS (TCS). Both plots are for $\scale^2 = 1.9\,\mathrm{GeV}^2$.}
    \label{fig::contrib}
\end{figure*}

{These corrections may prove to be non-negligible at currently available and foreseeable kinematics, and might be relevant for the correct interpretation of data.}
{An example is provided by the hadron tomography problem, in which the $t$-dependence of CFFs is Fourier-transformed into an impact-parameter dependence.} Fig.~\ref{fig::contrib} depicts the relative contribution of twist-4 {effects} for the dominant imaginary part of $\amp^{++}$ amplitude in DVCS and TCS, {\it i.e.}~$(\mathrm{Im}\amp^{++}|_{\mathrm{tw-2}}-\mathrm{Im}\amp^{++})/\mathrm{Im}\amp^{++}|_{\mathrm{tw-2}}$ ratio, for a pion target based on the model~\cite{Chavez:2021llq}. As expected, the effect does not depend significantly on  $\xi$, but it scales with $t$. For an exemplary point of $\xi = 0.2$, $t = -0.6\,\mathrm{GeV}^2$ and $\scale^2=1.9\,\mathrm{GeV}^2$ the inclusion of higher twist corrections leads to the change of the modulus of the amplitude by about $33\%$ for both DVCS and TCS. Note, however, that the kinematic twist-4 effects lead to a decrease in the modulus of the TCS amplitude, while they result in an increase in the DVCS amplitude.

\begin{figure*}[h]
    \centering
    \includegraphics[width=0.9\textwidth]{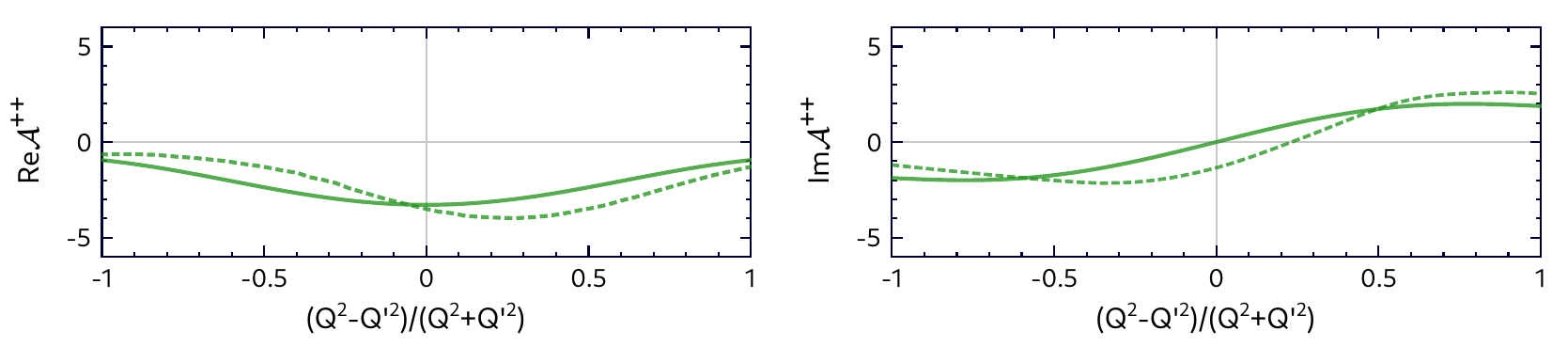}
    \caption{From~\cite{Martinez-Fernandez:2025gub}. Real and imaginary parts of $\amp^{++}$ amplitude as a function of $(Q^2-Q'^2)/(Q^2+Q'^2)$ ratio for $\scale^2 = 1.9\,\mathrm{GeV}^2$, $\xi=0.2$ and $t = -0.6\,\mathrm{GeV}^2$. Solid curves correspond to DDVCS evaluated at twist-2 accuracy. The dotted counterparts represent evaluations that include corrections up to twist-4 accuracy.}
    \label{fig::ratio}
\end{figure*}

Furthermore, asymmetries originating from amplitudes $\amp^{+-,\,0+,\,+0}$ open new ways to access the same GPDs. At LT and LO, the DVCS to TCS relation reads simply 
\be 
\amp^{++}_{\rm DVCS} = \left(\amp^{++}_{\rm TCS}\right)^*. 
\ee
Including higher twists, this relation does not hold anymore affecting the interpretation of data and the tests of GPD universality; see Fig.~\ref{fig::ratio} where the extreme right(left) corresponds to DVCS(TCS) while in-between DDVCS predictions are plotted for a pion target with GPD model~\cite{Chavez:2021llq}.

Finally, in addition to the kinematic twist corrections, there are genuine twist effects (genuine higher-twist distributions), which present a complex and separate task. This topic is beyond the scope of this section but, for the interested reader, a discussion can be found in Refs.~\cite{Anikin:2000em,Kivel:2000fg,Radyushkin:2000ap, Aslan:2018zzk}.

\subsection{DVMP at higher-order and higher-twist revisited }






\begin{figure*}[t]
\centering
\includegraphics[width=0.2\textwidth]{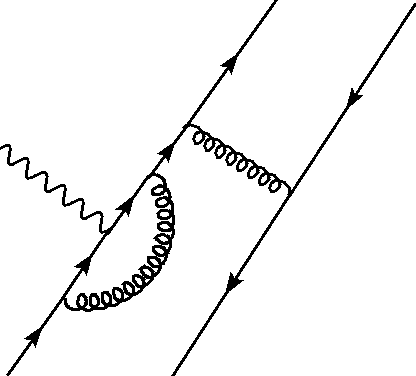}\quad
\includegraphics[width=0.2\textwidth]{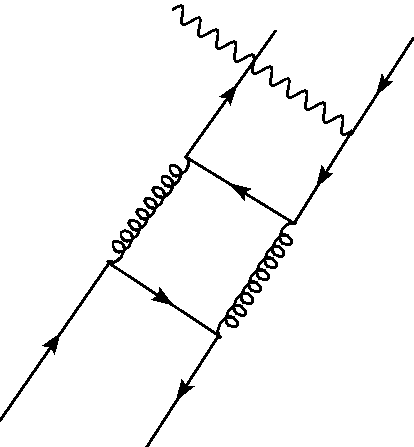}\quad
\includegraphics[width=0.2\textwidth]{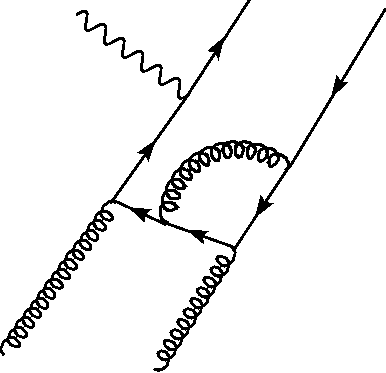}
\caption{
Representative partonic sub-diagrams that contribute
to DVV$_L^0$P at NLO.}
\label{fig:DVMPnloX}
\end{figure*}

The significance and complementarity of DVMP and DVCS
in accessing GPDs have long been recognized.
Considerable efforts are devoted to improving their theoretical descriptions
through implementing both higher-order pQCD corrections and power corrections.
While for DVCS the pQCD corrections up to NNLO
\cite{Braun:2020yib,Braun:2021grd,Ji:2023xzk},
and power corrections
\cite{Braun:2014sta,Guo:2021gru,Braun:2022qly}
have been worked out, 
DVMP corrections are known only to NLO contribution depicted in 
Fig.~\ref{fig:DVMPnloX},
see~\cite{Cuic:2023mki} and references therein.
The systematic inclusion of power corrections in DVMP remains challenging.
Recently, twist-3 contributions, which include both 2- and 3-body pion DAs,
have been determined~\cite{Kroll:2021ecb,Duplancic:2023xrt}.

The leading-twist (twist-2) DVMP amplitude arises from longitudinal photons, 
while transverse photons induce twist-3 effects. Although separating these 
contributions experimentally remains difficult, DV vector meson production at 
$Q^2 < 100~\text{GeV}^2$ 
and small 
$x_B$ 
is well described by the twist-2 
mechanisms 
\cite{Cuic:2023mki}. 
In contrast, DV pion production, measured only at 
$Q^2 < 10~\text{GeV}^2$ 
and moderate $x_B$, is dominated by 
transverse photons, {\it i.e.}, twist-3 effects~\cite{Goloskokov:2009ia, Goloskokov:2011rd, 
Duplancic:2023xrt}.

\begin{figure*}[t]
\centering
\includegraphics[width=0.8\textwidth]{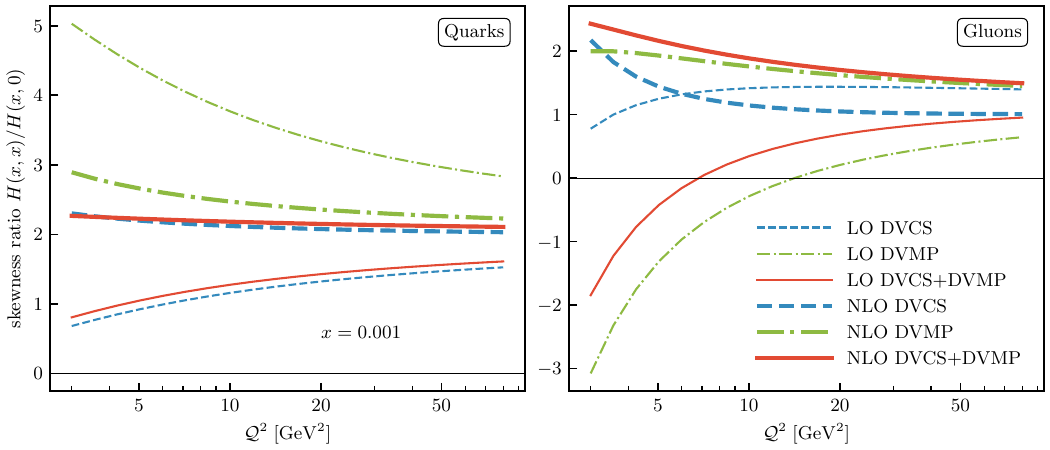}\\[2ex]
\caption{
Skewness ratio for the quark and gluon GPD $H$.
Depending on the processes used, the LO models (thin lines) show
large mutual differences of the resulting GPDs. The dependence
of NLO GPDs (thick lines) on the process is significantly smaller
\cite{Cuic:2023mki}.
}
\label{fig:DVMPnlo}
\end{figure*}

The DVMP amplitude 
$\gamma^* N \to M N'$ 
is expressed via transition form factors  given by convolutions of
hard-scattering kernels ($T^{a}$) with GPDs ($F^{a}$) and meson DAs ($\phi$).
While similar to Compton form factors in DVCS, transition form factors in DVMP 
depend additionally on the meson DA, 
making the description both richer in information
and more complex.
DVMP thus allows separation of GPDs by parity and flavor,
and provides LO access to gluon GPDs in DVV$_L^0$P (see Fig.~\ref{Fig_DVCS_DVMP}).

\newsavebox{\leftFig}
\newsavebox{\centerFig}
\newsavebox{\rightFig}
\sbox{\leftFig}{\includegraphics[width=0.35\textwidth]{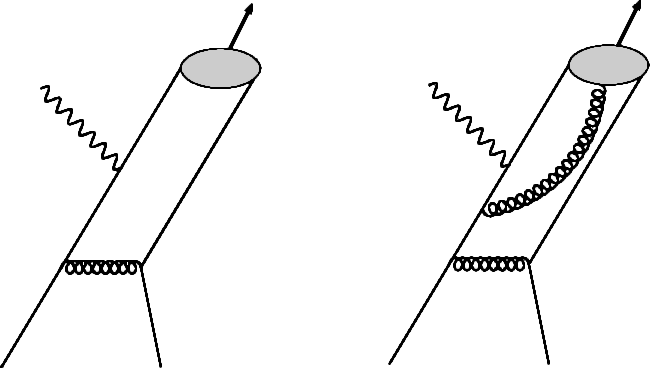}}
\sbox{\rightFig}{\includegraphics[width=0.4\textwidth]{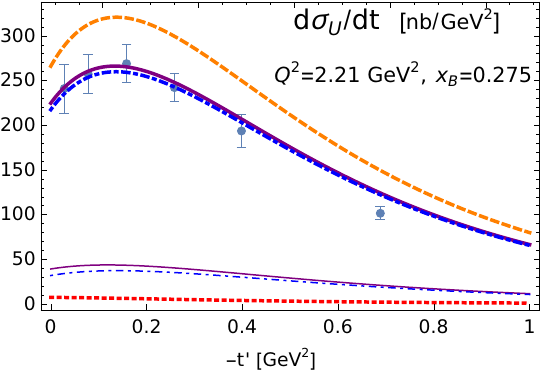}}
\newlength{\raiseFig}
\setlength{\raiseFig}{0.5\ht\rightFig - 0.35\ht\leftFig}
\begin{figure*}[t]
\centering
\begin{tabular}{ccc}
  \raisebox{\raiseFig}{\usebox{\leftFig}} 
&
  \raisebox{\raiseFig}{\usebox{\centerFig}} 
\hspace*{1.5cm}
&
  \usebox{\rightFig}
\end{tabular}
\caption{Left: Generic diagrams for 2- and 3-body subprocess amplitudes for DVMP.
Right: Differential DV$\pi^0$P cross section versus momentum transfer:
thick purple line unseparated $d\sigma_U/dt$ contribution,
blue dot-dashed $d\sigma_T/dt$ (twist-3), 
red dotted line $d\sigma_L/dt$ (twist-2) and 
the data taken from CLAS~\cite{CLAS:2014jpc}
(for details and other kinematical settings 
see Ref.~\cite{Duplancic:2023xrt}).
}
\label{fig:DVMPtw3}
\end{figure*}

The evolution of GPDs and DAs is conveniently implemented 
in the conformal moment representation \cite{Mueller:2005ed,Muller:2014wxa}, where conformal moments, 
analogous to Mellin moments in DIS, are defined via the Gegenbauer polynomials. 
In this basis, convolutions reduce to sums over conformal moments, 
which can be resummed using the Mellin-Barnes integral over the 
complex conformal spin $j$,
and {the} twist-2 transition form factor takes the form
\cite{Muller:2013jur}:
\begin{eqnarray}
&&
 {}^{a}\mathcal{T}(\xi, t, Q^2) 
= \frac{1}{2i}\int_{c-i \infty}^{c+ i \infty}\!
 dj \left[i\pm 
\left\{\begin{array}{c}
\tan \\
\cot
\end{array}\right\}
 \left(\frac{\pi j}{2}\right) \right]  
\nonumber \\
&&
\times  \xi^{-j-1}
\left[
T^a_{jk}(\mu)
\stackrel{k}{\otimes}
\phi_{k}(\mu)
\right]
F^a_j(\xi, t, \mu)
\, .
\label{eq:tff-conf}
\end{eqnarray}

A systematic framework for DVCS and DVMP based on the conformal 
partial wave expansion and the Mellin-Barnes techniques (CPaW formalism) 
was developed in~\cite{Mueller:2005ed,Mueller:2005nz},
and further detailed  in 
\cite{Kumericki:2007sa, Muller:2013jur}. 
This approach not only facilitates GPD evolution at NLO and beyond, 
but also enables flexible GPD modeling and efficient numerical 
implementations for global fits~\cite{gepard}.

In Ref.~\cite{Cuic:2023mki}, the CPaW formalism was applied, enabling for the first time the systematic extraction of GPDs from combined DVMP and DVCS data at twist-2 NLO accuracy.
The impact of NLO corrections on GPD shape
was investigated through global fits of GPDs to DIS, DVCS, and DV$\rho_L^0$P data at small 
$x_B$ and $Q^2 > 10$~GeV$^2$.
Rather than focusing solely on the magnitude of NLO corrections, 
the analysis highlighted their influence on the extracted GPDs, 
with NLO fits revealing GPD universality absent at LO, 
as evidenced by the skewness ratio shown in Fig.~\ref{fig:DVMPnlo}.
It was established that a simultaneous description of DIS, DVCS, and DVMP processes 
becomes feasible at NLO, providing insight into the proton structure through universal GPDs.
{A natural next step is to exploit DVMP as a source of information
on both nucleon and meson structure, through a combined study
of GPDs and meson distribution amplitudes.}

Experimental data for DV pion production 
(see~\cite{Duplancic:2023xrt} and references therein) 
indicate significant contributions from transversely polarized photons. 
Twist-3 calculations involving twist-2 {tensor} GPDs and 
twist-3 pion DAs have thus been proposed~\cite{Goloskokov:2011rd}, 
with satisfactory agreement achieved already in the Wandzura-Wilczek approximation (2-body pion Fock state).
Similarly, wide-angle pion production data require twist-3 effects, 
and calculations including the 3-body twist-3 contribution were needed 
\cite{Kroll:2021ecb}, 
yielding subprocess amplitudes valid also in the DV limit. Twist-3 pion DA parameters were 
constrained via the wide-angle analysis~\cite{Kroll:2018uvl}.

In Ref.~\cite{Duplancic:2023xrt}, DV$\pi^0$P was analyzed 
in the conventional momentum fraction representation
including both 2- and 3-body twist-3 contributions 
(see Fig.~\ref{fig:DVMPtw3}, left). 
The 3-body terms are essential for gauge invariance 
and influence 2-body ones via equations of motion. 
Endpoint singularities in the 2-body term were regularized 
using a modified perturbative approach with transverse momenta 
and the Sudakov factors, and alternatively with a dynamically generated gluon mass 
in a collinear approach. The study finds twist-3 contributions 
crucial for describing DV$\pi^0$ data (Fig.~\ref{fig:DVMPtw3}, right), 
except in COMPASS kinematics for relatively small $x_B$~\cite{COMPASS:2024hvm}
where twist-2 becomes relevant.
This suggests that the NLO corrections to twist-2 may be important at small $x_B$. 
Wide-angle pion production, also twist-3 dominated, 
provides constraints and complementary information on GPDs (large $t$), 
crucial for mapping the 3D partonic structure of the proton.

\subsection{Deep-exclusive processes beyond the leading order in QED}

Analysis of experimental observables for deep-exclusive reactions has to include processes beyond the leading order of QED, that involve radiation of extra photons and loop corrections. Emission of real photons by charged leptons, together with lepton vertex corrections necessary for cancellation of the infra-red (IR) divergences, lead to logarithmically-enhanced ($\times \log(Q^2/m_l^2)$) QED corrections to deep-exclusive cross sections at a level of tens of per cent for electrons and a few per cent for muons. Corrections to polarization asymmetries are normally smaller than for the unpolarized cross sections due to reduced sensitivity of soft-photon coupling to lepton's spin. These corrections are applied to  measurements in order to interpret the data in terms of hadronic form factors and/or GPDs; theoretical uncertainties of such corrections are constrained by detector acceptances and the (lack of) knowledge of hadronic response throughout the phase space of radiated photons.

While the approaches for QED corrections to DIS and elastic electron scattering were developed in 1960s 
\cite{mo1969radiative}, 
consistent treatment of exclusive electroproduction of mesons was put in place much later 
\cite{afanasev2002qed} 
for a new generation of data from JLab. Relevant theory developments for DVCS are presented in 
Refs.~\cite{mvdh2000, Akushevich:2017kct}.

In addition to logarithmically-enhanced QED corrections that are well understood from the theory prospective, yet another class of corrections is associated with photon's coupling to hadrons and it requires knowledge of hadronic structure for theory calculations. A well-known example of such QED corrections is two-photon-exchange that significantly limits accuracy of proton form factor measurements from unpolarized cross sections (see {\it e.g.}, ~\cite{afanasev2017two} 
for discussion). For electroproduction of pions, this problem was addressed in~\cite{afanasev2013two} 
using an approach with soft two-photon exchange that was recently re-applied for pion production in SIDIS 
\cite{lee2025soft}. 
Calculations within specific models were reported in 
Refs.~\cite{cao2020two, guo2022two}. 
Independent tests and experimental validation of the theoretical approaches for two-photon-exchange  calculations are possible due to the fact that corresponding corrections are charge-reversal-odd, {\it i.e.} they have opposite signs for negative vs positive leptons. Therefore, a prospective positron beamline at JLab~\cite{Afanasev:2019xmr} 
is expected to play a crucial role to address this problem.

We shall emphasize that appropriate methodology needs to be developed for analysis of QED corrections in deep-exclusive processes at new facilities such as the EIC.

\subsection{Securing the collinear factorization at high energy for heavy quarkonium photoproduction}

The Non-Relativistic QCD (NRQCD) factorization hypothesis~\cite{Bodwin:1994jh}, allows one to expand the amplitude of heavy quarkonium production in terms of relative velocity of heavy quarks in the bound-state $v^2$. The potential-model estimates give the value of $v^2\sim 0.3$ for charmonia and $v^2\sim 0.1$ for bottomonia, which makes the NRQCD velocity-expansion potentially reliable for the description of hadronization of heavy quarks into quarkonia. For the LO in $v^2$ and $\alpha_s${,} coefficient function of exclusive vector quarkonium photoproduction {reads}:
\be 
&&
 C_{g}^{\text{(0)}}\left(x,\xi;\mu_R \right) \nn \\ && =  \frac{x^2 \; {\cal C} }{(x + \xi -i\varepsilon)(x - \xi +i\varepsilon ) }  \langle{\cal Q}|\psi^\dagger(0)\sigma^i \chi(0)|0\rangle \varepsilon^{i}_{\gamma}. \nn \\ && \label{eq:quarkonium:Cg-LO}
\ee
{In here,} ${\cal C}=(4\sqrt{2}\pi \alpha_s(\mu_R) e e_{Q})/(m_Q^{3/2} N_c)$; $\varepsilon_\gamma^{i}$ is the polarization vector of the photon{.} {Up} to the LO in $v^2$, the long-distance matrix element (LDME) of NRQCD is 
\be 
&&
\langle{\cal Q}|\psi^\dagger(0)\sigma^i \chi(0)|0\rangle \nn \\ && = \sqrt{N_c} R_{\cal Q}(0) \varepsilon^{*i}_{{\cal Q}}/\sqrt{4\pi}+O(v^2).
\ee 
Here
$\varepsilon^i_{{\cal Q}}$ 
stands for the quarkonium polarization vector, 
$\psi(x)$ ($\chi(x)$) 
are the NRQCD heavy quark (antiquark) fields{,} and 
$R_{\cal Q}(0)$  
is the radial part of the potential-model wave-function of the state 
${\cal Q}$ 
at the origin. In the recent 
Ref.~\cite{Blask:2025jua} 
{the correction
at NLO in $v^2$ to the coefficient function}
has been computed,
which is proportional to the 
$O(v^2)$ 
LDME 
$\langle{\cal Q}|\psi^\dagger(0)\sigma^i {\bf D}^2 \chi(0)|0\rangle$, 
with 
${\bf D}$ 
being the covariant derivative. The genuine many-body relativistic effects, related to the admixture of 
$\ket{Q\bar{Q}g}$ 
and higher Fock-states in the physical quarkonium state 
$\ket{{\cal Q}}${,} will start to contribute at 
$O(v^4)$, and they can also be taken into account in the NRQCD expansion for the coefficient function.       

The NLO in $\alpha_s$ correction to the LO in $v^2$ coefficient function (\ref{eq:quarkonium:Cg-LO}) has been computed in Ref.~\cite{Ivanov:2004vd}. It was realized immediately
that at high photon-nucleon collision energies ($W_{\gamma p}$), where $\xi\ll 1$, {the contribution at NLO in $\alpha_s$} to the amplitude is dominated by the region $\xi\ll |x|\ll 1$, where the imaginary part of the NLO coefficient function behaves as:
\be
  &&  -\frac{i\pi c |x|}{2 \xi} \frac{{\alpha}_s(\mu_R)}{\pi} \ln \left( \frac{m_{Q}^2}{\mu_F^2} \right)
 f_i \langle{\cal Q}|\psi^\dagger(0)\sigma^j \chi(0)|0\rangle \varepsilon^{j}_{\gamma} \cr 
 \nn \\ && \equiv C_{i}^{\text{(1, asy.)}}(x,\xi;\mu_R,\mu_F), \label{eq:quarkonium:Cg-NLO:low-X} 
\ee
where $f_g=C_A$ and $f_q=2C_F$.

Being proportional to the $\ln(m_Q^2/\mu_F^2)$ and numerically dominant at $\xi\ll 1$ for the scales $\mu_F\sim$ few GeV, this contribution leads to {a} catastrophically large sensitivity of the $J/\psi$ photoproduction cross section to the choice of $\mu_F$ at high $W_{\gamma p}\gg M_{J/\psi}$, {see {\it e.g.} the discussion in
Ref.~\cite{Flett:2024htj}}.
This scale-sensitivity of the calculation is in fact so strong that it can even lead to unphysical conclusions, such as apparent dominance of the NLO quark GPD contribution in exclusive $J/\psi$ photoproduction at high energies, observed in Ref.~\cite{Eskola:2022vpi,Eskola:2022vaf} for some scale-choices. For the $\Upsilon$ {photoproduction} the problem is less severe, but the corrections discussed below are still important to achieve {precise} predictions.

Such instability of the NLO correction in collinear factorization (CF) of course calls into question the very applicability of the CF approach to this process. One may argue that at sufficiently high $W_{\gamma p}$ one should abandon the CF and use the dedicated factorization picture such as Colour-Glass Condensate (CGC, see {\it e.g.}, Ref.~\cite{Albacete:2014fwa} for a review). {Recently, the complete corrections at NLO in $\alpha_s$ and $v^2$} to the exclusive vector-quarkonium production amplitude in CGC formalism were computed in~\cite{Mantysaari:2022kdm}. 

  Within CF the systematic pathway towards the remedy of this situation has been first proposed in Ref.~\cite{Ivanov:2007je}, where it has been shown that at higher orders in $\alpha_s$ in the limit $\xi\ll |x| \ll 1$ the imaginary part of the coefficient function develops a series of corrections $\propto \alpha_s^n \ln^{n-1} (|x|/\xi)$. These corrections can be resummed with the help of High-Energy Factorization (HEF) formalism, proposed in Refs.~\cite{Collins:1991ty,Catani:1994sq} for the case of inclusive processes. The application of this formalism to the imaginary part of the exclusive production amplitude is possible, essentially {thanks to} the optical theorem. 

The resummation proposed in Ref.~\cite{Ivanov:2007je} gives the complete Leading Logarithmic Approximation (LLA), meaning that all the coefficients in front of terms $\propto \alpha_s^n \ln^{n-1} (|x|/\xi)$ are predicted correctly, including their $\mu_F$-dependence:
\begin{align}
    &C_i^{(n,\text{LLA})}(x,\xi;\mu_R,\mu_F) \cr
    &= \alpha_s^n(\mu_R) \frac{|x|}{\xi}  \ln^{n-1}\frac{|x|}{\xi} \sum\limits_{k=0}^n c_i^{(n,k)}(\mu_R) \ln^k\frac{m_Q^2}{\mu_F^2}. \label{eq:quarkonium:LLA-muF}
\end{align}
This becomes a problem if the usual phenomenological GPDs are used for the computation, with their scale-evolution kernels taken into account up to a fixed order $n_{\text{evol.}}=0$ for LO, 1 for NLO evolution etc. Then, the $\mu_F$-dependence of GPDs will compensate the $\mu_F$-logarithms in Eq.~(\ref{eq:quarkonium:LLA-muF}) with powers $k=n-n_{\text{evol.}},\ldots,n$, while the logarithms with $k<n-n_{\text{evol.}}$ are left un-cancelled. This 
results in the mismatch of the scale-dependence between the GPD and hard-scattering coefficient. Numerically it turns to  be as significant as the mismatch which the resummation was supposed to cure in a first place%
\footnote{See the Appendix B of Ref.~\cite{Lansberg:2021vie} for the discussion of this issue in a case of an inclusive process.}. To overcome this problem a Double-Logarthmic (DL) truncation of the full LLA HEF resummation was advocated in Ref.~\cite{Lansberg:2021vie}, which is equivalent to the LLA up to $O(\alpha_s^2)$ (NNLO) and is consistent with fixed-order evolution of GPDs up to NNLO. 

In Ref.~\cite{Flett:2024htj} this DL-truncated resummation has been applied to the process of exclusive vector quarkonium photoproduction. The resummed coefficient function was matched with the full NLO coefficient function of Ref.~\cite{Ivanov:2004vd} to provide a uniformly-accurate description across all values of $W_{\gamma p}$ starting from $W_{\gamma p}\gtrsim M_{{\cal Q}}$, where the resummation is not important{,} and going towards the region $W_{\gamma p}\gg M_{{\cal Q}}$ ($\xi\ll 1$), where the resummation starts to contribute. Significant reduction of the sensitivity to the scale $\mu_F$ is observed for the matched NLO$\oplus$DLA {results}. 

The resummed coefficient function  develops a ``hard Pomeron''-type behavior 
$$
C_i^{\text{(DLA)}}(x,\xi)\sim i\exp [(\alpha_s(\mu_R) N_c/\pi) \ln (|x|/\xi)]
$$ 
in the asymptotic regime $\xi\ll |x|\ll 1$.
{This, unfortunately, renders the matched coefficient function to be quite sensitive to the $\mu_R$-choice.
Combined with the residual $\mu_F$-uncertainty, this leads to {a} very significant scale-uncertainty band for the $d\sigma/dt(t=t_{\min})$.
}
Note however, that for the complete LLA the low-$x$ asymptotics would be 
$$
C_i^{\text{(LLA)}}(x,\xi)\sim i\exp [4\ln 2\cdot (\alpha_s(\mu_R) N_c/\pi) \ln (|x|/\xi)],
$$ 
which turns to be incompatible with the observed slope of the energy-dependence of the cross section. The $\mu_R$-sensitivity can be improved only via the inclusion of corrections beyond DLA. The formalism for this purpose is currently being developed for the inclusive quarkonium production case, see Refs.~\cite{Nefedov:2024swu, Nefedov:2024cuy}.

Also for the case of $J/\psi$ the matched DLA$\oplus$NLO predictions lie above the data, with only the lower limit of the uncertainty band being consistent with them. The description of the data is much better for the case of $\Upsilon(1S)$. Such a picture is consistent with the $O(v^2)$ corrections being larger for the case of $J/\psi$ rather than for $\Upsilon$, due to lower value of $v^2$ for the latter. In Ref.~\cite{Mantysaari:2022kdm}  the negative $O(v^2)$-correction to the $J/\psi$-photoproduction amplitude was obtained in the CGC calculation of this observable. Due to close links between CGC and HEF one naturally expects the $O(v^2)$-correction to decrease the cross section in case of DLA$\oplus$NLO CF computation. 
{However, this expectation contradicts the result of
Ref.~\cite{Blask:2025jua}, where the correction at NLO in $v^2$ to the
imaginary part of the LO CF was found to be positive.}

\subsection{{Machine learning methods for inverse problems in hadron structure studies}
}
\label{sec:theory_and_pheno_computing}

QCD global analyses of quantum correlation functions, such as PDFs and their higher-dimensional generalizations, the GPDs~\cite{Kriesten:2019jep, Kriesten:2021sqc}, have increasingly leveraged modern artificial intelligence and machine learning (AI/ML) fitting methodologies. Due to the complex nature of the inverse mapping from experimental data to the physics parameters of interest, many of the $x$-dependent shape features of hadron structure observables are difficult to trace back to input theoretical assumptions in the global fit. 
{This makes interpretability essential for comparing global fits obtained under different theoretical assumptions~\cite{Almaeen:2022imx}.}
A rigorous understanding of how input systematic theory choices shape the $x$-dependence of fitted PDFs/GPDs is therefore {one of the} key goals of femtographic studies.

\begin{figure}
    \centering
    \includegraphics[width=\linewidth]{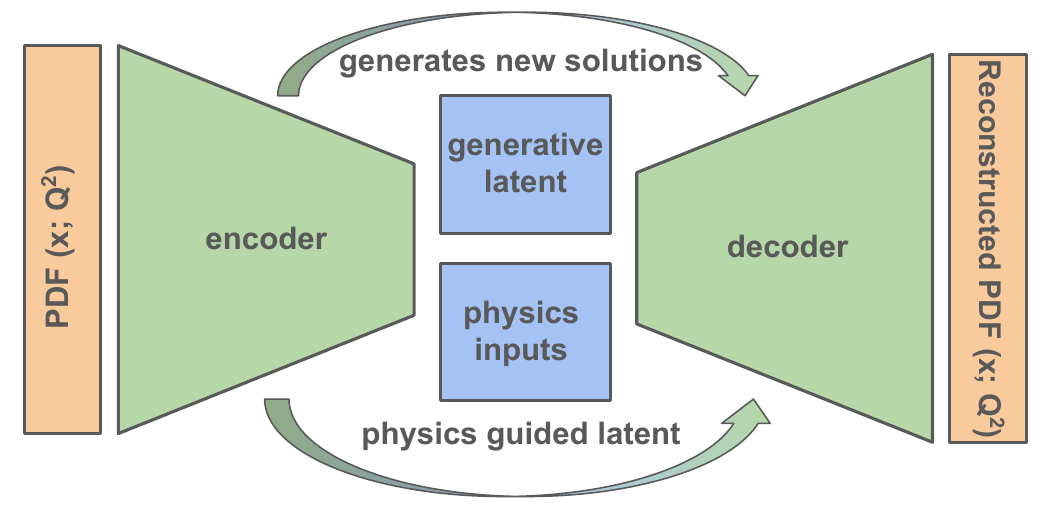}
    \caption{The VAIM takes in PDFs and physics observables such as lattice-calculable moments and reconstructs families of PDF solutions that solve the inverse problem.}
    \label{fig:vaim_infographic}
\end{figure}

Developments in lattice QCD provide increasingly complementary information to phenomenological analyses of experimental data in the form of pseudo- and quasi-PDF/GPDs for both quarks and gluons in hadrons~\cite{Ji:2013dva, Balitsky:2021bds}. Constructing lattice QCD-aware generative approaches bridges lattice data and phenomenology~\cite{Karpie:2019eiq, DelDebbio:2020rgv, Dutrieux:2024rem}. Some examples include:
\bi
\item
In Ref.~\cite{Kriesten:2023uoi}, a variational autoencoder inverse mapper (VAIM)~\cite{Almaeen:2021}, depicted in Figure~\ref{fig:vaim_infographic}, was shown to reconstruct PDFs from $x$-integrated information in the form of lattice-calculable Mellin moments;
\item
In a complementary application, VAIM maps the reduced pseudo-Ioffe-time distributions (RpITDs)~\cite{Radyushkin:2017cyf,Orginos:2017kos} to gluon PDFs~\cite{Kriesten:2025gti};
\item
GPD-sensitive observables, such as the Compton form factors (CFFs) entering the unpolarized DVCS cross section, have also been extracted using the VAIM architecture in an attempt to place constraints on partonic angular momentum~\cite{Almaeen:2024guo, Kriesten:2020wcx, Kriesten:2020apm, Grigsby:2020auv}.
\ei

The \texttt{XAI4PDF} framework is a machine learning model trained to classify PDFs according to the fitting assumptions that indirectly influence their $x$-dependent behavior. It utilizes attribution scores based on the guided backpropagation method~\cite{Lerma_2023} to generate human-readable saliency maps, which identify the input $x$-regions most relevant for the classification prediction. Once the model is trained, auto-differentiation is used to compute gradients of the classification score with respect to the input feature space --- the $x$-dependence. Guided backpropagation modifies differentiation through the rectified linear unit (ReLU) nonlinear activation function by applying a mask such that only positive gradients are propagated during the backward pass; for a detailed breakdown of the methodology see Ref.~\cite{Kriesten:2024are} and references therein. This procedure produces detailed maps that connect specific varying features of the PDF shape to the theoretical assumptions used in the underlying fit.

The \texttt{XAI4PDF} framework has the potential to complement generative-AI approaches like the VAIM by generating human-readable maps allowing for systematic benchmarking and uncertainty quantification~\cite{Kriesten:2024are}. While VAIM can be used to reconstruct families of PDF/GPD solutions to the inverse problem, \texttt{XAI4PDF} could be used to diagnose regions of the reconstructed quantum correlation functions which are most responsible for the differences among those families through calculated attribution scores.

The \texttt{XAI4PDF} framework, while currently applied to PDFs, can be naturally extended to GPDs where the input-space dimensionality increases, $(x;Q^{2}) \rightarrow (x,\xi,t;Q^{2})$, and the variations over this larger kinematic space are constrained by additional theory-based requirements like polynomiality, positivity constraints, {\it etc}. In this setting, the shape profile extracted from global fits is more model-dependent and less directly constrained than PDFs, making explainability tools particularly valuable for identifying which datasets or parameterization assumptions are most strongly associated with variations in specific regions of the GPD kinematics. This explainability diagnostic tool becomes especially important for understanding differences in derived $x$-integrated quantities, such as partonic angular momentum, extracted from GPD global analyses of exclusive processes.

\subsection{Convolution of the Chiral Effective Theory with the Light-Front Quark Model}
\label{sec:dynamical_models}

Phenomenological parametrizations of GPDs are essential for analyzing experimental data, offering flexible forms consistent with polynomiality, positivity, and known PDFs and form factors, though they often lack a direct link to nonperturbative QCD dynamics. In contrast, effective field–theory–based models--such as chiral perturbation theory, light-front holographic QCD, { light-front quark model (LFQM)} and the chiral quark–soliton model--provide a more dynamical, symmetry-guided picture of hadron structure. However, their predictive power is constrained by model-dependent assumptions, limited kinematic applicability, and difficulties in systematic improvement or direct comparison with experimental data. A promising strategy is to combine the empirical flexibility of phenomenological fits with the theoretical rigor of effective field theory (EFT)-inspired frameworks, supplemented by lattice-QCD inputs, to achieve a unified and physically transparent description of hadron structure.

In between the phenomenological models and lattice QCD simulations, various studies have been performed over the past 25 years~\cite{Ji:1997gm,Petrov:1998kf,Penttinen:1999th,Goeke:2001tz,Choi:2001fc,Choi:2002ic,Tiburzi:2001je,Theussl:2002xp,Boffi:2002yy,Schweitzer:2002nm,Scopetta:2003et,Ossmann:2004bp,Wakamatsu:2005vk,Mineo:2005vs,Pasquini:2006ib,Goeke:2008jz}. In particular, constraints from chiral EFT have been used to make predictions for various physical observables in the nucleon~\cite{Thomas:1983fh, Thomas:2000ny, Chen:2001nb, Myhrer:2007cf, Salamu:2014pka, Luan:2023lmt}. 
As a typical example, the sea quark flavor asymmetries in the nucleon have been investigated based on the observation that the long-range structure of the nucleon has contributions from the pseudoscalar meson cloud, associated with the model-independent leading nonanalytic behavior of observables expanded in a series in $M_\pi$. { In particular, the resolution~\cite{Ji:2013bca} of the lingering factor difference in the leading nonanalytic term coefficient of $M_\pi^2 \ \text{log} M_\pi^2$ provided convincing evidence of the link between the chiral EFT and QCD.
The result does not depend on the details of the ultraviolet regulator, since the nonanalytic structure is determined entirely by the infrared behavior of the underlying  theory.} Operationally, the sea quark distributions can be obtained through the convolution of a probability of the nucleon to split into a virtual meson and recoil baryon (``splitting function''), with the valence quark distribution of the meson.

In this respect, meson structure studies of the {LFQM} provide useful tools to study the nucleon structures via the convolution with the splitting functions computed by the { chiral EFT}.  The glimpse of the link between LFQM and QCD has been exhibited~\cite{Ma:2021yqx} in {a} toy version of QCD, namely QCD$_{(1+1)}$ at the large $N_c$ limit known as the 't~Hooft model~\cite{THOOFT1974461}. The solutions of the LFQM bound-state equation derived from the large $N_c$ QCD$_{(1+1)}$ was consistent with the Regge trajectories and the Gell-Mann--Oaks--Runner relation in the chiral limit~\cite{Ma:2021yqx}. It lends credence to the LFQM that one may extend its link to the real QCD with $N_c = 3$ in 3+1D. In recent development of LFQM,  particular emphasis has been placed on achieving both the consistency with the manifest covariance and the realistic phenomenological description of the meson structure and spectroscopy in the progress of LFQM~\cite{C13,C15,Choi:2017uos,C21,Arifi22}. 
Bakamjian-Thomas construction~\cite{BT53,KP91} appears a key ingredient of the LFQM to ensure the uniqueness of dynamical model predictions independent of the reference frame, the current components and the polarization vectors used in the computation of physical observables.  

{ While the splitting functions of mesons and baryons can be calculated from the chiral EFT, {their} moments are expanded as a series in the pseudoscalar meson mass ${\cal O}(M_\phi/\Lambda_{\chi})$ or small external momentum ${\cal O}(q/\Lambda_{\chi})$,} where $\Lambda_{\chi}\sim$ 1 GeV is the typical scale associated with the chiral EFT.
Traditional EFT calculations based on dimensional regularization have found it challenging to describe lattice results in the large-$Q^2$ region, or with pion mass 400~MeV~\cite{Young:2002ib, Leinweber:2003dg}, since the short-distance effect {arising} from the loop integral leads to a poor convergence of the chiral expansion~\cite{Donoghue:1998bs, Leinweber:2003dg}.
To improve the convergence, finite range regularization has been proposed and applied to the extrapolation of various quantities calculated by lattice QCD, including the vector meson mass, magnetic moments, magnetic and strange form factors, charge radii, and moments of PDFs and GPDs, from large pion masses to the physical mass~\cite{Young:2002ib, Leinweber:2003dg, Wang:2007iw, Allton:2005fb, Wang:1900ta, Wang:2012hj, Hall:2013dva, Shanahan:2012wh, Shanahan:2014uka}.
Here the three dimensional regulator is included in the loop integral and can be regarded as partially resumming the contribution from the higher order interaction.
In a similar vein, a {\it nonlocal} chiral effective Lagrangian has also been proposed~\cite{Wang:2010rib, He:2017viu, He:2018eyz}, in which the covariant regulator is automatically generated from the Lagrangian. 
One can obtain reasonable descriptions of the nucleon electromagnetic and strange form factors at large $Q^2$ using this method, which has also been applied to calculate the strange-anti-strange PDF asymmetry $s-\bar{s}$~\cite{Salamu:2018cny, Salamu:2019dok}, the sea quark TMD Sivers function~\cite{He:2019fzn}, and zero-skewness GPDs~\cite{He:2022leb} in the proton. 
More details about the nonlocal chiral effective theory can be found in { recent review articles~\cite{Wang:2022bxo, Wang:2025kzb}.
A framework for matching TMD PDFs onto chiral effective theory operators has also been discussed in terms of TMD hadronic distribution functions calculated in chiral perturbation theory~\cite{Copeland:2024wwm}. }

\begin{figure}[h]
\begin{center}
\includegraphics[width=0.40\textwidth]{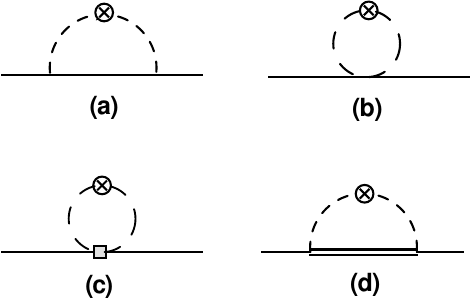}
\caption{One-loop diagrams for the proton to pseudoscalar meson (dashed lines) and octet baryon (solid lines) or decuplet baryon (double solid lines) splitting functions. The crossed circles ($\otimes$) represent the interaction with the external vector field from the minimal substitution, and the gray square (${\textcolor{gray}{\blacksquare}}$) 
denotes the last interaction term in Eq.~(\ref{eq:j1}). This figure is taken from Ref.~\cite{GHJMSW-2024}. } 
\label{fig:diagrams}
\end{center}
\end{figure}

In the one-loop approximation, the chiral Lagrangian 
can be expanded to reveal the interaction terms, given by
\begin{align}
\!\!\!\!\!\!
{\mathcal{L}}_{\rm int}
&= \frac{g_A}{2f_\pi}
  \left( \bar p\, \gamma^\mu \gamma_5 p\, \partial_\mu \pi^0
       + \sqrt2\, \bar p\, \gamma^\mu \gamma_5 n\, \partial_\mu \pi^+
  \right)  
\cr
&+ \frac{\cal C}{\sqrt{12} f_\pi}
  \left(
  - 2\, \bar p\, \Theta^{\nu\mu} \Delta_\mu^+\, \partial_\nu \pi^0
  \right. \cr
&+ 
  \left.- \sqrt2\, \bar p\, \Theta^{\nu\mu} \Delta_\mu^0\, \partial_\nu \pi^+
  \right. \cr
\qquad\qquad &+\left. \sqrt6\, \bar p\, \Theta^{\nu\mu} \Delta_\mu^{++}\, \partial_\nu \pi^-
  \right)
\cr
&+ \frac{i}{4 f_\pi^2}\, \bar p\, \gamma^\mu p\,
    (\pi^+ \partial_\mu \pi^-  -  \pi^- \partial_\mu \pi^+) \cr
&+ \frac{2i (b_{10}+b_{11})}{f_\pi^2}\, \bar p\, \sigma^{\mu\nu} p\,  
    \partial_\mu\pi^+\, \partial_\nu \pi^-  + {\rm H.c.},
\label{eq:j1}
\end{align}
where H.c. refers to the Hermitian conjugate, and $f_\pi=92.4$~MeV is the pion decay constant.
The axial vector coupling constant for the octet baryon is set to the standard value $g_A = 1.26$, and the octet-to-decuplet transition coupling ${\cal C}$ is chosen to be ${\cal C}=-\frac65 g_A$ from the SU$(6)$ spin-flavor symmetry.
Typically, in calculations of meson loop contributions to sea quark and antiquark asymmetries, assuming that the undressed proton has a flavor symmetric sea~\cite{Salamu:2014pka}, the meson coupling diagrams in Fig.~\ref{fig:diagrams} are the dominant source of differences between sea quark and antiquark PDFs and GPDs.

\begin{figure*}[tp]
\begin{center}
\includegraphics[scale=1]{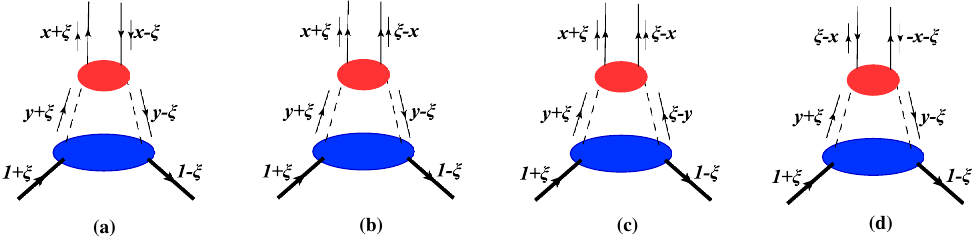}
\caption{Representation of the convolution formula in Eqs.~(\ref{eq:conv_a}), (\ref{eq:conv_b}), (\ref{eq:conv_c}), and (\ref{eq:conv_d}), respectively, with the \{dashed, thick solid, thin solid\} lines representing the \{pseudoscalar meson, proton, quark\}. The processes in diagrams (a) and (d) represent the DGLAP region for the quark and antiquark, respectively, while the processes in (b) and (c) contribute to the ERBL region. This figure is taken from Ref.~\cite{GHJMSW-2024}.}
\label{fig:convol_illu}
\end{center}
\end{figure*}

By combining various processes~\cite{GHJMSW-2024}, the contribution to the skewness nonzero GPDs from the diagram, {\it e.g.}, in Fig.~\ref{fig:diagrams}(a), can be expressed in the convolution form as

\begin{subequations}
\label{eq:convolution_sum}
\begin{align}
\label{eq:conv_a}
&H_q^{\rm (rbw)}(x,\xi,t)|\bigg|_{\xi<x} =  \nonumber \\[0.2cm]
&\hspace{1cm} \int_x^1 \frac{\dd{y}}{y} f_{\phi B}^{({\rm rbw})}(y,\xi,t)\, H_{q/\phi}\Big(\frac{x}{y},\frac{\xi}{y},t\Big), 
\\[0.2cm]
\label{eq:conv_b}
&H_q^{\rm (rbw)}(x,\xi,t)\bigg|_{x < \xi } =  \nonumber \\[0.2cm]
&\hspace{1cm}\int_\xi^1 \frac{\dd{y}}{y} f_{\phi B}^{({\rm rbw})}(y,\xi,t)\, H_{q/\phi}\Big(\frac{x}{y},\frac{\xi}{y},t\Big),\\[0.2cm]
\label{eq:conv_c}
&H_q^{\rm (rbw)}(x,\xi,t)\bigg|_{|x| < \xi} = 
\int_{-\xi}^{\xi} \frac{\dd{y}}{2y}\, \bigg[f_{\phi B}^{({\rm rbw})}(y,\xi,t)\nonumber \\[0.2cm]
&\hspace{1cm} \frac{1}{\pi} \int_{s_0}^\infty\!\! \dd{s}
     \frac{{\rm Im}\Phi_{q/\phi}{ \big( \frac12(1\!+\!\frac{x}{\xi}), 
                                        \frac12(1\!+\!\frac{y}{\xi}),
                                       s
                                  \big)}}{s-t+i\epsilon}\bigg],\, \nonumber \\[-0.1cm] \\
\label{eq:conv_d}
&H_q^{\rm (rbw)}(x,\xi,t)\bigg|_{\xi < -x } = \nonumber \\[0.2cm]
&\hspace{1cm}\int_{-x}^1 \frac{\dd{y}}{y} f_{\phi B}^{({\rm rbw})}(y,\xi,t)\, H_{q/\phi}\Big(\frac{x}{y},\frac{\xi}{y},t\Big), 
\end{align}
\end{subequations}
where $f_{\phi B}$ represents the nucleon-meson splitting function which can be computed by the chiral EFT as depicted in Fig.~\ref{fig:diagrams}. Here  
$H_{q/\phi}$ and $\Phi_{q/\phi}$ represent the meson's valence GPD and generalized distribution amplitude (GDA), respectively, which can be computed by the LFQM.
The total contribution of the rainbow diagrams in Figs.~\ref{fig:diagrams}(a) -- \ref{fig:diagrams}(d) to the GPDs is then obtained by summing the contribution of four subprocesses in Fig.~\ref{fig:convol_illu}.

\begin{figure}[tbp] 
\begin{center}
\includegraphics[width=0.48\textwidth]{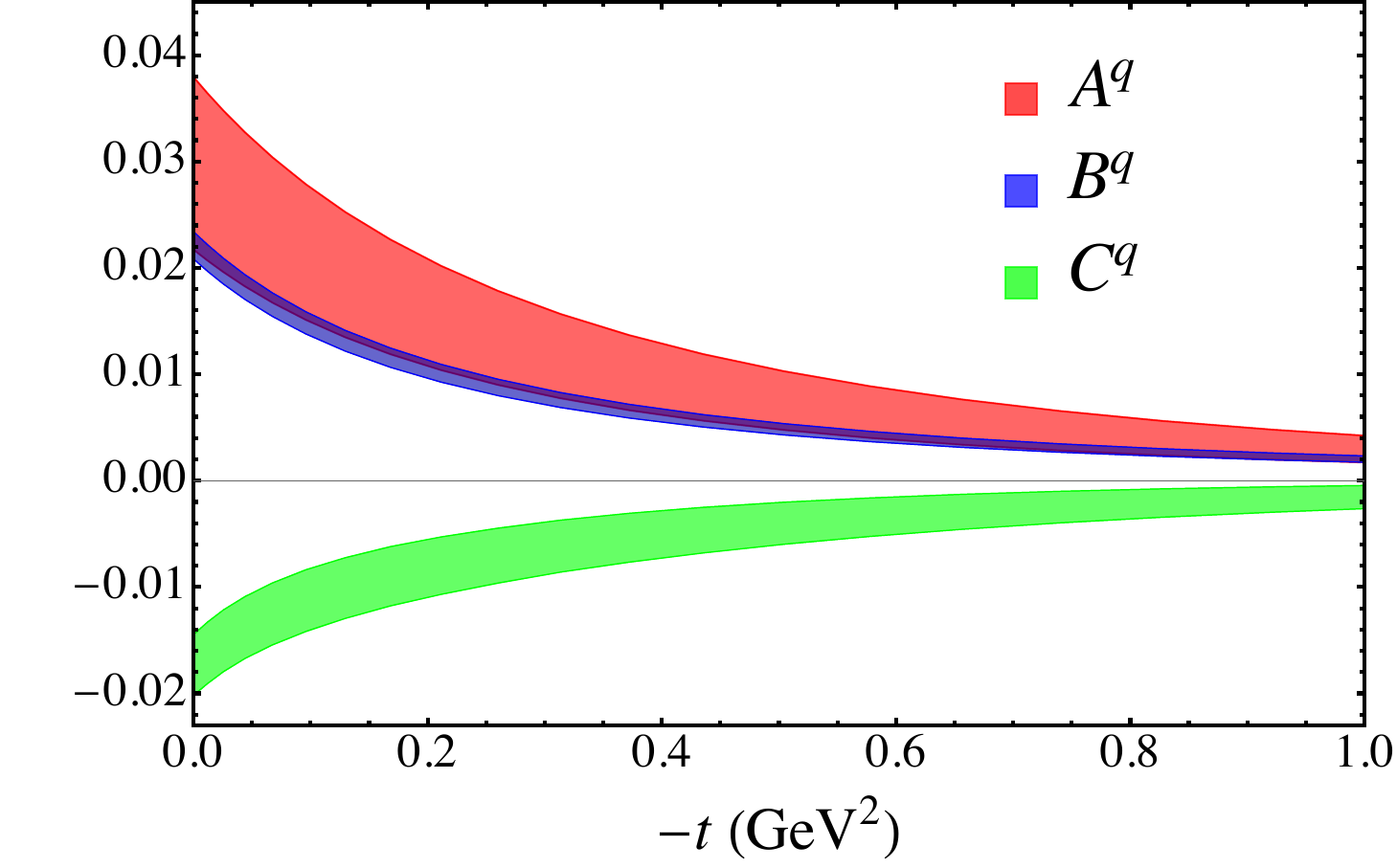}   
\caption{{GFFs} $A^q$ (red band), $B^q$ (blue band), and $C^q$ (green band) for quark flavor $q=u$ or $d$ as a function of $t$ arising from the pseudoscalar meson loop diagrams in Fig.~\ref{fig:diagrams}. The results for the $u$ and $d$ flavors for these diagrams are identical due to isospin symmetry. This figure is taken from Ref.~\cite{GHJMSW-2024}.}
\label{fig:GFFs_ud}
\end{center}
\end{figure}

From the calculated $H_q^{\rm (rbw)}$ and $E_q^{\rm (rbw)}$ GPDs~\cite{GHJMSW-2024}, one can compute the Dirac, Pauli, and gravitational form factors from the lowest two moments of the GPDs. Integrating the GPDs $H_q^{\rm (rbw)}$ and $E_q^{\rm (rbw)}$ over the quark light-cone momentum fraction $x$ {yields} the Dirac and Pauli form factors for a quark flavor $q$ (\ref{FF_1stmoment}). 
Note that  vanishing of the $\xi$ dependence in the integrals of the GPDs over $x$ provides a useful consistency check.

Furthermore, the gravitational form factors $A^q$, $B^q$ and $C^q$ can be obtained from the first moments of the $H_q$ and $E_q$ GPDs as
\begin{subequations}
\begin{eqnarray}
\int_{-1}^1\,\dd{x} x\, H_q^{\rm (rbw)}(x,\xi,t) &=& A^q(t) + 4\xi^2\, C^q(t), \\
\int_{-1}^1\,\dd{x} x\, E_q^{\rm (rbw)}(x,\xi,t) &=& B^q(t) - 4\xi^2\, C^q(t),
\label{eq:Gffs}
\end{eqnarray}
\end{subequations}
where the explicit dependence on the skewness $\xi$ now appears on the right hand side.
The numerical results for the GFFs $A^q(t)$, $B^q(t) \equiv 2J^q(t)-A^q(t)$, and $C^q(t) \equiv D^q(t)/4$ are shown in Fig.~\ref{fig:GFFs_ud} as a function of $t$~\cite{GHJMSW-2024}. See Eq.~(\ref{EMT_FF_decomposition}) and Eq.~(\ref{EMT_FF_decomposition_2}) for the definition of the form factors.
Because of isospin symmetry, the results for the $q=u$ flavor are identical to those for the $q=d$ flavor.
Despite the absolute value of the GPD $E_q$ being larger than that of the GPD $H_q$, the first moment of $E_q$ largely cancels between the positive-$x$ and negative-$x$ regions, leading to similar values for the $A^q$ and $B^q$ form factors.
In contrast, the $C^q$ form factor is negative and has a smaller magnitude compared with the other two form factors.

The extraction of GPDs directly from experimental data is of course the cornerstone of the quest to determine the 3-dimensional structure of the nucleon. Although pioneering studies have already been made by several groups worldwide on this effort~\cite{Diehl:2004cx, Kumericki:2009uq, Goldstein:2010gu, Guo:2022upw, Guo:2023ahv}, direct and model-independent extractions remain a formidable challenge. In this regard, the formulation presented here---convoluting the nucleon-meson splitting functions provided by the chiral EFT with the meson structure information, such as GPD and GDA, provided by the LFQM---can advance the global analysis framework toward a deeper understanding of hadron {structure}, incorporating lattice QCD calculations and experimental observations.

\subsection{Gravitational form factors of the nucleon in the chiral quark-soliton model}

The GFFs of the nucleon, see Sec.~\ref{Sec_GFF_intro},
convey essential information on its mass and spin. Since these fundamental properties are probed by external tensor fields (gravitons), 
the GFFs of the nucleon have been considered 
academically interesting objects~
\cite{Kobzarev:1962wt,Pagels:1966zza} 
for decades. However, GPDs opened a way for extracting information on the
GFFs~\cite{Muller:1994ses, Ji:1996nm, Radyushkin:1996nd}.
{They were extensively investigated within various theoretical
frameworks:
in the bag model~\cite{Ji:1997gm,Neubelt:2019sou},
in chiral perturbation theory~\cite{Belitsky:2002jp,Alharazin:2020yjv, Gegelia:2021wnj},
in lattice QCD~\cite{Hagler:2003jd,Gockeler:2003jfa,Shanahan:2018pib,Hackett:2023rif}, 
in chiral soliton approaches~\cite{Cebulla:2007ei, Kim:2012ts, Jung:2013bya, Jung:2014jja}, 
in chiral quark-soliton model~\cite{Goeke:2007fp, Kim:2020nug,
  Won:2022cyy, Won:2023ial, Won:2023cyd, Won:2023zmf}, 
in the light-front approaches and holographic QCD~\cite{Abidin:2009hr,Chakrabarti:2015lba,Kumar:2017dbf,Fujita:2022jus},
in the instanton QCD vacuum~\cite{Polyakov:2018exb}, 
in QCD sum-rules~\cite{Anikin:2019kwi,Azizi:2019ytx,Dehghan:2025ncw},
and in a classical model~\cite{Varma:2020crx}, 
among many others.}
  In this
Section, we briefly review the results from the chiral quark-soliton model 
($\chi$QSM).

The QCD EMT current, Eq.~(\ref{Def_EMT_current}), contains both the twist-2 and twist-4
contributions. Therefore, we will decompose the EMT current into the
twist-2 (spin-2) and twist-4 (spin-0) parts, which are associated with
the leading twist vector GPDs and twist-4 GPDs, respectively. The
twist-projected EMT currents are written as 
\begin{align}
  T_{\mu \nu}^{a} =  \bar{T}_{\mu \nu}^{a} + \hat{T}_{\mu \nu}^{a},
  \label{eq:tw2_tw4_split}
\end{align}
where the twist-2 ($\bar{T}_{\mu \nu}^{a}$) and twist-4 ($\hat{T}_{\mu
  \nu}^{a}$) parts are defined by 
\begin{align}
    \bar{T}_{\mu \nu}^{a}  
 = T_{\mu \nu}^{a}  - \frac{1}{4} g_{\mu \nu}  T_{\alpha}^{a, \alpha},
  \quad     \hat{T}_{\mu \nu}^{a}   
 = \frac{1}{4} g_{\mu \nu}  T_{\alpha}^{a, \alpha},
\end{align}
with $a=q,g$. 

{As given in (\ref{EMT_FF_decomposition}), 
the nucleon matrix elements of the EMT currents
are parametrized in terms of four independent Lorentz-invariant functions,
}
{\it i.e.}, 
$A^{a}$, $J^{a}$, $D^{a}$, and $\bar{C}^{a}$,
which convey information on the mass,
spin, and mechanical structure of the nucleon.
The SU(3) quark flavor structure of a generic GFF $\mathbb{F}^\chi$ can be 
denoted by the singlet ($\chi=0$), isovector ($\chi=3$) and octet ($\chi=8$), which are in relation with the quark components as 
follows
\be
&&
  \mathbb F^{0}=\mathbb F^{u} + \mathbb F^{d} + \mathbb F^{s}
  ,  \quad 
  \mathbb F^{3}=\mathbb F^{u} - \mathbb F^{d},   \nn \\ &&
  \mathbb F^{8}=\frac{1}{\sqrt{3}}\left(\mathbb F^{u}+\mathbb F^{d}
  -2\mathbb F^{s}\right). 
\label{eq:Decomp}
\ee
The singlet GFFs of a nucleon are given 
by the sum of all quark and gluon contributions 
\begin{align}
\sum_{a=q,g} \mathbb F^{a}(t) = \mathbb F(t).
\end{align}
The EMT current conservation constrains 
{that $\bar{C}(t) \equiv \sum_{a=q,g}\bar{C}^a(t)=0$}. 
However, each flavor component of $\bar{C}(t)$ 
does not necessarily vanish. Therefore, one should carefully consider the flavor decomposition of the GFFs.  

In fact, extracting the 
form factors $\bar{C}^a$ is challenging in most dynamical models
and lattice QCD due to the higher-twist (twist-4) corrections. 
To isolate this twist-4 term in a most systematic way,
we can decompose the symmetrized EMT current in terms of the twist-2 and
twist-4 components, as presented in Eq.~(\ref{eq:tw2_tw4_split}). 
With these twist-projected EMT currents employed,
we define their nucleon matrix elements as follows: 
\begin{align}
&   \mel{N(p',J'_{3})}
    {   \bar{T}_{\mu\nu}^{a}  ( 0 )  }
    {N(p,J_{3})} 
  \cr
  &= \bar{u}(p',J'_{3})
    \Bigg[
    A^{a}  ( t ) \frac{  P_{\mu} P_{\nu} }{  M_N } 
  + J^{a}  ( t ) \frac{  i  P_{ \{ \mu} \sigma_{\nu \} \alpha}
  \Delta^{\alpha} } { 2 M_N }  \cr
  &\qquad\qquad\qquad
  + D^{a}  ( t )  \frac{ \Delta_{\mu} \Delta_{\nu} -  \Delta^2 g_{\mu \nu} }{4M_N}      
 \cr
&\hspace{1cm} - g_{\mu\nu}  
    \left\{ \frac{t}{8 M_{N}} J^{a}(t) 
  - \frac{3t}{16 M_{N}} D^{a} (t) \right.\cr
 &\left.\qquad\qquad\qquad + \frac{M_{N}}{4}  
    \left(1- \frac{t}{4M_{N}^{2}}\right)
    A^{a} (t) \right\}
    \Bigg] 
    u(p,J_{3}),\cr
\end{align}
and
\begin{align}
&   \mel{N(p',J'_{3})}
    {   \hat{T}_{\mu\nu}^{a}  ( 0 )  }
    {N(p,J_{3})} \cr
& = \bar{u}(p',J'_{3})
    \Bigg[ g_{\mu\nu}  \bigg{\{} 
    M_{N} \bar{C}^{a}(t)
  + \frac{t}{8 M_N} J^{a}(t)  \cr
&- \frac{3t}{16 M_{N}} D^{a}(t) 
  + \frac{M_{N}}{4}  
    \left(1- \frac{t}{4M_{N}^{2}}\right)
    A^{a}(t) \bigg{\}}
    \Bigg] 
    u(p,J_{3}).\cr
\end{align}

The detailed formalism about the $\chi$QSM can be found in Refs.~\cite{Diakonov:1987ty, Diakonov:1997sj, Christov:1995vm}. Here,
we will sketch how the GFFs of the nucleon can be computed within the $\chi$QSM. The model is based on the effective low-energy QCD partition function
\begin{align}
    Z_{\mathrm{eff}} 
  &= \int 
    \mathcal{D} \psi^{\dagger} 
    \mathcal{D} \psi 
    \mathcal{D} U \cr
&  \exp(-
  \int d^{4} x \, 
    \psi^{\dagger} 
    \left(
    i\slashed{\partial} + i M U^{\gamma_{5}} + i \hat{m}
    \right) 
    \psi
  ), \\ 
  & U^{\gamma_5}(x) 
  \equiv U(x) \frac{1+\gamma_{5}}{2}  
  + U^{\dagger}(x) \frac{1-\gamma_{5}}{2}.
\end{align}
In the above, the second term describes the interaction of the quarks and the
chiral field $U = \exp(i \pi^{a} \lambda_{a})$ in the chiral-invariant manner. 
For the SU(2) subpart of the chiral field, we impose the hedgehog symmetry
\begin{align}
    U_{\mathrm{SU(2)}} 
  = \exp[i \hat{\bm{n}}\cdot  \bm{\tau} P(r)],
\end{align}
which is the minimal symmetry that aligns the spatial vector
with the isospin vector in the pion mean field $P(r)$,
where $\pi^{a}(\bm{r})= \hat{n}^{a} P(r)$, with
$\hat{n}^{a}=r^{a}/|\bm{r}|$ for $a=1,2,3$.
This symmetry ensures the invariance of the pion mean field under  
$\mathrm{SU}(2)_{\mathrm{flavor}} \otimes
\mathrm{SU}(2)_{\mathrm{spin}}$ rotations.  
To maintain the hedgehog symmetry in $\mathrm{SU}(3)_{\mathrm{flavor}}$,
we employ the trivial embedding of the strange component~\cite{Witten:1983tx},
\begin{align}
    U^{\gamma_{5}} 
  = \left( 
    \begin{array}{c c}
      U^{\gamma_{5}}_{\mathrm{SU(2)}} & 0 \\ 
      0                               & 1  
    \end{array}\right),
\end{align}
leaving the SU(2) as a subgroup: 
$\mathrm{SU}(2)_{\mathrm{flavor}}\otimes
\mathrm{SU}(2)_{\mathrm{spin}}\otimes \mathrm{U}(1)_{Y}\otimes
\mathrm{U}(1)_{Y_{R}}$. 
Here, $Y$ and $Y_{R}$ 
stand for the hypercharge and the right hypercharge, respectively. Since the pion mean field is identical to the solution of the classical equation of motion, we
can derive it by solving the equation of motion
self-consistently~\cite{Christov:1995vm}.  

Since the effective low-energy QCD partition function was obtained by integrating out the gluon field, the corresponding EMT current is derived as
\begin{align}
  T_{\mu\nu}^{\chi=0}(x) 
  = \frac{i}{4}   \bar{\psi} (x)
    \gamma_{ \{\mu} \overleftrightarrow{\partial}_{\nu\} }
    \lambda^{0} 
    \psi (x),
\label{eq:effectEMT}
\end{align}
where the flavor singlet matrix is defined as 
$\lambda^{0} = \mathrm{diag}(1,1,1)$. 
Note that Eq.~\eqref{eq:effectEMT} 
does not contain the gluon degrees of freedom. As discussed in Ref.~\cite{Won:2023zmf},
Eq.~\eqref{eq:effectEMT} can be regarded as an effective EMT operator:
\begin{align}
\sum_q T_{\mu \nu}^{q} (x) + T_{\mu \nu}^{g} (x) \big|_{\mathrm{QCD}}
  \xrightarrow{\mathrm{eff}} T_{\mu \nu}^{\chi=0}
  (x).
  \label{eq:total_eff}  
\end{align}
We have shown that the QCD sum rules, such as the mass and the spin relations, and the von Laue condition
are satisfied with this effective EMT current. However, when we look for corresponding flavor-tripet and octet EMT-like operators, which are required to carry out the
flavor decomposition, the situation becomes rather subtle. We can consider a naive construction of such operators by replacing the flavor-singlet matrix $\lambda^0$ 
in Eq.~\eqref{eq:effectEMT} 
with
$\lambda^3$ and $\lambda^8$
\begin{align}
T_{\mu\nu}^{\chi}(x) 
  = \frac{i}{4}   \bar{\psi} (x)
    \gamma_{ \{\mu} \overleftrightarrow{\partial}_{\nu\} }
    \lambda^{\chi} 
    \psi (x),
\label{eq:EMT_current_flav}  
\end{align}
and compute the GFFs with flavor decomposition~\cite{Wakamatsu:2005vk,
  Wakamatsu:2006dy, Won:2023ial, Won:2023cyd}.  
However, we can only derive the flavor-singlet EMT
current~\eqref{eq:effectEMT} as a N\"other current, and
we do not know yet how to derive the flavor-nonsinglet EMT currents in the effective theory. {This} implies that the current given in 
Eq.~\eqref{eq:EMT_current_flav} 
is not guaranteed to be consistent with the effective quark-gluon dynamics. Therefore, we have to come up
with a theoretically solid way to derive the effective flavor nonsinglet EMT operators from the corresponding QCD operators. We propose a plausible way to derive the effective flavor nonsinglet EMT operators from the instanton vacuum.

As shown in Refs.~\cite{Diakonov:1995qy, Balla:1997hf}, the QCD gluon operator can be converted into the effective quark operators in the effective theory while satisfying all low-energy theorems of QCD. We can generate the chiral-even twist-2 local operators by expanding the non-local vector current 
\begin{align}
&O^{\chi}_{\mu \nu_1 \ldots \nu_2}:=\bar{\psi} (x) \gamma_{ \{\mu}
                \overleftrightarrow{\cal D}_{\nu_1}
                \overleftrightarrow{\cal D}_{\nu_2} \ldots
                \overleftrightarrow{\cal D}_{\nu_n \} } \lambda^{\chi}
                \psi(x) \cr
                &\qquad\qquad\qquad\qquad\qquad\qquad\qquad\quad\;- \mathrm{traces}, \cr 
&O_{\mu \nu_1 \ldots \nu_2}^g:=-G_{ \{\mu \alpha}
       \overleftrightarrow{\cal D}_{\nu_1} \ldots
       \overleftrightarrow{\cal D}_{\nu_n} G_{\nu \} }^{\ \alpha} - \mathrm{traces}. 
\end{align}
The local currents in the leading and next-to-leading order are given by 
\begin{align}
&J_{\mu}^{\chi}=\bar{\psi} (x) \gamma_{ \mu }  \lambda^{\chi} \psi(x), \cr
&\bar{T}_{\mu \nu}^{\chi}= \frac{i}{4}\bar{\psi} (x) \gamma_{ \{\mu }
       \overleftrightarrow{\cal D}_{\nu \} }   \lambda^{\chi} \psi(x)
      - \mathrm{traces}, \cr
&\bar{T}_{\mu \nu}^{g} =
    -\frac{1}{2}G_{ \{\mu \alpha} G_{\nu \} }^{\ \alpha} - \mathrm{traces}, 
\end{align}
which will yield the first and second Mellin moments of the vector GPDs. They correspond respectively to the electromagnetic form factors and GFFs of the nucleon. 

In Ref.~\cite{Balla:1997hf}, it was shown that the effect of the gauge field in the covariant derivative is always suppressed 
in the dilute instanton configuration by its parametric dependence of $(M\bar{\rho})^{2}$ for the twist-2 local operators.
{Therefore,} it is consistent to keep only the ordinary derivative part [order unity $\sim O (1)$] in the twist-2
quark operators  
\begin{align}
\bar{\psi} (x) \gamma_{ \{\mu} 
  \overleftrightarrow{\cal D}_{\nu_1}
  \overleftrightarrow{\cal D}_{\nu_2} \ldots 
  \overleftrightarrow{\cal D}_{\nu_n \} }
  \lambda^{\chi} \psi(x) - \mathrm{traces}  
\cr 
\overset{\mathrm{eff}}{\longrightarrow}
\quad
\bar{\psi} (x)
    \gamma_{ \{\mu} 
    \overleftrightarrow{\partial}_{\nu_1} 
  \overleftrightarrow{\partial}_{\nu_2} \ldots
  \overleftrightarrow{\partial}_{\nu_n \} } \lambda^{\chi} \psi(x) -
  \mathrm{traces},  
\end{align}
and to set the twist-2 gluon operators equal to the null operators 
\begin{align}
&-G_{ \{\mu \alpha} \overleftrightarrow{\cal D}_{\nu_1} \ldots
  \overleftrightarrow{\cal D}_{\nu_n} G_{\nu \} }^{\ \alpha} -
  \mathrm{traces} \cr
  \overset{\mathrm{eff}}{\longrightarrow} \quad& 0 \quad
  + \quad O(M^{2} \bar{\rho}^{2}). 
\end{align}
Thus, we obtain the twist-2 effective flavor-nonsinglet EMT-like operators
\begin{align}
    &\bar{T}_{\mu\nu}^{\chi}(x) 
  = \frac{i}{4}   \bar{\psi} (x)
    \gamma_{ \{\mu} \overleftrightarrow{\partial}_{\nu\} }
    \lambda^{\chi} 
    \psi (x) - \mathrm{traces}, \cr 
    &\bar{T}_{\mu\nu}^{g}(x) 
  = 0.
\label{eq:EMT_current_tw2}
\end{align}
They imply that the use of the ordinary derivative operator is
validated with trace subtracted. Thus, we employ 
Eq.~\eqref{eq:EMT_current_tw2} to carry out the flavor decomposition
of the GFFs. Note that the essential difference lies in the subtraction
of traces (or twist classification). 

With the flavor-nonsinglet EMT-like effective operators constructed,  the GFFs of the nucleon within the $\chi$QSM are computed. The detailed formalism and results can be found in Refs.~\cite{Won:2022cyy, Won:2023ial, Won:2023cyd, Won:2023zmf}. Here, we will review the results and summarize the essential points from Ref.~\cite{Won:2023zmf}.
The results for the form factors $A^\chi(t)$  of the proton in the forward limit
($t=0$) are obtained to be 
\begin{align}
[\, \mathrm{SU}(3) \,] \;\;
    A^{0}_{p}(0)  
&=1, \;
    A^{3}_{p}(0)  
  = 0.25, \;
    A^{8}_{p}(0)  
    = 0.47,  \cr 
[\, \mathrm{SU}(2) \,] \;\;
    A^{0}_{p}(0)  
&=1, \; 
    A^{3}_{p}(0)  
  = 0.24.
\label{eq:massres}
\end{align}
Note that the twist-2 mass distribution of the nucleon is normalized to 
$3M_{\mathrm{cl}}/4$, 
where 
$M_{\mathrm{cl}}$ 
denotes the classical nucleon mass obtained from the $\chi$QSM. 
The missing contribution
$M_{\mathrm{cl}}/4$ 
must come from the twist-4 part of the EMT
current. As discussed in 
Eq.~\eqref{eq:EMT_current_tw2}, 
the gluon contributions to the leading-twist operators are suppressed by the packing fraction of the instanton
medium~\cite{Balla:1997hf, Polyakov:2018exb}. 
Thus, we can ignore their contribution to 
$A^\chi$ 
and also to the nucleon spin:   
\begin{align}
A^{g} =0, \quad J^{g}=0.
\end{align}
Using the results presented in Eq.~\eqref{eq:massres}, we can perform the flavor decomposition of the $A$ form factors 
\begin{align}
&[\, \mathrm{SU}(3) \,]\; 
    A^{u}_{p}(0)
 = 0.59, \, 
    A^{d}_{p}(0)
  = 0.35, \, 
    A^{s}_{p}(0)
  = 0.06,  \cr 
&[\, \mathrm{SU}(2) \,] \;  
    A^{u}_{p}(0)
 = 0.62, \, 
    A^{d}_{p}(0)
  = 0.38.
\end{align}
As expected, the up-quark contribution is dominant. While the down quark
contributes to $A$ by about 35~\%, the effect of the strange quark is
as large as 6~\%. Note that $A^q$ is not the portion of the nucleon mass.
As pointed out in Refs.~\cite{Lorce:2017xzd, Won:2023cyd}, 
the form factor $\bar{C}^q$ takes part in the 
the proton mass decomposition. 
In the $\chi$QSM, it is given by the quarks,
\begin{align}
M_p= \sum_q M_p^q = \sum_q (A^q(0) + \bar{C}^q (0)) M_p. 
\end{align}
$A^q(0)$ can rather be understood as the second Mellin
moments of the twist-2 unpolarized quark PDFs, which are the 
momentum {fractions} carried by the $u$-, $d$-, and $s$-quarks
in the proton: 
\begin{align}
\big[ \;\langle x \rangle_{u+\bar u} : \langle x \rangle_{d+ \bar d} : \langle x
  \rangle_{s+ \bar s} \;\big]
  = \big[\;59\% : 35\% : 6\%\;\big]. 
\end{align}

The mass radius of the proton is obtained as 
\begin{align}
    \langle r^{2}_{\mathrm{mass}} \rangle_{p} = 0.54~\mathrm{fm}^{2},  
\end{align}
which is equal to that in the flavor SU(2) symmetry~\cite{Polyakov:2018rew, Won:2022cyy}. 
{It is smaller than the charge radius of the proton computed within the same model framework, $\langle r^{2}_{e} \rangle_{p} = 0.78~\mathrm{fm}^{2}$ \cite{Kim:1995mr}.}

Similarly to the flavor decomposition of the $A$ form factor, 
we obtain the flavor-decomposed nucleon spin $J_p^q\equiv J_p^q\;(t=0)$,
\begin{align}
&[\, \mathrm{SU}(3) \,]\;\;
    J^{u}_{p} =  0.52, \;
    J^{d}_{p} = -0.06, \; 
    J^{s}_{p} =  0.04,   \cr 
&[\, \mathrm{SU}(2) \,] \;\; 
    J^{u}_{p} =  0.53, \; 
    J^{d}_{p} = -0.03.
\end{align}
As expected, the $u$-quark dominates over other quarks. 
These results are in agreement with findings 
from lattice QCD simulations~\cite{LHPC:2010jcs}. 
Comparing the results with those obtained from the SU(2)
$\chi$QSM~\cite{Goeke:2007fp,Won:2023cyd,Kim:2020nug},  
we find that the contribution of the $u$-quark to the total AM remains nearly unchanged, and the polarization of the $d$-quark
contribution is slightly enhanced. It indicates that the $s$ quark is polarized in the opposite direction to the $d$ quark, so that it almost cancels out each other. According to  Ji's
relation~\cite{Ji:1996ek}, 
the nucleon spin can be expressed as the sum of the intrinsic spin and the orbital angular momentum of the
quarks:  
\begin{align}
    J  = \frac{1}{2}   \sum_{q} \Delta q 
  + \sum_{q} L^{q}+J^{g}, \quad J^{g}=0.
\label{eq:Ji}
\end{align} 
Here, we consider only the quark contributions, 
since the gluon contributions are suppressed 
in the QCD instanton 
vacuum~\cite{Balla:1997hf, Diakonov:1995qy}. 
In the $\chi$QSM, the antisymmetric part of the $0k$ component of the EMT current yields the spin of the $s$-wave quarks, while the non-symmetric part is responsible for the $p$-wave quarks. Thus, the static quark spin and the relativistic motion of the quark yield the intrinsic spin and orbital angular momentum, respectively. Note that 50\% of the flavor-singlet angular momentum 
arises from the relativistic motion of the quarks inside the nucleon~\cite{Won:2022cyy}:
\begin{align}
    \frac{1}{2}
  = \frac{1}{2} \sum_{q} \Delta q + \sum_{q} L^{q} = 0.23 + 0.27.
\end{align}
However, there is a caveat. There are at least two reasons to examine the nucleon spin carefully: different UV divergence patterns between the total angular momentum, and the separate spin and the
orbital angular momentum~\cite{Kim:2023yhp}; lack of knowledge of the  matching between the QCD operator and the twist-3 effective
operator~\cite{Kim:2023pll}.

{Let us turn to the $D$-term form factor and the pressure distribution in the 3D Breit frame.} The pressure distribution must satisfy the von Laue stability condition (\ref{von_Laue_stability}) {due to the conservation of the energy-momentum tensor current}:
\begin{align}
    \int d^{3}r \,   
    p^{u+d+s}_{p}(r) 
  = 0.
  \label{eq:st_p}
\end{align}
As already mentioned, we have taken the effective operator 
\eqref{eq:EMT_current_tw2} derived from the QCD instanton vacuum and concentrate on the twist-2 part. However, the integral of the twist-2 part of the pressure distribution over $r$ is not zero:
\be
&&
    \int d^{3}r \, \bar{p}^{u+d+s}_{p}(r)
  = \frac{1}{4} M_{N}, \nn \\ && \int d^{3}r \, \hat{p}^{u+d+s}_{p}(r)
  = -\frac{1}{4} M_{N} .
\label{eq:hoho2}
\ee
It implies that the amount of energy $\frac{1}{4}
M_{N}$ in Eq.~\eqref{eq:hoho2} leaks away to the twist-4 quark and gluon part, of which the amount is $1/3$ times the normalization for the twist-2 energy distribution. Each quark contribution to the von Laue condition originates from the $\bar{C}$ form factor. Without knowing the twist-4 operator explicitly, we are
not able to deal with it. Thus, we can at most discuss the flavor decomposition of the twist-2 part. The integration over each flavor-decomposed pressure distribution is given as  
\be 
& &\int d^{3}r \, \bar{p}^{u}_{p}(r)
  = \frac{1}{4} M_{N} A^{u}_{p}(0), \nn \\ &&    \int d^{3}r \,
      \bar{p}^{d}_{p}(r) 
  = \frac{1}{4} M_{N} A^{d}_{p}(0),  \nn \\ &&
  \int d^{3}r \, \bar{p}^{s}_{p}(r)
  = \frac{1}{4} M_{N} A^{s}_{p}(0),
\ee
which shows that these are only proportional to the $A^{q}_{p}(0)$
form factors, which in turn are canceled out by the twist-4 part. 

On the other hand, the shear-force distribution comes from the off-diagonal part of the EMT. Thus, we can extract the $D$-term form factor from the shear-force distribution without being affected by the twist decomposition. 
We derive the flavor singlet, triplet, and octet $D$-term form factors 
by Fourier transform of the shear-force distributions as follows:  
\begin{align}
&[\, \mathrm{SU}(3) \,]\;\;
    D^{0}_{p}(0)
 = - 2.531, \; 
    D^{3}_{p}(0)
  = 0.063, \cr 
&\qquad\qquad    D^{8}_{p}(0)
  = - 0.697, \cr 
& [ \, \mathrm{SU}(2) \,]\;\;
    D^{0}_{p}(0)
 = - 2.531, \;
    D^{3}_{p}(0)
  = 0.295.
\end{align}
Note that the gluon contributions to the $D$-term form factor are suppressed at low normalization points $D^{g}=0$.


{In summary, the theoretical investigation of the nucleon GFFs is rather subtle. Both quark and gluon contributions must be taken into account, and their flavor decomposition requires flavor non-singlet EMT-like operators. In effective theories of QCD, a naive construction obtained by replacing the identity in flavor space with the Gell-Mann matrices leads to inconsistent results for the flavor decomposition of the
GFFs, and consequently for the mechanical properties as well.
}

\subsection{Gravitational form factors
and the equivalence principle 
}

The gravitational form factors of hadrons, while well-defined even in the absence of gravitational interactions, describe the interaction with classical gravity. The equivalence principle (EP), {see {{\it e.g.} \cite{Will:2005va} for an overview}}, plays a special role here, revealing an intriguing connection between the weakest and strongest fundamental interactions.

This connection can be approached from two perspectives. First, conservation laws provide an opportunity to derive the equivalence principle, transforming it from a fundamental principle into a low-energy theorem
\cite{Weinberg:1964ew}. Additionally, by performing flavor decomposition and extracting the separate contributions of quarks and gluons to GFFs, we gain unique insights into the individual interactions of quarks and gluons with gravity, that are otherwise unavailable. It is crucial to emphasize that this decomposition fundamentally relies on the non-perturbative nature of the QCD description of hadron structure, as the mixing of non-conserved terms in the EMT can be viewed as a correction to their non-perturbative values.

The derivation of {the EP} 
is most simple in the case of the leading approximation introduced by Einstein. 
Following Ref.~\cite{Teryaev:2016edw}, 
it is instructive to compare the gravitational and electromagnetic interactions in the classical limit ($\Delta \to 0$), the latter being described by the matrix element 
\begin{eqnarray}
  {\cal M}=\langle P'| J^{q}_{\mu} |P\rangle A^\mu(\Delta).
\label{A0}
\end{eqnarray} 
This matrix element at zero momentum transfer is fixed by the
fact that the interaction is due to the  local U$(1)$ symmetry,
whose  global counterpart produces the conserved charge:
\begin{eqnarray}
\langle P| J^{q}_{\mu} |P\rangle=2e^q P_\mu.
\label{ea}
\end{eqnarray}
Therefore, in the rest frame the interaction is completely defined by
the scalar potential:
\begin{eqnarray}
\label{0e}
  \left. { \cal M}_0 \right|_{\rm e.m.}=\langle P| J^{q}_{\mu} |P\rangle A^\mu = 2 M \cdot e^q  \phi (\Delta).
\label{AA}
\end{eqnarray}
At the same time, the interaction with the weak classical gravitational
field is:
\begin{eqnarray}
{ \cal M}=\frac{1}{2}\sum_{q,g} \langle P'| T^{q,G}_{\mu \nu} |P\rangle h^{\mu \nu} (\Delta),
\label{h4h}
\end{eqnarray}
where $h$ is a deviation of the metric tensor from its Minkowski value.
The relative
factor $1/2$, which will play a crucial role, arise from the fact
that the variation of the action with respect to the metric
produces an energy-momentum tensor with the coefficient $1/2$,
while the variation with respect to the classical source $A^\mu$
produces the current without such a coefficient. It is this
coefficient 
that guarantees the correct value for the
Newtonian limit, fixed by the  global translational invariance
\begin{eqnarray}
\sum_{a=q,g} \langle P| T^a_{\mu \nu} |P\rangle=2 P_\mu P_\nu.
\label{e}
\end{eqnarray}
Together with the approximation for $h$
(with factor of 2
having the geometrical origin) 
\begin{eqnarray}
\label{h}
h_{00}=2\phi (x)
\end{eqnarray}
this results in the rest frame expression:
\begin{eqnarray}
&&
\label{0g}
\left. {\cal M}_0 \right|_{\rm gravity}= \nn \\ && \sum_{a=q,g} \langle P| T^{a}_{\mu \nu} |P\rangle h^{\mu \nu}(\Delta)=
2 M \cdot M \phi (\Delta), 
\label{A}
\end{eqnarray}
where we deliberately used the same notation for gravitational and scalar
electromagnetic potentials and identified the one-particle state normalization factor
$2M$ to make the similarity between 
(\ref{0e}) 
and 
(\ref{0g})
explicit. 

{By comparing (\ref{0e}) and (\ref{0g})}
one observes that, due to conservation of electromagnetic current and energy-momentum tensor, the role played by the charge in the case of electromagnetic interaction is 
{taken }
by { mass} in the case of in gravitational interaction. 
{
This directly leads to the equivalence principle, since the gravitational charge is identified with the inertial mass, implying the universal charge-to-mass ratio:
\begin{equation}
\left.\frac{e}{M}\right|_{\rm e.m.}
 \; \longrightarrow \;
\left.\frac{M}{M}\right|_{\rm gravity}
=1.
\end{equation}
As a result, all bodies experience the same acceleration in a given gravitational field, irrespective of their composition or internal structure.
The 
{EP}
appears here as a low-energy theorem rather than a postulate.
The similarity with the electromagnetic case  clarifies 
the origin of such a theorem,  suggesting that the interaction 
with gravity is due to the { local}
counterpart of the  global symmetry, although it also can be proven
starting from the Lorentz invariance of the soft graviton
approximation \cite{Weinberg:1964ew}.}

{Moreover, the EP} 
allows one to describe the interaction with { quantum} spin, which might be even considered as a step in the direction of development of quantum gravity.  
As the gauge symmetry in gravity is more restrictive than in the electromagnetic case, it allows for the fixing of not only charges but also { moments} (terms linear in the momentum transfer $\Delta=P'-P$).

Let us discuss this crucial point in some detail. The situation with the terms linear in $\Delta=P'-P$ is different
for electromagnetism and gravity. 
In the electromagnetic case, such a term is determined by the specific dynamics and results in the anomalous magnetic moment. In contrast, the analogous contribution in the gravitational case is entirely fixed by the conservation of angular momentum.
{The reason is that the structure of the
Poincare group
is more rich than
that of the ${\rm U}(1)$ group.
Recall, that the restrictions for the low energy
limits of the interaction are coming from the fact, that the same group
is controlling both the interaction via local invariance,
and the conserved charges via the global invariance.
}


As a result, 
the Pauli form factor of the conserved EMT is zero, so 
there is no   Anomalous Gravitomagnetic Moment (AGM)~\cite{Kobzarev:1962wt}:
\be
\sum_{a=q,G} B^a(0)=0.
\label{Eq_principle}
\ee
This may be considered as a gravitational counterpart of Ji's sum rules.  Another manifestation of this property is the equality of the precession frequencies of classical and quantum rotators.  

The spin precession is naturally produced by the non-diagonal components 
$g_{0i}$ 
of the metric which are pronounced in terrestrial and orbital experiments due to Earth's rotation, see {\it e.g.}~\cite{Everitt:2011hp}.
The 
{EP} states that in a small freely falling local frame the effects of gravity are eliminated for any system including classical and quantum rotators.
However, spin precession due to spacetime curvature remains observable over extended regions. 
{We stress that we deal} with a small freely falling local frame. It is interesting to note, that the
{EP}
in the form of complete similarity of the behavior of classical and quantum rotators holds for {\it arbitrary} gravitational fields. This remarkable property immediately follows from the results obtained in~\cite{Obukhov:2013zca}. 
 
{A} crucial role is played by the quantum measurement, which is usually considered to have a marginal or philosophical role. Indeed, the spin rotation with the same frequency as a classical rotator is trivial if spin is considered as a usual (pseudo)vector.
At the same time, {a} quantum measurement performed in a non-inertial rotating frame makes the equality of classical and quantum frequencies a non-trivial manifestation of the equivalence principle. 

The Extended Equivalence Principle abbreviation (ExEP), which may also be interpreted as Exact EquiPartition,
is formulated as a requirement for quark and gluon contributions to AGM
vanishing separately:
\be
B^a(0) =0; \,\ a=q,\,g.
\ee
incorporated to Ji's sum rules
obviously results in the equal partition of momentum and (total) angular momentum between quarks and gluons:
{
\be
J^q: J^g=A^q(0): A^g(0).
\ee
}
{This property is violated in perturbative QCD, where quark and gluon contributions mix under renormalization, but it has been conjectured that nonperturbative phenomena, such as confinement and spontaneous chiral symmetry breaking, may restore it approximately \cite{Teryaev:2016edw}.}

A similar situation occurs for other structures that are strictly forbidden by conservation laws for conserved EMT. While they are allowed to be nonzero for separate quark and gluon contributions, their respective values tend to be small.
One should first mention the ``cosmological constant'' 
\begin{eqnarray}
\langle P| T^{a}_{\mu \nu} |P\rangle = {\bar C}^a g_{\mu \nu}.
\label{Cbar}{}
\end{eqnarray}
It is important to note that the extraction of pressure in proton ~\cite{Burkert:2018bqq} 
actually assumes the smallness of ${\bar C}$ for quarks, providing the zero integral of pressure weighted with radius squared (being the necessary condition of equilibrium), while the negative integral{,} weighted with the fourth power of radius and related to the $D$-term, ensures that this equilibrium is stable. 

Indeed, zero ${\bar C}$ may be associated with the Feynman Principle of virtual work 
$$
\int p\, dV =0.
$$
Its violation for quarks means that the equilibrium is achieved only if the interaction with gluons is taken into account. As soon as the 
stability condition $D < 0$ is considered for quarks separately (which is implied by the available data), the smallness of $\bar C$ allows one to consider the stable mechanical equilibrium of quarks separately from gluons.  

At the same time, substantial $\bar{C}$ would mean that the equilibrium of quarks cannot be achieved if their interaction with gluons is not taken into account. In that case, one may also expect the gluon's role in maintaining the stability of that equilibrium to be considerable.
Still, one cannot exclude that the equilibrium (equality to zero of total $\bar{C}$) requires both quark and gluon contributions, while the stability (inequality $D <0$) may be achieved separately.

Another interesting structure is provided by the analog of shear viscosity 
$\mu$
corresponding to the T-odd {(dissipative)} term in the EMT of the classical fluid  \cite{LandauLifshitzFluid1987}:
\begin{eqnarray}
T_{visc}^{\alpha \nu} = 
{-}\mu
\frac{\partial v ^{\alpha}}{\partial x_\nu}|_{(\rm symmetric,\, traceless)}.
\end{eqnarray}
{The quantum analog may be obtained if the velocity role is played by the average momentum of hadrons while the derivative is substituted by the 
their relative momentum $\Delta$, as 
implied by 
the Fourier transform.} 

It is instructive to consider the production of $\pi \eta$ pairs with exotic quantum numbers $J^{PC} = 1^{-+}$~\cite{Teryaev:2022pke,Song:2025zwl}:
\begin{eqnarray}
\langle \pi \eta (P, \Delta) |T^a_{\alpha \nu}|0 \rangle_{\mu_F^2}= E^a (M^2_{\pi \eta}, \mu_F^2) P_{\alpha} \Delta_{\nu},
\label{EMT_GDA_exotic}
\end{eqnarray} 
{where the new invariant form factor $E^a$
depends on the $\pi \eta$ invariant mass 
$M^2_{\pi \eta}$ and on factorization scale $\mu_F^2$.}
{ The contribution (\ref{EMT_GDA_exotic}) 
corresponds} to the crossing of the (naive) T-violating contribution into the EMT hadronic matrix element from {the direct (GPD) channel, 
$\langle h'  |T^a_{\alpha \nu} | h \rangle$, to the cross-conjugate (GDA) channel, $\langle h'  \bar{h}| T^a_{\alpha \nu} | 0 \rangle$.}
The smallness of the such term relates the smallness of {the}
corresponding GDA (also observed  in the perturbative limit 
\cite{Song:2025zwl})
and smallness of {the} shear viscosity in the holographic approach.
The latter originates from extra-dimensional gravity
{\cite{Kovtun:2004de}}, establishing in this way some gravity/QCD interplay. The other ExEP manifestations may provide stability of confinement against the strongest gravitational fields in four dimensions~\cite{Teryaev:2016edw}, 
such as those produced by black holes.

{
Although gravitational form factors can be studied independently of gravity itself, their gravitational interpretation provides a connection between hadron structure and the equivalence principle. 
This connection becomes particularly intriguing when the 
{EP}
is extended to the level of quarks and gluons. 
The ExEP postulates vanishing of the quark and gluon anomalous gravitomagnetic moments, 
$B^q(0)$ and 
$B^g(0)$, implying that the partition of the nucleon momentum and total angular momentum between quarks and gluons is nearly identical. While this property is violated in perturbative QCD, it has been conjectured that nonperturbative dynamics may drive the system toward the extended-equivalence-principle limit. 
Whether this behavior reflects a deeper property of QCD, and whether it has implications beyond hadron structure, for example, for the behavior of confined systems in extremely strong gravitational fields, remains an open question and calls for further detailed investigation. 
}


\subsection{Probing quark orbital angular momentum with GTMDs}
\label{sec:oam_gtmds}
{GTMDs} represent the most comprehensive two-parton correlation functions for describing the internal structure of nucleons. The study of GTMDs is motivated by several key factors. First, quantities such as one-dimensional {PDFs}, {FFs}, and their three-dimensional extensions---{TMDs} and 
GPDs---can all be obtained as specific kinematical projections of GTMDs. GTMDs therefore serve as the \textit{``mother distributions''} of these lower-dimensional functions~\cite{Meissner:2009ww}. {Importantly, while GTMDs provide a more general underlying framework, the phenomenological extraction of nucleon structure is currently driven primarily by GPDs, which are directly accessible in hard exclusive processes such as deeply virtual Compton scattering and exclusive meson production, as discussed in detail in this whitepaper. In this sense, GTMD-based considerations are most impactful when they reduce to or constrain GPD matrix elements and their moments. This provides a direct connection between partonic phase-space correlations and experimentally measurable GPD observables, particularly in the context of nucleon tomography and spin decomposition via Ji’s sum rule.}  
On the other hand, GTMDs contain somewhat richer physical information than either TMDs or GPDs, as certain GTMDs---such as $F_{1,4}$ and $G_{1,1}$---do not survive in the TMD or GPD limits. Exploring the physics encoded in these ``lost'' components provides unique insights into partonic correlations that cannot be accessed otherwise.  
Third, for specific kinematic configurations, the Fourier transform of GTMDs can be related to the Wigner distributions~\cite{Belitsky:2003nz}. The Wigner distributions, familiar from phase-space formulations of quantum mechanics and widely used in atomic, molecular, and optical physics, enable five-dimensional imaging of the nucleon. Furthermore, certain GTMDs directly probe nontrivial correlations between the orbital motion of partons and the spin of either the nucleon or the parton itself~\cite{Lorce:2011kd,Bhattacharya:2024sno}:  
\begin{align}
\langle \vec{L}_{q/g} \cdot \vec{S}_{N} \rangle &\sim  F^{q/g}_{1,4} \, , \qquad
\langle \vec{L}_{q/g} \cdot \vec{S}_{q/g}\rangle \sim G^{q/g}_{1,1} \, .
\label{e:spin_orbit_correlations}
\end{align}
These compelling considerations highlight the central role of GTMDs as the ``ultimate'' parton correlation functions, encompassing the complete spatial and momentum information of partons within hadrons. Below, we identify a physical process that can provide the \textit{first} experimental access to quark orbital angular momentum (OAM) by exploiting its connection to GTMDs, as indicated by the first relation in Eq.~(\ref{e:spin_orbit_correlations}).

\begin{figure}[b]
\begin{center}
\includegraphics[width=6cm]{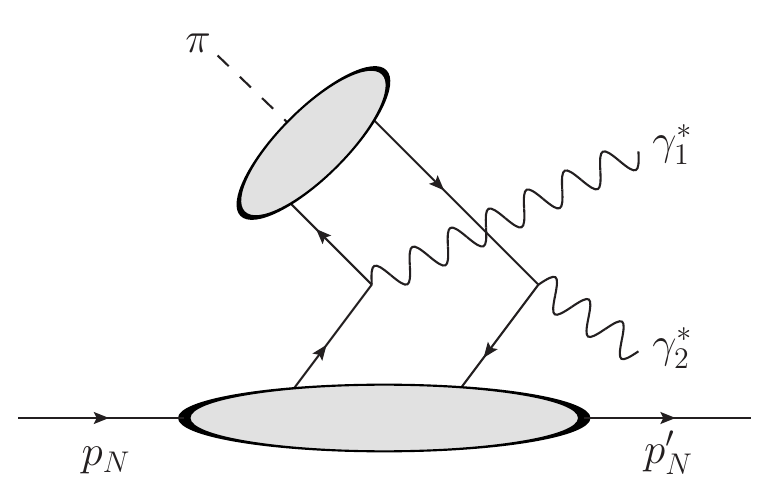}

\vspace{0.7cm}

\includegraphics[width=6cm]{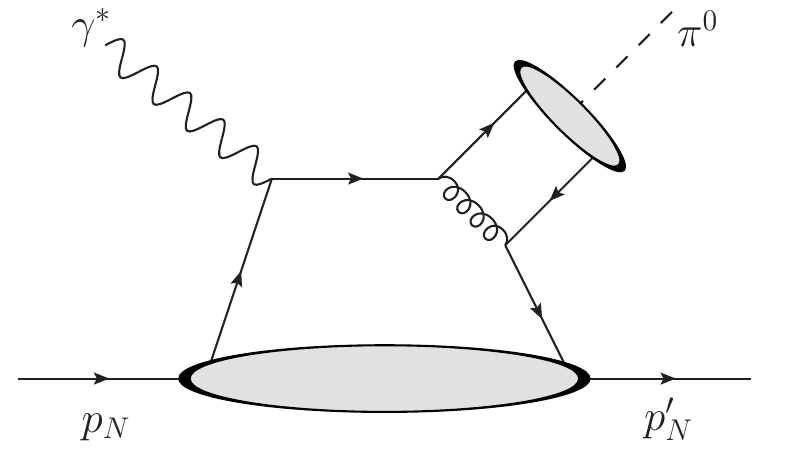}
\caption{Top: Exclusive pion--nucleon double Drell--Yan process. Bottom: Exclusive $\pi^0$ production in electron--proton collisions.}
\label{fig:process}
\end{center}
\end{figure}

In Ref.~\cite{Bhattacharya:2017bvs}, the authors identified for the first time a physical process sensitive to quark GTMDs: the exclusive double Drell--Yan process in pion--nucleon collisions (see top panel of Fig.~\ref{fig:process}). 
That work also
demonstrates the existence of an observable directly sensitive to the quark OAM and spin--orbit correlations. However, two major challenges arise. 
The suggested process suffers from an extremely low event rate, with a cross section 
$\sim \alpha_{\rm em}^{4}$. Moreover, GTMDs can be accessed only in the ERBL region, whereas the quark OAM density,
\begin{align}
L^{q/g}(x, \xi) \sim \int d^2 k_\perp \, k^2_\perp \,  F^{q/g}_{1,4}(x, \xi, k_\perp, \Delta_\perp),
\end{align}
relates to the total OAM $L^{q/g}$ only in the forward limit 
$\xi = 0$.
Consequently, one faces the nontrivial task of extrapolating $F^{q/g}_{1,4}$, which is experimentally accessible only in the ERBL region, to the forward limit relevant for OAM extraction.

Motivated by these challenges, our recent work~\cite{Bhattacharya:2023hbq} proposed an alternative approach to access GTMDs through exclusive $\pi^0$ production in electron--proton collisions,
\be
e(l)  + p(p_N, \lambda)
\rightarrow e(l') +\pi^0(p_\pi) +     p(p'_N, \lambda') \, ,
\label{DVMP_Sh}
\ee
with a longitudinally polarized proton target. This framework avoids several of the limitations associated with the double Drell--Yan process. {Below we argue that the amplitude of the reaction (\ref{DVMP_Sh}) involves a linear combination of (moments of) {GTMDs} and GPDs. This interplay is itself physically significant, as it connects two complementary descriptions of hadron structure and therefore offers a valuable opportunity to simultaneously constrain both GTMD and GPD dynamics within a unified exclusive framework.}

We employ the standard kinematical variables:
photon virtuality is $Q^2 = -q^2 = -(l - l')^2$ and
$\gamma^*p$ center-of-mass energy $W^2 = (p_N + q)^2$.
The skewness parameter $\xi$ is defined as in (\ref{Def_xi})
and $t = (p'_N - p_N)^2 = -\frac{4\xi^2 M^2 + \Delta_\perp^2}{1 - \xi^2}$. We neglect the pion mass ($p_\pi^2 \approx 0$) for simplicity.

At leading order in perturbative QCD, four Feynman diagrams contribute (see the bottom panel of Fig.~\ref{fig:process}), and the schematic form of the amplitude $A$ is expressed as
\begin{align}
& A \propto \int dx \int d^{2}k_\perp \,\cr 
& H(x, \xi, z, k_\perp, \Delta_\perp) \,
f^q(x, \xi, k_\perp, \Delta_\perp) \,
\int dz \, \phi_\pi(z),\cr
\end{align}
where $H(..)$ denotes the hard part, $f^q(..)$ represents the soft part from the proton, and $\phi_\pi(z)$ is the leading twist-2 pion 
DA. We perform the collinear twist expansion of the hard part in powers of $k_\perp$ and $\Delta_\perp$:
\begin{align}
H(k_\perp, \Delta_\perp) &=
H(0,0)
+ \frac{\partial H(k_\perp,0)}{\partial k_\perp^\mu}\bigg|_{k_\perp=0} k_\perp^\mu \cr
&+ \frac{\partial H(0,\Delta_\perp)}{\partial \Delta_\perp^\mu}\bigg|_{\Delta_\perp=0} \Delta_\perp^\mu
+ \ldots
\label{e:exp}
\end{align}

We consider the exclusive  production of $\pi^0$ by the transversely polarized virtual photon.
The leading (twist-2) term $H(0,0)$ vanishes for this process, as a result of the conservation of angular momentum along the direction of the virtual-nucleon beam. Therefore, the first nonzero contribution arises at twist-3 order. Ensuring electromagnetic gauge invariance during this expansion is essential; we preserve it using the special propagator technique. Picking up one factor of $k_\perp$ from the hard part yields an amplitude proportional to ($k^2_\perp$)-weighted moments of GTMDs, whereas a factor of $\Delta_\perp$ leads to a GPD contribution. {Consequently, the amplitude is of twist-3 nature and involves a convolution of GTMD moments (related to quark OAM) with GPDs, thus linking two complementary descriptions of hadron structure within a single observable.}
Note that the $k_\perp$ dependence of the pion DA is neglected here since it is itself of twist-3 nature. Including it would make the amplitude proportional to a twist-3 GTMD moment times a twist-3 pion DA, which is power-suppressed compared to the current twist-3 $\times$ twist-2 contribution. The case involving a twist-3 pion DA coupled to {tensor} GPDs has been studied by the GK group~\cite{Duplancic:2023xrt} and is omitted here for simplicity.

The resulting polarization-dependent cross section is
\begin{align}
&\frac{d \sigma_{{T}}}{dt \, dQ^2 \, dx_B \, d\phi} =\nonumber \\[0.2cm]
&\frac{(N_c^2-1)^2 \, \alpha_{\rm em}^2 \alpha_s^2 \, f_\pi^2 \, \xi^3 \Delta_\perp^2}
{
N_c^4 (1-\xi^2) Q^{10}(1+\xi)}
\left[ 1 + (1-y)^2 \right]
\nonumber \\[0.2cm]
& \times
\lambda \sin(2\phi) \, 2a \,
{\rm Re} \!\left[
\left( i{\cal F}_{1,4} + i{\cal G}_{1,4} \right)
\left( {\cal F}_{1,1}^* + {\cal G}_{1,1}^* \right)
\right],
\label{e:asym}
\end{align}
where $a = {2(1-y)}/[{1+(1-y)^2}]$ and 
$f_\pi$ 
is the pion decay constant. This expression reveals that the GTMD moment ${\cal F}_{1,4}$---and hence the quark OAM---produces a distinctive $\sin(2\phi)$ modulation in the longitudinal single target-spin asymmetry. The azimuthal angle $\phi = \phi_{l_\perp} - \phi_{\Delta_\perp}$ encodes the correlation between the scattered lepton and the recoil proton.

The corresponding unpolarized cross section reads
\begin{align}
&\frac{d \sigma_{{T}}}{dt \, dQ^2 \, dx_B \, d\phi} = \nonumber \\[0.2cm]
&\frac{(N_c^2-1)^2 \, \alpha_{\rm em}^2 \alpha_s^2 \, f_\pi^2 \, \xi^3 \Delta_\perp^2}
{
N_c^4 (1-\xi^2) Q^{10}(1+\xi)}
\left[ 1 + (1-y)^2 \right]
\nonumber \\[0.2cm]
& \times \Bigg\{
\bigg[
|{\cal F}_{1,1}+{\cal G}_{1,1}|^2
+ |{\cal F}_{1,4}+{\cal G}_{1,4}|^2  \cr 
&\hspace{1cm}
+ 2 \frac{M^2}{\Delta_\perp^2} |{\cal F}_{1,2}+{\cal G}_{1,2}|^2
\bigg]
\nonumber \\[0.1cm]
& \hspace{0cm}
+ \cos(2\phi) \, a
\left[
- |{\cal F}_{1,1}+{\cal G}_{1,1}|^2
+ |{\cal F}_{1,4}+{\cal G}_{1,4}|^2
\right]
\Bigg\}.\cr
\label{e:unpol}
\end{align}
Interestingly, the unpolarized cross section exhibits a clear $\cos(2\phi)$ modulation. Two noteworthy features emerge:
\bi
\item the quark Sivers function contributes through ${\rm Im}[F_{1,2}]|_{\Delta=0} = -f_{1T}^\perp$, even without target polarization, mirroring the behavior of the gluon Sivers function identified in Ref.~\cite{Boussarie:2019vmk};  
\item the worm-gear function enters via ${\rm Re}[G_{1,2}]|_{\Delta=0} = g_{1T}$, again for an unpolarized target. 
\ei
These helicity-flip terms can be isolated by taking the $\Delta_\perp \to 0$ limit. Importantly, since both polarized and unpolarized terms are of twist-3 nature, the unpolarized contribution does not dilute the asymmetry signal.

\begin{figure}
\centering
\includegraphics[scale=0.30]{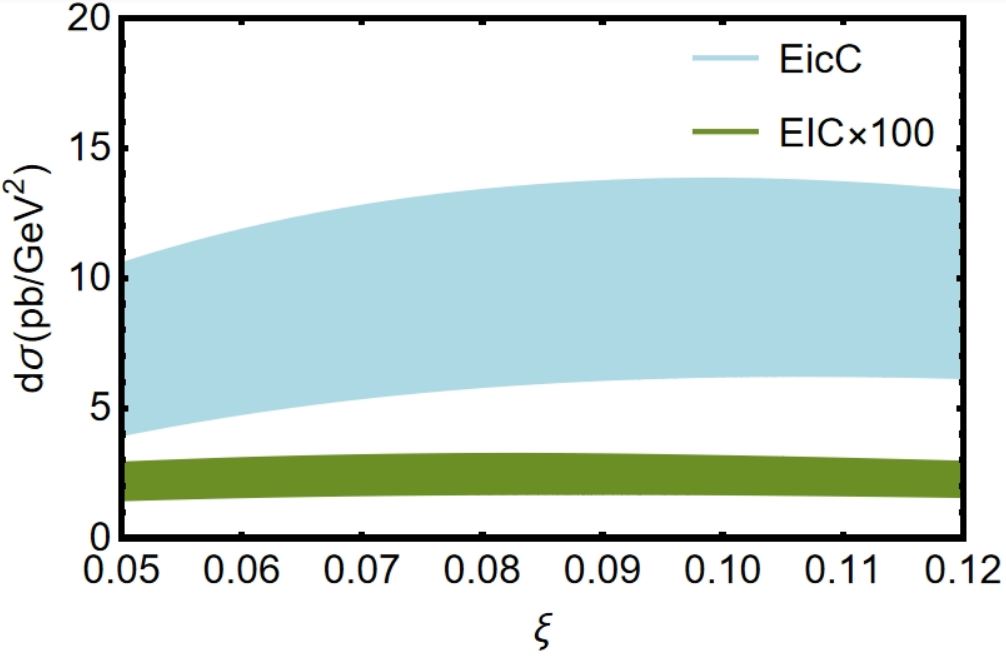}
\includegraphics[scale=0.30]{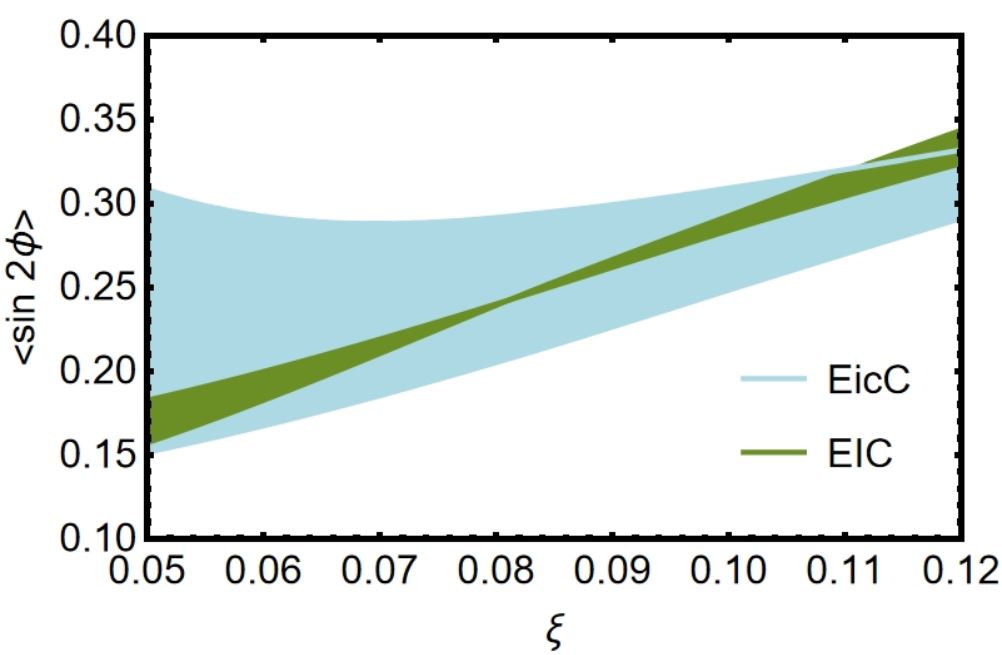}
\caption{Top: Unpolarized cross section for EIC kinematics ($Q^2 = 10~\textrm{GeV}^2$, $\sqrt{s_{ep}} = 100~\textrm{GeV}$) and EicC kinematics ($Q^2 = 3~\textrm{GeV}^2$, $\sqrt{s_{ep}} = 16~\textrm{GeV}$). The EIC curve is scaled by a factor of 100. Bottom: average $\langle \sin(2\phi) \rangle$ as a function of $\xi$. The variable $t$ is integrated over $[-0.5~\textrm{GeV}^2, -4\xi^2 M^2/(1-\xi^2)]$. This Figure is taken from Ref.~\cite{Bhattacharya:2023hbq}; for a detailed discussion of the uncertainty bands, we refer the reader to that work.}
\label{plot}
\end{figure}

The relevant GTMD moment entering Eq.~(\ref{e:asym}) is
\begin{align}
{\cal F}_{1,4} &=
\int_{-1}^{1} dx \,
\frac{
x \xi
\int d^2 k_\perp \, k_\perp^2 \,
F_{1,4}^{u+d}(x, \xi, \Delta_\perp, k_\perp)
}{
M^2 (x+\xi - i\epsilon)^2 (x-\xi + i\epsilon)^2
}\cr
&\times\int_{0}^{1} dz \,
\frac{\phi_\pi(z)(1+z^2-z)}{z^2 (1-z)^2}.
\end{align}
The hard part diverges as $z \to 0, 1$, signaling endpoint singularities that challenge collinear factorization. Following Ref.~\cite{Goloskokov:2007nt}, we regularize those singularities by modifying the $z$-integration limits to
$\int_{\langle p_\perp^2 \rangle /Q^2}^{1 - \langle p_\perp^2 \rangle /Q^2} dz$,
with $\langle p_\perp^2 \rangle = 0.04~\textrm{GeV}^2$ extracted from the CLAS data~\cite{Bhattacharya:2023hbq}. We adopt the asymptotic pion DA $\phi_\pi(z) = 6z(1-z)$.  
Discontinuities in the derivative of quark GPDs at $x = \pm \xi$, combined with double poles in the integrals, can also induce divergences. To regulate them, we shift
$$
\frac{1}{(x-\xi + i\epsilon)^2} \to \frac{1}{(x-\xi - \langle p_\perp^2 \rangle /Q^2 + i\epsilon)^2},
$$ 
and analogously for $x=-\xi$. These refinements highlight the crucial role of intrinsic (quark) transverse motion in preserving factorization.

To minimize backgrounds, we focus on the large-$\xi$ region (thus suppressing gluon contributions) and the large-$|t|$ region (thus reducing Primakoff effects, which scale as $1/t$). Notably, the interference between the Primakoff process and gluon GTMD $F_{1,4}$ also produces a $\sin(2\phi)$ modulation, serving as an observable background~\cite{Bhattacharya:2023yvo}.
The unpolarized cross section and $\sin(2\phi)$-weighted asymmetry are shown in Fig.~\ref{plot}. {The cross section is sizable in the kinematic regime relevant for the EicC, while it is comparatively smaller at EIC energies, consistent with its $1/Q^{10}$ scaling behavior. This strong suppression with increasing $Q^2$ suggests that it may be particularly interesting to study this reaction in the kinematic range accessible at JLab. In fact, existing measurements with longitudinally polarized targets are already available in this regime, which might be useful.} The $\sin(2\phi)$ asymmetry {on the other hand} is significant at both facilities, establishing exclusive $\pi^0$ electroproduction as a promising channel for the first experimental extraction of the quark OAM distribution. 

{We note in passing that Ref.~\cite{Bhattacharya:2026qnd} proposed potential experimental signatures of the gluon GTMDs $F_{1,4}^g$ and $G_{1,1}^g$ through exclusive heavy (axial-) vector meson production in lepton--proton collisions within the framework of collinear twist-3 factorization. In particular, characteristic $\cos (2\phi)$ and $\sin (2\phi)$ azimuthal modulations were shown to provide sensitivity to gluonic OAM and spin--orbit correlations, thereby opening a new avenue to probe the nucleon spin structure at the future EIC. This process again highlights the interplay between GTMDs and GPDs within a single observable, reflecting their complementary roles in describing hadron structure.}

In summary, we proposed a novel method to extract the quark OAM associated with the Jaffe--Manohar spin sum rule by measuring the azimuthal angular correlation 
$\sin(2\phi)$ 
in exclusive 
$\pi^0$ 
electroproduction at the EIC and EicC. This observable provided a clean and sensitive probe of quark OAM for several compelling reasons. 
First, the azimuthal asymmetry was not a mere power correction, since both the unpolarized and the longitudinally polarized cross sections contributed at twist--3, ensuring leading sensitivity to quark OAM. Second, this process enabled, for the first time, direct access to the quark GTMD 
$F_{1,4}$ 
in the DGLAP region---the key distribution underlying the partonic OAM density. 
Beyond OAM extraction, our study also revealed another remarkable feature: the quark components of 
$F_{1,2}$ and $G_{1,2}$, 
corresponding respectively to the Sivers and worm--gear functions, contributed to the unpolarized cross section in the forward limit. This finding was particularly striking, as these functions are traditionally accessed only through transversely polarized targets. 
We computed the differential cross section within the collinear higher--twist expansion framework and carried out detailed numerical analyses. The results exhibited a sizable $\sin(2\phi)$ modulation whose magnitude and sign were directly governed by the quark OAM distribution. Within the kinematic range accessible to both EIC and EicC, this observable could be thoroughly explored, paving the way for the first experimental determination of the Jaffe--Manohar quark OAM distribution.

\section{Recent updates from lattice QCD
}
\label{sec:latticeQCD}

Lattice QCD studies have explored 
GPDs for many years through form factors and generalized {FFs}, employing the {OPE}.
More recently, focus has shifted to direct calculations at the physical pion mass (see Ref.~\cite{Constantinou:2020hdm} for details).
Similar to {PDFs}, extracting information becomes increasingly difficult at higher Mellin moments and momentum transfers between initial and final states due to signal suppression.
However, advancements in PDF calculations have spurred significant progress in developing new methods to access the $x$- and $t$-dependence of GPDs ($t=-Q^2$).
Applying these new methods to GPD calculations in lattice QCD presents several challenges.
Unlike collinear PDFs, extracting $x$-dependent GPDs is more complex due to the required momentum transfer ($Q^2$) between the initial and final states.
Furthermore, GPDs are historically defined in the Breit frame, where momentum transfer is equally distributed between these states.
This necessitates separate, computationally expensive calculations for each momentum transfer value.
Despite these challenges, progress has been made in extracting $x$-dependent GPDs using lattice QCD. The most recent development is a new Lorentz-invariant parametrization that bypasses the need for a symmetric kinematic frame calculation. This methodology is already established theoretically and has been confirmed through numerical simulations {\cite{Bhattacharya:2022aob,Bhattacharya:2023jsc,Bhattacharya:2025yba,Chu:2025kew}}. 

The following sections provide an overview of the current status and recent developments in lattice QCD studies of GPDs and related quantities. 
Sec.~\ref{sec:lattice_Breit} summarizes pioneering lattice calculations in the Breit frame, and discusses their comparison with traditional form-factor analyses and the first tomographic reconstructions.
Sec.~\ref{sec:latt_syst} focuses on investigating systematic uncertainties, presenting renormalization-group and renormalon-resummed improvements that enhance the precision and convergence of LaMET extractions.
Sec.~\ref{sec:asymmetric} introduces the newly developed asymmetric-frame approach and its Lorentz-invariant formulation, which enable efficient determination of the $t$-dependence and access to higher Mellin moments within accessible computational resources.
Finally, Sec.~\ref{sec:latt_twist3} extends the discussion beyond the leading-twist approximation, highlighting recent lattice explorations of twist-3 PDFs and GPDs that probe quark–gluon correlations and transverse-spin structure.

\subsection{Breit-frame lattice GPD calculations  }
\label{sec:lattice_Breit}    

\begin{figure}[h!]
\centering
\includegraphics[scale=0.475]{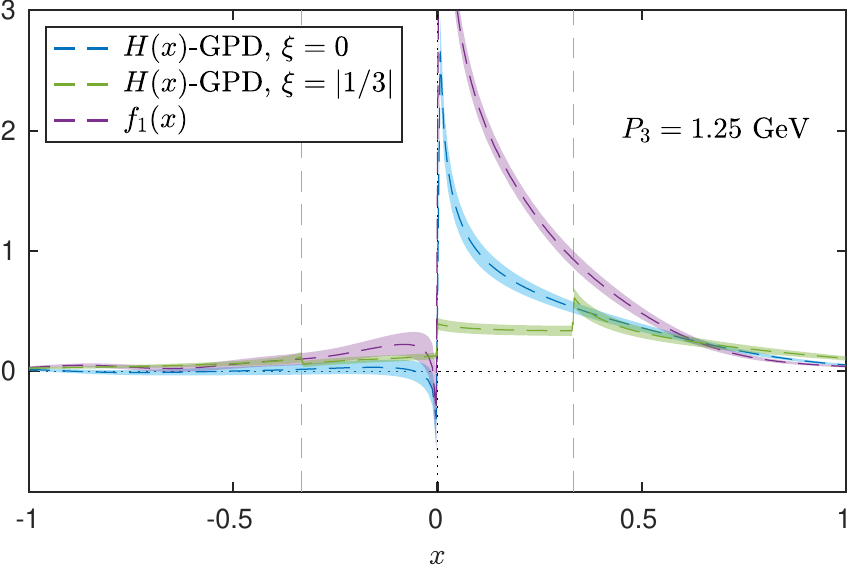}       \includegraphics[scale=0.475]{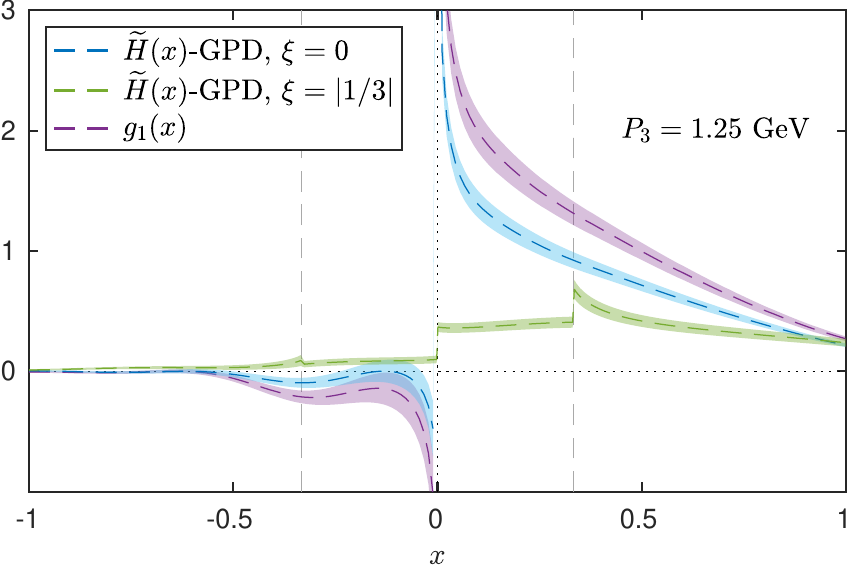}
\includegraphics[scale=0.475]{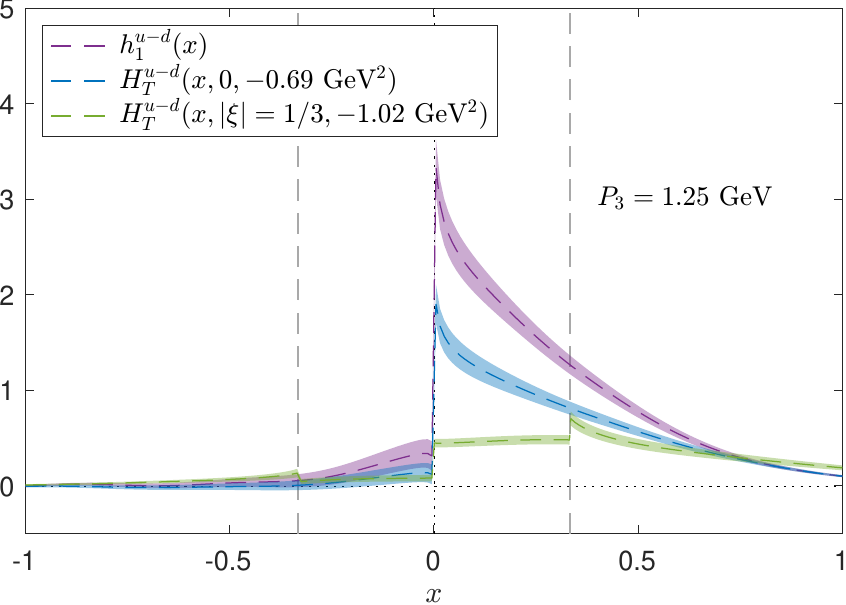}
\caption{{Top}: $H$-GPD for $\xi=0$ (blue band) and $\xi=|1/3|$ (green band), as well as the unpolarized PDF (violet band) for $P_3=1.25$~GeV. The area between the vertical dashed lines is the ERBL region. {Middle}: The same as the {top} panel, but for $\widetilde{H}$. {bottom}: The same as the {top} panel, but for $H_T$. This
figure is taken from Refs.~\cite{Alexandrou:2020zbe,Alexandrou:2021bbo}.}
\label{fig:ETMC_GPDs}
\end{figure}

The first lattice calculations of $x$-dependent GPDs used the LaMET approach, investigated the pion valence-quark GPD at zero skewness and multiple momentum transfers with a pion mass of $M_\pi \approx 310$~MeV~\cite{Chen:2019lcm}.
While these results showed reasonable agreement with traditional local-current form-factor calculations at similar pion masses, the large uncertainties prevented definitive conclusions about different models of the GPD's kinematic dependence.
Regarding the proton, the first calculation uses LaMET to study the Bjorken-$x$ dependence of isovector nucleon GPDs and presented the {vector} ($H$, $E$) and {axial-vector} ($\widetilde{H}$, $\widetilde{E}$) GPDs for both zero and nonzero skewness at three values of the momentum boost, using an ensemble of a pion mass of $M_\pi \approx 260$~MeV~\cite{Alexandrou:2020zbe}.
The results for nonzero skewness are important as $\xi\ne0$ gives rise to the nontrivial ERBL region, and, in addition, allows direct access to $\widetilde{E}$ GPD. 
{We note that the momentum-space matching becomes unreliable near $x=\xi$, preventing direct extraction of this region from LaMET. However, this limitation does not arise in the same way in short-distance factorization, where the GPD is constrained through coordinate-space information. This emphasizes the need of complementarity between the two approaches.}
The work of Ref.~\cite{Alexandrou:2020zbe} was extended to the {tensor} GPDs~\cite{Alexandrou:2021bbo} using the same ensemble. Fig.~\ref{fig:ETMC_GPDs} shows results from Ref.~\cite{Alexandrou:2020zbe,Alexandrou:2021bbo} for the three leading GPDs, $H,\,\widetilde{H},\,H_T$.

\begin{figure}[tb]
\includegraphics[width=0.45\textwidth]{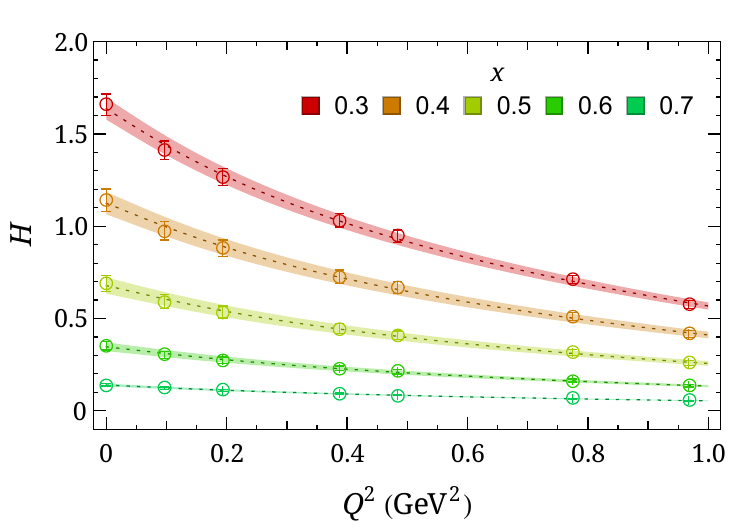}
\includegraphics[width=0.45\textwidth]{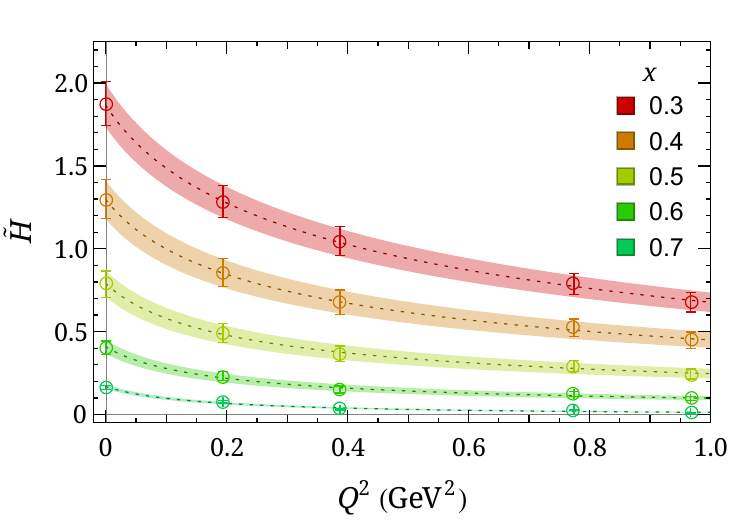}
\includegraphics[width=0.45\textwidth]{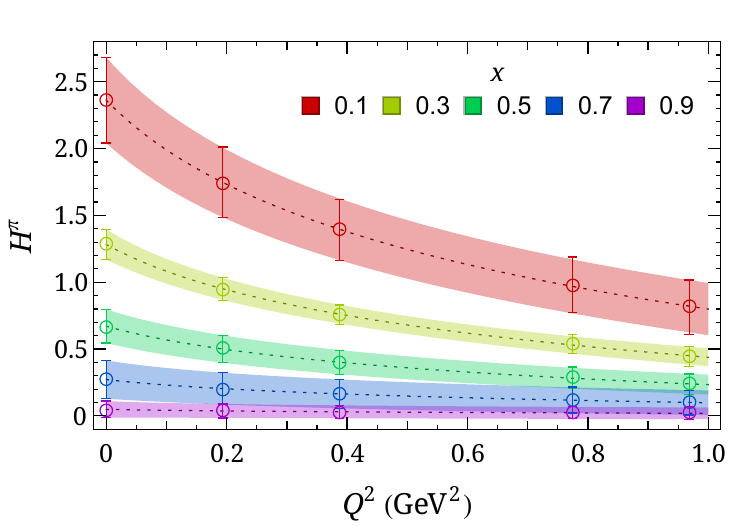}
\caption{
Nucleon isovector $H$ (left),  
and $\tilde{H}$ (middle) and pion valence-quark (right) GPDs
at $\xi=0$ with $z$-expansion to $Q^2$ at selected $x$ values. This
figure is taken from Refs.~\cite{Lin:2020rxa,Lin:2021brq,Lin:2023gxz}.
\label{fig:3DGPD}}
\end{figure}

Refs.~\cite{Lin:2020rxa,Lin:2021brq} reported the first lattice QCD calculations of {vector and axial-vector}
nucleon GPDs at the physical pion mass with boost momenta around 2.0~GeV and multiple momentum transfers.
Later, the first LQCD $x$-dependent pion valence-quark GPD calculated directly at physical pion mass using LaMET with next-to-next-to-leading order (NNLO) perturbative matching correction was also reported~\cite{Lin:2023gxz}.
The $z$-expansion was used to interpolate the $Q^2$ dependence of the light-cone GPD functions, and as shown in Fig.~\ref{fig:3DGPD}, the fit describes the lattice data well. The interpolated $Q^2$ dependence was used to make the first tomography study using lattice QCD calculations at a physical pion mass.

\begin{figure*}[tb]
\centering
\includegraphics[width=0.62\columnwidth]{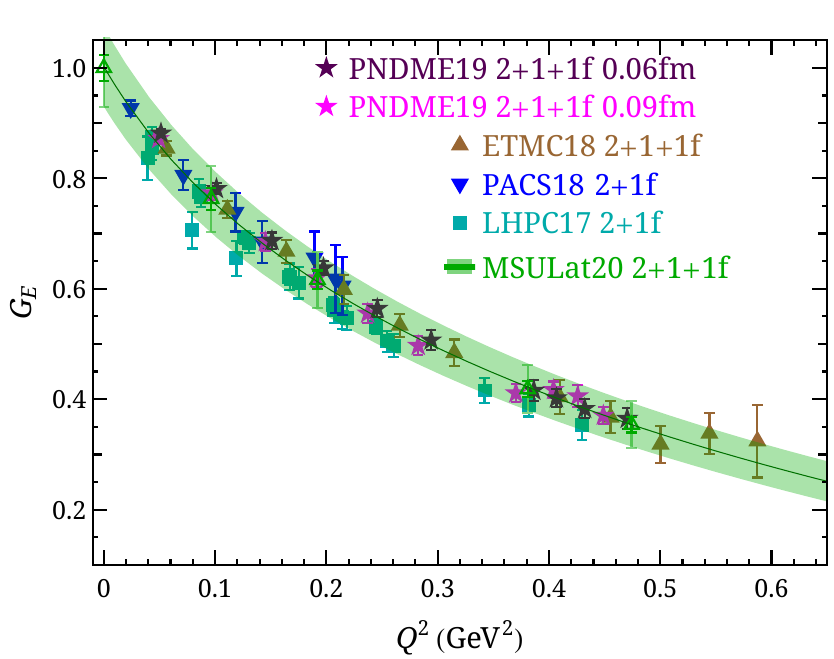} 
\includegraphics[width=0.62\columnwidth]{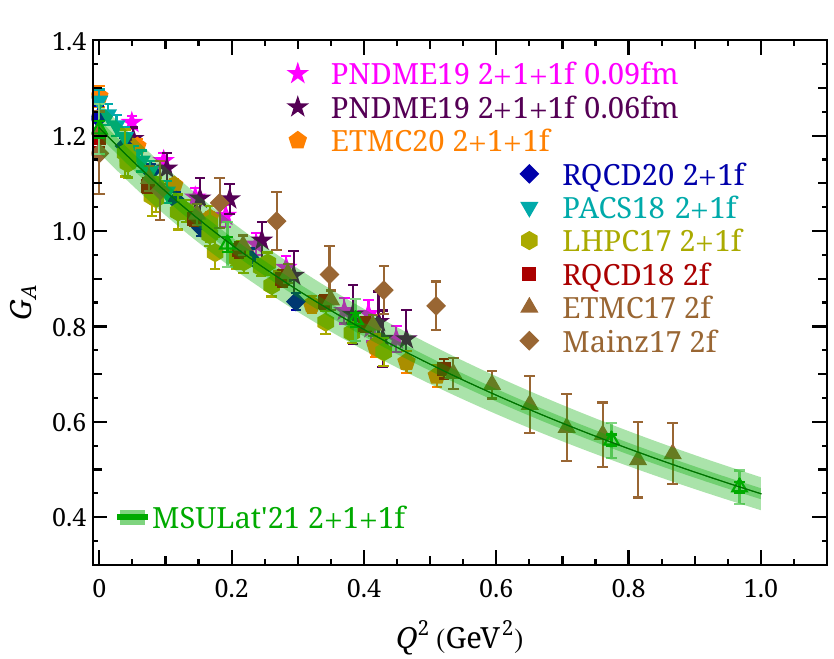}
\includegraphics[width=0.62\columnwidth]{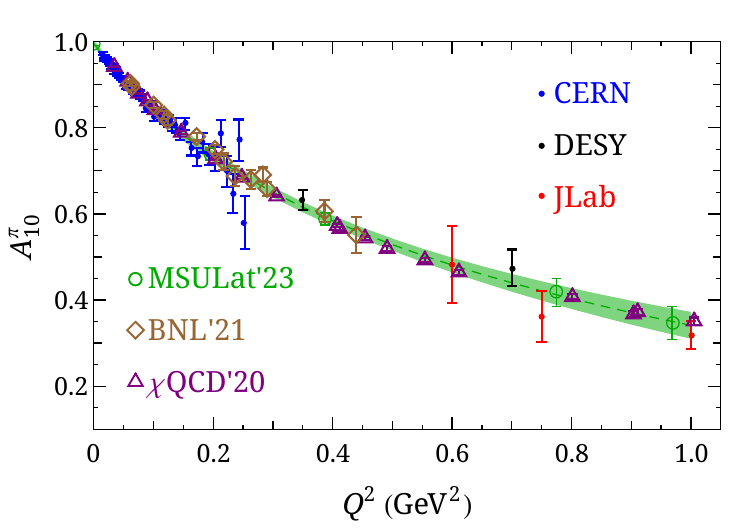}
\includegraphics[width=0.62\columnwidth]{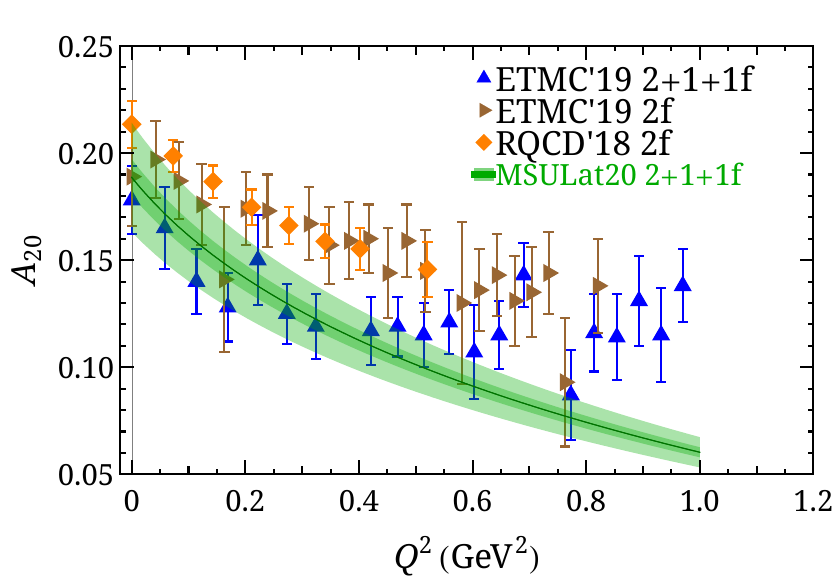}
\includegraphics[width=0.62\columnwidth]{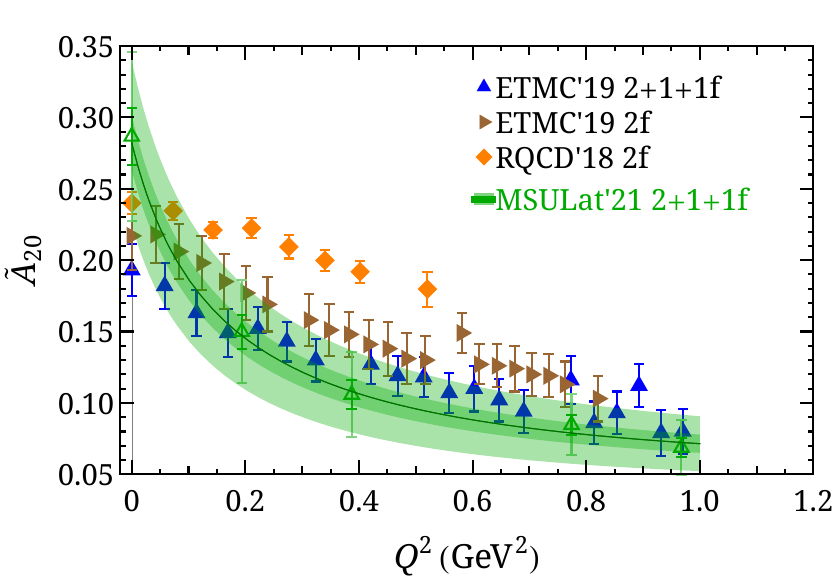}
 \includegraphics[width=0.62\columnwidth]{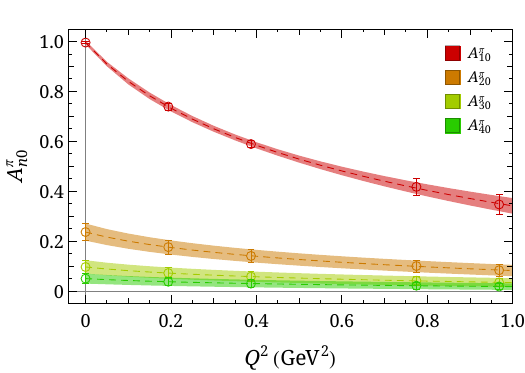}
\caption{
\label{fig:LatGFF}
(top) The nucleon isovector electric (left), axial (middle) and pion (right) form factor results obtained from $x$-dependent GPDs~\cite{Lin:2020rxa,Lin:2021brq,Lin:2023gxz} (labeled as green bands in the plots) as functions of transfer momentum $Q^2$, and comparison with other lattice works calculated near physical pion mass~\cite{Green:2014xba,Rajan:2017lxk,Hasan:2017wwt,Capitani:2017qpc,Alexandrou:2017hac,Bali:2018qus,Alexandrou:2018sjm,Jang:2018djx,Shintani:2018ozy,RQCD:2019jai,Alexandrou:2020okk,Wang:2020nbf,Gao:2021xsm}, and experimental data~\cite{NA7:1986vav,JeffersonLab:2008jve,JeffersonLab:2008gyl,Horn:2007ug, JeffersonLabFpi-2:2006ysh,JeffersonLabFpi:2000nlc} in the case of pion. 
(bottom) The nucleon isovector generalized FFs $\{A,B\}_{20}(Q^2)$ and $\tilde{A}_{20}(Q^2)$ obtained from $x$-dependent GPDs~\cite{Lin:2020rxa,Lin:2021brq} (labeled as green bands in the plots) 
compared with other lattice results calculated near physical pion mass as functions of {the momentum transfer $Q^2$}~\cite{Bali:2018zgl,Alexandrou:2019ali}.
In the rightmost plot, the lowest four {vector} pion generalized FFs $A_{n0}$ with $n\in[1,4]$ obtained from taking the moment integral using the pion GPD function~\cite{Lin:2023gxz}.
This
figure is taken from Refs.~\cite{Lin:2020rxa,Lin:2021brq,Lin:2023gxz}.
}
\end{figure*}

To validate the reliability of their lattice-derived tomography results, Refs.~\cite{Lin:2020rxa,Lin:2021brq,Lin:2023gxz} calculated moments by integrating the lattice-extracted $H$,
$\tilde{H}$
and $H^\pi$ GPDs over $x$. These moments were then compared with results from earlier lattice calculations that employed traditional form factors and generalized form factors at or near the physical pion mass.
The zero-skewness limit of the GPD function is related to the Mellin moments by taking the $x$-moments~\cite{Ji:1998pc,Hagler:2009ni}, which gives the nucleon {vector and axial-vector} generalized FFs
$A_{ni}(Q^2)$, 
$\tilde{A}_{ni}(Q^2)$, 
and pion generalized FFs $A_{ni}^\pi(Q^2)${.} 
When $n=1$ in the nucleon class, we get the Dirac and Pauli electromagnetic, and axial form factors $F_1(Q^2) = A_{10}(Q^2)$, $F_2(Q^2) = B_{10}(Q^2)$,
$G_A(Q^2) = \tilde{A}_{10}(Q^2)$;
{when} $n=2$, generalized FFs $A_{20}(Q^2)$, $B_{20}(Q^2)$ and $\tilde{A}_{20}(Q^2)$. Similarly for pion form factor $F_\pi$ ($n=1$) and GFFs $A_{n0}^\pi(Q^2)$.
The upper panels of 
Fig.~\ref{fig:LatGFF} 
present the nucleon and pion 
$n=1$ 
moments from $x$-dependent GPDs evaluated at the physical pion mass~\cite{Lin:2020rxa,Lin:2021brq,Lin:2023gxz}, 
together with earlier LQCD OPE determinations reported by other collaborations.
To enable comparison with a broader set of lattice-nucleon studies, the Sachs electric 
($G_E$) 
and magnetic 
($G_M$) 
form factors are reconstructed using the 
$F_{1,2}$ obtained via the $x$-integral in 
$G_E(Q^2)=F_1(Q^2)+q^2F_2(Q^2)/(2 M_N)^2$ 
and 
$G_M(Q^2)=F_1(Q^2)+F_2(Q^2)$.
In each panel, the green point and green band denote results from $x$-dependent GPDs, while the OPE-based calculations from other groups appear as markers in various colors.
Overall, the single–lattice-spacing results display good consistency across methods and collaborations. 
The lower panels of Fig.~\ref{fig:LatGFF} show the $n=2+$ moment from $x$-dependent GPDs alongside generalized FFs obtained using OPE approaches at or near the physical pion mass~\cite{Alexandrou:2019ali,Bali:2018zgl} for the nucleon.
Even within the same OPE framework and collaboration, the two data sets for $A_{20}$ exhibit noticeable tension,
indicating that the systematic effects for these generalized FFs can be intricate.
Since the blue points correspond to finer lattice spacing, larger spatial volume, and larger 
$M_\pi L$, 
they are expected to have reduced systematic uncertainties.
Ref.~\cite{Lin:2020rxa} moment 
$A_{20}(Q^2)$ 
agrees more closely with OPE-based results at smaller momentum transfer 
$Q^2$, 
whereas 
$B_{20}(Q^2)$ 
shows better agreement with OPE calculations at higher 
$Q^2$.
Comparing the 
$N_f=2$ data from Ref.~\cite{Alexandrou:2019ali} with the 
$N_f=2$ results from Ref.~\cite{Bali:2018zgl}, one finds agreement in the magnitudes of 
$A_{20}$ 
and 
$B_{20}$, 
but with differing slopes, likely reflecting distinct analysis choices and remaining systematics.
Both Refs.~\cite{Lin:2020rxa} and Ref.~\cite{Alexandrou:2019ali} determinations are based on a single ensemble. Future investigations incorporating additional lattice artifacts, such as explicit lattice-spacing dependence, will be important for fully accounting for the differences observed.

\begin{figure}[tb]
\begin{center}
\includegraphics[width=0.99\columnwidth]{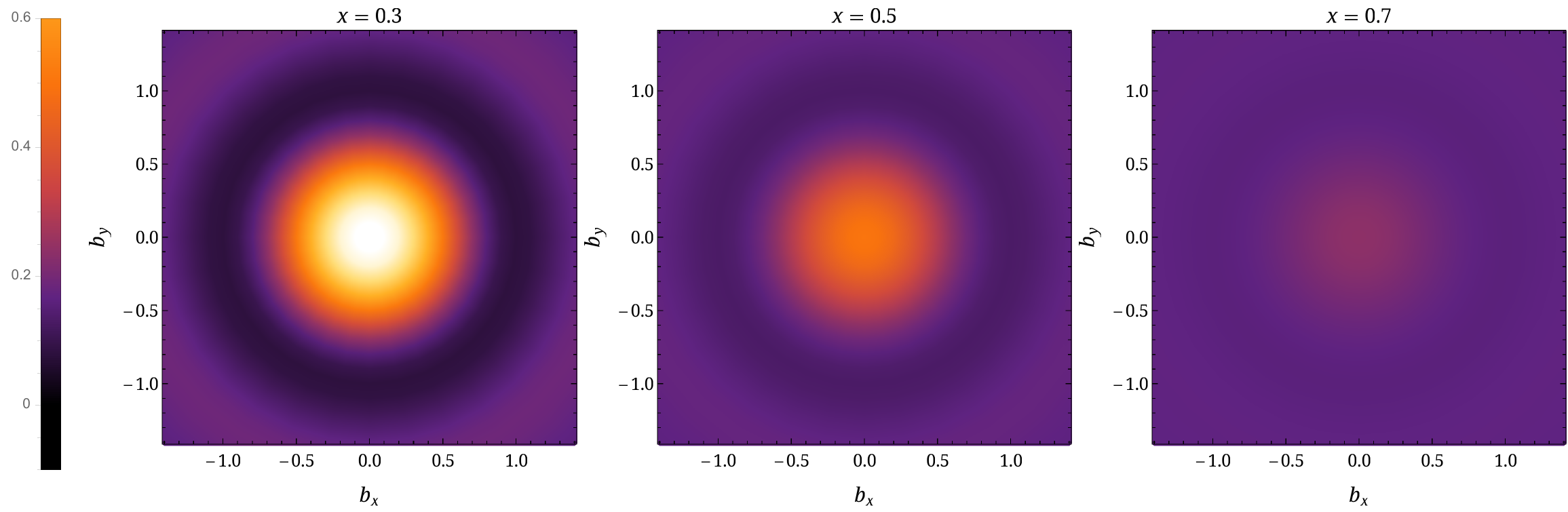}\\
\includegraphics[width=0.99\columnwidth]{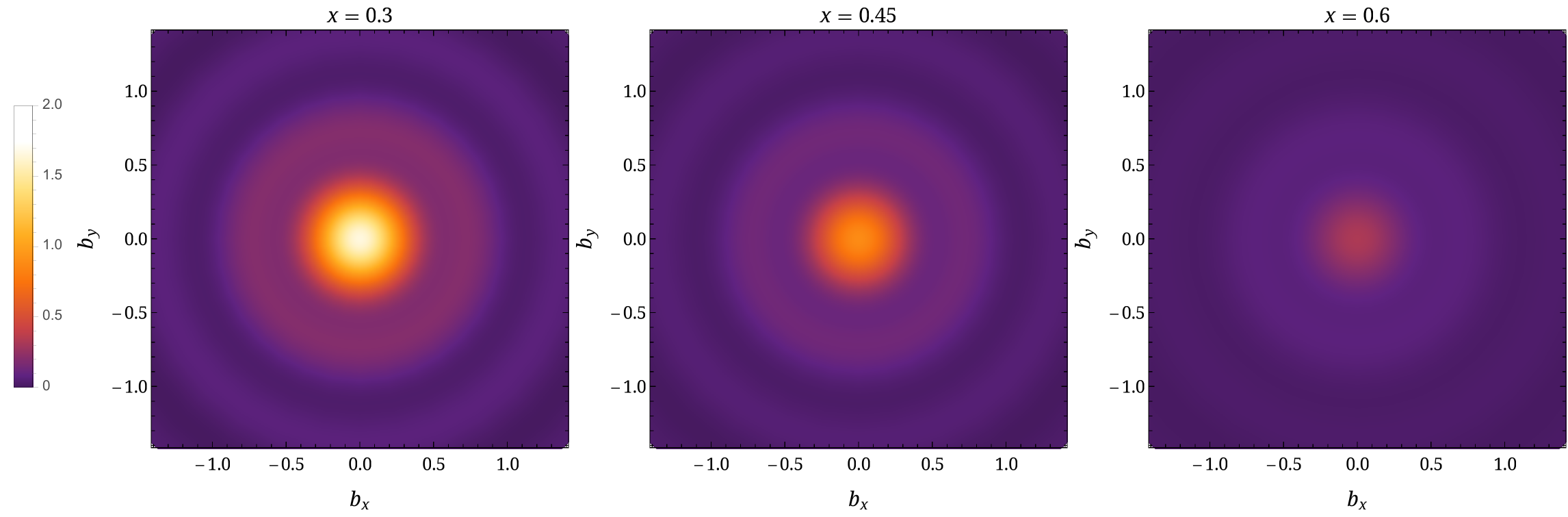}\\
\includegraphics[width=0.99\columnwidth]{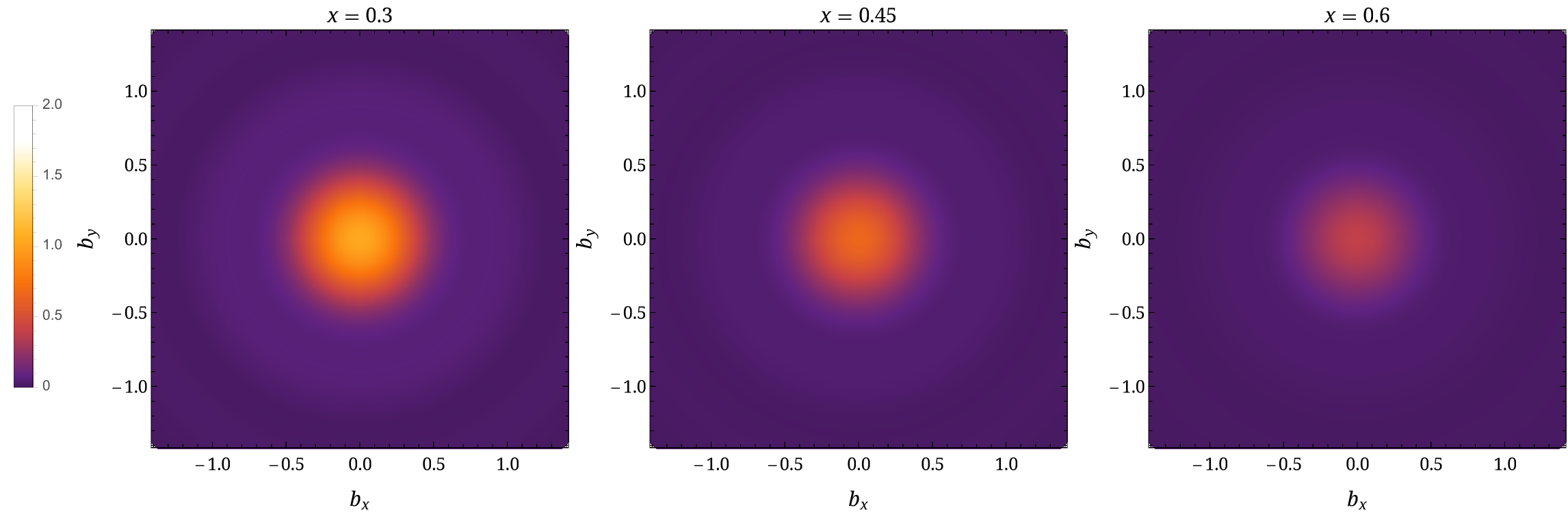}
\end{center}
\caption{
Unpolarized and polarized nucleon (upper and middle) and pion (bottom) tomography from lattice QCD: three-dimensional impact-parameter--dependent parton distribution as a function of $x$ and $b$ using lattice $H$, $\tilde{H}$ and $H^\pi$ GPD functions at physical pion mass. This
figure is taken from Refs.~\cite{Lin:2020rxa,Lin:2021brq,Lin:2023gxz}.}
\label{fig:b-density}
\end{figure}

Ref.~\cite{Lin:2020rxa} performed the Fourier transform of the
$Q^2$-dependent vector GPDs $H(x,\xi=0,Q^2)$ to obtain impact-parameter–dependent distributions~\cite{Burkardt:2002hr}
\be 
&&
\{q,
\Delta q
,q^\pi \}(x,\mathbf{b}) \nn \\ && =
\int \frac{ d^2{\mathbf{\Delta}}}{(2\pi)^2} \{H,\tilde{H},H^\pi\}(x,\xi=0,Q^2=
\mathbf{\Delta}^2) e^{i\mathbf{\Delta}\,\cdot \, \mathbf{b} },  \nn \\ &&
\label{eq:impact-dist}
\ee
with $b \equiv |\mathbf{b}|$ indicating the transverse distance from the center of momentum.
Fig.~\ref{fig:b-density} displays the first LQCD results for the impact-parameter–dependent 2D distributions at $x=0.3$, 0.45 and 0.6 for unpolarized and polarized nucleon and pion.
These distributions describe the probability density of a parton with momentum fraction $x$ located at transverse distance $b$, 
providing the first $x$-dependent nucleon tomography from LQCD.
Similar tomography results for the {axial-vector} GPD, $\tilde{H}(x,\xi=0,Q^2)$, can be found in Ref.~\cite{Lin:2021brq}, as well as the pion~\cite{Lin:2023gxz}.

\subsection{Investigation of methodology-specific systematic uncertainties} 
\label{sec:latt_syst}
Since the numerical studies discussed above, the LaMET framework has seen advances including renormalization-group resummation (RGR)~\cite{Su:2022fiu} and leading–renormalon resummation (LRR)~\cite{Zhang:2023bxs}.
RGR is designed to resum logarithms arising from the mismatch between the intrinsic physical scale and the final renormalization scale of the parton.
The strategy is to choose an intermediate scale at which the logarithms vanish and then evolve to the target scale using the renormalization group.
This can be applied both to the renormalization of the bare matrix elements and to the perturbative matching.
LRR targets divergences associated with the infrared renormalon (IRR)~\cite{Zichichi:1979gj}, whose impact can remain even when using RGR alone.
Its first application--to the pion PDF~\cite{Zhang:2023bxs}--demonstrated that the combination of LRR with RGR substantially reduces systematic effects.
These improvements have been employed for {DA}s~\cite{Holligan:2023rex} and PDFs~\cite{Su:2022fiu,Zhang:2023bxs,Holligan:2024umc,Gao:2023ktu}, and so far have been applied to nucleon isovector $H$ and $E$ GPDs from Ref.~\cite{Holligan:2023jqh} on a physical-pion-mass ensemble, covering momentum transfers $Q^2=[0, 0.97]~\text{GeV}^2$ at $\xi=0$ and $Q^2
=
0.23\text{ GeV}^2$ at $\xi=0.1$, renormalized in the $\overline{ \rm MS}$ scheme at $\mu=2.0$~GeV, with two- and one-loop matching, respectively.

\begin{figure*}[htp]
 \centering
 \includegraphics[width=0.45\linewidth]{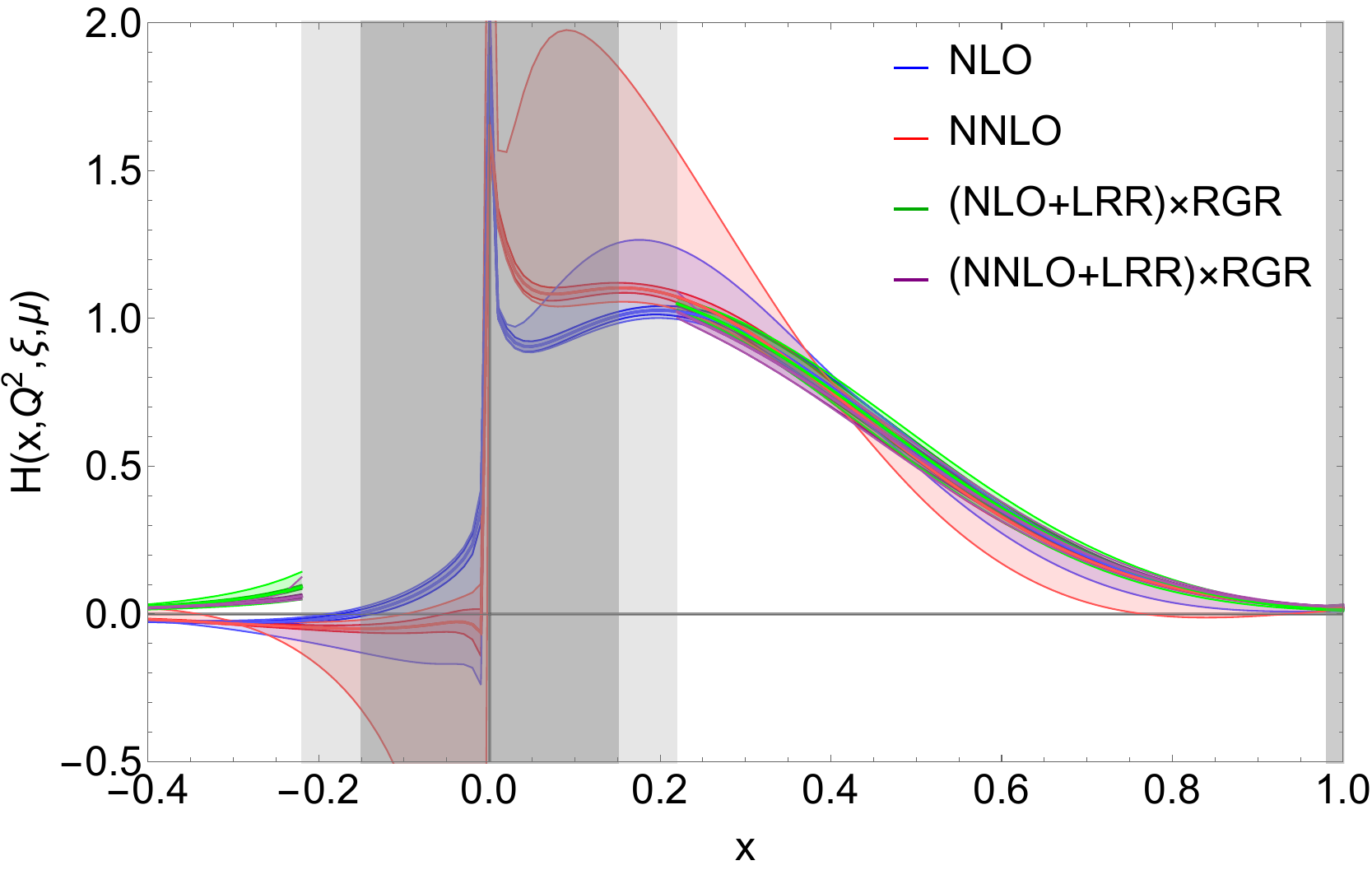}
 \includegraphics[width=0.45\linewidth]{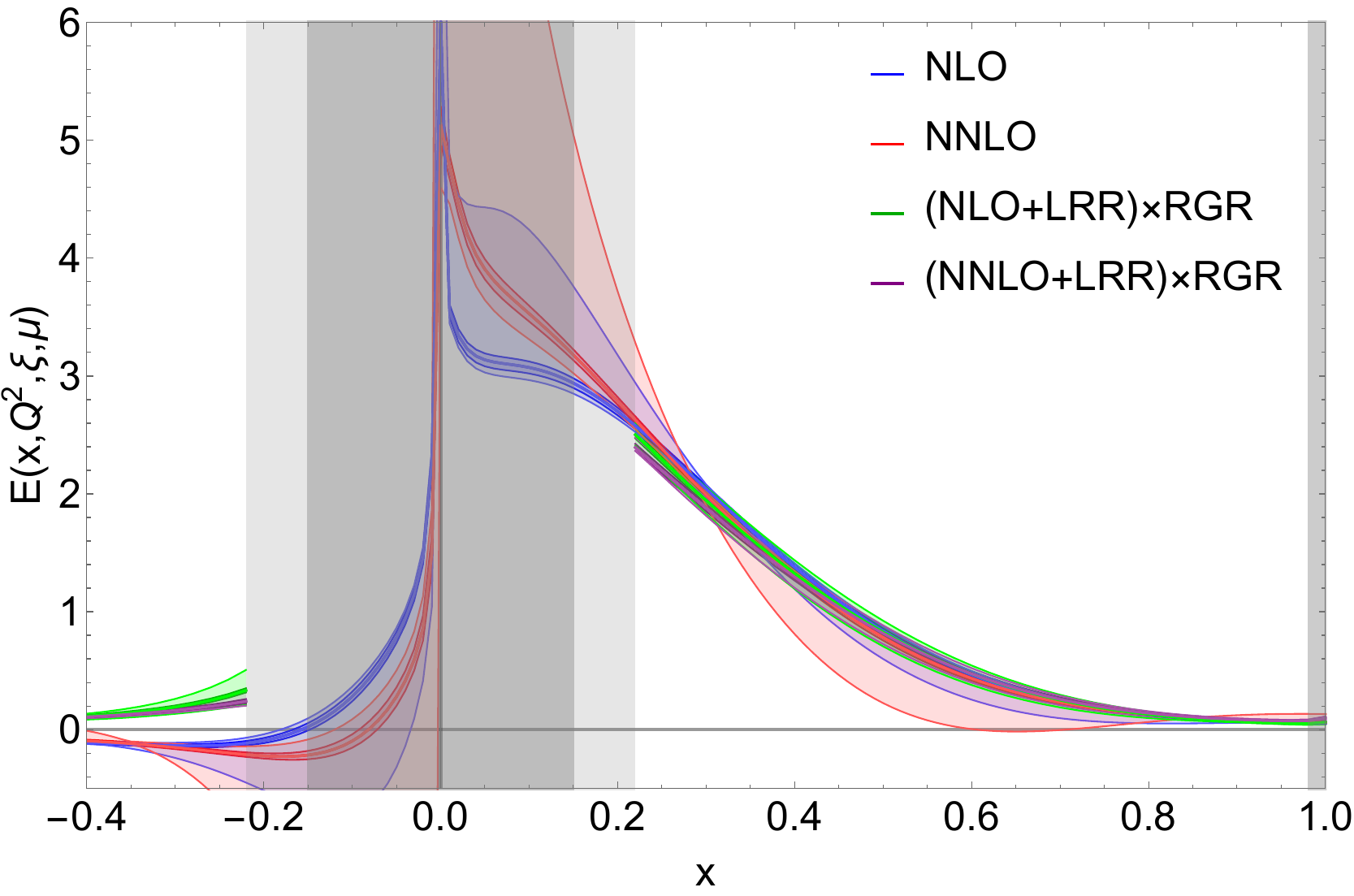}
 \caption{
 Lightcone {nucleon isovector} $H$ and $E$ GPDs (left and right, respectively) with NLO (blue), NNLO (red), NLO+LRR  (green) and  NNLO +LRR (purple) evaluated at $Q^2=0.39\text{ GeV}^2$ and $\xi=0$~\cite{Holligan:2023jqh}.
 The inner bands are statistical errors;
 the outer bands are combined statistical and systematic errors, derived from the scale variation.
 The dark-gray regions are the $x$-values at which the LaMET calculation breaks down.
 In addition, when RGR is applied, the matching formula breaks down for 
$|x|\lesssim 0.2$, 
 which is shaded in light gray. This
figure is taken from Ref.~\cite{Holligan:2023jqh}.}
 \label{fig:xi0GPD-Q2-0p39}
\end{figure*}

The first NNLO GPD analysis incorporating RGR~\cite{Su:2022fiu} together with LRR was reported in Ref.~\cite{Holligan:2023jqh}.
Fig.~\ref{fig:xi0GPD-Q2-0p39} shows representative results for the {nucleon vector-isovector}
GPDs 
$H$ 
and 
$E$  
at 
$Q^2=0.39\text{ GeV}^2$~\cite{Holligan:2023jqh}.
The central values across all four schemes are broadly consistent, indicating that the main benefit of RGR and LRR is a reduction in systematic uncertainties and improved convergence in the matching procedure, as reflected by the close proximity of the NLR and NNLR results.
While moving from NLO to NNLO tends to increase the systematic bands, applying LRR (NLR $\rightarrow$ NNLR) mitigates this growth,
underscoring the need for systematic control to keep pace with higher-order matching and renormalization.
Similar improvements are also observed at nonzero skewness in the same study; more detailed results for other momentum transfers and the $\xi\neq 0$ case can be found in Ref.~\cite{Holligan:2023jqh}.

\subsection{New lattice GPD results from asymmetric frame }
\label{sec:asymmetric}

As discussed above, calculations of GPDs on the lattice were initially performed in the Breit frame, wherein the momentum transfer is symmetrically distributed between the initial and final hadron state. In such a setup, the all-to-all propagator required for the determination of the appropriate matrix elements needs to be separately computed for each value of 
$t$.
Thus, the full mapping of 
$t$-dependence of the distribution poses a formidable computational task.
However, recently a new frame of reference was introduced~\cite{Bhattacharya:2022aob} 
to circumvent this problem, {\it i.e.}, the asymmetric frame, which ascribes the whole momentum transfer to the source state. Since the sink momentum is then fixed, several values of 
$t$ 
can be obtained in a single calculation. Below, we outline the principles of this approach and discuss some selected results.

The method relies on parameterizing the underlying matrix elements in terms of Lorentz-invariant (LI) amplitudes. For example, in the vector case for a spin-$1/2$ hadron, the matrix element can be parametrized as
\be
\label{eq:parametrization_vector}
&&
F^{\mu}(z,P,\Delta)  = \bar{u}(p_f,\lambda') \bigg [ \dfrac{P^{\mu}}{M} A_1 + M z^{\mu} A_2 \nn \\ && + \dfrac{\Delta^{\mu}}{M} A_3 + i M \sigma^{\mu z} A_4 + \dfrac{i\sigma^{\mu \Delta}}{M} A_5 
+ \dfrac{P^{\mu} i\sigma^{z \Delta}}{M} A_6
\nonumber \\
&& 
+ M z^{\mu} i\sigma^{z \Delta} A_7 + \dfrac{\Delta^{\mu} i\sigma^{z \Delta}}{M} A_8  \bigg ] u(p_i, \lambda),
\ee
where the amplitudes are denoted by $A_i$ ($i=1,\ldots,8$), $p_{i/f}$ are hadron's initial/final momenta, and $\lambda/\lambda'$ stand for the corresponding polarization variables, $P=(p_i+p_f)/2$ is the average momentum, $\Delta=p_f-p_i$ -- momentum transfer, $M$ -- hadron's mass, and $z$ -- Wilson line's length in the insertion operator.
While Ref.~\cite{Bhattacharya:2022aob} 
handled only the vector case, further work extended this to the axial vector 
\cite{Bhattacharya:2023jsc} 
and tensor 
\cite{Bhattacharya:2025yba} 
sectors.
These two additional parameterization contain, respectively, 
8 amplitudes 
$\widetilde{A}_i$ 
and 12 amplitudes 
$A_{Ti}$.
GPDs can then be expressed in terms of these amplitudes. The formalism is also available for the pion {vector} GPD, and a recent implementation can be found in Ref.~\cite{Ding:2024saz}.

As hinted above, the primary motivation for introducing this formalism was computational efficiency. However, an important byproduct is the development of alternative definitions of quasi-GPDs {\cite{Bhattacharya:2022aob}} that can be employed, which still converge to the same light-cone GPDs upon matching.
For example, apart from the standard (frame-dependent) definition of 
$H/E$ 
{vector} GPDs, it was shown that an alternative LI definition can be used, with modified contributions of some amplitudes and/or contributions of some of them vanishing altogether. This means that some amplitudes have contributions that start at a higher order than the leading twist, {\it i.e.}, different definitions of quasi-GPDs have nonequivalent explicit power corrections. The latter would suggest that alternative definitions may evince faster convergence to physical GPDs due to the absence of explicit power corrections. However, there are also implicit power corrections in amplitudes that contribute to twist-2. Hence, convergence must be established numerically for each case, since it is dictated by the underlying non-perturbative dynamics. For example, Refs.~\cite{Bhattacharya:2022aob,Cichy:2023dgk} has shown that the rate of convergence of the 
$H$ 
GPD is basically the same for the standard definition and the LI one, while the latter is significantly faster-converging for 
$E$.

\begin{figure}[t!]
    \centering
\includegraphics[scale=0.43]{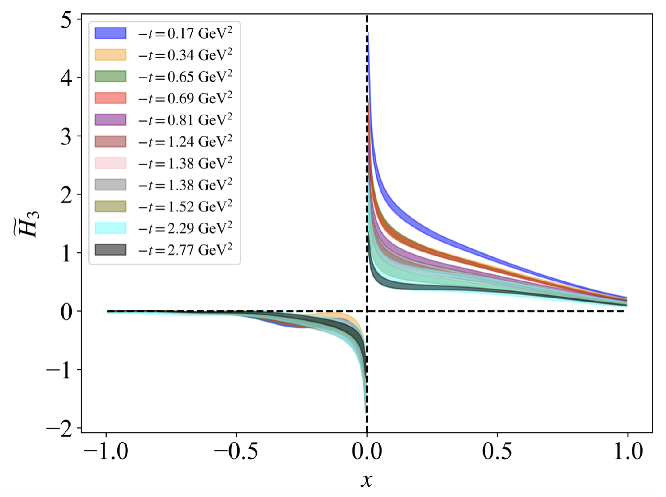}
\includegraphics[scale=0.37]{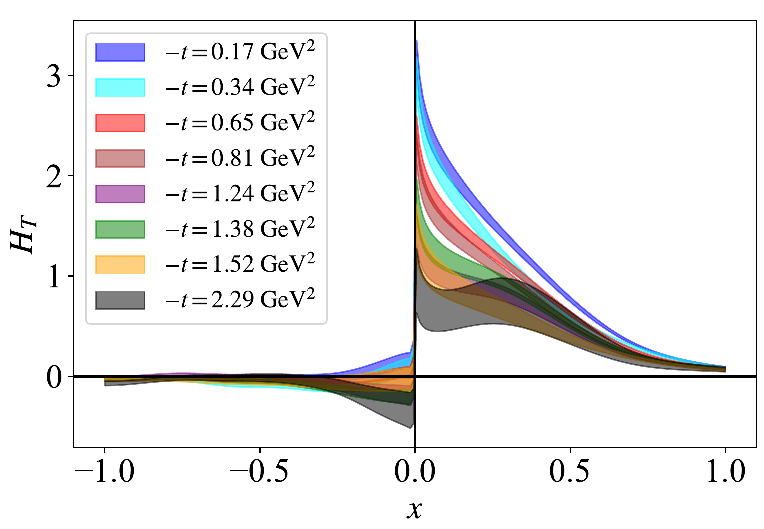}
    \caption{The momentum-transfer squared dependence of lattice-extracted GPDs, all given in the $\overline{\rm MS}$ scheme at 2~GeV and obtained on the lattice with $P_3=1.25$~GeV.
    Top: {axial-vector} GPD $\widetilde{H}$ (denoted by $\widetilde{H}_3$ to reflect the usage of the standard definition); results extracted from the symmetric frame 
    ($-t=0.69,1.38,2.77$~GeV$^2$) and the asymmetric frame (all other values of $-t$)~\cite{Bhattacharya:2023jsc}.
    Bottom: the {tensor} GPD $H_T$, only the asymmetric frame. This
figure is taken from Refs.~\cite{Bhattacharya:2022aob,Bhattacharya:2025yba}.}
    \label{fig:GPDs_asymmetric}
\end{figure}

The numerical work accompanying the development of the general amplitudes-based framework for vector, axial vector, and tensor GPDs concentrated on establishing proofs-of-principle of the methodology, {\it i.e.}, on numerical checks of the frame independence of amplitudes and on the feasibility of obtaining data in a wide range of $t$ values, in the zero-skewness case.
The former was established by comparisons of the amplitudes extracted from the asymmetric frame and from the standard symmetric frame (with a separate lattice calculation).
The latter, in turn, resulted in determining the $t$-dependence of GPDs in a range between around $-t=0.17$~GeV$^2$ and 
$-t=2.29$~GeV$^2$.
While such large values of $-t$ may be significantly affected by $t$-dependent power corrections, they pose no difficulty for the lattice from the point of view of the signal.
As an example, Fig.~\ref{fig:GPDs_asymmetric} shows the $t$-dependence of nucleon's $H$-type polarized GPDs -- {axial-vector} $\widetilde{H}$~\cite{Bhattacharya:2023jsc} and {tensor} $H_T$~\cite{Bhattacharya:2025yba}.
Both were obtained in the same setup of $N_f=2+1+1$ twisted-clover fermions, at the {lattice spacing} $a\approx0.093$ fm {(translating to a UV cutoff $hc/a$ of around 2.1 GeV)}, 260 MeV pion mass, and $L\approx3$ fm as spatial extent.
The nucleon was boosted to $P_3\approx1.25$~GeV and the results were matched to light-cone GPDs.
In the {axial-vector} case, the GPDs shown are from the standard definition (denoted $\widetilde{H}_3$; converging slightly faster than the proposed alternative one), and the plot also includes three values of $-t$ obtained in the symmetric frame.
The {tensor} GPD $H_T$ is shown using the alternative LI definition and only data from the asymmetric frame.
Very recently, the zero-skewness asymmetric-frame formalism was extended to $\xi\neq0$~\cite{Chu:2025kew}.

\begin{figure}[t!]
    \centering
\includegraphics[width=0.99\columnwidth]{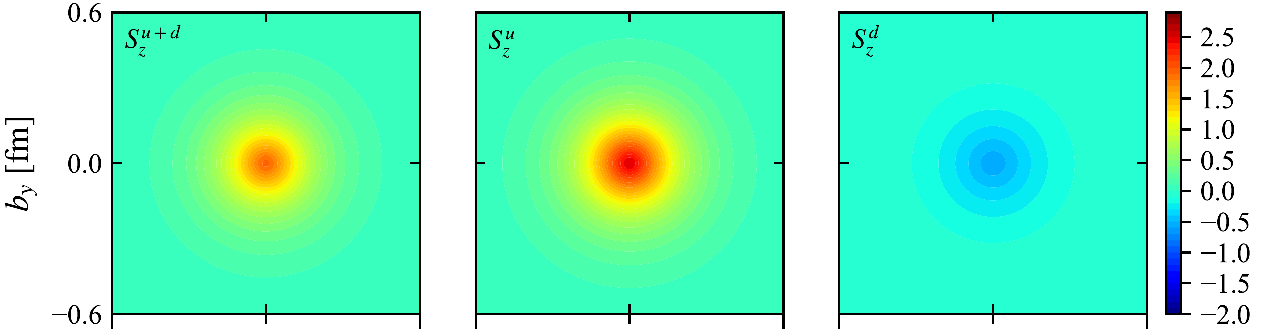}
\includegraphics[width=0.99\columnwidth]{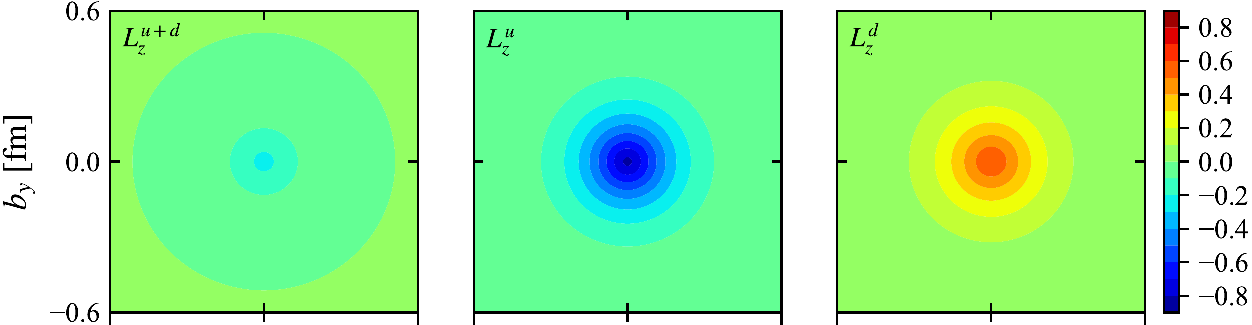}
\includegraphics[width=0.99\columnwidth]{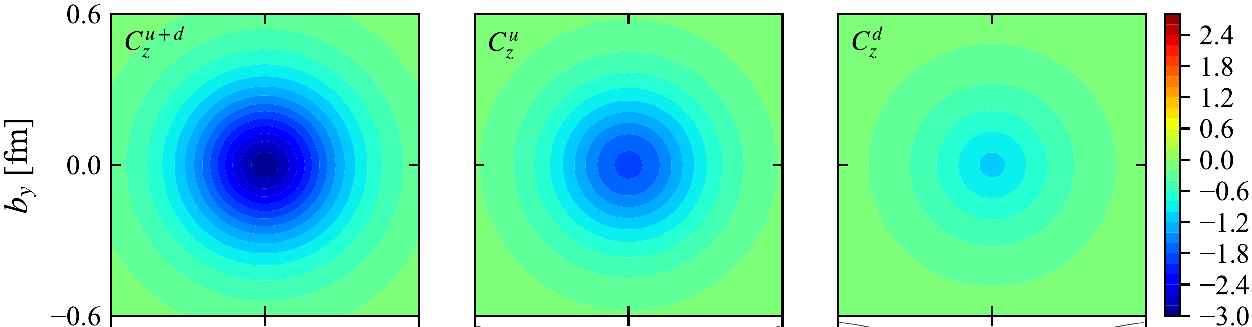}
    \caption{Tomographic pictures of: light quark helicity density (top), light quark orbital angular momentum (middle), and light quark spin-orbit correlation density (bottom). Left for $u+d$, middle for $u$, and right for $d$. This
figure is taken from Ref.~\cite{Bhattacharya:2024wtg}.}
    \label{fig:GPDs_asymmetric_moments}
\end{figure}

The framework of asymmetric frames was also employed to extract Mellin moments of GPDs in the {vector}~\cite{Bhattacharya:2023ays} and {axial-vector}~\cite{Bhattacharya:2024wtg} cases.
This thread used short-distance factorization based on non-local operator product expansion.
Thanks to the latter, one can reach high orders of the Mellin moments, limited only by statistical precision.
In practice, moments up to $A/B/\widetilde{A}_{5,0}$ could be extracted before the values become suppressed below statistical errors ($\widetilde{E}$ and, thus, $\widetilde{B}_{n,0}$ cannot be accessed at zero skewness due to a vanishing kinematic coefficient).
Note that this is already a much higher order than can be reached with standard lattice methodology from before the advent of novel techniques employing non-local operators.
As an example, Fig.~\ref{fig:GPDs_asymmetric_moments} displays tomographic pictures of the moments~\cite{Bhattacharya:2024wtg}, obtained upon Fourier transform of the $t$-dependence to impact-parameter space.
In the upper panel, the light quark helicity density is shown, related to the first moment of the $\widetilde{H}$ {axial-vector} GPD.
The middle panel presents the orbital angular momentum of $u$ and $d$ quarks, expressed by the difference of second moments of {vector} $H$ and $E$ GPDs (giving the total angular momentum) and the helicity density.
Finally, the bottom panel shows the light quark spin-orbit correlation density, {\it i.e.}, the difference of the second moment of $\widetilde{H}$ and the first moment of $H$, upon neglecting the mass-suppressed{,} and irrelevant at this level of precision{,} contribution from two of the {tensor} GPDs. Another recent work on extracting moments of proton GPDs can be found in Ref.~\cite{HadStruc:2024rix}, as well as for the pion $H$ GPD~\cite{Gao:2025inf}.

\begin{figure*}[t!]
    \centering
\includegraphics[width=0.93\textwidth]{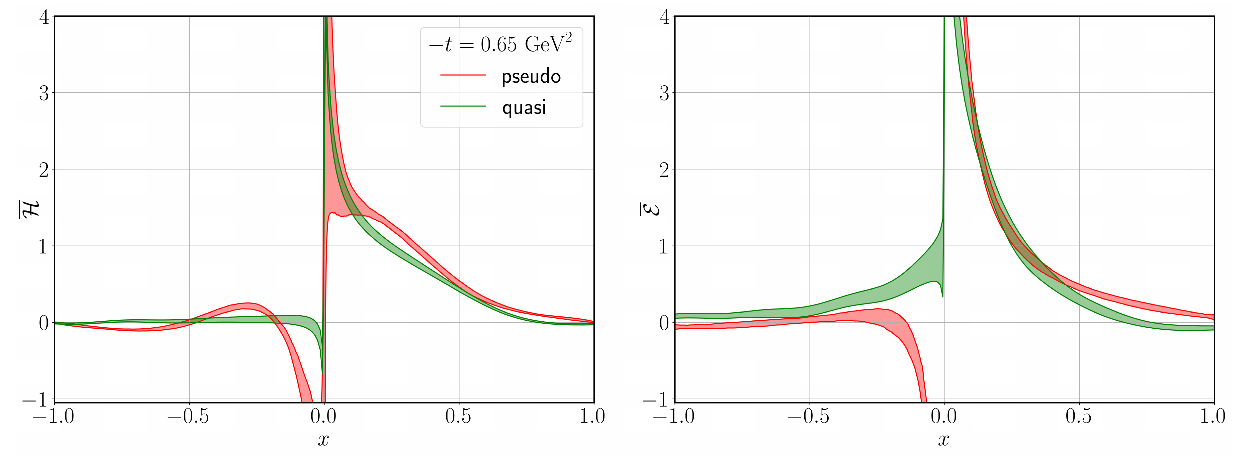}
    \caption{{Vector GPDs} $H$ (left) and $E$ (right) reconstructed using the pseudo-distribution (red bands) and the quasi-distribution approaches (green bands). In both cases, $-t=0.65$~GeV$^2$. This
figure is taken from Ref.~\cite{Bhattacharya:2024qpp}.}
    \label{fig:GPDs_quasi_pseudo}
\end{figure*}

Short-distance factorization can also be used to obtain matched Ioffe-time distributions (rather than moments) and reconstruct the $x$-dependence of GPDs, often referred to as the pseudo-GPD approach.
The same lattice data as used in Ref.~\cite{Bhattacharya:2022aob}, complemented with additional calculations at other values of $P_3$, {were} input in this case~\cite{Bhattacharya:2024qpp}.
The reconstruction of $x$-dependence followed by imposing a fitting Ansatz for the light-cone GPDs, having carefully assessed a range of Wilson line lengths that can be treated as short distances.
It is then natural to compare the outcome of this approach with that from quasi-distributions (factorized in momentum space), as shown in Fig.~\ref{fig:GPDs_quasi_pseudo}.
In the quark part ($x>0$), we observe qualitatively similar dependence for both GPDs, with quantitatively largest differences around $x=0.25$ for $H$ and $x\gtrsim0.5$ for $E$.
The antiquark part ($x<0$) yields no definite conclusions and should be considered unreliable. Obviously, systematic effects are different in both approaches, and notably, the use of the fitting Ansatz in the pseudo-GPD method may introduce a model bias. Thus, a more promising prospect is to combine the quasi- and pseudo-distribution methods in a single analysis, as argued in Ref.~\cite{Ji:2022ezo}, rather than using them independently, as discussed below.

\begin{figure}[t!]
    \centering
\includegraphics[width=0.99\columnwidth]{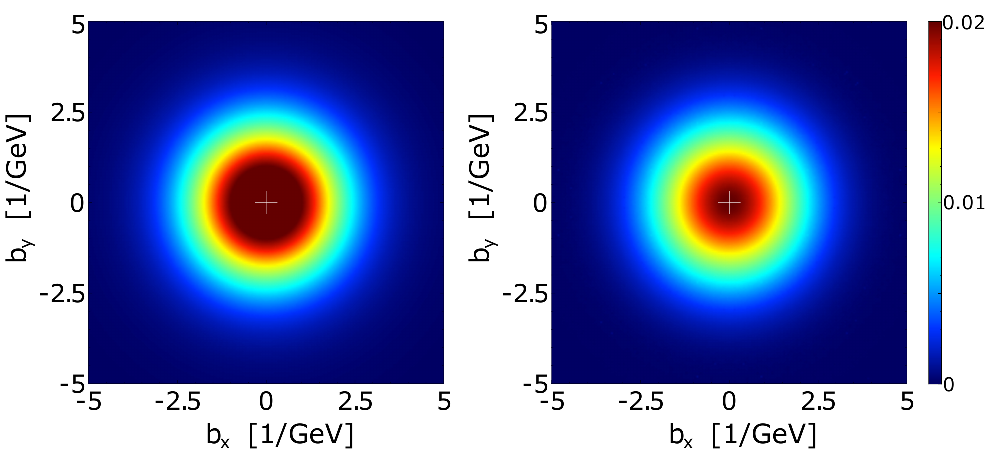}\\
\includegraphics[width=0.99\columnwidth]{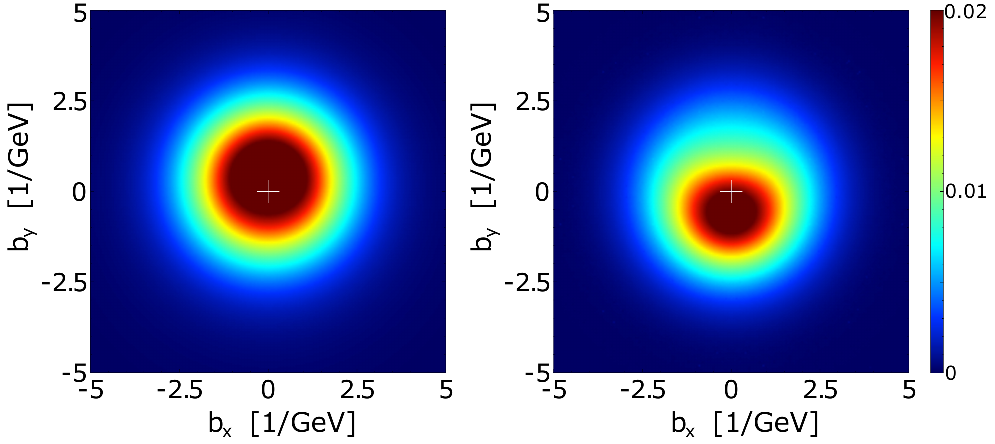}
\caption{Examples of tomographic pictures for an unpolarized (top) and a transversely polarized (bottom) proton at $x=0.2$, for valence up (left) and valence down (right) quarks.
The color scale is the same in all plots, and the up quark plots are rescaled by 2 for better comparison. This
figure is taken from Ref.~\cite{Cichy:2024afd}.}
    \label{fig:GPDs_lat_tomography}
\end{figure}

Another application of asymmetric-frame data is an exploratory work on integrating lattice results with experimental data for elastic scattering~\cite{Cichy:2024afd}.
In this approach, a specific ratio was defined to reduce the contamination of lattice data with previously unaccounted-for systematic effects.
Using such a ratio in conjunction with experimental data, tomographic images of the nucleon were extracted, along with information on the total angular momentum carried by valence quarks.
In addition, a new type of a ``shadow term'' was defined and shown to be useful in estimating model uncertainties and studying nucleon tomography beyond the standard bell shape. 
An example of tomographic pictures is shown in Fig.~\ref{fig:GPDs_lat_tomography}.

The final and most recent application is a new framework based on artificial neural networks (ANNs)~\cite{Chu:2025jsi}.
ANNs are used as a tool for a robust reconstruction of the 
$x$ dependence and the $t$ dependence, currently at zero skewness. Moreover, the ANN framework is employed to combine information from quasi- and pseudo-distribution approaches, exploiting their complementary strengths that originate from the factorization of the lattice object into its light-cone counterpart in either momentum or coordinate space.

The introduction of the asymmetric-frame approach significantly boosted the prospects of extracting GPDs from the lattice.
The formalism yields significantly more efficient simulations in terms of computational time and is flexible with respect to the extracted quantities.
There is also no obstacle in handling higher-twist GPDs, see the following subsection. Obviously, future work needs to analyze different sources of systematic effects that may be hidden at various steps of the multi-step procedure to go from bare lattice data to final distributions and physical observables, {\it e.g.}, those related to angular momentum or mechanical properties.

\subsection{Beyond leading twist }
\label{sec:latt_twist3}

\begin{figure}[h!]
  \hspace*{-0.2cm} 
\centering
  \includegraphics[scale=0.48]{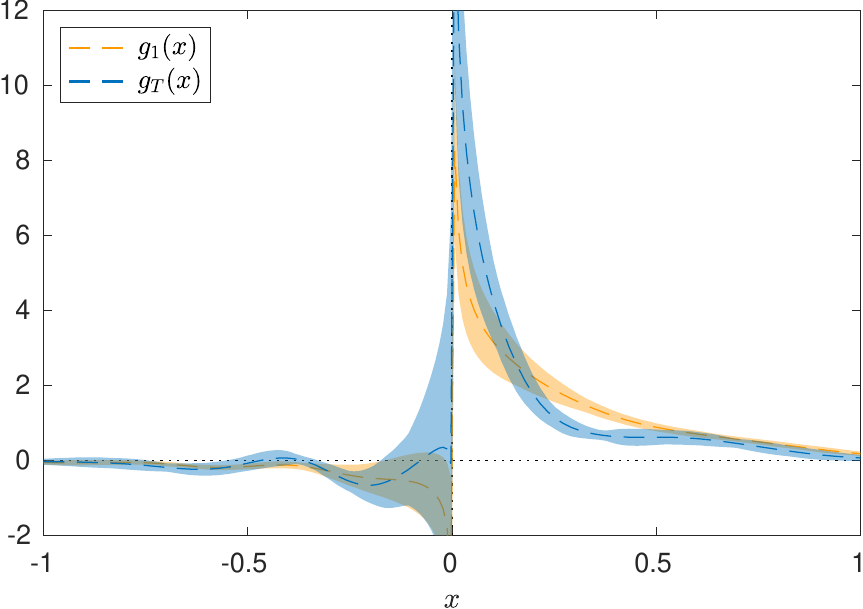}
 	\caption{Comparison of $x$-dependence of $g_T$ (blue band) and $g_1$ (orange band) at $P_3=1.67$~GeV. This
figure is taken from Ref.~\cite{Bhattacharya:2020cen}.}
 	\label{fig:gT_g1}
 \end{figure}

\begin{figure}[h!]
    \centering
    \includegraphics[scale=0.48]{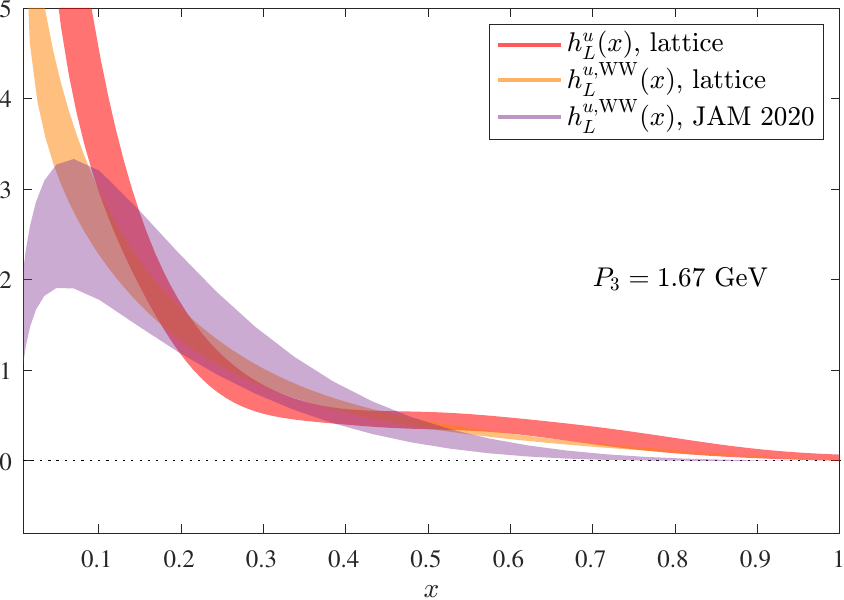}\hspace{0.1cm}
   \includegraphics[scale=0.48]{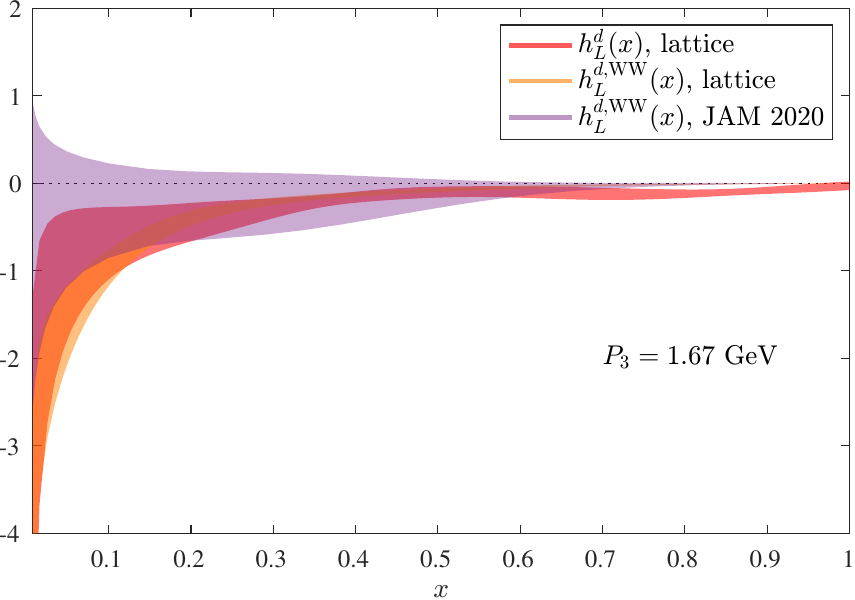} 
\caption{Test of the Wandzura-Wilczek approximation for up (left panel) and down (right panel) distributions, at $P_3=1.67$~GeV. For the separate flavors, we show $h_L(x)$ (red) with $h_L^{\rm WW}(x)$ (orange) extracted from lattice QCD. Results for $h_L^{\rm WW}(x)$ from the JAM collaboration~\cite{Cammarota:2020qcw} (violet) are also included for comparison. This
figure is taken from Ref.~\cite{Bhattacharya:2021moj}.}   
\label{fig:hL_WW_u_d}
   \end{figure}

The so-far discussion in this chapter is restricted to the leading twist (twist-2) of GPDs from lattice QCD. However, the twist-3 contributions, although lacking a probabilistic interpretation, encode a critical insight into quark-gluon correlations and quark transverse spin structure. Lattice QCD has explored the first $x$-dependent results for two twist-3 PDFs using the LaMET approach on $N_f=2+1+1$ twisted mass clover fermion ensembles (pion mass of 260 MeV), with momentum boosts of 0.83, 1.25, and 1.67~GeV. The first one is the chiral-even $g_T(x)$, which is the twist-3 counterpart of the helicity PDF. The first and only lattice calculation was reported in 2020~\cite{Bhattacharya:2020cen}, demonstrating the feasibility of the methodology, with results that validate 
{that} the magnitude of the twist-3 contributions is as sizable as the leading twist (see Fig.~\ref{fig:gT_g1}). Additionally, the Wandzura–Wilczek approximation was tested and found to hold well over a broad $x$ range, while also allowing for up to 40$\%$ genuine twist-3 effects. Also, the chiral-odd twist 3 PDF $h_L(x)$, the counterpart of the transversity PDF, was also computed within the framework of LaMET\cite{Bhattacharya:2021moj}. Among the various results is the flavor decomposition, which offers interesting qualitative conclusions on the role of up- and down-quark for $h_L$. The individual-quark flavor was also explored within the Wandzura–Wilczek approximation (see Fig.~\ref{fig:hL_WW_u_d}). More details can be found in Refs.~\cite{Bhattacharya:2020cen,Bhattacharya:2021moj}. 

\begin{figure*}[h!]
\includegraphics[scale=0.4]{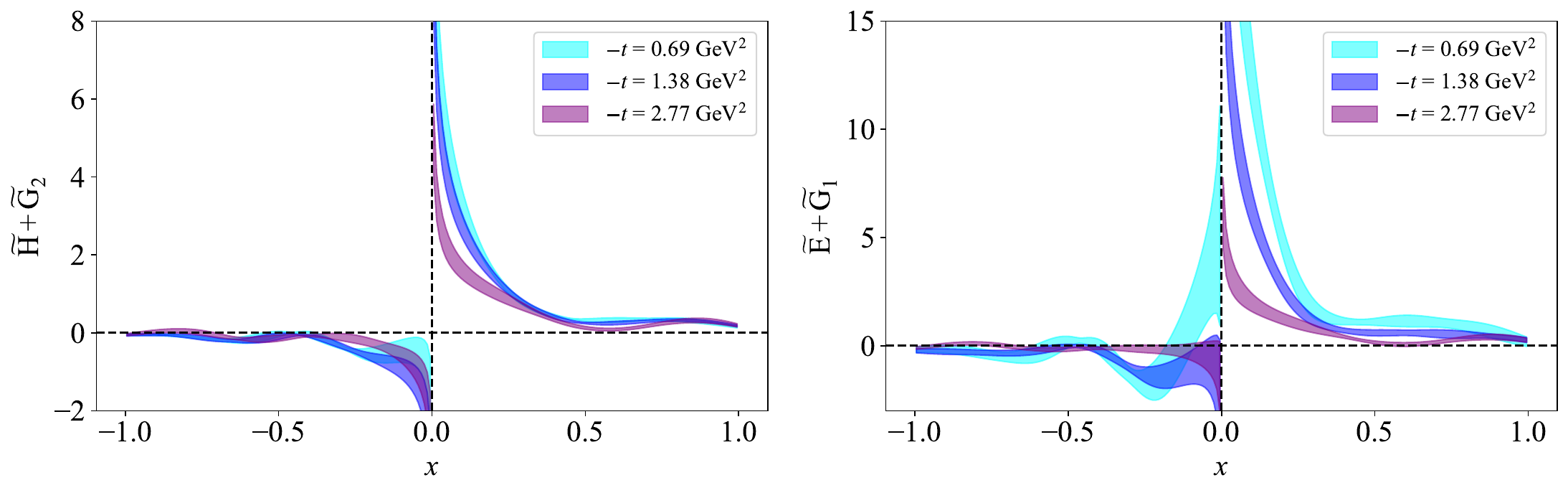}\\
\includegraphics[scale=0.4]{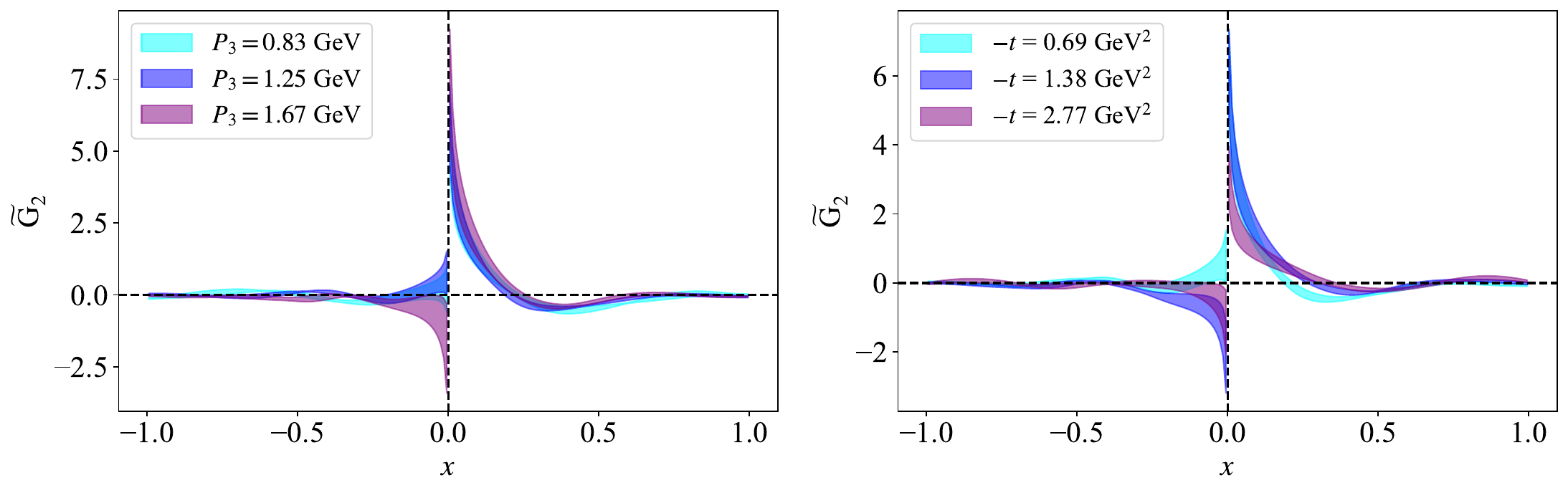}\hspace*{-14.85cm}
\includegraphics[scale=0.4]{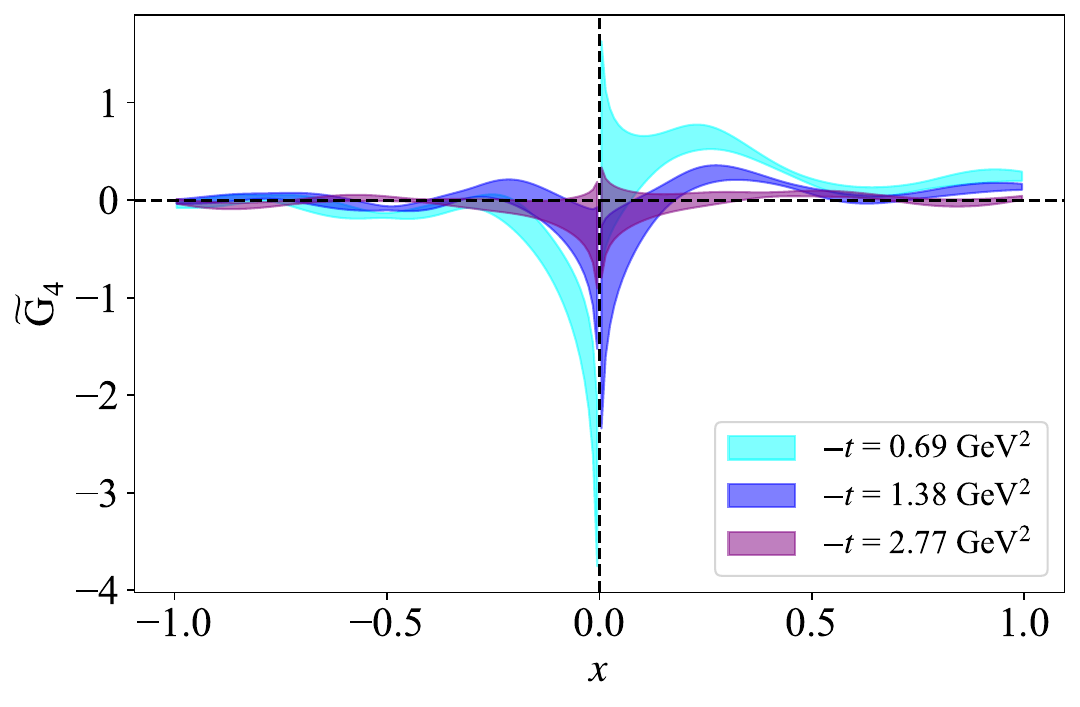} 
\vskip -0.3cm
\caption{$\widetilde{H}+\widetilde{G}_2$ (upper left), $\widetilde{E}+\widetilde{G}_1$ (upper right), $\widetilde{G}_4$ (lower left), and the indirectly obtained $\widetilde{G}_2$ (lower right) at $P_3=1.25$~GeV for various values of $-t$. Results are given in the $\overline{\rm MS}$ scheme at a scale of 2~GeV. The bands correspond to the statistical errors and the systematic uncertainty due to the $x$-dependent reconstruction. This
figure is taken from Ref.~\cite{Bhattacharya:2023nmv}.}
\label{fig:G1G2G4_final}     
\end{figure*}

The first lattice QCD determination of twist 3 GPDs for the proton was delivered for the chiral even axial twist 3 GPDs using LaMET in the symmetric frame (Breit frame)~\cite{Bhattacharya:2023nmv}. The same ensemble was used as for the $g_T$ and $h_L$ calculation, where the pion mass is 260~MeV. Due to the enhanced gauge noise for matrix elements with momentum transfer, the momentum boost is focused on 1.25~GeV. The values of the momentum transfer accessed through the symmetric frame are $-t =$ 0.69, 1.38, and 
2.76~GeV$^2$, all at zero skewness.
Besides extracting the three twist-3 axial GPDs that are nonzero at zero skewness (see Fig.~\ref{fig:G1G2G4_final}), some consistency checks have been performed. This includes examining the local limit, the Burkhardt–Cottingham–type sum rules, and Efremov–Teryaev–Leader–type sum rules. These were validated mainly within uncertainties. Currently, this calculation is extended to the asymmetric kinematic frame {outlined} in Sec.~\ref{sec:asymmetric}, which gives access to a wide range of values for $-t$ at a reduced computational cost.

\section{
Development of phenomenology}
\label{sec:theory_and_pheno}

Recent years have witnessed remarkable progress in the study of GPDs, driven by advances in data analysis, theoretical understanding, and experimental techniques.
Ongoing efforts in data analysis, see Sec.~\ref{sec:theory_and_pheno_pheno}, have not only improved the extraction of GPD information from existing data on exclusive processes but also helped to identify current limitations. Those will be addressed through new experiments, the involvement of lattice QCD calculations, and the exploration of novel processes.
The latter are exemplified in Secs.~\ref{sec:photon-meson_pair_photoproduction} and~\ref{Sub_sec_UPC}, which discuss the exclusive photoproduction of a photon-meson pair, and measurements sensitive to GPDs in ultra-peripheral collisions, respectively.
In addition, we review recent progress in the phenomenology of non-perturbative quantities other than GPDs, that are accessible in exclusive reactions.
Specifically, Sec.~\ref{sec:pheno:TGPDs} addresses transition GPDs, Sec.~\ref{sec:theory_and_pheno_tda} focuses on transition distribution amplitudes (TDAs) encoding partonic correlations in the backward kinematic regime, and Sec.~\ref{sec:pheno:GDAs} examines generalized distribution amplitudes (GDAs) probed in $e^+e^-$ annihilation.

\subsection{Progress in data analysis} 
\label{sec:theory_and_pheno_pheno}

In total, thousands of experimental points for DVCS are currently available. Many groups interested in the phenomenology of GPDs are analyzing these data~\cite{Kumericki:2015lhb,Moutarde:2018kwr,Guo:2023ahv, Cuic:2023mki, Kriesten:2020apm, Guo:2025muf, Panjsheeri:2025vpa, Goharipour:2025lep}. This effort is facilitated by open-source frameworks like PARTONS~\cite{Berthou:2015oaw} and Gepard~\cite{gepard}, as well as a new open database for GPD physics~\cite{Burkert:2025gzu}. In addition, novel computing methods based on machine learning techniques are being developed, aiming to reduce the model bias in extracted quantities (see {\it e.g.}, Refs.~\cite{Kumericki:2011rz,Dutrieux:2021wll,Grigsby:2020auv} and Sec.~\ref{sec:theory_and_pheno_computing}). A rigorous phenomenology of DVCS is now possible thanks to the precise description of this process, including recent calculation of NNLO~\cite{Braun:2022bpn} and higher-twist corrections~\cite{Braun:2022qly} (see also Sec.~\ref{sec:advences_in_theory_DVCS_ht}).

The extraction of GPDs from DVCS data remains, however, far from satisfactory. The origin of this problem lies in the multidimensional nature of GPDs, necessitating constraints from a much larger sample than in the case of, for instance, 1D PDFs. In addition, the DVCS signal is always mixed with the BH signal, reducing the statistical power of the available data in fits. Moreover, DVCS only probes a specific combination of GPDs, see Eq.~\eqref{Def_CFF}, meaning that this process provides only partial access to GPDs. Last but not least, the deconvolution of GPDs from CFFs is an ill-defined problem (see Sec.~\ref{sec:DVCS_and_CFFs}), meaning that access to full information on GPDs is even further restricted.

All the aforementioned issues motivate new measurements, as well as the search for alternative and complementary sources of information about GPDs. The community is eagerly awaiting the electron-ion collider (EIC) being constructed at BNL~\cite{AbdulKhalek:2021gbh}, the electron-ion collider EicC designed at HIAF, China~\cite{Anderle:2021wcy}, and an anticipated upgrade of JLab~\cite{Accardi:2023chb}. The prospects for these new machines are discussed in Sec.~\ref{sec:new_experiments}. From the point of view of GPD phenomenology, it is important to note that each of them will cover a complementary kinematic domain (with some very welcome overlap), which is important for mapping CFFs across the full kinematic range. Such mapping is important \textit{per se} for the extraction of GPDs, but also because a good knowledge of CFFs over the entire domain of $\xi$ is needed in analyses employing dispersion relations.
Another class of experiments utilizes hadron-hadron collisions, where GPDs are probed in an unprecedented low-$x_B$ domain via UPCs (see Sec.~\ref{Sub_sec_UPC} for more details).

{Equally important is the access to different and independent observables (also for processes other than DVCS), which involve different coefficient functions and thus provide complementary constraints on the underlying GPDs. The simultaneous analysis of such observables is crucial for reducing model dependence and achieving a reliable global extraction of GPDs.}
The best example is DVMP, where the produced mesons act as filters for the types of active partons. Recently, a global fit has for the first time enabled the systematic extraction of GPDs from combined DVMP and DVCS data at NLO accuracy~\cite{Cuic:2023mki}. Exclusive photoproduction of heavy vector mesons, on the other hand, provides a clean probe of gluons. Data for both processes already exist, but their interpretation within the GPD framework is more challenging than in the case of DVCS. Specifically, for light mesons, factorization is proven only for longitudinally polarized virtual photons~\cite{Collins:1996fb}. Additionally, describing meson formation requires non-perturbative distribution amplitudes, introducing another source of uncertainty in the phenomenology of this process. For heavy mesons, NLO calculations are known to lack stability at high energies. A recent attempt to address this problem includes resumming soft gluon contributions using the so-called high-energy factorization~\cite{Flett:2024htj}. 
{For TCS, where experimental data have recently become available~\cite{CLAS:2021lky}, phenomenological analyses are exploiting its complementarity with DVCS for combined global studies~\cite{Grocholski:2019pqj}.}

Other recently considered processes sensitive to GPDs, for which, however, data have not yet been collected, include DDVCS
\cite{Belitsky:2002tf,Guidal:2002kt}, and processes with two new particles in the final state (the $2 \to 3$ processes), such as a pair of photons~\cite{Pedrak:2017cpp, Pedrak:2020mfm,Grocholski:2021man,Grocholski:2022rqj} or a photon- meson pair~\cite{Boussarie:2016qop, Duplancic:2018bum, Duplancic:2022ffo,Duplancic:2023kwe,Crnkovic:2025man}. 
DDVCS is particularly interesting from the perspective of its description in the language of GPDs, as it can be considered a unification of DVCS and TCS. At the same time, DDVCS can be used to directly probe GPDs in kinematic regions not accessible with these two processes. Recent developments regarding DDVCS include a new description utilizing the Kleiss-Stirling spinor techniques
\cite{Kleiss:1985yh}, which have been employed in new impact studies
\cite{Deja:2023ahc}. 
This marks an important step toward new experimental proposals~\cite{Alvarado:2025huq}. On the other hand, 
$2 \to 3$ 
processes present an interesting case for studying factorization theorems, as demonstrated by Refs.~\cite{Qiu:2022pla,Nabeebaccus:2024mia}. These processes also provide unique information about GPDs. For instance, the exclusive production of two photons gives access to the $C$-odd combination of GPDs, 
Eq.~(\ref{C_odd_GPDs}), whereas processes like DVCS, TCS, and DDVCS only probe the $C$-even combination 
({\it cf}. Eq.~(\ref{C_even_GPDs})). More information about ``novel'' processes sensitive to GPDs can be found in Secs.~\ref{sec:photon-meson_pair_photoproduction}, \ref{Sub_sec_UPC}.

Another alternative source of information about GPDs comes from the lattice-QCD side, made possible in particular by the development of the {LaMET}~\cite{Ji:2020ect} and the short-distance operator expansion (SDE)~\cite{Radyushkin:2019mye} frameworks. Progress in this field is discussed elsewhere (see Sec.~\ref{sec:latticeQCD}), however, it is worth mentioning here the first attempts at meaningfully incorporating lattice-QCD results into GPD phenomenology, as seen in works such as Refs.~\cite{Guo:2022upw,Cichy:2024afd}.

In the end, we emphasize that the limited access to GPDs through currently measured processes does not prevent meaningful physics from being conducted within the GPD formalism. In particular, relying on a few reasonable and well-tested assumptions~\cite{Moutarde:2018kwr}, at low-$x_B$ one can directly relate the slope of the DVCS cross section as a function of the $t$ variable with the nucleon tomography. This method of accessing nucleon tomography has been used in the past 
using the results of 
HERA experiments~\cite{H1:2005gdw,H1:2009wnw,ZEUS:2008hcd}, 
{of JLab Hall A and CLAS data \cite{Dupre:2016mai,Dupre:2017hfs},}
and, most recently, by COMPASS~\cite{COMPASS:2018pup}. In addition, the method will be used by EIC experiments to perform one of their flagship measurements~\cite{Aschenauer:2025cdq}. Other interesting quantities can be extracted from DVCS amplitudes based on the dispersion relation techniques. In particular, one can extract the so-called subtraction constant, directly related to the $D$-term, and therefore to the mechanical properties. This relation has been utilized in analyses like~\cite{Burkert:2018bqq,Kumericki:2019ddg,Duran:2022xag,Moutarde:2019tqa}. Another interesting quantities one can access using dispersion relations are the Froissart-Gribov projections~\cite{Semenov-Tian-Shansky:2023ysr}. This technique has only very recently been developed for \mbox{spin-$\nicefrac{1}{2}$} targets and allows to quantify the response of a hadron target to probes carrying various angular momenta.

\subsection{Exclusive photoproduction of a photon-meson pair}
\label{sec:photon-meson_pair_photoproduction}

\newsavebox{\leftfig}
\newsavebox{\rightfig}
\sbox{\leftfig}{\includegraphics[width=0.45\columnwidth]{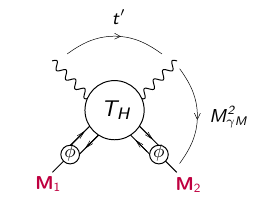}}
\sbox{\rightfig}{\includegraphics[width=0.45\columnwidth]{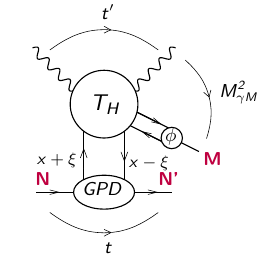}}
\newlength{\raisefig}
\setlength{\raisefig}{0.5\ht\rightfig - 0.35\ht\leftfig}
\newlength{\arrowraise}
\setlength{\arrowraise}{0.5\ht\rightfig}
\begin{figure}[!ht]
\centering
\begin{tabular}{ccc}
  \raisebox{\raisefig}{\usebox{\leftfig}} &
\hspace{-0.55cm}
  \raisebox{0.5\ht\rightfig}{$\longrightarrow$} 
\hspace{-0.55cm}
&
  \usebox{\rightfig}
\end{tabular}
\caption{Left: Brodsky-Lepage factorization for $\gamma M_1 \to \gamma M_2$ process.
Right: factorization of $\gamma N \to \gamma M N'$ process (similar to the TCS, see Fig.~\ref{Fig_TCS_DDVCS_MBIHER}), where $-t \ll M_{\gamma M}^2$.}
\label{fig:fact-gMgM}
\end{figure}

Photon-meson photoproduction,
$\gamma N \to \gamma M N'$,
has been proposed as a promising alternative~\cite{Boussarie:2016qop,Duplancic:2018bum},
offering complementary access to GPDs. 
Similar to DVMP, this process probes quark flavor and includes gluon contributions at leading order, with the additional advantage of accessing {tensor} GPDs already at leading twist in transverse vector meson production.

Unlike DVCS and DVMP, which are $2 \to 2$ processes, 
photon-meson photoproduction is a $2 \to 3$ process. 
It is characterized by more complex kinematics and a more involved 
leading-order hard-scattering amplitude. 
While  $2 \to 2$ processes primarily provide ``moment-type'' constraints,  at leading order 
effectively probing GPDs at $x = \pm \xi$, 
the additional complexity of photon-meson photoproduction 
provides a more detailed sensitivity to the $x$-dependence of GPDs~\cite{Qiu:2023mrm}. 
An analogous situation arises in DDVCS, where the presence of an additional scale leads to a similar effect.
Other $2 \to 3$ processes have also been proposed. 
These include meson-meson photoproduction~\cite{Ivanov:2002jj,Enberg:2006he,Siddikov:2022bku}, 
which also provides access to {tensor} GPDs; 
diphoton photoproduction~\cite{Pedrak:2017cpp,Pedrak:2020mfm,Grocholski:2021man,Grocholski:2022rqj},
which has been analyzed at NLO; 
and the pion-nucleon to photon-photon process~\cite{Qiu:2022bpq,Qiu:2024mny}, 
which represents the crossed version of photon-meson photoproduction.

Photon-meson photoproduction has gained significant interest lately.
The vector meson case, {\it i.e.},
$\gamma \rho_L$ and $\gamma \rho_T$ photoproduction
\cite{Boussarie:2016qop},
as well as quark contributions to the pseudoscalar meson case,
{\it i.e.}, $\gamma \pi$ photoproduction
\cite{Duplancic:2018bum,Qiu:2023mrm},
have been investigated at leading order, 
and the feasibility of experimental
measurements at JLab, COMPASS, EIC, as well as in UPC collisions
at the LHC, has been discussed in detail
\cite{Duplancic:2022ffo,Duplancic:2023kwe,Qiu:2023mrm}.
Photoproduction of photon–heavy-meson pairs 
has also been analyzed
\cite{Siddikov:2024blb},
where gluon contributions were found to be leading.

The factorization of the hard-scattering amplitude
for photon-meson photoproduction
can be understood in two steps (see Fig.~\ref{fig:fact-gMgM}).
First, following Brodsky–Lepage large-angle factorization
\cite{Lepage:1980fj}, the amplitude for
$\gamma M_1 \rightarrow \gamma M_2$
factorizes into a convolution of meson DAs $\phi$
and a hard-scattering kernel $T_H$ for
$\gamma(q \bar{q}) \rightarrow \gamma(q \bar{q})$.
This holds if the invariant mass $M_{\gamma M}^2$
and momentum transfer $t'$ are large,
corresponding to large scattering angles
or large transverse photon momentum.
Second, when the momentum transfer to the nucleon, $t$, is small,  
the incoming meson DA can be replaced by a nucleon GPD,  
yielding a factorized description similar to TCS 
(with the photon of virtuality $Q'^2 > 0$ replaced  
by a photon-meson pair of invariant mass $M_{\gamma M}^2$ -- see Fig.~\ref{Fig_TCS_DDVCS_MBIHER}.
For $C = +1$ states, quark $(q \bar{q})$ Fock states
can be replaced by gluon ones $(gg)$.

{The resulting photoproduction amplitude exhibits a 
much richer kinematics dependence compared to DVCS or DVMP.
It can be expressed
in terms of the skewness $\xi$, momentum transfer $t$,
the center-of-mass energy squared $s$, and an additional
dimensionless parameter $\alpha$ defined as
\be
1-\alpha = -t'/M_{\gamma M}^2.
\label{Def_alpha_Crnkovic}
\ee
This parameter is related to the scattering angle $\theta$ of the outgoing photon in the center-of-mass frame of the
$\gamma (\bar{q}_1 q_2) \to \gamma (\bar{q}_1 q_2)$
or
$ \gamma q_1 \to  \gamma q_2 (\bar{q}_1 q_2)$
hard subprocesses, see  Fig.~\ref{fig:fact-gMgM}, as 
$ \cos \theta = 2 \alpha-1$.}

The same hard subprocess amplitude $T_H$ describes
the leading-twist subprocesses
on both sides of Fig.~\ref{fig:fact-gMgM},
provided that one takes into account that the incoming quark momentum fraction given by
$(x + \xi)/(2\xi)$ extends beyond the usual range of meson momentum fractions
and that the appropriate $i \epsilon$ prescription is retained in the propagators%
\footnote{
This is analogous to the relation between the meson transition form factor and DVCS, and between the meson electromagnetic form factor and DVMP.
}.

The factorization for photon-meson photoproduction
involving quark GPDs has recently been proven~\cite{Qiu:2022pla}.
In contrast, the contributions of gluon GPDs
to the photoproduction of light neutral pseudoscalar mesons
pose challenges, 
{since standard collinear factorization may break down due to soft rescattering effects.}
Breaking of factorization and potential remedies
have been discussed in Refs.~\cite{Nabeebaccus:2023rzr,Nabeebaccus:2024mia}.

\begin{figure}[!ht]
\centering
\begin{tabular}{c}
\includegraphics[width=0.45\textwidth]{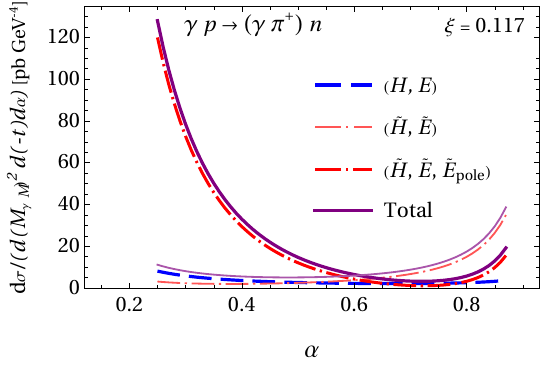}\\
\includegraphics[width=0.45\textwidth]{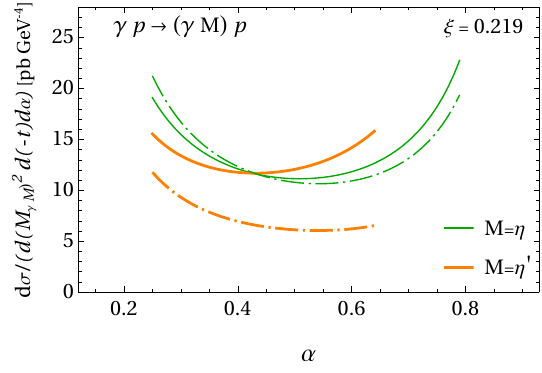}
\end{tabular}
\caption{Full differential cross section for $\gamma p \to \gamma \pi^+ n$ (top panel) 
and $\gamma p \to \gamma \eta (\eta') p$ (bottom panel) {as
functions of the dimensionless parameter
$\alpha$ (\ref{Def_alpha_Crnkovic}).} 
In the lower graph, the dashed lines represent quark-only contributions, 
while the solid lines also include the two-gluon contributions to
$\eta$ ($\eta'$).
{This figure is taken from Ref.~\cite{Crnkovic:2025man}.}} 
\label{fig:res-gMgM}
\end{figure}

Current efforts focus on systematizing results for all relevant
channels, developing fast numerical codes incorporating various
DA and GPD models with QCD evolution, and extending the analysis to
neutral pseudoscalar meson photoproduction.
In \cite{Crnkovic:2025man}, the focus is on the 
$\gamma N \to \gamma \eta (\eta') N$ 
reaction channel.
It was shown that, in contrast to the two-gluon exchange contributions associated with gluon GPDs, the gluonic ($gg$) components of the $\eta (\eta')$ distribution amplitudes do not lead to endpoint or integration singularities. Therefore, the gluonic content of the $\eta (\eta')$ mesons can be straightforwardly incorporated within the factorized framework, unlike the case of gluon GPDs in the nucleon sector discussed above.
The gluon contributions are, as expected, more pronounced in $\gamma \eta'$ photoproduction -- the dashed lines on the lower panel of Fig.~\ref{fig:res-gMgM} denote only the quark contributions, and a comparison with the solid lines illustrates the impact of the gluonic component.
Similarly to the case of pions,
the cross sections are of the order of picobarns.
On the upper panel of Fig.~\ref{fig:res-gMgM}, the differential cross section for $\gamma N \to \gamma \pi^+ N$ is analyzed in terms of axial and vector GPD contributions. For this process, the $\pi$-pole contribution becomes significant at moderate to large values of $\xi$ \cite{Crnkovic:2025man}.

\subsection{Measurements sensitive to GPDs in ultra-peripheral collisions}
\label{Sub_sec_UPC}

GPDs can also be constrained through measurements of exclusive processes in ultra-peripheral collisions (UPCs). In UPCs, hadrons interact with one  another at a distance that is large, {\it i.e.}, greater than the sum of their respective radii. This results in a suppression of colored hadronic interactions and a dominance of colorless exchanges, such as the interaction through photons, pomerons, and potentially odderons~\cite{Albrow:2010yb}. 

\begin{figure}[!ht]
\begin{center}
\includegraphics[width=0.4\textwidth]{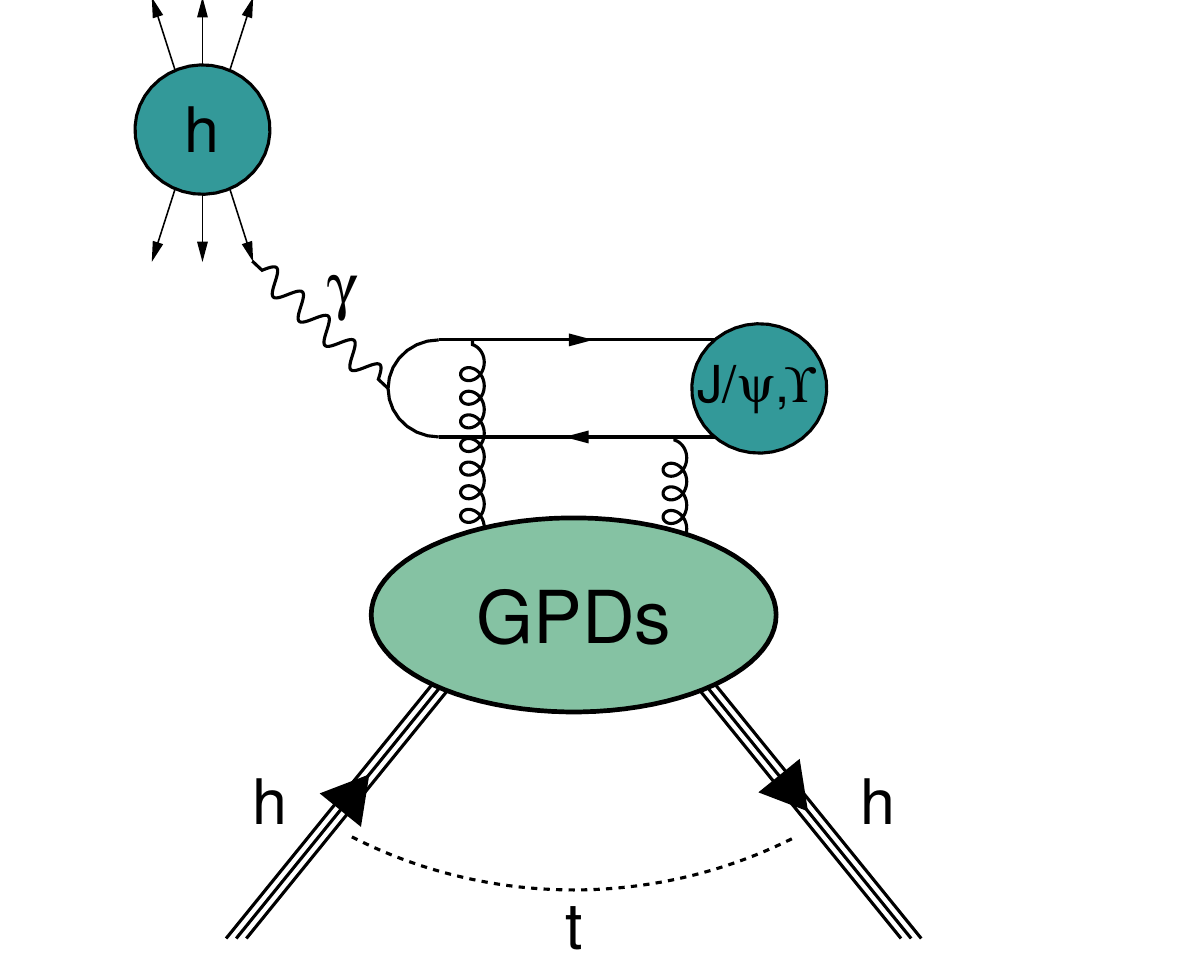}
\caption{Schematic diagram representing the exclusive photoproduction of quarkonia in ultra-peripheral hadron-hadron collisions. Here, $h$ stands for proton or ion.}\label{fig:UPC_GPD}
\end{center}
\end{figure}

Measurements sensitive to GPDs in UPCs have been performed so far through the study of exclusive quarkonium production, 
mainly through exclusive photoproduction of a 
$J/\psi$ 
or an 
$\Upsilon$.  
In such a case, one of the beam hadrons emits a quasi-real photon, which serves as a probe of the other beam hadron.  The quasi-real photon fluctuates into a quark-antiquark pair, which itself interacts, at leading order, through gluon exchange with the probed hadron and evolves into a final-state quarkonium. The large mass of the charm or bottom quark provides the hard scale needed for the factorization of the studied process. A corresponding schematic diagram is shown in Fig.~\ref{fig:UPC_GPD}. 
 
The advantage of UPCs with respect to the study of GPDs lies in their high-energy reach, or  equivalently in their low-$x_B$  reach, where $x_B \simeq {Q^2}/{W_{\gamma h}^2}$, with $W_{\gamma h}$ standing for the center-of-mass energy of the $\gamma h$ system, see Fig.~\ref{fig:UPC_GPD}. 
At the LHC,  presently the highest-energy hadron collider, one can span the $x_B$
region down to $x_B=10^{-6}$. Moreover, by comparing GPDs constrained from data collected in electron-hadron scattering and in UPCs, a crucial test of the universality of GPDs and  the  understanding of perturbative QCD corrections can be obtained~\cite{Mueller:2012sma,Moutarde:2013qs}. At the moment, however, exclusive measurements in UPCs have not been used yet to constrain GPDs. 

Exclusive quarkonium production has been measured in proton-antiproton collisions~\cite{CDF:2009xey} at the Tevatron, in proton-proton~\cite{LHCb-PAPER-2012-044,LHCb-PAPER-2013-059,LHCb-PAPER-2014-027,LHCb-PAPER-2015-011,LHCb-PAPER-2018-011,LHCb:2024pcz}, proton-lead~\cite{ALICE:2014eof,ALICE:2018oyo,CMS:2018bbk,ALICE:2023mfc} and lead-lead~\cite{ALICE:2012yye,ALICE:2015nmy,CMS:2016itn,ALICE:2019tqa,LHCb-PAPER-2020-043,ALICE:2021gpt,ALICE:2021tyx,LHCb-PAPER-2022-012,CMS:2023snh,ALICE:2023jgu,ATLAS:2025uxr,CMS:2025bsz} 
collisions at the LHC, and in deuteron-gold~\cite{STAR:2021wwq} and gold-gold~\cite{PHENIX:2009xtn} 
collisions at RHIC. {Typically, the quarkonium is reconstructed through its decay products detected in the central region, while requiring the absence of any additional particle activity in that region. Furthermore, a veto of signal detection in the (far-)forward and (far-)backward detectors can improve the exclusivity of the event.}

\begin{figure}[!ht]
\begin{center}
\includegraphics[width=0.99\columnwidth]{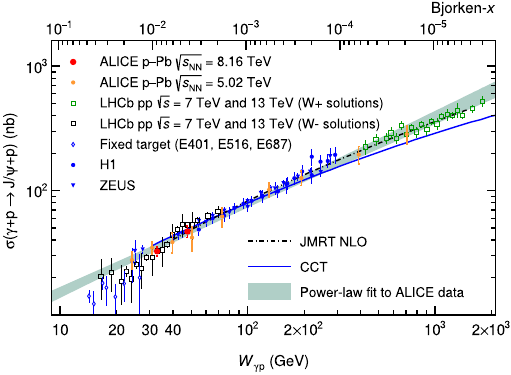}
\caption{Cross section for exclusive $J/\psi$ photoproduction off the proton as a function of the photon-proton center-of-mass energy (lower abscissa axis) and of Bjorken-$x$ (upper abscissa axis) for data collected at colliders in lepton-proton, proton-proton and proton-lead interactions, as well as at fixed-target  experiments. Figure taken from~\cite{ALICE:2023mfc}.}\label{fig:UPC_comp}
\end{center}
\end{figure}

Proton-proton collisions at the LHC currently provide access to the lowest possible 
$x_B$ 
values. However, in these collisions, one cannot unambiguously identify which of the two protons serves as the photon emitter. Depending on which of the two beam protons emits a photon, the emitted photon has lower or higher energy, corresponding to a different configuration of the event kinematics. In the first approximation, this problem is absent when analyzing proton-ion collisions,  particularly when using heavy ions. The flux of photons emitted by a beam particle is 
proportional to the square of its atomic number. Hence, in collisions with a proton and a heavy ion, such as gold at RHIC and lead at the LHC, the heavy ion acts, at first order, as the photon emitter, and the ambiguity is lifted.  The probed $x_B$ region in such collisions (at the LHC) extends down to $10^{-5}$. 
Figure~\ref{fig:UPC_comp} 
presents the cross section for exclusive $J/\psi$ photoproduction off the proton, obtained from measurements in lepton-proton scattering at the HERA collider experiments~\cite{ZEUS:2002wfj,H1:2005dtp,H1:2013okq}, proton-proton collisions at 
LHCb~\cite{LHCb-PAPER-2012-044,LHCb-PAPER-2013-059,LHCb-PAPER-2018-011}\ and proton-lead collisions at ALICE~\cite{ALICE:2014eof,ALICE:2018oyo,ALICE:2023mfc} as well as from measurements at fixed-target experiments~\cite{Binkley:1981kv,Denby:1983az,E687:1993hlm}. As can be seen, the photoproduction cross sections extracted from the different data sets are compatible with one another in the kinematic regions of common coverage. This is in agreement with measurements probing the same underlying physics. 

In order to study nuclear GPDs, ion-ion collisions provide the necessary information. Gold-gold collisions and lead-lead collisions have been studied extensively at, respectively, RHIC and the LHC. As do the measurements in proton-proton collisions, most of the measurements in ion-ion collisions suffer from an ambiguity in the identity of the photon emitter. However, recently both the ALICE and CMS experiments used their zero-degree calorimeters (ZDCs) installed in the far-backward and far-forward regions along the beamline~\cite{ALICE:2023jgu,CMS:2023snh} in order to resolve this issue. The method is based on the detection in the ZDCs of neutrons emitted by one or both of the interacting ions, after they mutually excite each other through Coulomb interaction~\cite{Baltz:2002pp,Guzey:2013jaa,Guzey:2016piu}. The probability of Coulomb excitation by the interacting beam particles depends on the impact parameter, increasing with decreasing impact parameter. At the same time, the photon flux exhibits an impact-parameter-dependent energy dependence, with, {\it e.g.}, the photon flux for high-energy photons increasing at smaller impact parameter. Hence, by performing measurements of exclusive quarkonium production for event categories corresponding to an absence of neutron emission, neutron emission by one of the two beam particles only, and neutron emission by both beam particles, one can unambiguously extract the exclusive quarkonium photoproduction cross sections corresponding to the emission of low- and high-energy quasi-real photons. 

While high-energy hadron-hadron collisions provide access to the low-$x_B$ region, the gaseous fixed target at the CERN LHCb experiment, SMOG2~\cite{2707819}, allows probing the high-$x_B$ region. Here, interactions of a proton or lead-ion beam with a proton target or heavier targets, consisting of deuteron, nitrogen, oxygen and the noble gases ranging from helium to xenon,  can be studied. It has been demonstrated that data collection using the SMOG2 system can be performed in parallel with data taking in proton-proton collisions with high efficiency and reconstruction resolution, and this without affecting the quality of data collection in beam-beam collisions~\cite{PhysRevAccelBeams.27.111001}. At present, plans are under development to install a transversely polarized target at the LHCb experiment~\cite{Hadjidakis:2018ifr, Aidala:2019pit}, {for LHC Run 5}. Such a target will allow access to the GPDs $E$ and $\widetilde{E}$ in the high-$x_B$ region, in a way that is complementary to that probed in lepton-proton interactions.

\subsection{Non-diagonal hard exclusive reactions and transition GPDs}
\label{sec:pheno:TGPDs}

Beyond the standard diagonal GPDs, the concept has been extended to transition GPDs~\cite{Goeke:2001tz,Guichon:2003ah}, that describe matrix elements between ground-state nucleons and hadronic resonances. This extension provides a new window into the internal structure of excited baryons and their connection to fundamental QCD symmetries, see Ref.~\cite{Diehl:2024bmd} for a review. Experimental access to transition GPDs is provided by
non-diagonal form of DVCS and DVMP  processes,
involving nucleon transitions into a resonance.
This allows one to probe not only the ground-state structure but also the spatial and dynamical properties of baryon resonances. As in the nucleon case, these quantities unify aspects of form factors and parton distributions, but now they encode resonance excitations, providing access to new observables, such as transition {GFFs}~\cite{Ozdem:2019pkg}, 
particularly the nucleon-to-resonance transition angular momentum~\cite{Kim:2023xvw}.

\begin{figure}[h]
\centering
\includegraphics[width=0.8\columnwidth]{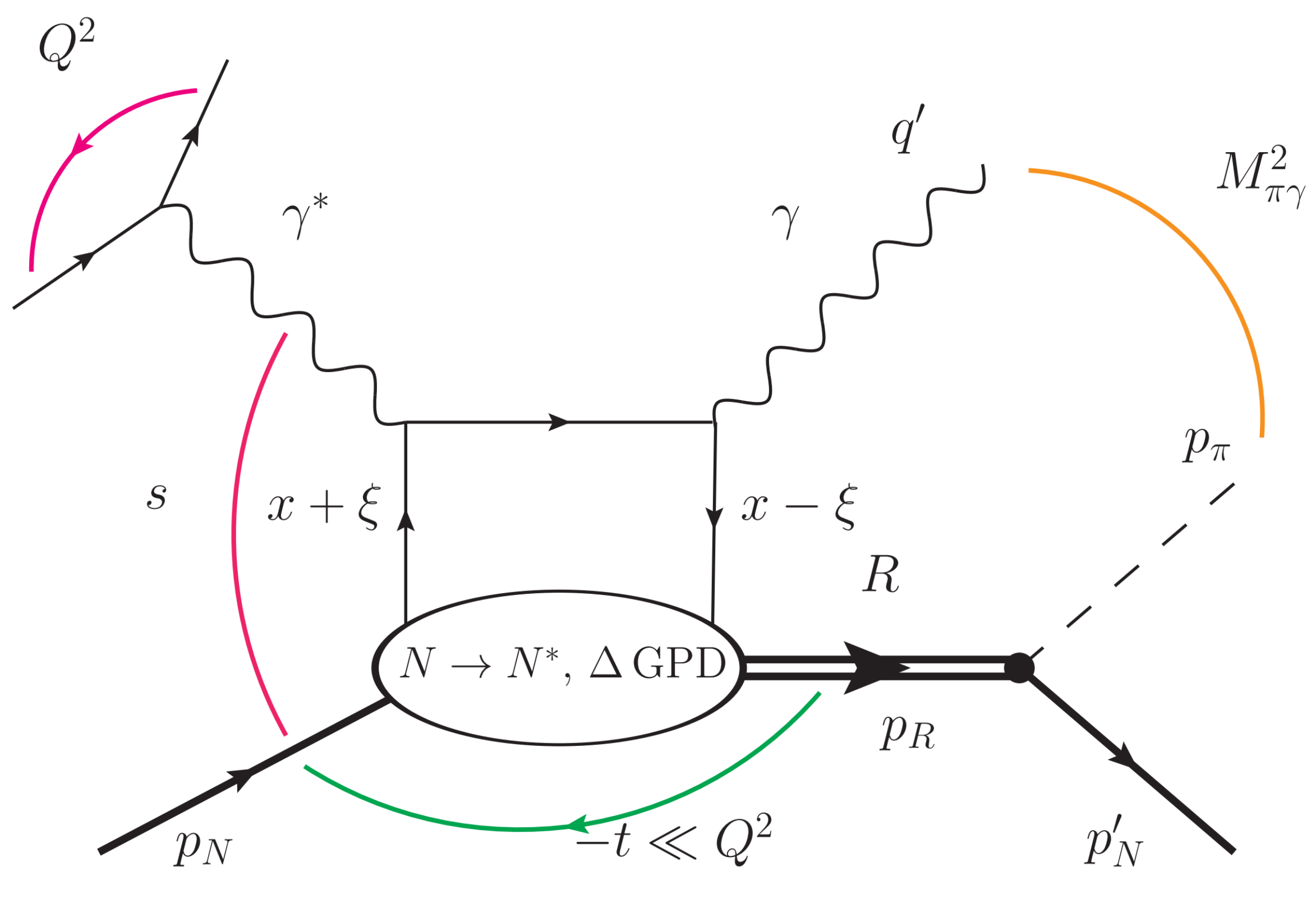} \ \ \ 
\includegraphics[width=0.8\columnwidth]{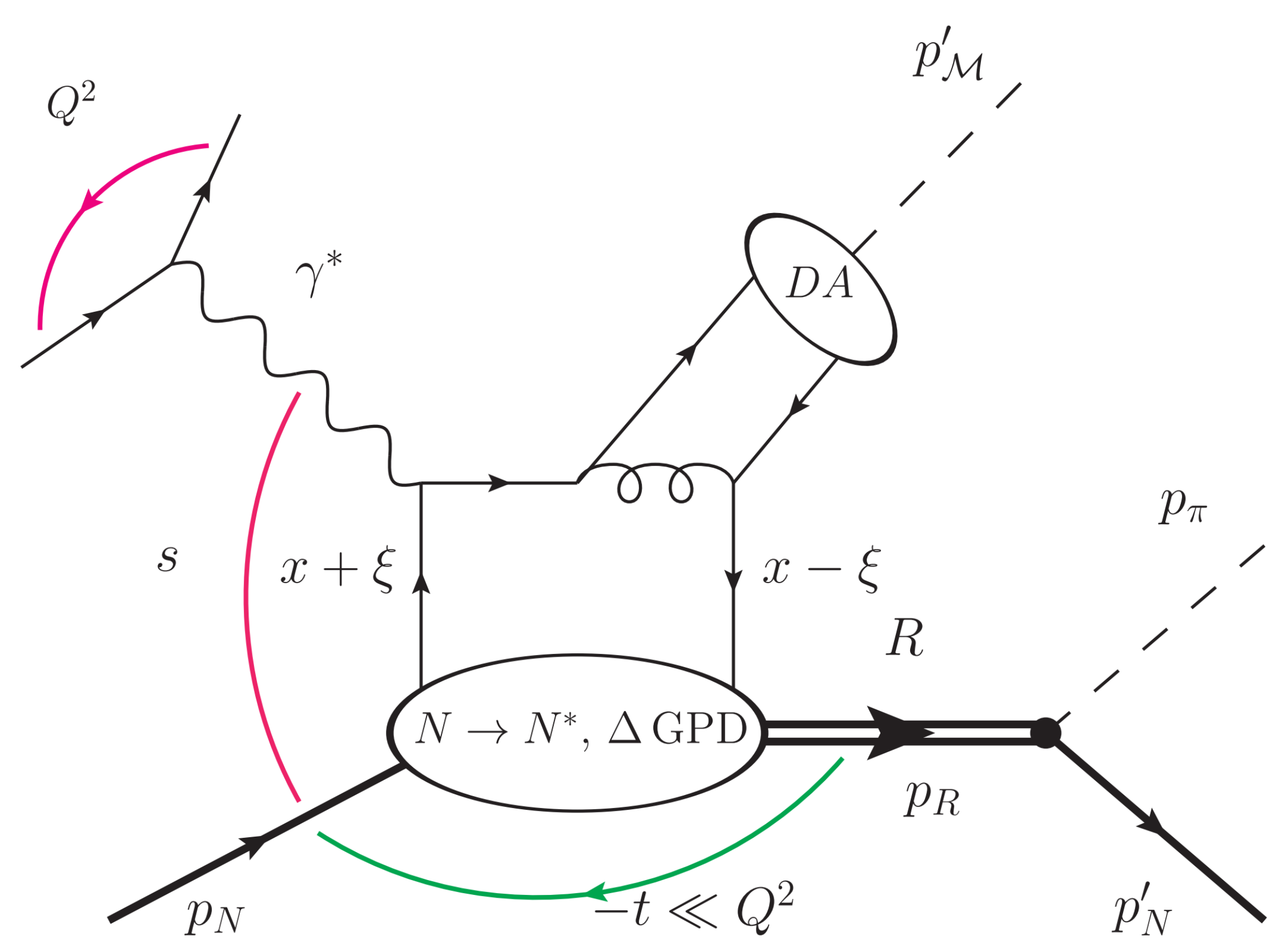}
\caption{Non-diagonal DVCS and DVMP reactions involving a transition from nucleon to an excited nucleon state $R=N^*,\, \Delta$. }
\label{fig:NDDVCS}
\end{figure}

For the simplest case, the 
$N\to\Delta(1232)$
transition, a total of $16$ independent twist-2 GPDs are identified~\cite{Kim:2024hhd,Kroll:2022roq,Kim:2026kdo}, linked to transition form factors such as the Jones–Scadron multipoles and the Adler axial form factors, see~\cite{Pascalutsa:2006up}. In Ref.~\cite{Semenov-Tian-Shansky:2023bsy} the GPD framework for non-diagonal DVCS has been extended to the second resonance region $\pi N$ resonances  
${\rm P}_{11}(1440)$, ${\rm D}_{13}(1520)$ and
${\rm S}_{11}(1535)$. The cross section estimates
for the kinematical conditions of JLab 12~GeV were performed using a phenomenological parametrization
with the first Mellin moments of transition GPDs normalized to known electromagnetic transition form factors~\cite{Ramalho:2023hqd}, and also accounting for the contribution of pion exchange in the $t$-channel. 
Non-diagonal DVMP involving the production of $\pi-\Delta(1232)$ final state has been recently addressed in Ref.~\cite{Kroll:2022roq}. Additionally, {an}
application of the
nucleon-to-hyperon transition GPD formalism
for hard exclusive reactions involving strangeness production, such as
$\gamma^* N \to \{\Lambda, \, \Sigma\} K$,
was presented in~\cite{Kroll:2019wug}.

An interesting opportunity lies in  introducing transition GPDs between a nucleon and the $\pi N$ system, as initially suggested in Ref.~\cite{Polyakov:2006dd}.
The $N \to \pi N $ GPDs are expected to provide comprehensive information about the spectrum of the $\pi N$ system produced through non-local QCD probes in hard exclusive reactions. A notable aspect of these GPDs is their potential to connect the chiral regime (pion produced at the threshold, $W_{\pi N} \sim M_N+m_\pi$, 
where $W_{\pi N}$ stands for the center-of-mass energy of the ${\pi N}$ system) 
with the resonance regime ($W_{\pi N} \sim M_\Delta, \, M_{N^*}$). In the chiral regime, $N \to \pi N$ GPDs can be systematically computed using chiral perturbation theory, see  {\it e.g.}, Ref.~\cite{Alharazin:2023zzc}.
Furthermore, by employing dispersive techniques based on the  Watson final state interaction theorem
{\cite{Watson:1954uc}}, one can extend these results into the resonance regime.  A toy example of such a connection for spinless hadrons was recently discussed in the $\pi \to \pi \pi$ case in Ref.~\cite{Son:2024uxa}.

The first dedicated measurements of non-diagonal hard exclusive reactions at JLab 12~GeV have already demonstrated the feasibility of accessing transition GPDs by measuring the  
$e p \to e' \pi^- \Delta^{++}$  channel~\cite{CLAS:2023akb}. 
Future data sets will explore DVCS with $N \to \Delta$ and higher $N^\ast$ final states.
The proposed JLab 22~GeV energy upgrade will significantly extend the accessible kinematics in $Q^2$ and $W$.
Complementary opportunities exist at other facilities: 
COMPASS/AMBER with high-intensity muon and meson beams, 
J-PARC and FAIR for hadron-induced reactions, 
the future EIC and EicC for high-energy exclusive scattering at small~$x$, 
and LHC ultraperipheral collisions for probing transition GPDs in the diffractive regime in the gluon-dominated mode.

\subsection{Backward hard exclusive reactions and TDAs}
\label{sec:theory_and_pheno_tda}

\subsubsection{ $u$-channel exclusive meson electroproduction and nucleon-to-meson TDAs}
 
The theoretical framework for understanding exclusive meson electroproduction in the backward region in the framework of collinear QCD factorization is now quite developed~\cite{Pire:2021hbl, Pire:2025wbf}. Although no complete proof of factorization exists in this new domain~\cite{Gayoso:2021rzj}, there are rather solid hints from recent experimental studies at JLab~\cite{CLAS:2017rgp, JeffersonLabFp:2019gpp,CLAS:2023akb},
supporting the definition of a new class of nonperturbative objects called transition distribution amplitudes (TDAs).

\begin{figure}[!ht]
\begin{center}
\includegraphics[width=0.35\textwidth]{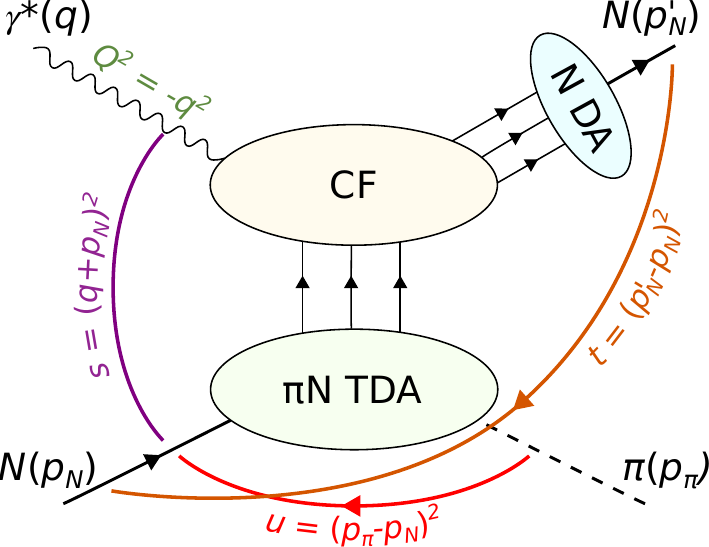}
\end{center}
\caption{Kinematical quantities and the collinear factorization mechanism for $\gamma^* N \to \pi N$ in the near-backward kinematical regime (large $Q^2$, $s$;
fixed $x_B$; $|u| \sim 0$). The lower blob, denoted $\pi N$ TDA, depicts the nucleon-to-pion  transition distribution amplitude; the $N$ DA blob depicts the
nucleon distribution amplitude; CF denotes the hard subprocess amplitude (coefficient function). Figure from Ref.~\cite{Pire:2025wbf}.}
\label{Fig_Factorization}
\end{figure}

The golden channels for measuring TDAs are meson ($\cal M$) deeply exclusive electroproduction 
\begin{equation}
    \gamma^*(q)+ N(p_N) \to {\cal M}(p_{\cal M})+ N(p'_N)
\end{equation}
in the nearly-backward  kinematics defined by small $|u|$, large $Q^2=-q^2$ and $s=(p_N+q)^2$ with fixed $x_B$ (see Fig.~\ref{Fig_Factorization} for the kinematics).  This reaction indeed was the first to bring evidences of existence of the backward cross section peak at large $Q^2$, which could be the signal for the manifestation of the QCD hard scattering mechanism~\cite{CLAS:2017rgp}. A basic prediction of the QCD picture -- namely the dominance of transverse virtual photon polarization -- could not initially be tested. This test was successfully conducted a few years later in the study of backward $\omega$-meson electroproduction~\cite{JeffersonLabFp:2019gpp}. 

 
 {In contrast to the case of GPDs, the experimental information 
  on processes involving TDAs 
  \cite{CLAS:2017rgp, JeffersonLabFp:2019gpp,CLAS:2023akb}
  remains much less complete, while the number of the relevant independent invariant functions is larger.} In the specific case of nucleon-to-pion transitions, one has eight independent 
 nucleon to pion transition distribution amplitudes ($ \pi N$ TDAs) $H_s^{\pi N}$ defined as (here for the proton-to-$\pi^0$ case):
\be
&& 4 
(P \cdot n)^3 \int\left[\prod_{j=1}^3 \frac{d \lambda_j}{2 \pi}\right] e^{i \sum_{k=1}^3 x_k \lambda_k(P \cdot n)} \nn \\ && 
\left\langle\pi^0\left(p_\pi\right)\right| \widehat{O}_{\rho \tau \chi}^{\,uud}\left(\lambda_1 n, \lambda_2 n, \lambda_3 n\right)\left|N^p\left(p_N, s_N\right)\right\rangle \nn \\
&& =\delta\left(x_1+x_2+x_3-2 \xi\right) i \frac{f_N}{f_\pi M_N} \sum_s\left(s^{\pi N}\right)_{\rho \tau, \chi} \nn \\ && \times H_s^{\pi N}\left(x_1, x_2, x_3, \xi, u ; \mu^2\right),
\label{Def_piN_TDA}
\ee
where the tri-local light cone operator 
is defined as
\be 
&&
\widehat{O}_{\rho \tau \chi}^{uud} 
(\lambda_1n,\lambda_2n,\lambda_3n) \nn \\ &&  =\varepsilon_{c_1 c_2 c_3}
  u^{c_1}_{\rho}(\lambda_1 n)
  u^{c_2}_{\tau}(\lambda_2 n)  
  d^{c_3}_{\chi}(\lambda_3 n).
  \label{Def_3q_operator}
\ee
Here, $c_{1,2,3}$ stand for the color group indices  and $\rho$, $\tau$, $\chi$
denote the Dirac indices of the quark field operators;
$f_\pi=93$~MeV and $f_N=5.0 \times 10^{-3} $~GeV$^2$  determines the normalization of the nucleon light-cone wave function at the origin.  
The sum in 
(\ref{Def_piN_TDA}) 
stands over the set of eight leading-twist-$3$ 
Dirac structures
\be
&&
(s^{\pi N})_{\rho \tau, \chi} \nn \\ && =\left\{ (v_{1,2}^{\pi N} )_{\rho \tau, \chi}, (a_{1,2}^{\pi N})_{\rho \tau, \chi}, (t_{1,2,3,4}^{\pi N})_{\rho \tau, \chi}\right\}\,, 
\ee
which are constructed from  the fully covariant components,
see  Ref.~\cite{Pire:2011xv}. Invariant TDAs
$H_s^{\pi N}$ in (\ref{Def_piN_TDA}) are functions
of three momentum fraction variables $x_i$, the skewness variable $\xi$
defined with respect to the $u$-channel momentum transfer $\xi =- \frac{(p_{\cal M}-p_N)\cdot n}{(p_{\cal M}+p_N)\cdot n}$ 
and the invariant momentum transfer 
$u = (p_N-p_{\cal M})^2$, as well as the factorization scale $\mu^2$.
The variables $x_i$ are subject to the momentum conservation constraint $\sum_i x_i=2 \xi$. 

A new modeling strategy has been recently developed~\cite{Pire:2025wbf, Pire:2025enz} to construct the
TDAs  from the factorized Ansatz for corresponding spectral densities (quadruple distributions) 
\be
F(\sigma,\, \rho,\, \omega,\, \nu)= f (\sigma, \rho) h(\sigma,\, \rho,\, \omega,\, \nu),
\label{Fact_Anz_Quad}
\ee
as~\cite{Pire:2010if}:
\be
\begin{aligned}
H(w, v, \xi)= & \int_{-1}^1 d \sigma \int_{-1+\frac{|\sigma|}{2}}^{1-\frac{|\sigma|}{2}} d \rho \int_{-1+|\sigma|}^{1-\left|\rho-\frac{\sigma}{2}\right|-\left|\rho+\frac{\sigma}{2}\right|} \nn \\
& \times d \omega \int_{-\frac{1}{2}+\left|\rho-\frac{\sigma}{2}\right|+\frac{\omega}{2}}^{\frac{1}{2}-\left|\rho+\frac{\sigma}{2}\right|-\frac{\omega}{2}} d \nu \, \delta(w-\sigma-\omega \xi) \nn \\
& \times \delta(v-\rho-\nu \xi) F(\sigma, \rho, \omega, \nu), 
\end{aligned}
\\
\label{eq:spectral_represention_wv}
\ee
where $w \equiv x_3-\xi$ and $v \equiv \frac{1}{2} \sum_{k, l=1}^3 \varepsilon_{3 k l} x_k$
are the so-called quark-diquark coordinates.

\begin{figure}[!ht]
\begin{center}
\includegraphics[width=0.99\columnwidth]{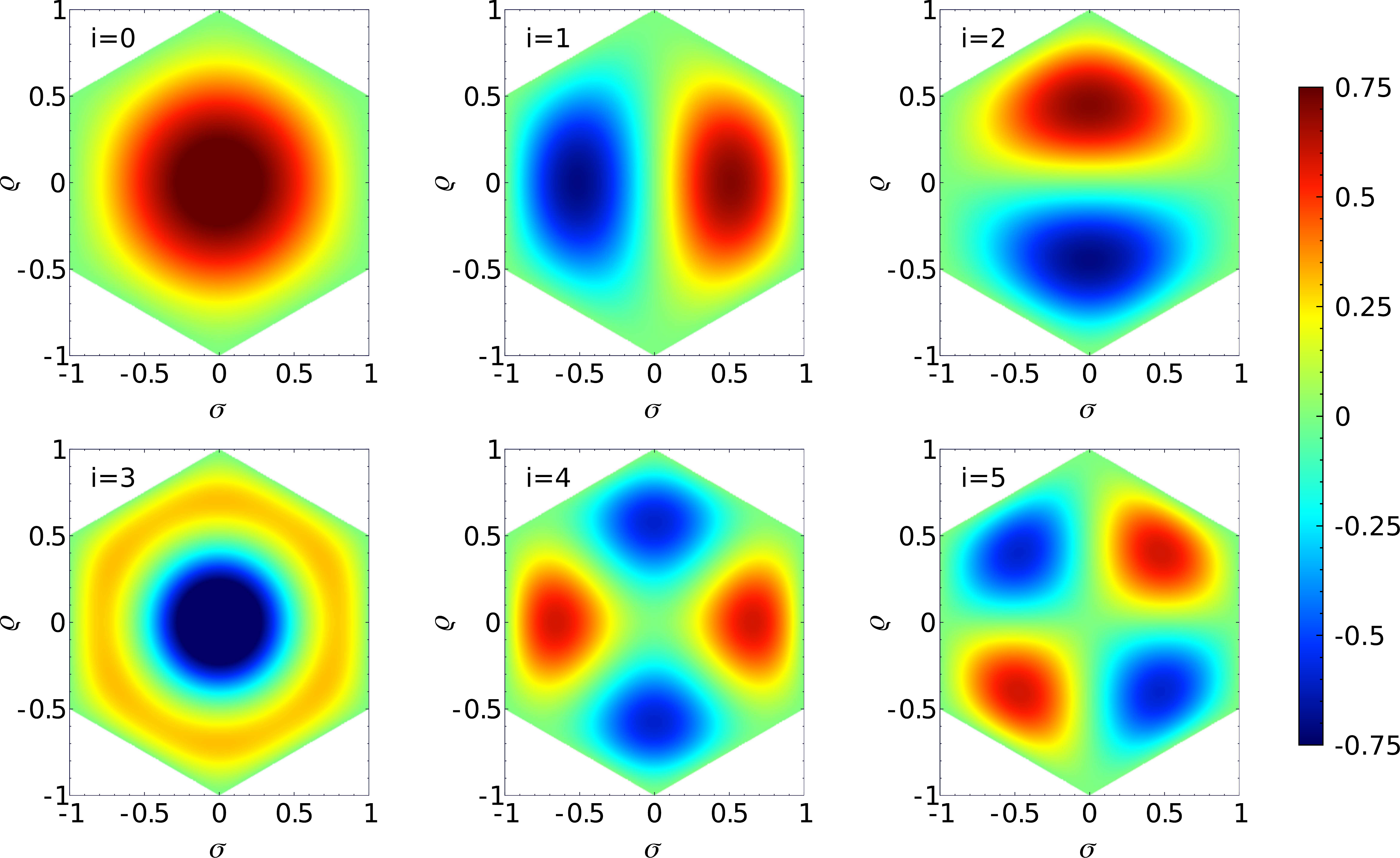}
\end{center}
\caption{Lowest-order ($i=0,\ldots,5$) orthogonal polynomials on the hexagon, multiplied by the corresponding weight function given by Eq.~\eqref{eq:tda_weight_function}. Figure from Ref.~\cite{Pire:2025wbf}.}
\label{fig:tda_fig_Hex}
\end{figure}

With the use of the factorized Ansatz (\ref{Fact_Anz_Quad}), 
the forward (zero skewness) limit of TDA (\ref{eq:spectral_represention_wv}) 
is expressed through 
$f(\sigma, \rho)$ 
defined on the hexagon 
$\sigma \in [-1;\,1]; \; \rho \in [-1+|\sigma|/2; \, 1-|\sigma|/2]$, {see Fig.~\ref{fig:tda_fig_Hex}}. Since  this forward limit is not related to any measured quantity, we construct a flexible parametrization of the forward function developed on a basis of orthogonal polynomials on the hexagon,  which has been devised in the context of hexagonal optical elements. The weight function defining the orthogonal basis is chosen as
\be
&&
W(\sigma, \rho) \sim \left(1-\sigma^2\right)^d\left((\rho-1)^2-\frac{\sigma^2}{4}\right)^d \nn \\ && \times \left((\rho+1)^2-\frac{\sigma^2}{4}\right)^d \,,
\label{eq:tda_weight_function}
\ee
which ensures good convergence at the boundaries of the domain of definition, controlled by the parameter $d$. The first few orthogonal functions are shown in Fig.~\ref{fig:tda_fig_Hex} for $d=1$.

The normalized profile function is written as a 3-body generalization of Radyushkin's factorized double distribution Ansatz profile~\cite{Musatov:1999xp}: 
\be 
&&
h(\sigma,\,\rho,\, \omega,\, \nu) = 
\frac{\Gamma(3b+3)}{2^{5b+2} \Gamma(1+b)^3} \nn \\ &&
\times \left(1+2 \nu -\omega -2 \left|\rho -\frac{\sigma }{2}\right|\right)^b
\nn \\ &&
\times
\left(1-2 \nu -\omega -2 \left|\rho +\frac{\sigma}{2}\right| \right)^b (1 -|\sigma |+\omega)^b
\nn \\ &&
\times \left(1-\frac{1}{2} \left(\left|\rho -\frac{\sigma }{2}\right|+\left|\rho+\frac{\sigma }{2}\right|+|\sigma |\right) \right)^{-3b-2}, \nn \\ &&
\label{eq:profile_function}
\ee
where the parameter $b$ controls the strength of the skewness dependence of the resulting TDA. This allows for spanning a quite large space of parameterizations and enables the extraction of information on TDAs from experimental data.

More data from various reactions are urgently needed to confirm the proposed partonic picture and to extract the nucleon-to-meson TDAs, or at least determine some of their features, in order to study the baryonic and mesonic content of nucleons. The dedicated experiment 
for backward $\pi^0$ production is currently under preparation~\cite{Li:2020nsk}.
The nucleon-to-photon TDAs have also been defined, in particular for the description of  backward DVCS and TCS~\cite{Pire:2022fbi}.
In Ref.~\cite{Pasquini:2024qxn} nucleon-to-photon TDAs were estimated in a model based on light cone hadronic wave functions. Photoproduction experiments, as well as processes from pion~\cite{Pire:2016gut} or antiproton~\cite{Lansberg:2012ha} beams, have been theoretically investigated. They are definitely a potential source of future progress in this field, both with lepton pair and with charmonium production~\cite{Pire:2022kwu}.

\subsubsection{$\gamma \pi$ transition distribution amplitudes and  backward DVCS in the Sullivan process}
\label{sec:bDVCS}

An additional opportunity to explore hadronic structure through near-backward hard exclusive reactions is provided in the meson sector. The pion-to-photon 
($\pi \gamma$) 
TDAs exhibit features analogous to GPDs, as they are matrix elements of the same operators~\eqref{Def_operator}.
Specifically, they share the same support region, \((x,\xi)\in[-1,1]\otimes[-1,1]\), and their interpretation in terms of parton–hadron scattering amplitudes can be constructed~\cite{Pire:2004ie}.

\begin{figure}[!ht]
\begin{center}
 \includegraphics[width=0.8\columnwidth]{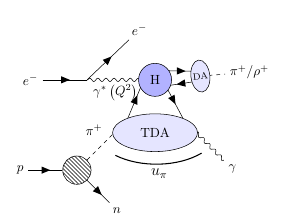} \includegraphics[width=0.8\columnwidth]{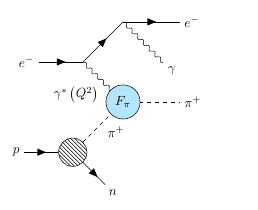}
\caption{\label{Fig_bDVCS}Top panel: Backward DVCS in the Sullivan process. Bottom panel: the corresponding Bethe-Heitler process. This figure is adapted from Ref.~\cite{Castro:2025rpx}.}
\end{center}
\end{figure}

There are four leading twist 
$\pi \gamma$ TDAs: one vector, one axial and two transversity TDAs. In the $\gamma^{\ast} \pi \to \gamma \pi$ process, shown separately in Fig.~\ref{Fig_bDVCS} for the QCD-induced sub-process and its Bethe-Heitler counterpart, only the axial quark TDA $A^{\pi\gamma}$ contributes. It is defined (omitting the Wilson lines required to ensure gauge invariance of the definition) as~\cite{Pire:2004ie,Lansberg:2006fv}
\be 
&&
\label{eq:AxialTDADef}
\frac{e}{f_{\pi}}\left({\cal E}\cdot\Delta\right)A^{\pi\gamma}(x,\xi,u,\mu)=\frac{1}{2}\int\frac{d \kappa}{2\pi}e^{ix (P \cdot n) \kappa }
\nn \\ &&
\langle\gamma(q',\lambda'_\gamma) |\bar{\psi}^{d}(-\kappa  n /2)\gamma^{+}\gamma_{5}\psi^{u}(\kappa n/2)|\pi(p_{\pi})\rangle,
\nn \\ &&
\ee
where $P=(p_\pi+q')/2$, $\Delta=q'-p_\pi$, and  ${\cal E}(q',\lambda'_\gamma)$ is the polarization vector of the final state photon. 

One can relate the axial $\pi \gamma$ TDA (\ref{eq:AxialTDADef}) 
to the axial form factor measured in the pion weak radiative decay $\pi^{+}\rightarrow \ell^{+}\nu_{\ell}\gamma$:
    \begin{equation}\label{eq:FFsDef}
         \int_{-1}^1 dx  A^{\pi\gamma} (x,\xi,u,\mu)  =   \frac{f_\pi}{m_{\pi}} F_{A} (u) \,, 
    \end{equation}
thus imposing a constraint on possible parametrizations. There are various approaches to modeling $\pi \gamma$ TDAs. In Ref.~\cite{Castro:2025rpx}, 
they are first computed in the DGLAP region through the overlap of light-front wave functions, following the procedure already used for GPDs, with models for both the pion and photon light-front wave functions.
The corresponding double distributions are then derived explicitly, allowing the determination of the TDAs in the ERBL domain. Consequently, the resulting models automatically satisfy the polynomiality constraints.

\begin{figure}[!ht]
        \centering
        \includegraphics[width=0.99\columnwidth]{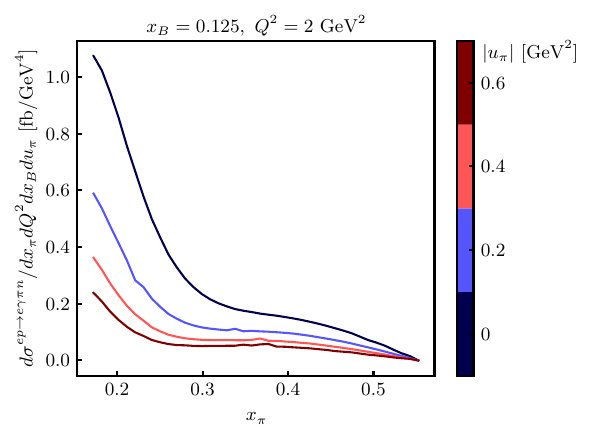} 
        \includegraphics[width=0.99\columnwidth]{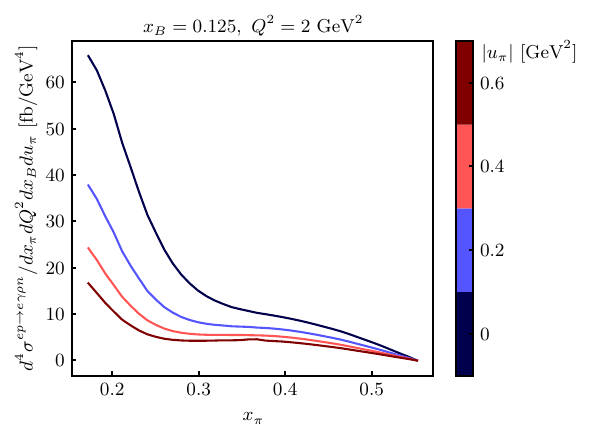}
\caption{\label{fig_CSaxial} Differential cross section for $ep\rightarrow e\gamma\pi n$ (upper-panel) and $ep\rightarrow e\gamma\rho n$ (lower-panel) at typical JLab 22~GeV kinematics, $Q^{2}=2~\textrm{GeV}^{2}$, $x_{B}=0.125$ as a function of the incoming pion momentum fraction $x_{\pi}$, for a number of $u_{\pi}$ values. This figure is taken from Ref.~\cite{Castro:2025rpx}.}
\end{figure}

The calculation of the scattering amplitude for the process $ep \rightarrow e\gamma\pi n$ proceeds in the usual way. As illustrated in Fig.~\ref{Fig_bDVCS}, two sub-processes contribute in this case, as in conventional DVCS. In the backward kinematic regime, the Bethe–Heitler amplitude is found to be negligible relative to the QCD contribution. The resulting estimates for the cross sections are shown in Fig.~\ref{fig_CSaxial} for JLab 22~GeV kinematics~\cite{Accardi:2023chb}, for both $ep \rightarrow e\gamma\pi n$ and $ep \rightarrow e\gamma\rho n$ processes. The latter is sensitive to the vector TDA. The study opens a novel avenue to access $\pi \gamma$ TDAs in the backward DVCS channels. 
The cross section estimates suggest that a first signal could be observed in JLab experiments at current energies, and that a productive program could be pursued with dedicated experiments at higher energies, although more precise feasibility studies are clearly required.

\subsection{Meson pair production in $e^- e^+$ annihilation and GDAs}
\label{sec:pheno:GDAs}

It has been recognized since the early works on GPDs~\cite{Muller:1994ses, Diehl:1998dk, Polyakov:1998ze} that the related concept of di-hadron generalized distribution amplitudes (GDAs) provide access to the quark and gluon structure of mesons. GDAs are defined as the Fourier transforms of the matrix element between the vacuum and the $A \bar A$ di-hadron state of the same operators, {Eqs.}
(\ref{Def_operator})
and
(\ref{Def_operator_G}), 
involved in the definition of the GPDs. For the di-pion chiral-even case, they read
\be
&&
\phi^q(z,\zeta,s) =\frac{1}{2} \int \frac{d \kappa}{2 \pi} e^{i x(P \cdot n) \kappa} \nn \\ && \left\langle \pi(p_\pi) \pi(p'_\pi)  \right| \bar{q}(-\frac{1}{2} \kappa n) n \cdot \gamma q(\frac{1}{2} \kappa n)\left|0\right\rangle;   \nn \\ &&
\phi^g(z,\zeta,s) =\frac{1}{P \cdot n} \int \frac{d \kappa}{2 \pi} e^{i x(P \cdot n) \kappa} \nn \\ && \left\langle \pi(p_\pi) \pi(p'_\pi) \right| n_\alpha G_a^{\alpha \mu}(-\frac{1}{2}\kappa n) G_{a \mu}^\beta(\frac{1}{2} \kappa n) n_\beta\left|0\right\rangle, \nn \\ && 
\ee
where $z$ denotes the light-cone momentum fraction of the quark, $\zeta$ is the light-cone momentum fraction of the final-state  meson, $P = p_\pi+p'_\pi$ and $s=P^2$. The golden reaction~\cite{Diehl:2000uv} to access GDAs is the crossed ($s\leftrightarrow t$) process of DVCS, namely
\be
\gamma^*(q) +\gamma(q') \to \pi (p_\pi)+ \pi(p'_\pi),
\label{dipion}
\ee  
in the kinematics where $s \ll Q^2 = -q^2$.
The experimental evidence of the validity of this approach had to wait for the BELLE data~\cite{Belle:2015oin} on $\pi^0 \pi^0$ production, which was successfully analyzed in~\cite{Kumano:2017lhr}, allowing for a first meaningful  leading order extraction of the di-pion chiral even quark GDA.

Recent theoretical progress on the di-pion GDAs includes the calculation~\cite{Lorce:2022tiq,Pire:2023kng,Pire:2023ztb} of kinematical higher twist corrections to the leading order amplitudes of the process~(\ref{dipion}), and the related reaction $\gamma^*(q)  \to \gamma(q')~\pi (p_\pi) \pi(p'_\pi)$. 
It is phenomenologically very interesting
to access these processes in a
broad enough range of $s$ to be able to perform the impact parameter picture interpretation~\cite{Pire:2002ut} of GDAs. A way to access the very elusive chiral-odd GDAs has recently been proposed~\cite{Bhattacharya:2025awq}.
Accessing baryon-antibaryon GDAs has also been recently discussed~\cite{Han:2025mvq,Han:2025eao} which are crucial to check the relations between (nucleon) GPDs and GDAs.

Future progress in this domain is expected on both {the} theoretical and {the} experimental sides. On the one hand, the development of the non perturbative description of mesons with various theoretical tools has recently emerged. There are still too few applications of these tools to the determination of di-meson GDAs, but the first attempts~\cite{Dorokhov:2011ew, Mezrag:2023nkp} are promising. Lattice studies of GDAs are still missing but one can hope that the lattice community will tackle soon this challenge. On the other hand, the high luminosity of the current $e^- e^+$ colliders BELLE 2 and BES III opens the way to the detailed study of many channels with scattering amplitudes which depend on di-meson GDAs. 

GDAs are also building pieces of electroproduction amplitudes such as di-meson hard electroproduction when the di-meson invariant mass is moderate~\cite{Warkentin:2007su}. They are also playing an important role in the understanding of various $B$-meson decays~\cite{Chen:2002th,Krankl:2015fha,Yan:2025ocu}.

\subsection{Double Deeply Virtual Compton Scattering}
\label{Sec_DDVCS_prospects}
As mentioned in Sec. \ref{ddvcs_solid}, the study of 
DDVCS represents a crucial step forward in accessing the three-dimensional structure of the nucleon through {GPDs}. Unlike conventional DVCS or TCS, DDVCS involves a spacelike incoming photon and a timelike outgoing photon, enabling an independent variation of {both photon virtualities mapping the ERBL region of GPDs through} the parameters $\xi'$ and $\xi$ defined in Eq. (\ref{xip_sca}) and (\ref{xi_sca}) respectively. Given its unique access to the GPD phase space, particular attention has been given to this process from both theoretical and experimental perspectives. In this context, recent phenomenological studies assess the phase-space coverage and experimental reach of forthcoming measurements at JLab and the future EIC, including projections for beam and target spin-dependent observables supporting the approved experimental programs SoLID$\mu$~\cite{SoLIDu} and $\mu$CLAS12~\cite{uCLAS12}.

\begin{figure}[!ht]
        \centering
        \includegraphics[width=0.99\columnwidth]{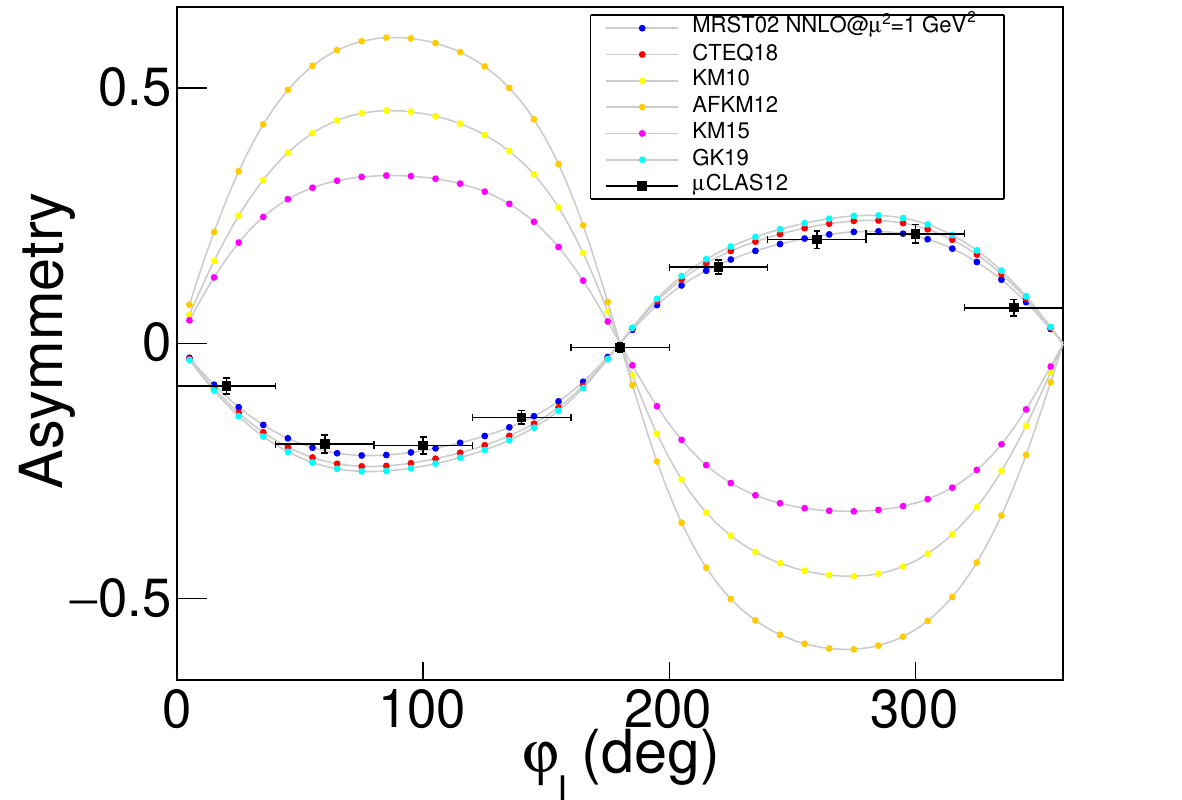}
        \includegraphics[width=0.99\columnwidth]{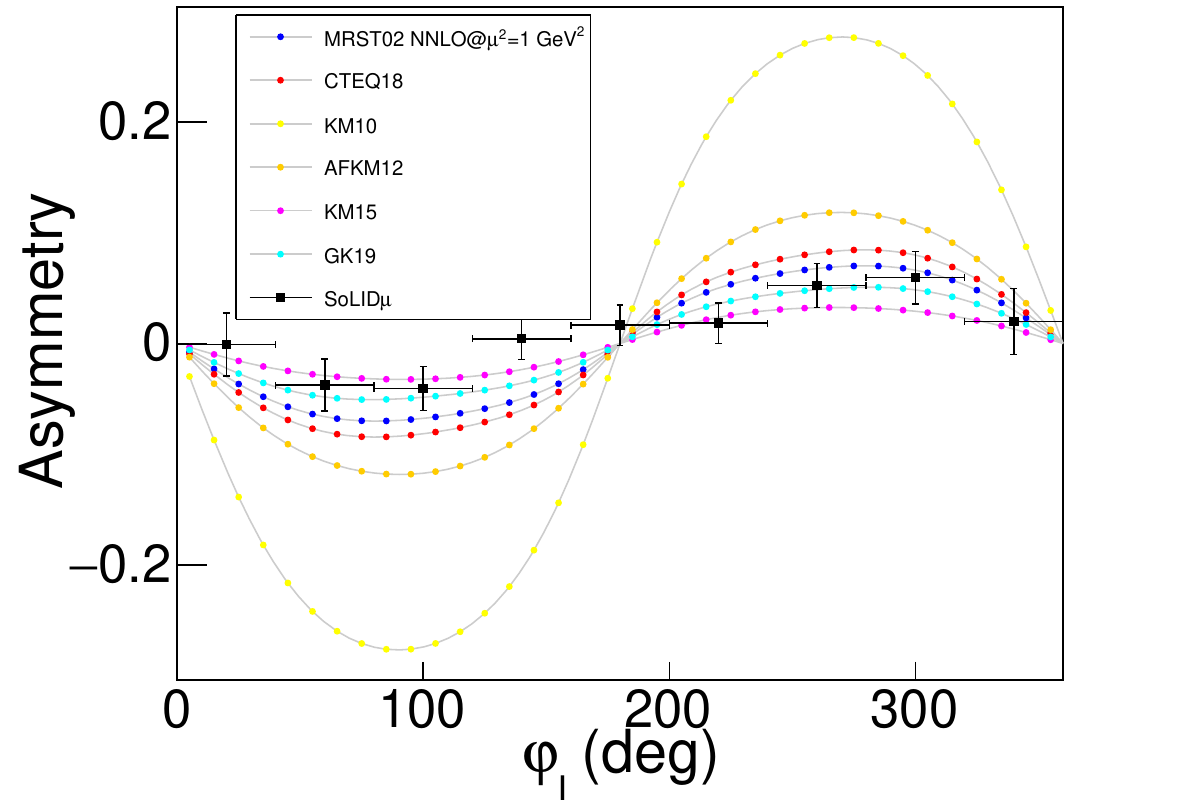}
        \includegraphics[width=0.99\columnwidth]{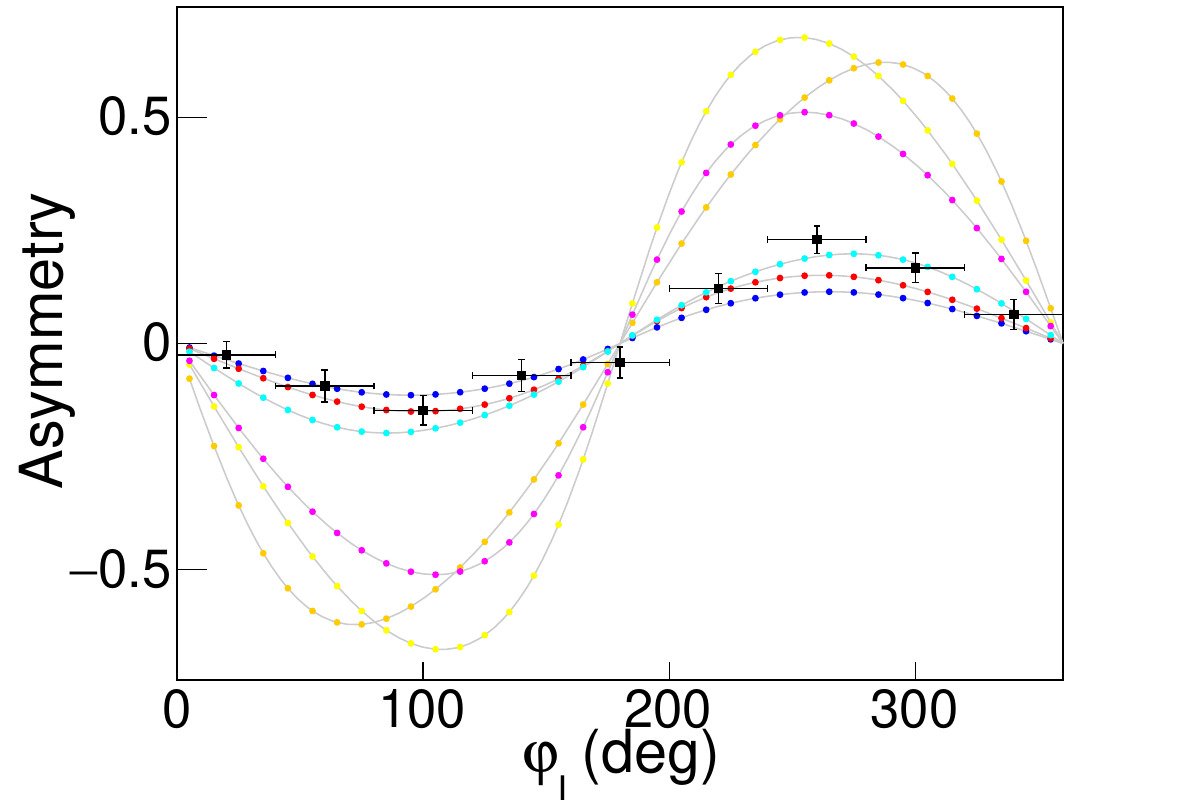}
\caption{{Examples of BSA} projections for the 11 GeV (top) and 22 GeV (middle) configurations at JLab and the EIC (bottom), assuming 100 days, 200 days, and one year of data taking, respectively~\cite{Alvarado:2025huq}. Bin boundaries are given by $\xi'<-0.056$ and $\xi>0.45$ (left), $-0.053<\xi'<0.01$ and $\xi<0.214$ (middle), and, $\xi'>-0.005$ and $\xi<0.00417$ (right).}
\label{ddvcs_bsa_11_22_eic}
\end{figure}

Assuming the foreseen acceptance of the SoLID$\mu$ and $\mu$CLAS12 spectrometers, the phase space coverage of the approved experiments was studied in~\cite{Alvarado:2025huq} for both the 11~GeV and 22~GeV beam configurations, following the ongoing discussions for a CEBAF energy upgrade, and for a nominal luminosity of $\mathcal{L}=10^{37}\;\text{cm}^{-1}\cdot\text{s}^{-1}$. The physics case is driven by the access to GPDs through BSA measurements, whose projected statistical precision suggests a significant discriminating power among GPD models while enabling three-dimensional exploratory measurements across the DDVCS phase space, as shown in Fig.~\ref{ddvcs_bsa_11_22_eic}. Comparable sensitivity was obtained for the Target Spin, the Double, and the Beam charge asymmetry projections, assuming $80\%$ polarized NH$_{3}$ targets.

Currently, only the GK and VGG models offer explicit support for DDVCS computations, although the latter is based on an early implementation that has become outdated. The KM model, in contrast, is currently limited to DVCS kinematics. To enable a more comprehensive phenomenological exploration, an extension of the KM framework was implemented in~\cite{Alvarado:2025huq} to produce DDVCS predictions. Nevertheless, further studies on the analyticity of the Compton tensor are required to link the real and imaginary parts of CFFs through a consistent dispersion relation. Furthermore, target-spin–dependent observables are found to be strongly suppressed at EIC kinematics, a feature likely driven by the absence of sea-quark and gluon contributions in existing GPD frameworks, which are expected to dominate the GPD dynamics at higher center-of-mass energies. Hence, it is foreseen that continued model development and refinement will be essential to achieve meaningful interpretations once DDVCS data becomes available.

\begin{figure}[!ht]
        \centering
        \includegraphics[width=0.8\columnwidth]{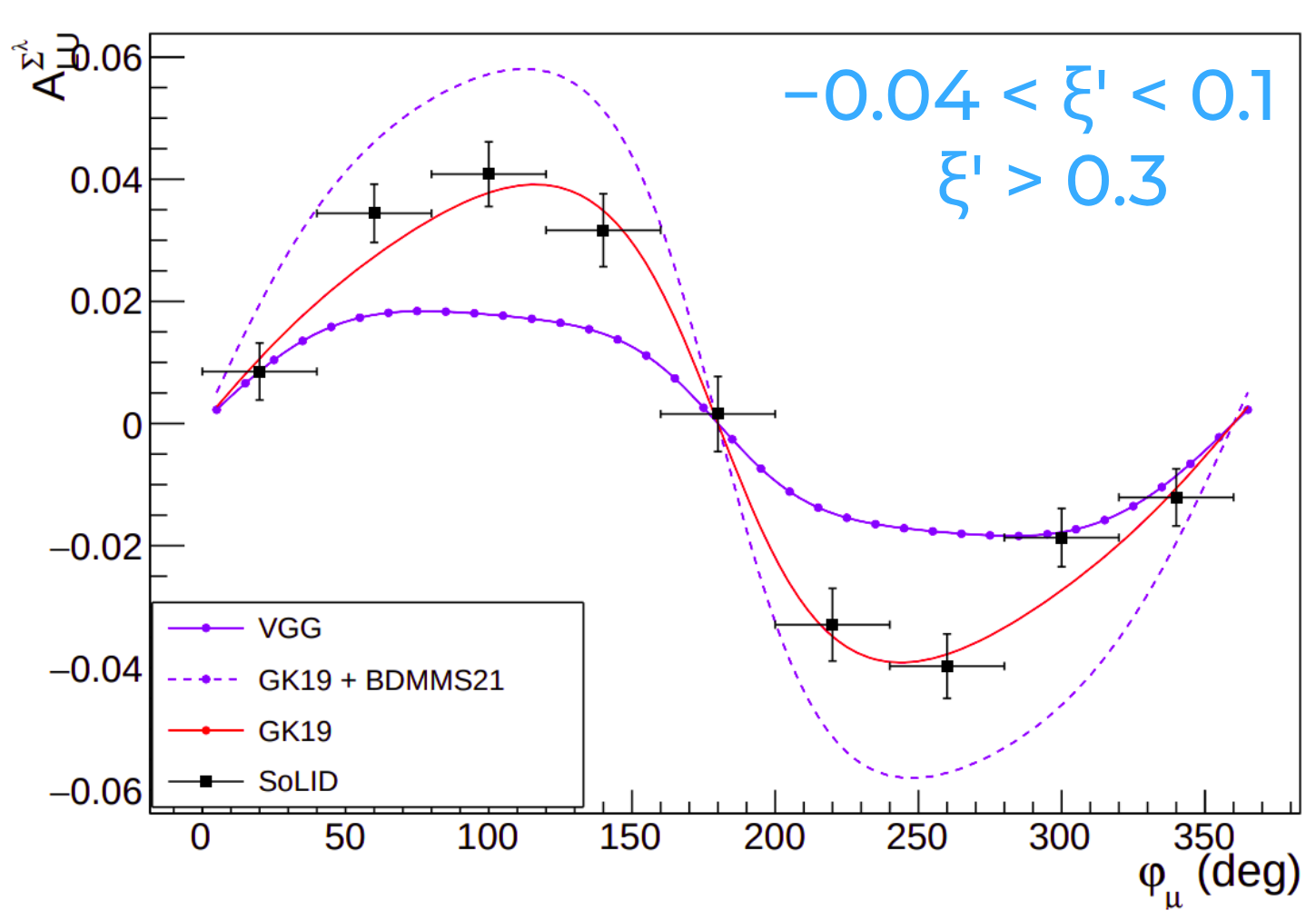}
        \includegraphics[width=0.8\columnwidth]{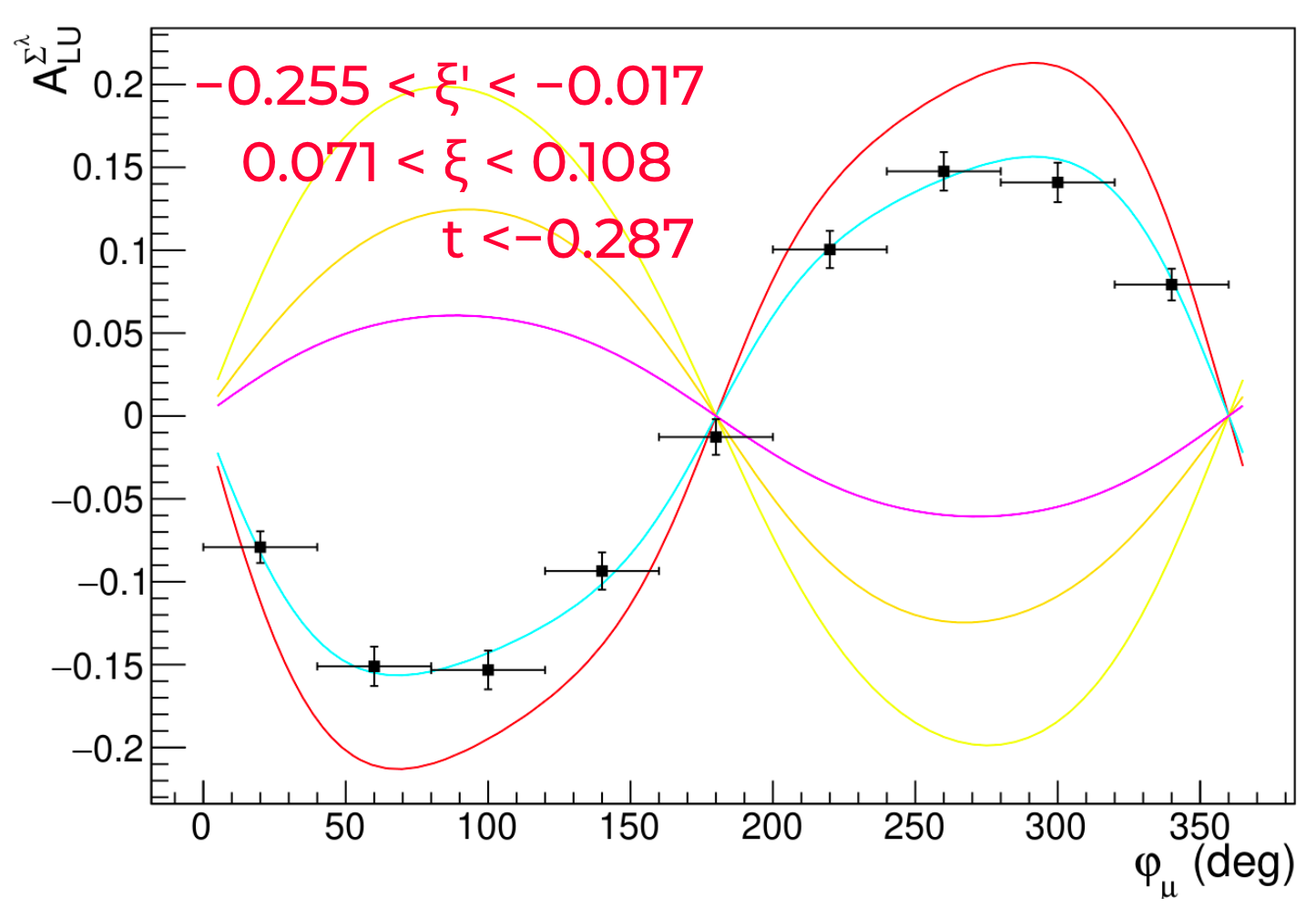}
        \includegraphics[width=0.83\columnwidth]{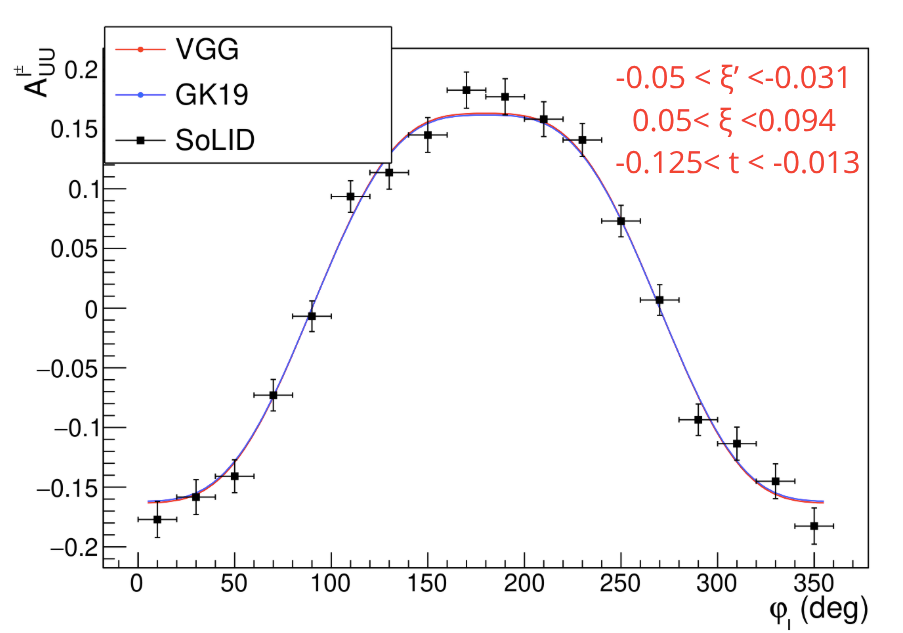}
\caption{Example BSA projections for the SoLID$\mu$ experiment proposal in the $\xi'>0$ (top) and $\xi'<0$ (middle) regions. Example of a $\mu$CA projection (bottom) for the SoLID$\mu$ experiment proposal. {Shadow GPD contributions are described by the BDMMS21 model.}}
\label{ddvcs_solid_prop}
\end{figure}

The experimental prospects for DDVCS are particularly promising for SoLID$\mu$ and $\mu$CLAS12. In particular, it is foreseen the first-time observation of 
{a sign change for the BSA}
when transitioning between the $\xi' > 0$ and $\xi' < 0$ kinematic regions, a distinctive feature linked to the analytic structure of CFFs. {Moreover, the phase space coverage extends to $\xi$ values up to 0.4, granting access to a kinematic region where the BSA is predicted to receive significant contributions from shadow GPDs~\cite{Moffat:2023svr,uCLAS12,SoLIDu}, thus providing the first experimental constraints on such models.} An example of the model-predicted BSA and the expected statistical precision is shown in Fig.~\ref{ddvcs_solid_prop}. 

In addition, discussions have recently emerged regarding the access to the real part of the CFFs through {angular asymmetries.} The muon charge asymmetry ($\mu$CA), a measurement proposed for the SoLID$\mu$ experiment~\cite{Belitsky:2003fj}, {compares the $\mu^{+}$ and $\mu^{-}$ detection rates at a fixed azimuthal decay angle $\varphi_{\ell}$ through the ratio $(N^{\mu^{-}}(\varphi_{\ell}) - N^{\mu^{+}}(\varphi_{\ell}))/(N^{\mu^{-}}(\varphi_{\ell}) + N^{\mu^{+}}(\varphi_{\ell}))$, where $N^{\mu^{\pm}}$ denotes the acceptance-corrected experimental yield of charged $\mu^{\pm}$ particles. A representative $\mu$CA experimental projection is shown in Fig.~\ref{ddvcs_solid_prop}, featuring the its large model-predicted amplitude and the expected statistical precision.} This observable would enable experimental sensitivity to the real part of the CFF, providing anticipated results ahead of positron beam configurations and opening a new approach for constraining GPD models in the near term.

In summary, the DDVCS program represents a major opportunity to advance the experimental and theoretical understanding of GPDs beyond the reach of DVCS and TCS. In the coming years, the combined efforts at SoLID$\mu$ and $\mu$CLAS12 will explore the double-virtual regime for the first time, enabling tests of the analytic properties of the Compton amplitude and providing stringent constraints on GPD models. Continued theoretical developments, including the implementation of a dispersion relation and the incorporation of sea-quark and gluon dynamics, will be essential to fully exploit the physics potential of these measurements. Together, these advances will establish DDVCS as a key channel for probing the three-dimensional structure of the nucleon.

\section{Prospective experiments and future experimental facilities  }
\label{sec:new_experiments}

\subsection{Proposed GPD studies with SoLID}

The SoLID~\cite{JeffersonLabSoLID:2022iod} would maximize the science return of the 12~GeV energy upgrade of the 
CEBAF with an attractive combination of large acceptance (full azimuthal $\phi$ coverage) and high luminosity of $10^{37}\,$cm$^{-2}\cdot$s$^{-1}$ ($>100\times$CLAS12, $>1000\times$EIC) capabilities. The SoLID GPD program is particularly well suited to SoLID's open geometry and high luminosity, and is intended to make use of these strengths to provide unique data difficult to obtain in any other manner. 

\subsubsection{DVCS with polarized targets}

DVCS has been the golden channel to experimentally study GPDs~\cite{Ji:1996nm,Belitsky:2001ns}. 
To allow for a full flavor decomposition to extract the GPDs of individual quarks, it is desired to collect precise data over a more complete phase space {than previous measurements} and with different experimental observables. It is especially crucial to conduct measurements with a transversely polarized target, which is essential to access the poorly known GPD $E$. SoLID enables the measurement of DVCS on transversely polarized $^3$He with 11~GeV longitudinally polarized electron beam, where the single-spin asymmetry ($A_{UT}$) and the double-spin asymmetry ($A_{LT}$) provide great sensitivity to decouple different CFFs in the polarized neutron-DVCS reaction. A run-group measurement, in parallel with the already approved SIDIS experiment on longitudinally and transversely polarized $^3$He (E12-10-006), is under exploration. In combination with the DVCS measurement using longitudinally polarized proton targets (E12-11-108), one can perform flavor decomposition to isolate the CFFs of $u$ and $d$ quarks.

\subsubsection{Timelike Compton scattering at high luminosity}

Timelike Compton Scattering (TCS) is the photoproduction of a virtual timelike photon ($Q^{\prime 2} > 0$) on a nucleon, where the final-state virtual photon immediately decays into a lepton pair:
\be 
\gamma(q) + p(p_N) && \to \gamma^{*}(q')+p(p^{\prime}_N)  \nn \\ && \to \ell^{-}(k) + \ell^{+}(k') + p(p^{\prime}_N) \, .
\label{eq:TCS}
\ee 
TCS provides a direct probe of nucleon GPDs, accessing regions of phase space that are complementary to those explored in DVCS (Fig.~\ref{Phy_Spa}). It can provide valuable information for GPD extraction~\cite{CLAS:2021lky} and offer an important test of the universality of GPDs and the QCD factorization approach.

The JLab 12 GeV upgrade enables access to the TCS process  above the resonance production region. The first TCS measurement on the proton was published by CLAS12~\cite{CLAS:2021lky}. Both the beam polarization and forward-backward asymmetries were measured to be nonzero, providing strong evidence for the contribution of the quark-level mechanisms parameterized by GPDs to this reaction. The comparison of the measured polarization asymmetry with DVCS-data-constrained GPD model predictions for the imaginary and real parts of $H$ points toward the interpretation of GPDs as universal functions. The measured forward-backward asymmetry is reproduced significantly better by the VGG model with the $D$-term than without it, suggesting the sensitivity of TCS to the $D$-term. 

The SoLID TCS experiment E12-12-006A~\cite{E12-12-006A} is the perfect next-stage experiment after the experimental demonstration of the existence of the TCS process. With luminosity 100 times larger {than CLAS}, SoLID will allow us to perform a mapping of the $t$, photon virtuality and skewness dependence of TCS. This is essential for understanding factorization, higher-twist effects, and NLO corrections. The experiment will push the TCS study to a precision era.

\subsubsection{Double DVCS with muons} \label{ddvcs_solid}

\begin{figure}[t!]
 \begin{center}
\includegraphics[width=0.48\textwidth]{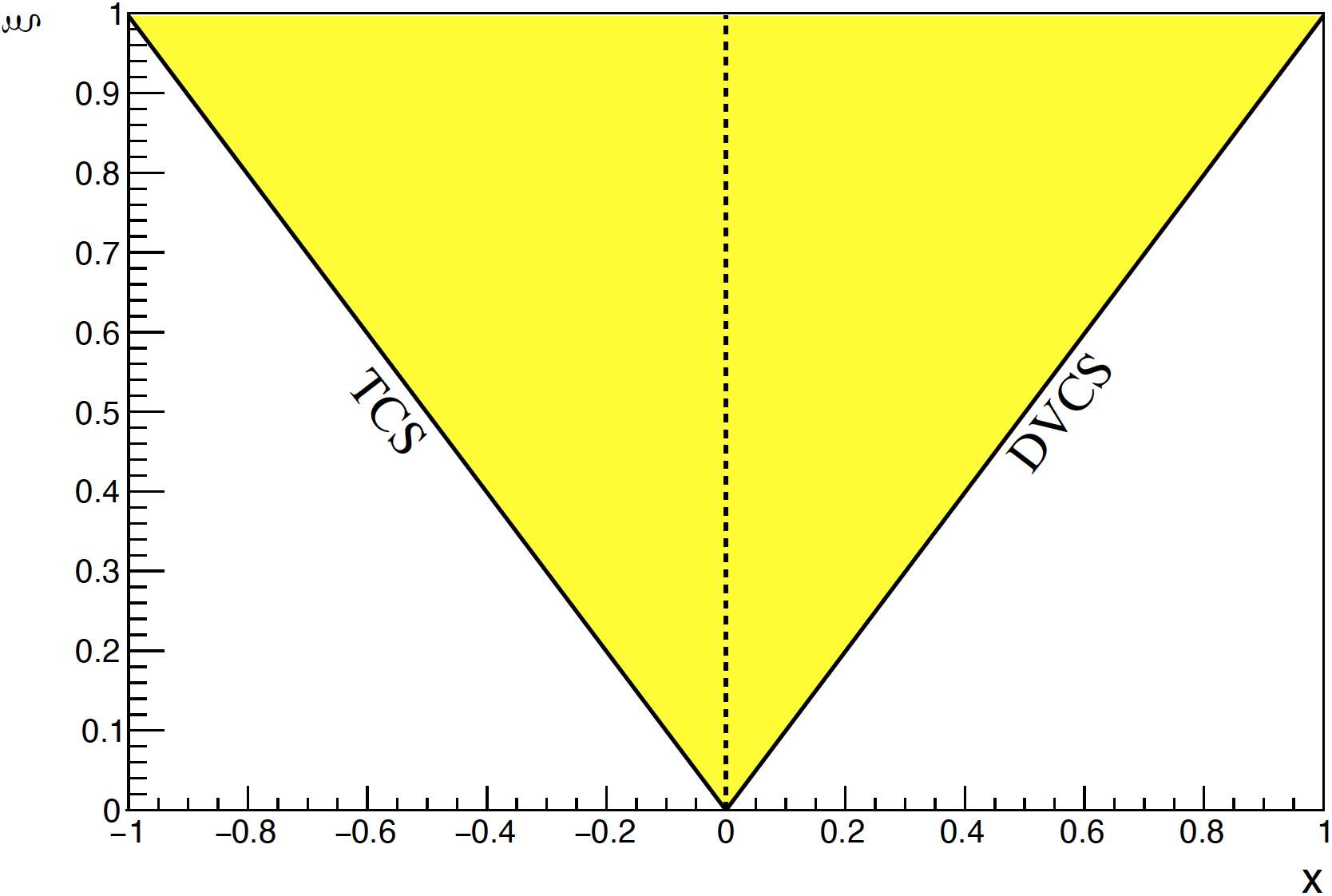}
\caption{GPD coverage of the physical phase space of the imaginary part of the CFFs: the yellow area represents the DDVCS domain, bounded on the one side by the TCS, and on the other side by the DVCS lines. This figure is taken from Ref.~\cite{Zhao:2020th}
}
\label{Phy_Spa}
\end{center}
\end{figure}
Double DVCS~\cite{Guidal:2002kt,Belitsky:2002tf,Deja:2023ahc} provides a unique tree-level, pointwise sensitivity to the ERBL kinematic region of GPDs, enabling direct investigation of the $(x,\xi)$-dependence. At the leading twist-2 and leading $\alpha_s$-order, the DDVCS process corresponds to the absorption of a space-like photon by a parton of the nucleon, followed by the emission from the same parton of a time-like photon decaying into a $\ell \bar{\ell}$-pair (middle panel of Fig.~\ref{Fig_TCS_DDVCS_MBIHER}). 
The scaling variables associated with this process are defined as 
\begin{eqnarray}
\xi' & = & \frac{Q^2-Q'^2+t/2}{2Q^2/x_{B}-Q^2-Q'^2+t} \, , \label{xip_sca} \\
\xi  & = & \frac{Q^2+Q'^2}{2Q^2/x_{B}-Q^2-Q'^2+t} \, , 
\label{xi_sca}
\end{eqnarray}
representing the generalized Bjorken variable ($\xi'$) and the skewness ($\xi$). When the final photon becomes real, the DDVCS process turns into DVCS, which corresponds to the restriction $\xi'=\xi$ in the Bjorken limit. In the limit where the initial photon becomes real, DDVCS turns into the TCS process, corresponding to the restriction $\xi'=-\xi$ 
in the Bjorken limit. In this sense, the DDVCS reaction generalizes both the DVCS and TCS processes, as shown in Fig.~\ref{Phy_Spa}. 

\begin{figure}[t!]
\centering
\includegraphics[width=0.475\textwidth]{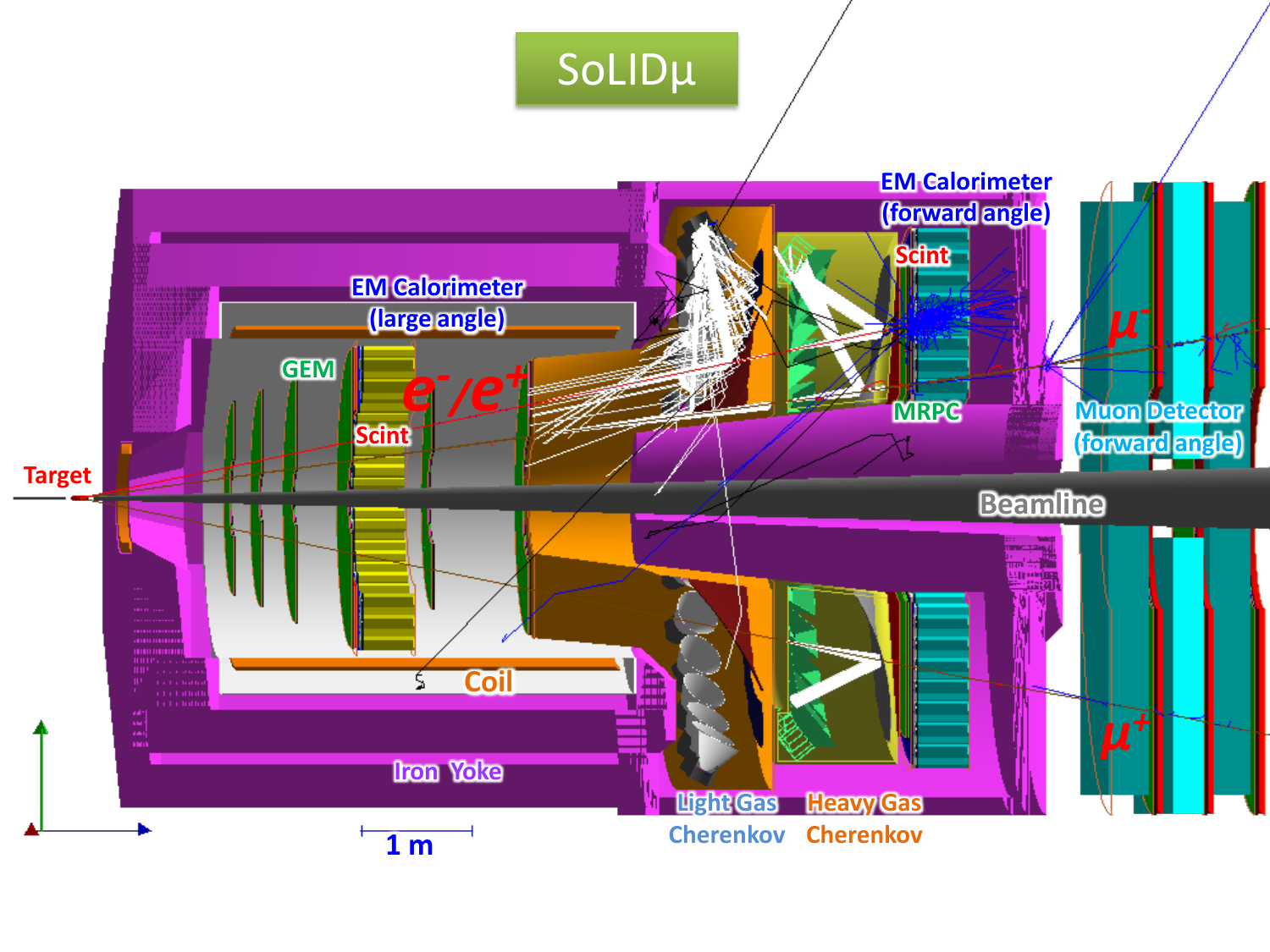}
\caption{\small{SoLID${\mu}$ spectrometer concept, including SoLID and the added forward angle muon detector.}}
\label{fig:solid}
\end{figure}
To achieve these measurements, SoLID is to be complemented with a dedicated muon detector (Fig.~\ref{fig:solid}) to constitute the SoLID$\mu$ spectroimeter. The Forward Angle Muon Detector, located after the downstream end-cap, consists of three layers of iron slabs instrumented with tracking capabilities and scintillators. This configuration provides significant coverage of the DDVCS muons and allows the efficient investigation of the $(\xi,\xi')$ space for $Q^2 \le 10$~GeV$^2$ and $-t < 2$~GeV$^2$. The experiment~\cite{solid_ddvcs} was approved in Summer 2025 to operate over a period of 110 days with a 15~cm long unpolarized liquid hydrogen target and a 3~$\mu$A beam intensity at the luminosity $10^{37}$cm$^{-2}\cdot$s$^{-1}$. It will enable the measurement of the BSA and probe the expected sign change of the imaginary parts of the CFFs $\{ \mathcal{H}, \mathcal{E} \}$ around $Q^2$=$Q'^2$. This represents a strong testing ground for the universality of the GPD formalism~\cite{Anikin:2017fwu}. 

\subsubsection{Deep exclusive meson production with SoLID}

A special kinematic regime is probed in Deep Exclusive Meson Production reactions, where the initial hadron emits a quark-antiquark or gluon pair.  This has no counterpart in the
usual parton distributions and carries information about the $q\bar{q}$ and $gg$ components of the hadron wavefunction.  
Because quark helicity is conserved in the hard scattering regime, the produced meson acts as a helicity filter~\cite{Goeke:2001tz}. In particular, leading order QCD predicts that vector meson production is sensitive only to the vector GPDs, $H$ and $E$, whereas pseudoscalar meson production is sensitive only to the axial-vector GPDs, 
$\tilde{H}$ and $\tilde{E}$.  

The GPD $\tilde{E}$ remains one of the least constrained nucleon GPDs~\cite{Cuic:2020iwt}. It is related to the pseudoscalar nucleon form factor $G_P(t)$, whose extraction is difficult because it plays only a minor role at the momentum transfers characteristic of nucleon $\beta$-decay. Moreover, the contribution of $\tilde{E}$ into standard DVCS observables is usually kinematically suppressed. This makes DVCS alone poorly suited for constraining $\tilde{E}$ especially compared with channels such as pseudoscalar meson production, where pion-pole effects can enhance sensitivity \cite{Goeke:2001tz}. Since the forward limit of $\tilde{E}$ does not reduce to a known parton distribution function, measurements sensitive to $\tilde{E}$ provide access to genuinely new aspects of nucleon structure that cannot be inferred from inclusive scattering data alone. Ref.~\cite{Frankfurt:1999fp} 
identified the single spin asymmetry for exclusive
$\pi^{\pm}$ production from a transversely polarized nucleon target as the most sensitive observable to probe $\tilde{E}$. The experimental access to $\tilde{E}$ is through the azimuthal distribution of the emitted pions. The $\sin(\phi-\phi_s)$ asymmetry, where $(\phi-\phi_s)$ 
is the angle between the target polarization vector and the reaction plane, is related to the parton-helicity-conserving part of the scattering process and is sensitive to the interference between $\tilde{H}$ and $\tilde{E}$ 
\cite{Frankfurt:1999fp,Diehl2005}. Refs.~\cite{Frankfurt:1999fp,Belitsky2004} 
note that ``precocious scaling'' is likely to
set in at moderate $Q^2\sim 2-4$~GeV$^2$ for this observable, as opposed to the absolute cross section, where scaling is not expected until $Q^2>10$~GeV$^2$.
\begin{figure}[t!]
\centering
\includegraphics*[width=0.99\columnwidth]{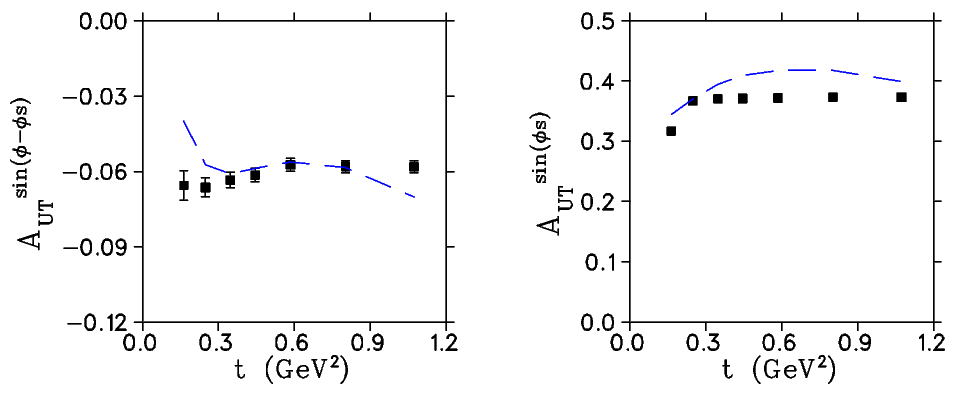}
\hspace{0.2cm}
\caption{\small{
Projected uncertainties for $A_{UT}^{\sin(\phi-\phi_s)}$ and $A_{UT}^{\sin(\phi_s)}$ in the $\vec{n}(e,e'\pi^-)p$ reaction from a transversely polarized $^3$He target and SoLID. $Q^2>$4 GeV$^2$ and $W>2$ GeV restrictions are applied. 
The dashed curve represents the input asymmetry into the simulation, and the data points represent the extracted asymmetry moment values from an unbinned maximum likelihood (UML) analysis of simulated SoLID data. This figure is taken from Ref.~\cite{JeffersonLabSoLID:2022iod}.}}
\label{fig:demp_proj}
\end{figure}

SoLID, in conjunction with a polarized $^3$He target, can be used to probe $\tilde{E}$. Since polarized $^3$He is an excellent proxy for a polarized neutron, the reaction of interest is essentially $\vec{n}(e,e'\pi^-)p$. The only previous data are from HERMES~\cite{HERMES:2009gtv},
for average values $\langle x_B \rangle =0.13$, $\langle Q^2 \rangle =2.38$~GeV$^2$. Although the observed 
$\sin(\phi-\phi_s)$ asymmetry moment is small, the HERMES data are consistent with GPD models based on the dominance of 
$\tilde{E}$ over $\tilde{H}$ at low $-t=-(q-p_{\pi})^2$ 
\cite{Goloskokov:2009ia}. SoLID will permit measurements of the $\sin(\phi-\phi_s)$ modulation of the transverse target spin asymmetry at higher $Q^2$ and $x_B$, with much smaller statistical errors over a wider range of $t$. Fig.~\ref{fig:demp_proj} shows the E12-10-006B 
\cite{atpi_proposal} projections for the two most important transverse single spin asymmetry moments, assuming detection of triple-coincidence $\vec{^3{\rm He}}(e,e'\pi^-p)pp$ 
events. The $\sin(\phi-\phi_s)$ moment (left) provides access to $\tilde{E}$ and is the primary motivation of the measurement. There is growing theoretical interest in the 
$\sin(\phi_s)$ moment (right), as it provides access to the tensor GPD $H_T$. 

\subsubsection{$J/\psi$ production at SoLID}

\begin{figure}[b!]
 \begin{center}
  \includegraphics[width=0.99\columnwidth]{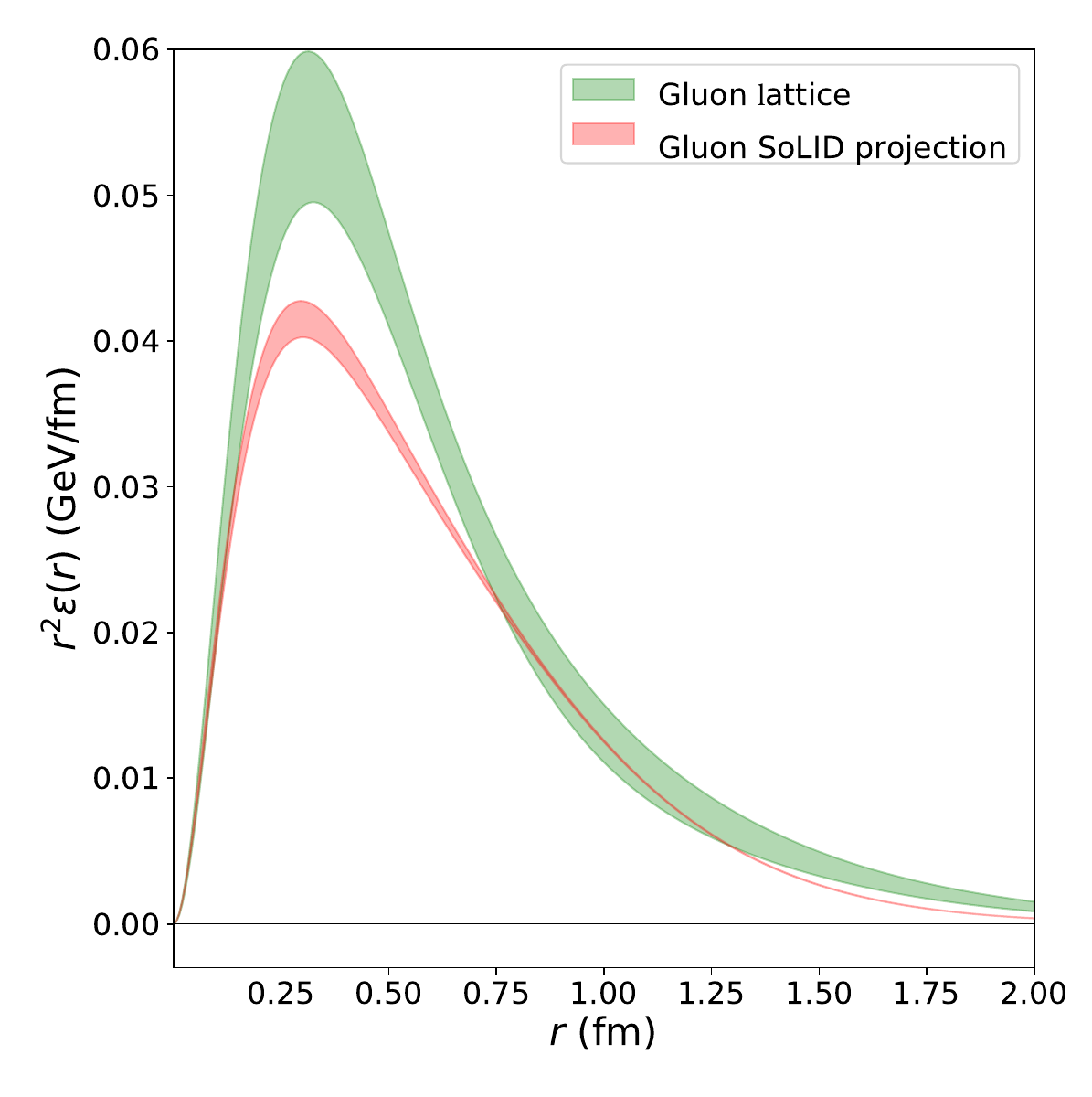}
  \caption{The impact of the SoLID $J/\psi$ experiment on the precision of the knowledge of the mass density (pink) compared to the latest lattice QCD calculations. The difference between Lattice and the simulated SoLID curve accounts for different in $A^g(0)$ between lattice and  simulations.}
  \label{fig:simg-solid}
 \end{center}
\end{figure}
Measurements of $J/\psi$ photo-production near threshold at JLab will improve dramatically with the realization of the Solenoidal Large Intensity Device~\cite{JeffersonLabSoLID:2022iod}. 
Furthermore, electro-production studies near threshold--beginning at a virtual-photon four-momentum 
\( Q^{2} \approx 1.5\ \text{GeV}^2\) and decreasing to about 
\(Q^{2} \approx 0.5\ \text{GeV}^2\) at the highest invariant mass--will also become feasible with an 11~GeV incident electron beam.

In Fig.~{\ref{fig:simg-solid}} we show the impressive impact of an extraction of the gluon mass density from simulated precision data of SoLID using the cross sections obtained from the hQCD model using the parameters obtained from the $J/\psi$-007 experiment. The simulated data were obtained assuming a 3$\mu$A incident electron beam striking a 15 cm long liquid hydrogen target leading to a luminosity of 10$^{37}$ cm$^{-2}\cdot$s$^{-1}$ for a period of 50 days. The photo-production of $J/\psi$ configuration of SoLID uses a 3-fold coincidence where the $J/\psi$ $e^+e^-$ decay-pair together with the recoiling proton are detected. A similar impact can be shown for the scalar density.

\subsection{Prospective experiments with JLab at 22~GeV}

The 
CEBAF at JLab is a world-class facility at the forefront of nuclear physics research. With its 12~GeV program~\cite{Arrington:2021alx}, CEBAF has provided an unprecedented window into the fundamental structure of matter, particularly in the realm of hadron physics. 

To push the boundaries of knowledge and address some of the most profound questions in modern physics, the community is preparing for a new era of discovery. A significant effort is underway to develop the scientific case and technical feasibility for a staged upgrade of CEBAF, which would include a major energy increase to 
22~GeV.  

{
This upgrade can
be realized by leveraging recent advancements in accelerator technology, allowing the energy reach to be extended within the existing tunnel footprint and using the existing CEBAF 
Superconducting Radio-Frequency
(SRF) cavity system. The proposal involves replacing the highest-energy arcs with Fixed Field Alternating Gradient (FFA) arc~\cite{Bogacz:2024phg} and increasing the number of recirculations. This novel permanent magnet technology will also lead to significant energy savings and lower operating costs.  }

\begin{figure}[h!]
    \centering
\includegraphics[width=0.99\columnwidth]{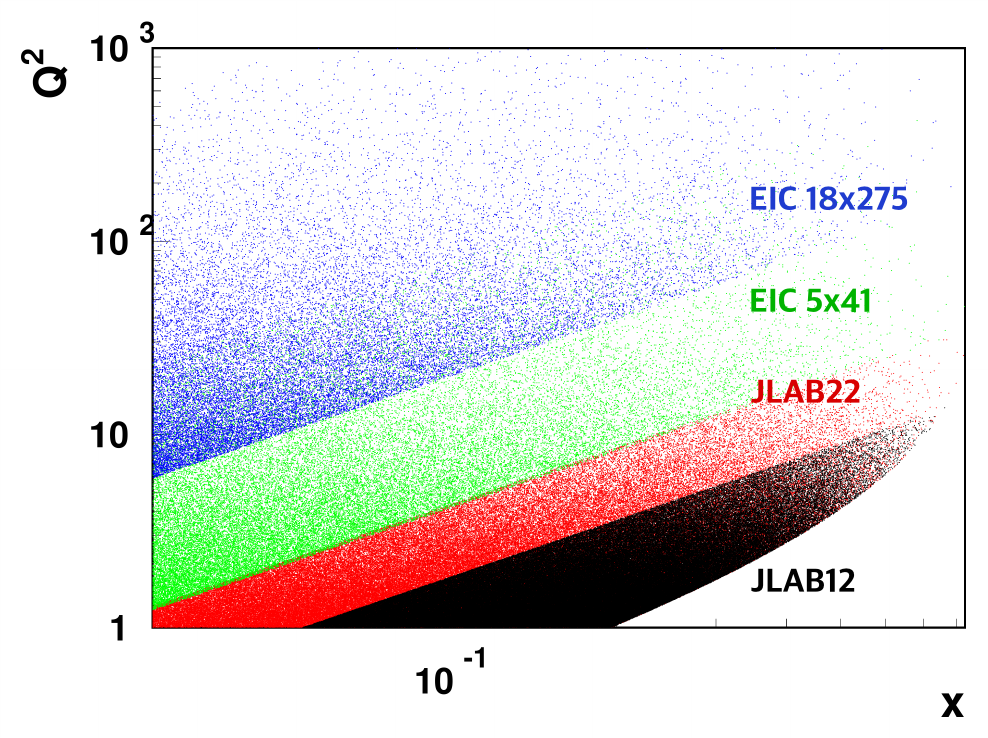}
    \caption{Kinematic coverage $Q^2$ vs $x$ of JLab at 12~GeV,  at 22~GeV, and EIC for the two energies configurations (low and high). Figure courtesy of Harut Avagyan.} 
    \label{fig:phase-space}
\end{figure}

{
A central motivation for the upgrade is the substantial extension of the accessible kinematic range, particularly in $Q^2$. The higher beam energy will provide a new window between the current 12~GeV program and the future EIC, as illustrated in Fig.~\ref{fig:phase-space}. This overlap is essential for connecting precision measurements in the valence region with the broader kinematic coverage of the EIC, thereby enabling a more complete understanding of QCD across different energy scales.
}

A 22~GeV program 
{would}
offer unparalleled opportunities to explore the intricacies of QCD in the valence quark regime with high-precision, high-luminosity experiments. Furthermore, it will enable investigations into the transition region toward sea quark dominance and allow access to the study of heavier hadrons with diverse structures (the so-called XYZ states). Operating at the ``luminosity frontier'' with large acceptance detectors and high-precision spectrometers, CEBAF at 22~GeV will provide unique insights into the nature of QCD and the emergence of hadron structure, exploring various facets of the non-perturbative dynamics.A white paper with the initial science case has been published \cite{Accardi:2023chb}, followed by the most recent refinements detailed in \cite{Accardi:2026slw}.


One of the flagship opportunities enabled by the upgrade is the study of the gluonic structure of the nucleon through near-threshold 
photo- and electroproduction production of 
$J/\psi$ 
and higher-mass charmonium states 
($\chi_c$ and $\psi(2S)$). 
{Recent experimental results from JLab show the feasibility of these studies \cite{Duran:2022xag},\cite{GlueX:2019mkq},\cite{CLAS:2026lls} and more precise measurements at 11 GeV are planned in Hall B, Hall C, Hall D, and with the SoLID detector in Hall A ~\cite{Chen:2014psa}}.
These measurements  provide crucial insights into the proton's gluonic structure through the study of  the  gluon GFFs. 
A recent study 
\cite{Pentchev:2025qyn} 
shows that the gluon GFFs extracted from the available 
$J/\psi$  data using two diametric approaches yield compatible results, and these data are also on the same scale as lattice calculations. Therefore, more comprehensive experimental and theoretical studies are needed to better understand the reaction mechanism. Furthermore, with an energy upgrade, an order of magnitude increase in the polarization figure-of-merit in GlueX will allow for measurements of polarization observables essential for disentangling reaction mechanisms.




%
%
{The increased beam energy is equally important for the study of hard-exclusive reactions.}
In addition to high luminosity, the higher beam energy would push the momentum transfer $Q^2$ into a region where the theoretical framework for GPDs is most applicable. It  would therefore be a game changer for DDVCS measurements, {\it cf.} Sec.~\ref{Sec_DDVCS_prospects}.
The detectors at JLab, such as an upgraded CLAS12 and SoLID, would be able to handle the high rates, making it possible to measure this rare process. Simulations show that it would provide the necessary kinematic reach and enable access to a resonance-free region between 2 and 3~GeV, which is ideal for studying the scaling and evolution of GPDs.

{Beyond nucleon imaging, the upgrade will also substantially advance studies of meson structure, particularly through measurements of pion and kaon electromagnetic form factors.}
Measuring these
form factors reveals the internal spatial distribution of their charge and momentum, providing insights into the dynamics of the strong force, including confinement and chiral symmetry breaking.
For a spin-$0$ particle, like the pion, its form factor is directly extracted from the longitudinal cross section ($\sigma_L$). The corresponding high-precision measurements are typically carried out in Hall C that is currently the world's only facility capable of performing $L/T$ separations over a wide kinematic range. 
Experiments with the 12~GeV CEBAF program
%
have already made significant contributions to this field.
%
%
The 22~GeV CEBAF upgrade would extend the kinematic range of these measurements, providing even more precise data, and greater overlap with EIC data, enabling critical cross-checks with its future findings.

{The higher beam energy will also greatly expand the reach of elastic and transition form-factor studies.}
Extending nucleon, pion, and resonance transition form factor measurements to momentum transfers $Q^2$ of approximately 30~GeV$^2$ is a pivotal goal of the CEBAF 22~GeV upgrade. In fact, data at $Q^2\sim 30$~GeV$^2$ will serve as a critical test of theoretical models describing the transition from a non-perturbative, confinement-dominated behavior to a more perturbative, asymptotically free nature of the strong force. Furthermore, these studies are essential for elucidating the mechanism of the emergence of hadron mass, particularly within the Dyson-Schwinger approach.


{
In summary, a 22~GeV CEBAF would substantially extend the scientific reach of the current 12~GeV program by enabling precision studies of gluonic structure, hard-exclusive reactions, meson structure, and elastic and transition form factors at unprecedented momentum transfers. Operating at the luminosity frontier, it would provide a unique bridge between present JLab measurements and the future EIC, thereby strengthening the worldwide effort to understand the emergence of hadron structure from QCD.
}

\subsection{DVCS at EIC 
} \label{sec:DVCS}

\begin{figure}[!ht]
\centering
\includegraphics[width=0.99\columnwidth]{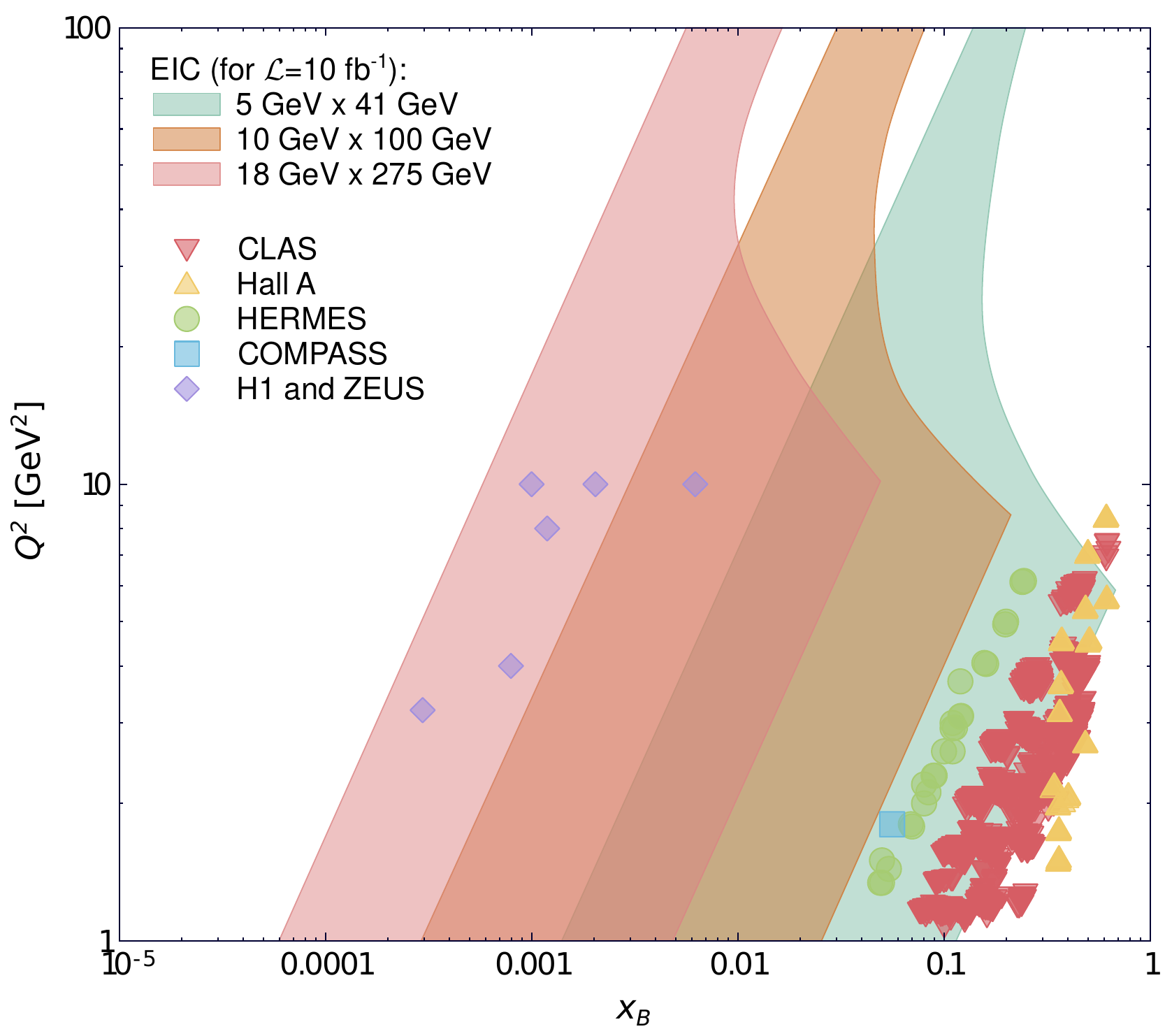}
\caption{Coverage of the $(x_B$,~$Q^2)$  region by ePIC and existing DVCS data for a proton target. Figure reproduced from Ref.~\cite{Aschenauer:2025cdq}.}
\label{fig:Aschenauer:2025cdq:xBVsQ2AndxBVstOverQ2AndData}
\end{figure}

New impact studies for the measurement of DVCS at the EIC have been recently presented in Ref.~\cite{Aschenauer:2025cdq}. They provide a detailed assessment of the impact that future EIC measurements will have on nucleon tomography and CFF extraction, incorporating detector effects, radiative corrections, and background contributions. Using the latest ePIC detector design and state-of-the-art simulation tools, including the EpIC MC generator~\cite{Aschenauer:2022aeb}, the analysis demonstrates that the EIC’s capabilities will enable precise access to GPD-sensitive observables across a broad range in $x_B$ and $Q^2$ (see Fig.~\ref{fig:Aschenauer:2025cdq:xBVsQ2AndxBVstOverQ2AndData}). The foreseen measurements will fill longstanding gaps in the kinematic plane, particularly at small $x_B$, and allow for meaningful tests of GPD evolution.

In the new study, backgrounds from QED radiation and exclusive $\pi^0$ contamination were carefully evaluated. QED effects are shown to be localized in extreme regions of inelasticity $y$, which are mostly excluded by standard cuts. The background from exclusive $\pi^0$ production was quantified via full MC simulation, showing sub-percent contamination at the lowest beam energy configuration and even less at higher energies. The excellent performance of the ECAL and the suppressed cross section for exclusive $\pi^0$ production at small $x_B$ are key to achieving this low background, confirming the experimental feasibility of clean DVCS measurements at the EIC.

\begin{figure}[!ht]
\centering
\includegraphics[width=0.99\columnwidth]{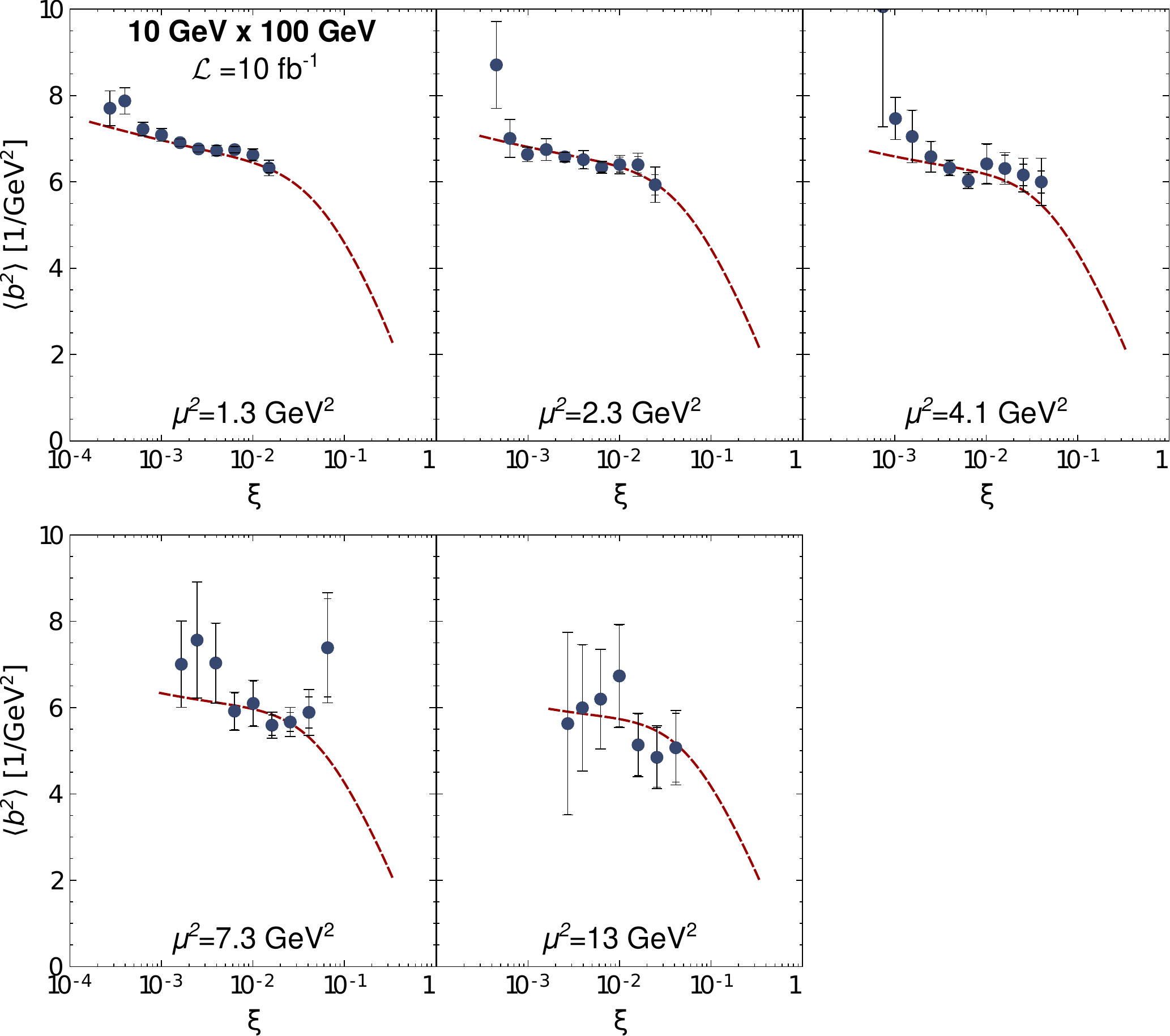}
\caption{Projections for the average transverse sizes of $q^{\mathrm{DVCS}}(\xi,b)$ distributions (nucleon tomography as probed by DVCS), $\langle b^2 \rangle = ({\int d b\,  b^2\,q^{\mathrm{DVCS}}(\xi,b)})/({\int d b\, q^{\mathrm{DVCS}}(\xi,b)})$, for EIC as a function of $\xi$ for $\mu^2 \equiv Q^2$ bins and $\mathcal{L}=10\,\mathrm{fb}^{-1}$. The reference values obtained from the GK model are denoted by the red dashed curves. Figure reproduced from Ref.~\cite{Aschenauer:2025cdq}.}
\label{fig:Aschenauer:2025cdq:nt_sum}
\end{figure}

\begin{figure}[!ht]
\centering
\includegraphics[width=0.99\columnwidth]{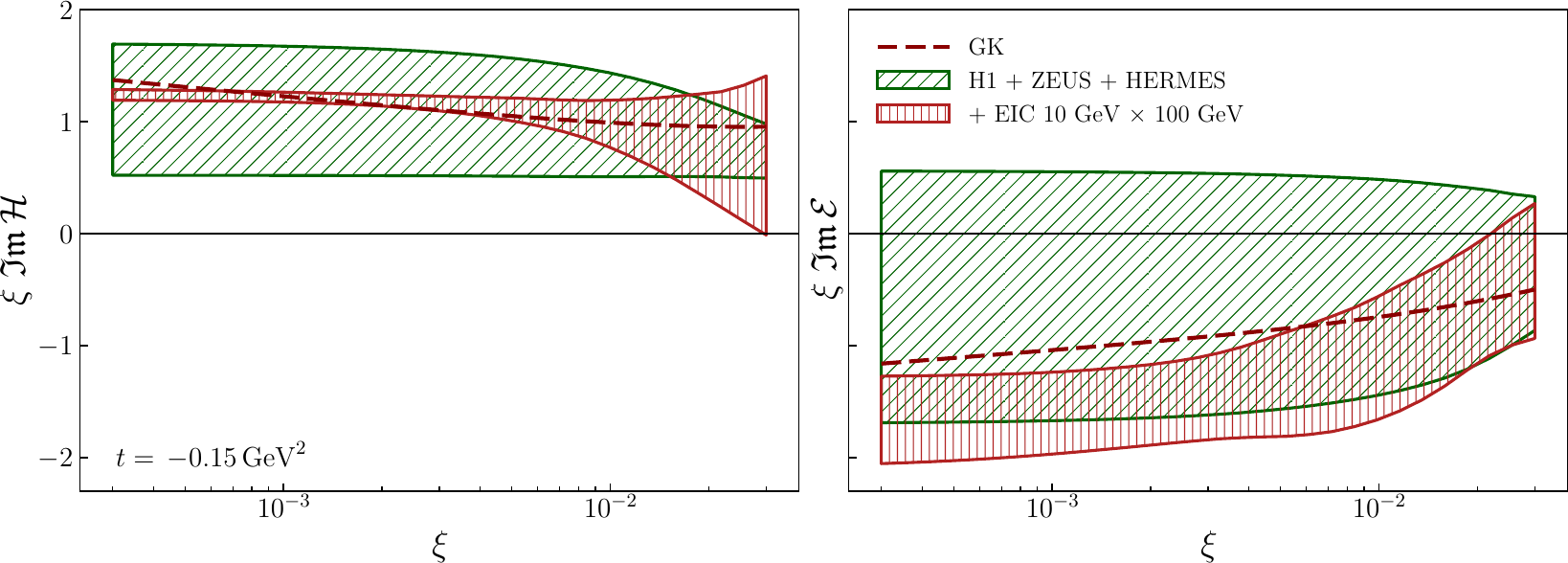}
\includegraphics[width=0.99\columnwidth]{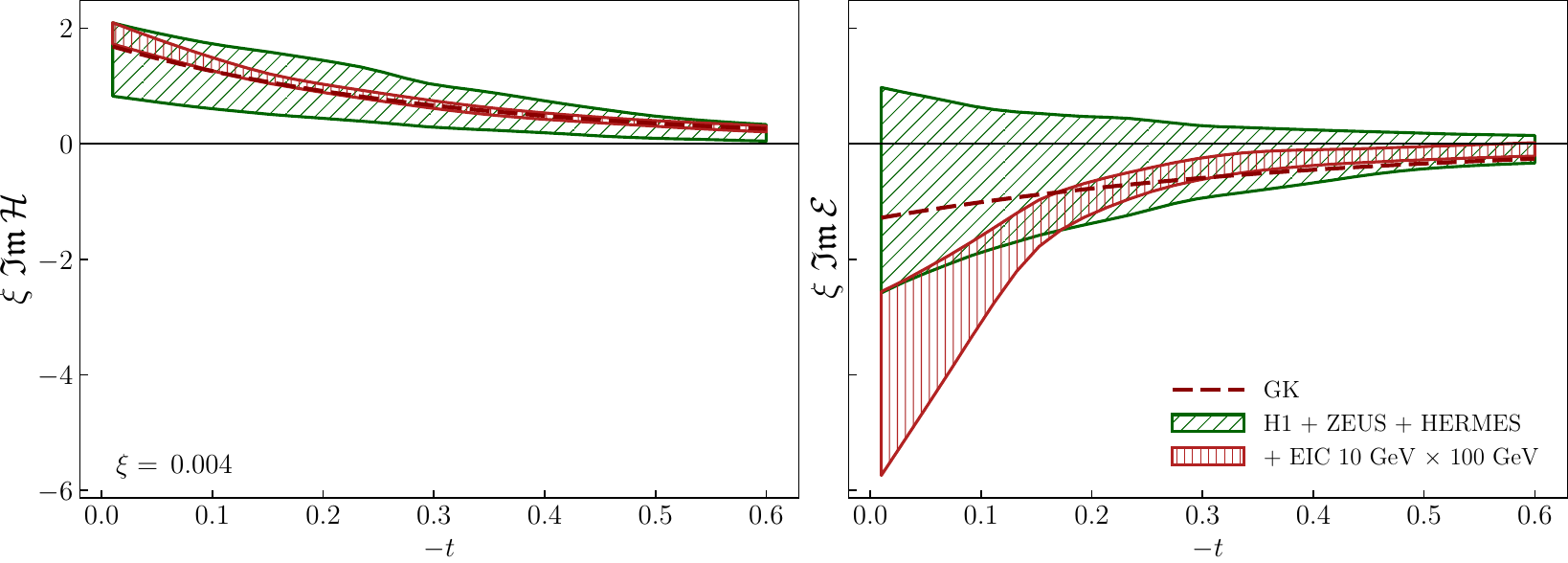}
\caption{CFFs $\im\mathcal{H}$ (left) and $\im\mathcal{E}$ (right) extracted from $A_{LU}$ asymmetries (projections for EIC for $\mathcal{L}=10\,\mathrm{fb}^{-1}$) as a function of $\xi$ (first row)
    and $t$ (second row), as extracted by training an ensemble of neural nets to only old HERA data
(green slanted dashes) and additionally to simulated EIC data (red vertical dashes) at $Q^2 = 4\,\mathrm{GeV}^2$.
GK model values are plotted for comparison (red dashed line). Figure reproduced from Ref.~\cite{Aschenauer:2025cdq}.}
\label{fig:Aschenauer:2025cdq:nncff}
\end{figure}

The study includes a demonstration of extracting nucleon tomography information and CFFs from unpolarized cross sections and $A_{LU}$ asymmetries, respectively. The main results are presented in Figs.~\ref{fig:Aschenauer:2025cdq:nt_sum} and~\ref{fig:Aschenauer:2025cdq:nncff}. One can see that nucleon tomography at the EIC will benefit from wide kinematic coverage and high-precision measurements, enabling detailed mapping of the proton’s spatial and spin structure across varying $x_B$ and $Q^2$. The analysis demonstrates that evolution effects impacting tomographic extractions, currently poorly understood, can be systematically studied using the EIC's extended lever arm in $Q^2$. 

{
The extraction of CFFs from $A_{LU}$ asymmetries is based on a framework incorporating flexible functional forms close to non-parametric methods to reduce model bias and improve uncertainty propagation. At EIC kinematics the asymmetry is mostly sensitive to the interference between DVCS and BH, and can be used to constrain  the imaginary parts of the CFFs $\mathcal{H}$ and $\mathcal{E}$, see Eq.~\eqref{eq:1}.
The sensitivity to the CFF $\mathcal{E}$ is due to the excellent statistics and long lever arm in $t$ available at the EIC.
}

The impact studies highlight the significant constraining power of EIC measurements, particularly on $\mathcal{H}$, and also on $\mathcal{E}$, which are central to the Ji sum rule analysis. 
The study provides a clear benchmark for the precision and robustness of future extractions, and reinforces the EIC’s role as a critical facility for GPD and 3D nucleon structure studies.

\subsection{Capabilities for partonic imaging with the ePIC experiment}

The EIC science program has a robust component devoted to exclusive physics, with an emphasis on partonic imaging, as seen in both the EIC white paper~\cite{accardi2014electronioncolliderqcd} and the EIC Yellow Report~\cite{AbdulKhalek:2021gbh}. In order to make a clean event selection of exclusive production off a proton possible, it is desirable to tag the scattered beam proton, which is nearly collinear to the outgoing hadron beam, in addition to the reconstruction of the centrally produced, new final-state particles, such as a photon in DVCS (see Sec.~\ref{sec:DVCS}) or a vector meson. In addition to the partonic imaging program in electron-proton collisions, there
is an increasing interest in elucidating the partonic structure of the nucleus, using an electron beam~\cite{PhysRevD.104.114030, TU2020135877, Jentsch:2021qdp, FRISCIC2021136726}.  

\begin{figure}[h]
    \centering
   \includegraphics[width=0.99\columnwidth]{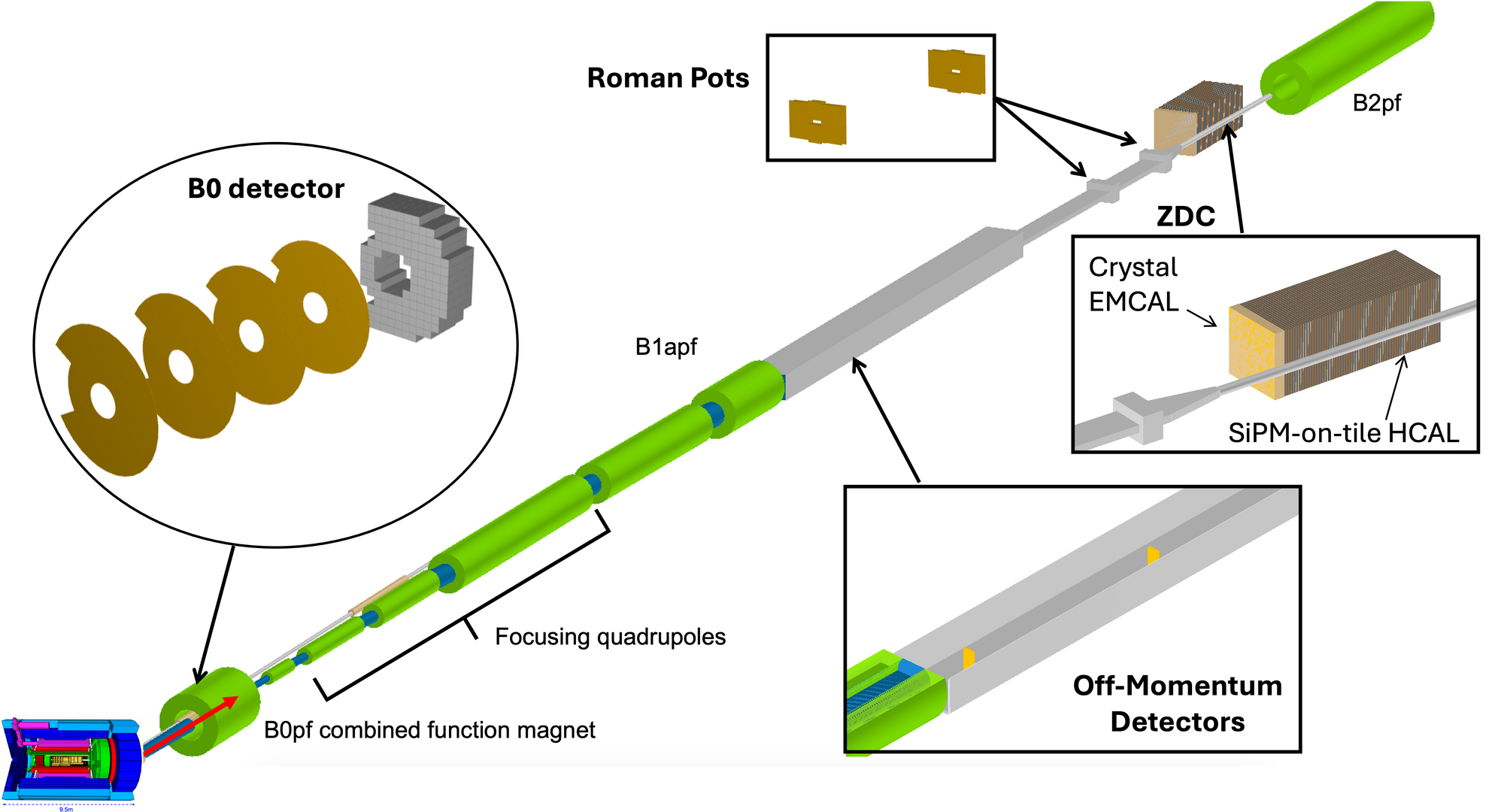}
    \caption{The lattice of magnets to steer and focus the outgoing hadron beam is depicted by the green cylinders in the figure. Each of the four subsystems is directly integrated into the beamline. Starting from the interaction point and moving from left to right, the B0 detector, the off-momentum detectors, the Roman pots, and the zero-degree calorimeter are indicated. The image was generated using the ePIC software framework~\cite{ref:ePIC_dd4hep}.}
    \label{fig:far_forward_detectors}
\end{figure}

In order to enable the full breadth of the proposed exclusive physics program, a detector system integrated with the outgoing hadron beam, the so-called ``far-forward'' detector system, has been included in the ePIC design from the very beginning. Four independent subsystems cover the region $\eta$ > 4.5, which falls well outside the acceptance of the ePIC central detector. These subsystems are represented in Fig.~\ref{fig:far_forward_detectors}, while Table~\ref{tab:FFDetectors_acceptance} summarizes the geometric acceptance for far-forward scattered protons and neutrons. The various subsystem components are detailed below.

\begin{table*}[htp]
\centering
\begin{tabular}{ c c c c }
\hline
Detector & Used for & $\theta$ acceptance [mrad] & $x_{L}$ acceptance  \\
\hline
 B0 tracker    & $p$ & $5.5 - 20.0$  & N/A \\
 Off-Momentum  & $p$ & $0.0 - 5.0$   & $0.45 - 0.65$  \\
 Roman Pots    & $p$ & $0.0 - 5.0$   & $0.6 - 0.95 ^*$ \\
 Zero-Degree Calorimeter & $n$ & $0.0 - 4.0$   & N/A \\
 \hline
\end{tabular}
\caption{ Summary of the geometric acceptance with the current ePIC design for far-forward scattered protons and neutrons in polar angle $\theta$ and longitudinal momentum fraction $x_{L} = {p_{z, p'}}/{p_{z, beam}}$, with $p_{z, p'}$ 
the longitudinal scattered-proton momentum and $p_{z, beam}$ the average proton-beam momentum. *The acceptance of the Roman pots at high values of $x_{L}$ depends on the accelerator beam optics, which can be changed to enhance either luminosity or acceptance.}
\label{tab:FFDetectors_acceptance}
\end{table*}

\subsubsection*{The B0 detector} 

\begin{figure}[h]
    \centering
    \includegraphics[width=0.99\columnwidth]{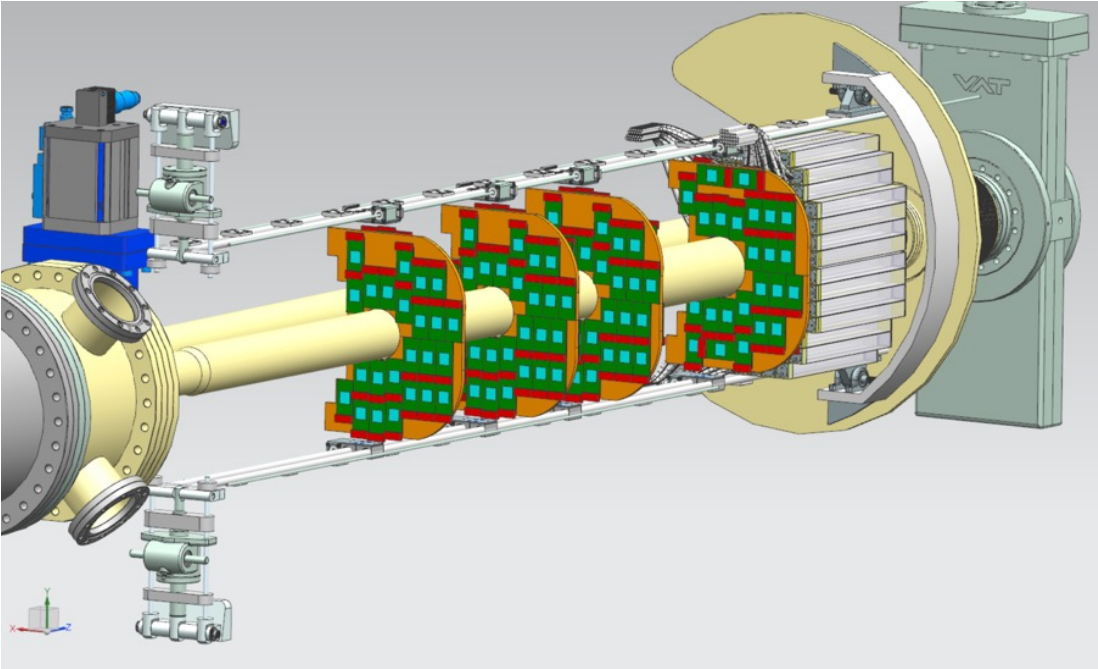}
    \caption{The B0 tracking detector layers (left) followed by the electromagnetic calorimeter crystals (right), inside the B0 magnet. The outer magnet substructure is not represented in the picture. }
    \label{fig:B0_detectors}
\end{figure}

The B0 detector, depicted in Fig.~\ref{fig:B0_detectors}, consists
of a tracking system and an electromagnetic calorimeter, embedded inside the first dipole magnet (B0pf), when exiting the ePIC barrel detector along the hadron-beam direction. The tracking system consists of four layers of AC-LGAD silicon with $\sim 15-20\,\mu\mathrm{m}$ spatial resolution, and provides high-resolution timing ($\sim 35\,\mathrm{ps}$) needed to correct for the momentum smearing induced by the 25~mrad rotation of the proton beam bunches. The system will enable the reconstruction of charged particles, in particular protons,  with polar scattering angle $5.5 \; \rm{mrad} < \theta < 20.0 \; \rm{mrad}$. For a proton beam of 100~GeV, this corresponds to scattered protons with transverse momenta in the range $0.6 \; {\rm GeV} < p_{T} < 2 \; \rm{GeV}$. It should be noted that the acceptance is not azimuthally symmetric due to the presence of the electron beam components in the same magnet bore, making symmetric tracking coverage impossible.

Behind the tracking detector, ${ \rm PbWO}_{4}$ crystals paired with silicon photomultipliers create a compact, high-resolution electromagnetic calorimeter to aid in the reconstruction of low-energy photons from nuclear de-excitation and hard photons from processes such as backward DVCS.

\subsubsection*{The Roman pots} 

\begin{figure}[h]
    \centering
    \includegraphics[width=0.99\columnwidth]{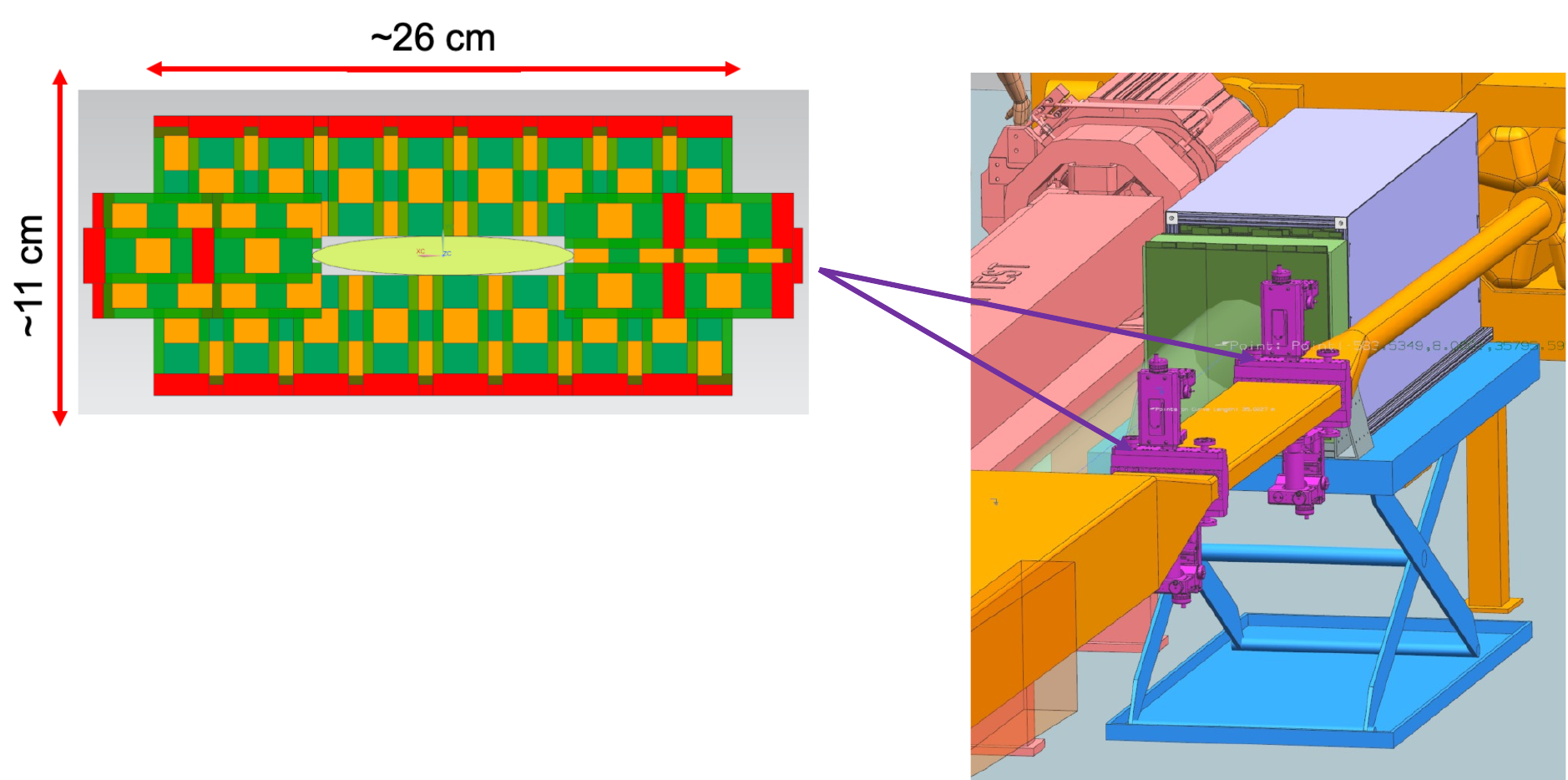}
    \caption{Depiction of the Roman-pots subsystem and its support structure in the present ePIC design. The active detector area of the Roman pots consists of staves, with AC-LGADs mounted on both sides. These same staves are used for both the B0 tracker and off-momentum detector subsystems.}
    \label{fig:roman_pots}
\end{figure}

The Roman pots are designed to maximize the geometric acceptance provided by the magnetic apertures and beam optics of the outgoing hadron beam, especially at very low $p_{T} \sim 0.2$~GeV (inner detector edge at a distance of $1\,\mathrm{cm}$ or less from the hadron beam core). These detectors are comprised of the same AC-LGAD silicon technology as that of the B0 spectrometer. However, due to the absence of a magnet, a different tracking approach is required. This approach makes use of the knowledge of proton transport through the hadron lattice magnets. The Roman pots are organized into two stations, at around 32 and 34 meters from the interaction point. The layout of the AC-LGAD sensors in the detector planes in addition to the support structure in the current ePIC CAD model are shown in Fig.~\ref{fig:roman_pots}. 

\subsubsection*{The off-momentum detectors} 

\begin{figure}[h]
    \centering
    \includegraphics[width=0.8\columnwidth]{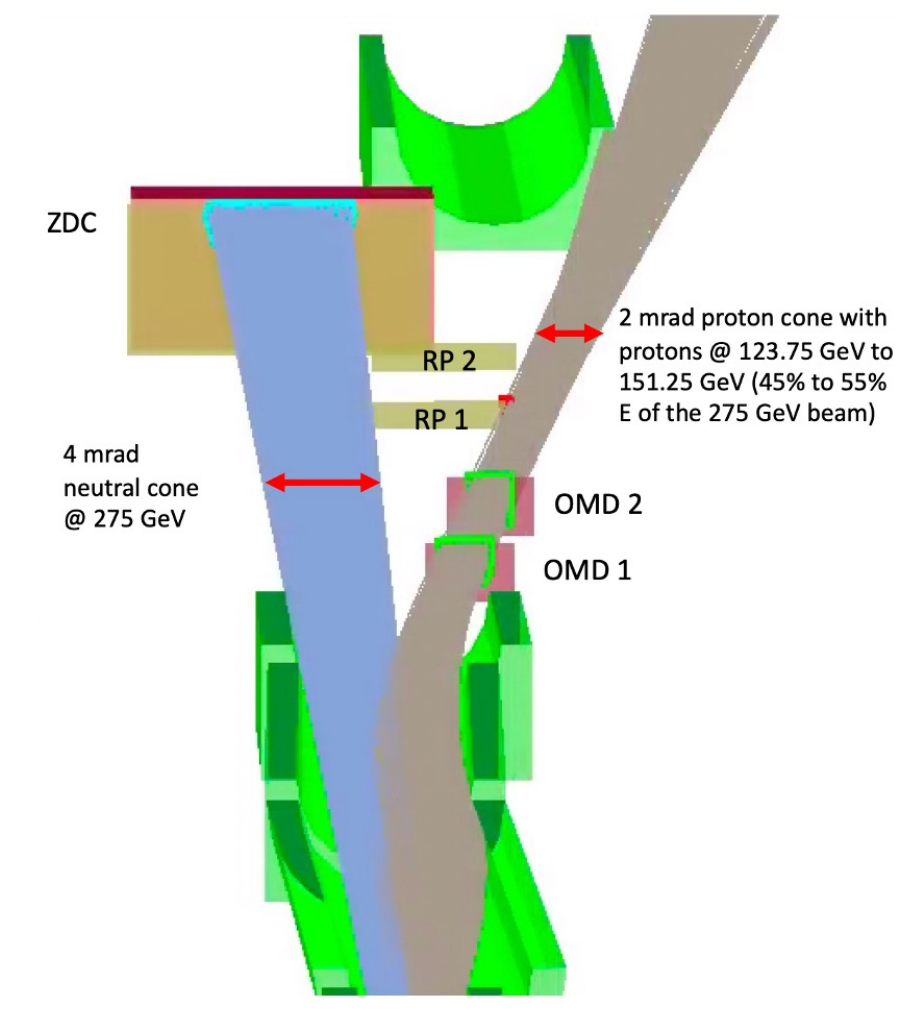}
    \caption{Full EICROOT simulation~\cite{ref:EICROOT} showing the neutron cone (blue), going towards the zero-degree calorimeter, and the off-momentum proton profile being steered off-axis towards the off-momentum detector planes.}
    \label{fig:offMomBeamProfile}
\end{figure}

The role of the Off-Momentum Detectors (OMDs) lies primarily 
in the reconstruction of protons with a large loss in 
longitudinal momentum, of around 40\% or more with
respect to the nominal beam momentum. The detection of particles with such large momentum loss is of crucial
importance for studies with nucleus collisions, in particular for 
spectator tagged DIS~\cite{Jentsch:2021qdp} and spectator-tagged vector-meson production~\cite{TU2020135877}. These detectors function identically to Roman pots, but are situated further away from the beam to capture protons that are bent away from the beam core due to their lower rigidity, as illustrated in Fig.~\ref{fig:offMomBeamProfile}.

\subsubsection*{The zero-degree calorimeter} 

\begin{figure*}[h]
    \centering
    \includegraphics[width=0.95\linewidth]{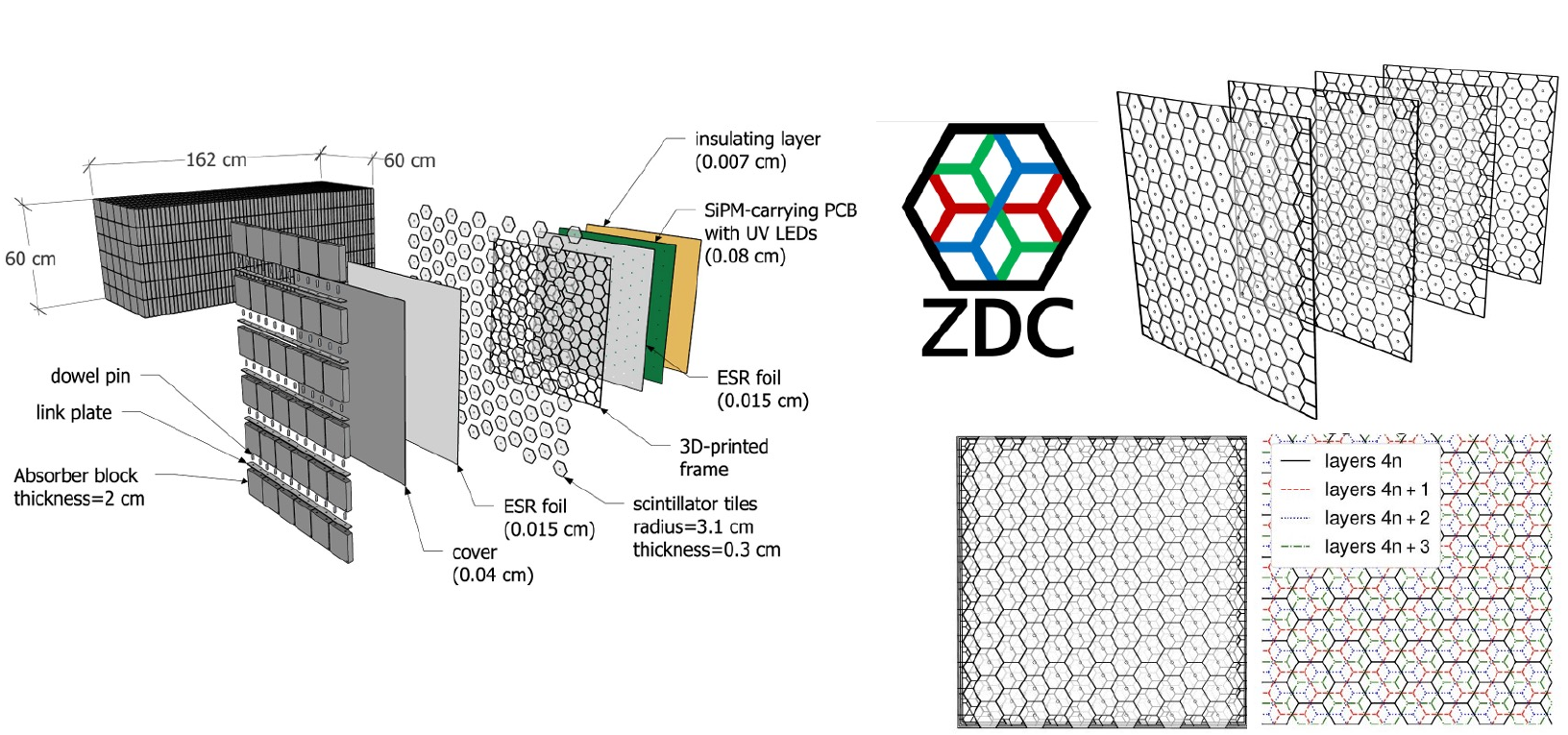}
    \caption{Exploded view of the ePIC SiPM-on-tile calorimeter. 
    The scintillator is etched into small cells, each embedded with a SiPM.
    These cells are then staggered transversely to create sub-cells, which enable finer pixelation that can take advantage of modern computing techniques to achieve an imaging calorimeter with far fewer channels than a conventional silicon imaging calorimeter. The algorithm that takes advantage of this approach can be found in Ref.~\cite{PAUL2024169044}.}
    \label{fig:ZDC}
\end{figure*}

The Zero-Degree Calorimeter (ZDC) is comprised of a  PbWO4 crystal calorimeter and a calorimeter consisting
of alternating layers of scintillating and absorber material. The latter calorimeter, named “SiPM-on-tile.”, is an imaging calorimeter 
made of etched cells of scintillating tiles, iron absorber material, and a silicon photomultiplier (SiPM). 
It is depicted in Fig. ~\ref{fig:ZDC}.

The combination of all four far-forward subsystems allows for an extended coverage for tagging and reconstructing protons (and charged particles in the B0) as well as neutral final states at far-forward rapidity. The overall performance of the far-forward detector system  depends on the individual subsystems and is summarized in Table~\ref{tab:FF_detector_performance}.

\begin{table*}[h]
\centering
\begin{tabular}{ c c c c }
\hline
Detector & Energy Resolution & Position/Angular Resolution & $p_{T}$ Resolution  \\
\hline
 B0 tracker    & N/A & 20 $\mu$m & $<$ 7\% for $p_{T} \sim 1$~GeV \\
 B0 EMCAL      & $\Delta E/E = \frac{1-3\%}{\sqrt{E}} \oplus 2\%$ & 1-2 cm & N/A \\
 Off-Momentum  & N/A & 20 $\mu$m & 5\% for $p_{T} \sim 1$ GeV \\
 Roman Pots    & N/A & 20 $\mu$m & 5\% for $p_{T} \sim 1$ GeV \\
 ZDC EMCAL     & $\Delta E/E = \frac{1-3\%}{\sqrt{E}} \oplus 2\%$ & 1-2 cm & N/A \\
 ZDC HCAL      & $\Delta E/E = \frac{50\%}{\sqrt{E}} \oplus 5\%^{*}$ & $< \frac{1 \rm{mrad}}{\sqrt{E}}$ & N/A \\
 \hline
\end{tabular}
\caption{ Summary of the detector resolutions for the far-forward subsystems. Full discussion of the requirements for the various far-forward subsystems, with references, can be found in~\cite{ref:ePIC_FF_wiki}. *The ZDC SiPM-on-Tile hadronic energy resolution can be improved via a graph neural network energy reconstruction.}
\label{tab:FF_detector_performance}
\end{table*}

For the measurement of exclusive photon or meson production off a proton, the scattered proton can be detected in the B0 or in the Roman pots, depending on the beam-energy configuration and the scattered-proton momentum. When reconstructing the scattered proton in the far-forward system, a center-of-mass-energy--dependent gap is present in the Mandelstam-$t$ distribution, because the coverage of the far-forward system is incomplete, especially at the lower hadron beam energies {(e.g. for 100 GeV protons). This can be corrected for by relying on the kinematic constraints imposed by the exclusivity of the specific process. It should be noted that this imposes an additional difficulty on the full event reconstruction due to potential contamination from misidentification of different processes, for example mistaking exclusive $\pi^{0} \rightarrow \gamma \gamma$ production for DVCS when the two photons merge in a calorimeter cluster, or a photon is missed. However, due to high-granularity calorimetry across the ePIC fiducial acceptance, this type of misidentification has been demonstrated to be manageable~\cite{Aschenauer:2025cdq}.}

When studying diffractive production of a meson or photon off a nucleus, the (virtual) photon exchanged between the beam electron and nucleus can interact either with the nucleus as a whole or with one of the nucleon constituents inside the nucleus. The former is referred to as coherent production and implies the nucleus to remain intact, while the latter is referred to as incoherent production and implies the break up of the nucleus. With the exception of the lightest nuclei, the scattered nucleus  in coherent production stays too close to the beam line to be detected in the far-forward system. For incoherent production, nucleus debris can be detected in the far-forward system. 

In general, both for coherent and for incoherent production, 
the reconstruction of kinematic variables is based on the scattered lepton and the newly created particles detected in the ePIC central detector. The coherent cross section differential in the Mandelstam-$t$ variable shows as a function of $t$ a diffraction pattern characterized by an alternate succession of maxima and minima, which correlate with the size of the nucleus under study. For the incoherent cross section, such pattern is absent. At low Mandelstam $t$, coherent production dominates, but quickly, at increasing values of $t$, incoherent production becomes largely dominant. Hence, if one is interested in coherent production, it is vital to heavily reduce the incoherent contribution. Signal detection in the far-forward system is used to veto events from incoherent production. Ongoing studies on exclusive meson production off nuclei, for which figures will be released in the near future by the ePIC collaboration, indicate that using the far-forward detector system the incoherent contribution can be reduced such as to resolve the first two maxima and the first minimum of the coherent signal in the lowest-$t$ region, 
while in the region of larger $t$, the incoherent signal remains dominant. 

Besides the reduction of the incoherent contribution, it is also important to resolve the diffractive pattern of minima and maxima from coherent production. Here, the momentum resolution of the scattered beam electron forms the  limiting factor, smearing out the diffractive pattern. However, recently, a new method has been introduced that allows to restore the diffractive pattern~\cite{Kesler:2025ksf}. In this method, the $t$ distribution is projected along the direction perpendicular to the plane formed by the initial and scattered beam lepton. This allows to reduce the effect of the electron-momentum resolution.

\subsection{Proton and $\pi^+$ DVCS at EicC}
While the EIC at BNL will cover a broad kinematic domain extending deep into the small-$x$ gluon–dominated regime, the proposed Electron–Ion Collider in China (EicC) is designed to focus on the sea-quark region at moderate energies. Together, these facilities establish a global program in hard exclusive reactions that will decisively advance our understanding of the inner structure of hadrons.
The white paper for EicC was released in 2021~\cite{Anderle:2021wcy}, 
while the Chinese version was published a year earlier~\cite{CAO:2024fdz}. 
It is proposed to experimentally investigate the multi-dimensional structure of the nucleon in the sea quark region with the aim of bridging the kinematic coverage gap between EIC and JLab, as clearly shown in Fig.~\ref{fig:kinematic_coverage}.
The 3D imaging in terms of GPDs through exclusive processes is one of the most vital goals of the EicC~\cite{Cao:2023wyz,Huang:2025wdq}. 
Other central scientific topics are the precise measurement of TMDs~\cite{Zeng:2022lbo}, electro- and photo-production of heavy quarkonium~\cite{Wang:2023thy,Cao:2019gqo} and exotic hadronic states~\cite{Cao:2023rhu,Cao:2020cfx,Yang:2020eye,Cao:2019kst}.
The electron beam with a polarization degree of around 80\% is designed to collide with the proton beam with a polarization of about 70\% under the center of mass energies of 15 to 20~GeV and the luminosity of (2–3) $\times 10^{33}$ cm$^{-2}$ s$^{-1}$.
This project proposes to upgrade the High-Intensity heavy-ion Accelerator Facility (HIAF)~\cite{Yang:2013yeb} in Huizhou, China, \\red{which became operational late last year.} Additionally, a polarized electron ring will be installed as part of the upgrade.
Motivated by the requirements of the physics objectives, a central detector system is designed with the coverage of charged particles within pseudo-rapidity $-3.5 <\eta < 3.5$, and the electromagnetic calorimeter for electron and photon detection covering a range of $-3 < \eta <3$.
The far-forward detectors, including Roman pots, off-momentum detector, and zero-degree calorimeter (ZDC), are particularly relevant for the extremely forward-going proton required by the GPDs physics.
{The specified central and far-forward detectors constitutes the baseline setup of the EicC detection system, as detailed in the forthcoming Conceptual Design Report.}  

\begin{figure}[htb]
\centering\includegraphics[width=0.35\textwidth]{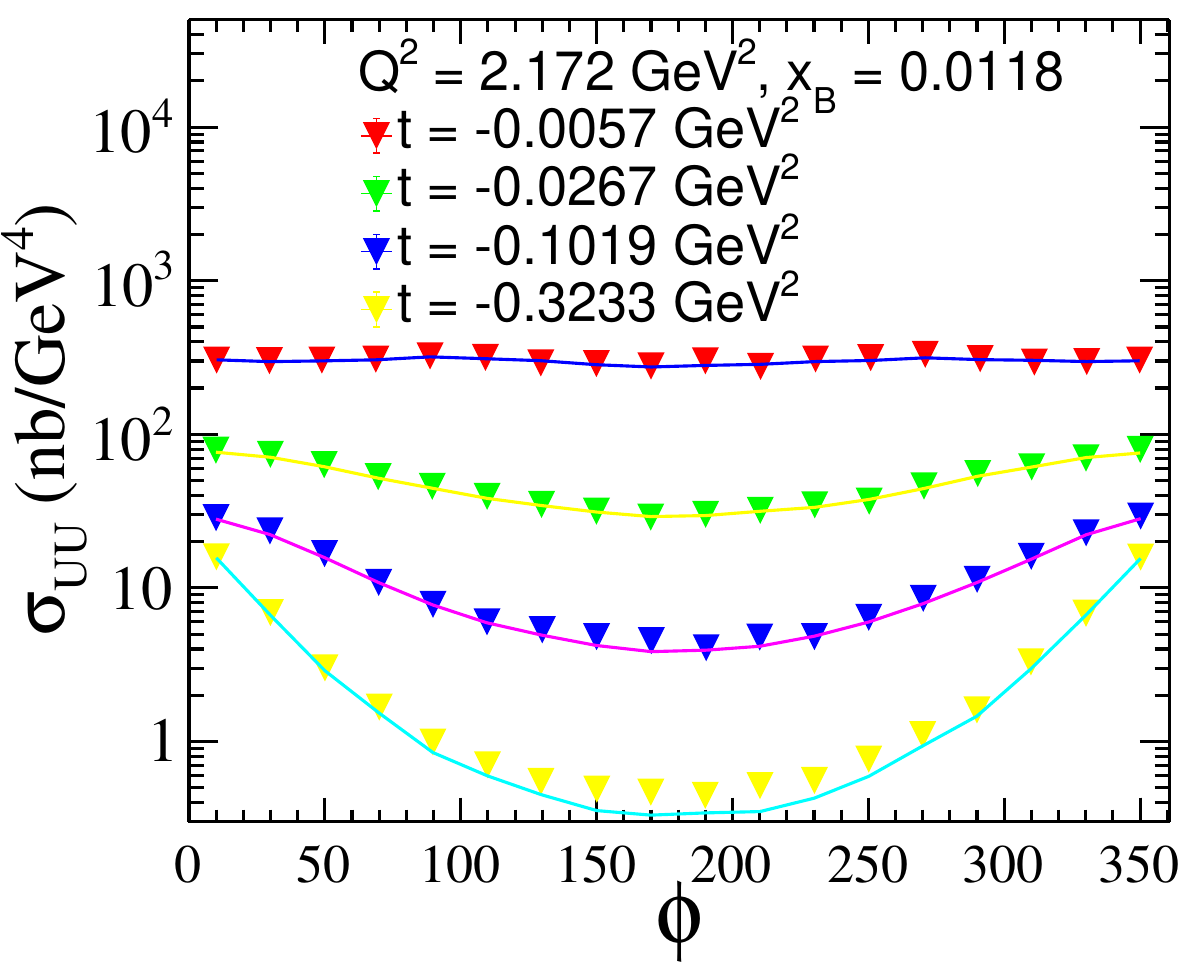}\\
\centering\includegraphics[width=0.35\textwidth]{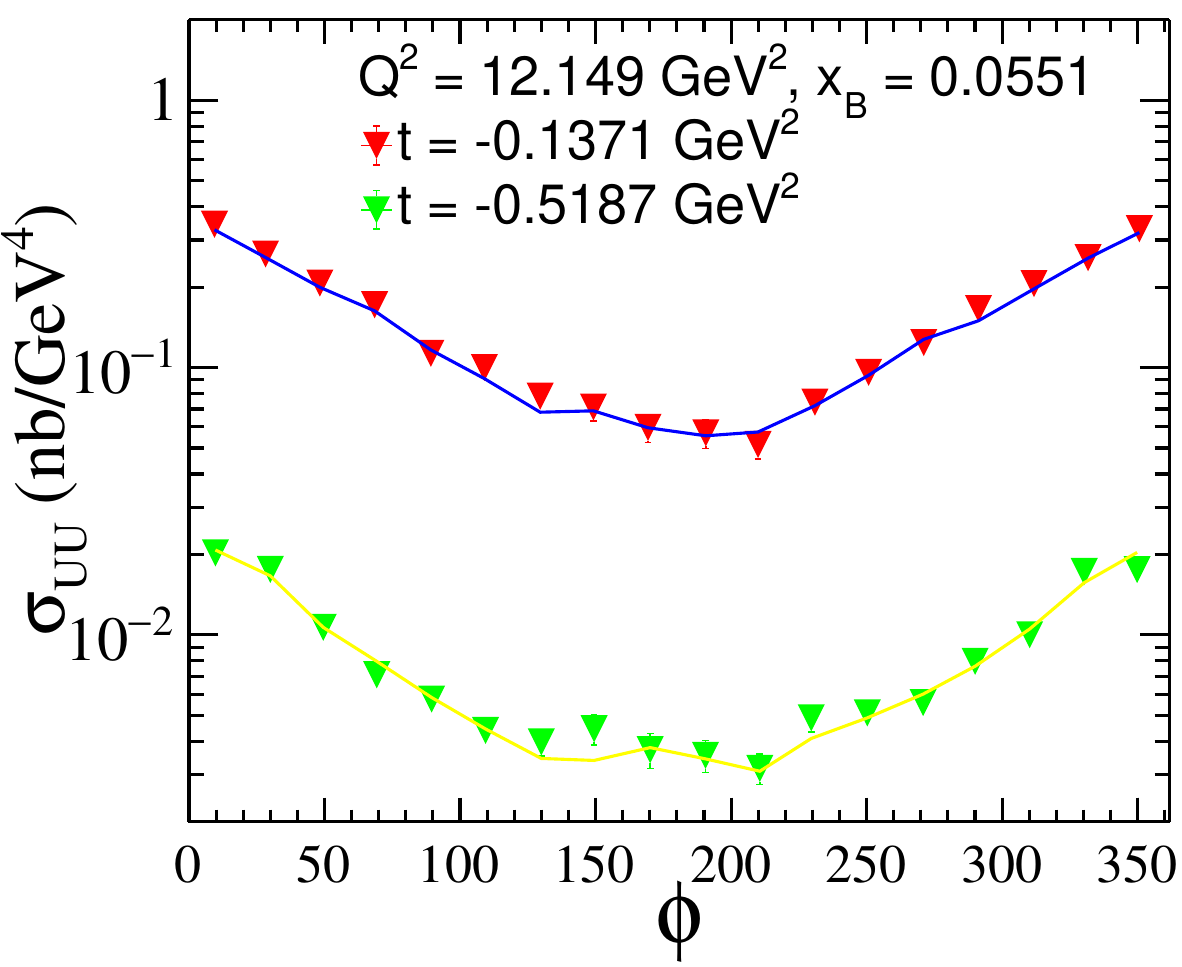}
\caption{
Projected data of unpolarized cross section {of proton DVCS} in two selected ($x_B$, $Q^2$) bins of the EicC kinematical region within {the} adapted MILOU generator~\cite{Perez:2004ig,Cao:2023wyz}.
{The colored lines represent} Goloskokov-Kroll (GK) model~\cite{Goloskokov:2005sd,Goloskokov:2006hr}.
}
\label{fig:EicC_pDVCS_proj}
\end{figure}

\begin{figure}[htb]
\centering\includegraphics[width=0.45\textwidth]{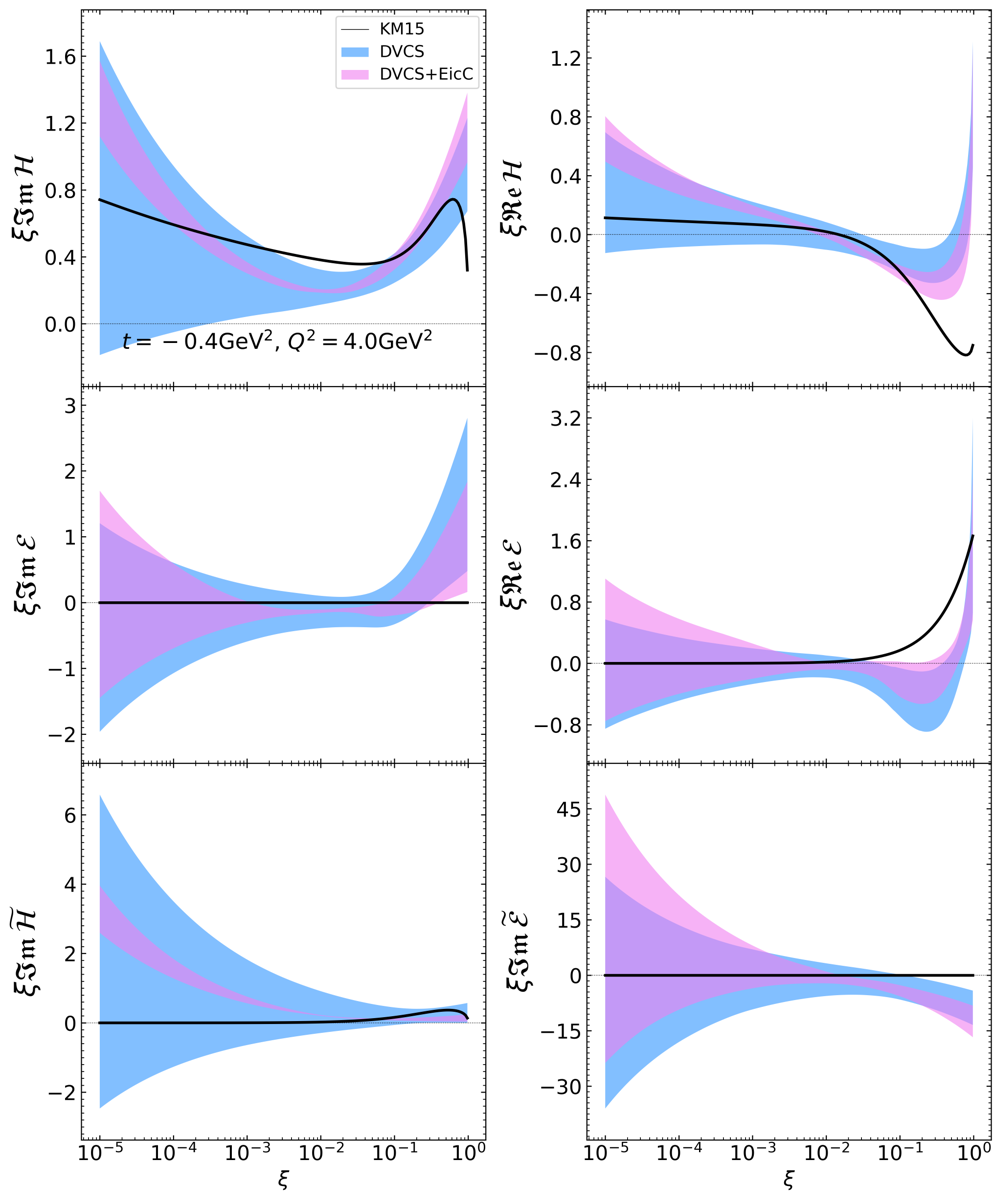}
\caption{
The extraction of CFFs versus skewness $\xi$ at $Q^2 = $ 4.0 GeV$^2$ and $-t =$ 0.4 GeV$^2$.
The blue and red error bands are uncertainties before and after including the pseudodata of asymmetries at EicC. Figure is adapted from Ref. \cite{Huang:2025wdq,Huang:2026eai} by further enabling 
dispersion relations.
}
\label{fig:EicC_pDVCS_imp}
\end{figure}

The proton DVCS is the most extensively studied exclusive process relevant to GPDs from both experimental and theoretical sides.
Pseudo-data of unpolarized cross sections in two typical bins {($0.0102 < x_B < 0.0139$, $1.63  < Q^2 < 2.64$ GeV$^2$ and $0.0475 <  x_B < 0.0646$, $6.98  < Q^2 < 18.45$ GeV$^2$)} in the EicC kinematical region are shown in Fig.~\ref{fig:EicC_pDVCS_proj}.
The simulation takes into account the acceptance and resolution of the state-of-art design of detectors within the adapted MILOU generator~\cite{Perez:2004ig,Cao:2023wyz}.
The estimated statistical uncertainties of asymmetry measurements range from 1.0\% $\sim$ 5.0\% {for all 69 kinematic bins within the range of $1.0 < Q^2 <  80.0$ GeV$^2$, $0.003 < x_B < 0.3$, and $0.02 < -t < 1.0$ GeV$^2$} during a year of data-taking with a luminosity of $2 \times 10^{33}$ cm$^{-2}$ s$^{-1}$ \cite{Huang:2025wdq}.
The systematical errors are expected to not exceed or be comparable to the statistical ones.
{The impact of these pseudo-data of all asymmetries ($A_{UT}$, $A_{UL}$, $A_{LU}$, $A_{LL}$, and $A_{LT}$) on the extraction of CFFs is shown in Fig. \ref{fig:EicC_pDVCS_imp} with the help of Gepard neural network~\cite{Kumericki:2011rz,Kumericki:2019ddg,Cuic:2020iwt} in a similar framework in Fig.~\ref{fig:flavsep}.
The kinematic cuts $Q^2 > 1.5~\text{GeV}^2,  -{t}/{Q^2} < 0.2$ are applied to all data sets with an aim to suppress the higher-twist contribution.
Re$\widetilde{\mathcal{H}}$ and Re$\widetilde{\mathcal{E}}$ are set to be vanishing.
Dispersion relations are implemented for the CFFs ${\mathcal{H}}$ and ${\mathcal{E}}$ \cite{Huang:2026eai}, unlike in a previous impact study \cite{Huang:2025wdq}.} 
The global analysis concludes that EicC can fairly improve the precision of all CFFs down to $\xi \sim 10^{-4}$, {with substantial improvements in the sea-quark region.}

At the leading twist, pseudoscalar meson production is sensitive to the axial-vector GPDs. {The $\pi^0$ DVMP is advantageous at EicC where the accessible regime of $Q^2>$ 10.0~GeV$^2$ will allow us to investigate the applicability of the factorization theorem \cite{Chen:2026vff}.} On the other hand, polarized deuterons and helium-3 beams are available for flavor separation of CFFs.

\begin{figure}[htb]
\centering\includegraphics[width=0.8\columnwidth]{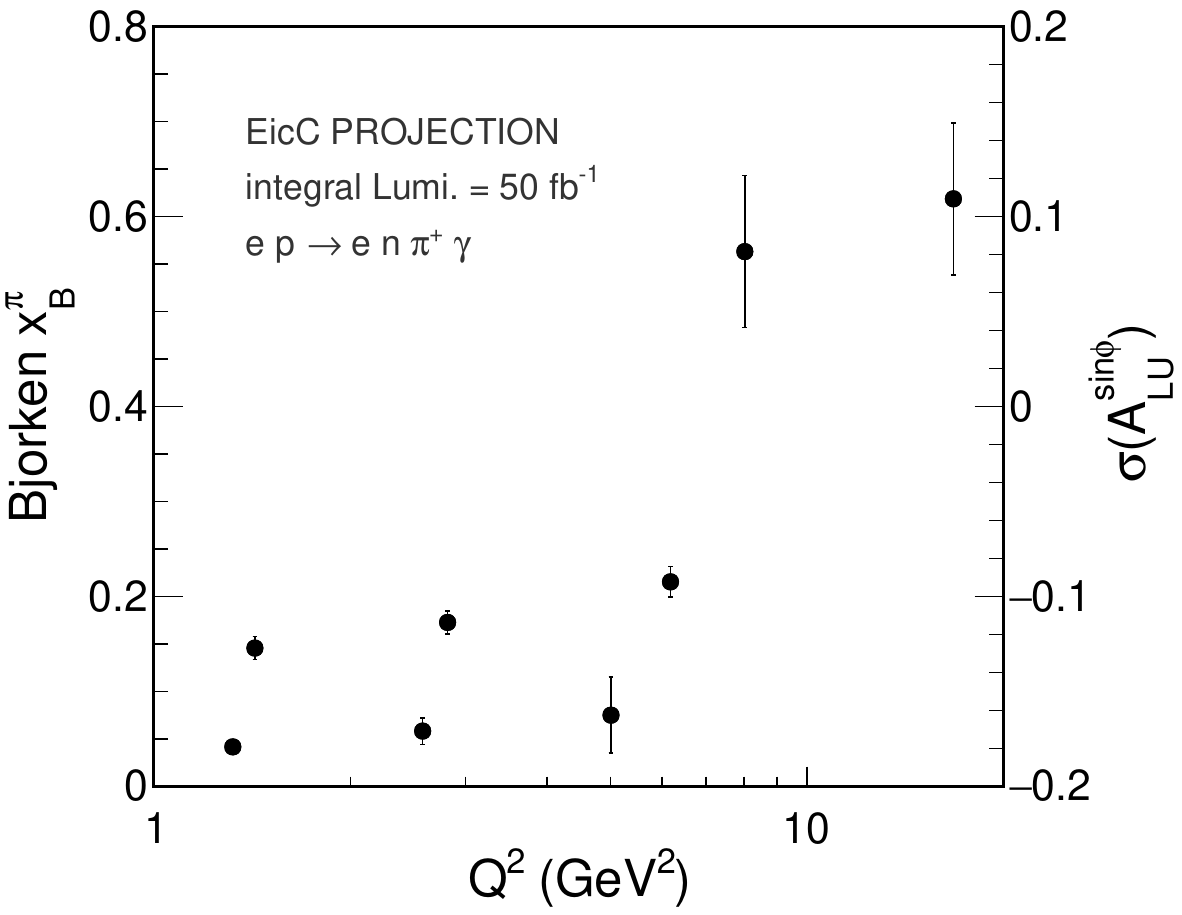}
\caption{
Projected statistical uncertainties of beam spin asymmetry with longitudinal polarized electron beam in the EicC kinematical region within generator~\cite{Chavez:2021koz}.
}
\label{fig:EicC_piDVCS_proj}
\end{figure}

As highlighted in the sections above, both pions and kaons play a central role in our understanding of strong interactions.
In addition to exploring nucleon's structure, EicC has the potential to measure the structure of those very important mesons in QCD.
The $\pi^+$ and $K^+$ electromagnetic form factor up to $Q^2 = 20 \sim 25$~GeV$^2$ can be extracted through reactions $p(e,e'\pi^+)n$ (with $-t \lesssim 0.6$~GeV$^2$) and $p(e,e'K^+)n$ (with $-t \lesssim 0.9$~GeV$^2$) at EicC \cite{Lu:2025bnm}, a {remarkable} extension of the JLab $Q^2$ range in 
Fig.~\ref{fig:fpi_proj}.
Similar processes with additionally detected real photons, {\it e.g.} $p(e,e'\gamma\pi^+)n$ and $p(e,e'\gamma K^+)n$, can be used to access the multidimensional structure of the pion and the kaon~\cite{Amrath:2008vx,Chavez:2021koz}.
For the first reaction, the interference term
between Compton scattering and the BH process reads~\cite{Amrath:2008vx}:
\be \label{eq:BSAsin}   
&&
\sum_{\textrm{spin}}\mathcal{M}^*_{\textrm{DVCS}}\mathcal{M}_{\textrm{BH}} \propto F_\pi(t_\pi) \nn \\ && \times \bigg[ \frac{1-y_\pi+y_\pi^2/2}{y_\pi \sqrt{1-y_\pi}} \textrm{Re} \mathcal{H}_{\pi} \cos\left(\phi_{\textrm{Trento}}\right) \nn \\ && + h_e \frac{1-y_\pi/2}{\sqrt{1-y_\pi}}
    \textrm{Im} \mathcal{H}_{\pi} \sin\left(\phi_{\textrm{Trento}}\right) \bigg],
\ee
with $y_\pi$ being the fraction of the beam energy carried by the virtual photon in the subprocess $e \pi \to e\pi \gamma$, and $h_e$ being the helicity of the electron beam.
Measurements of the unpolarized cross section and electron-beam-spin asymmetry probe the real and imaginary parts of the CFF 
$\mathcal{H}_{\pi}$, respectively. 
{It is shown that quarks and gluons interfere destructively and that gluon contribution to $\pi^+$ structure could be identified by a reduction in the magnitude of the $\pi^+$ DVCS beam spin asymmetry at EicC energies}~\cite{Chavez:2021koz}.
The projected statistical uncertainties of beam-spin asymmetry with a longitudinally polarized electron beam are shown in Fig.~\ref{fig:EicC_piDVCS_proj}, following the simulation of Ref.~\cite{Chavez:2021koz}.
The sensitivity, being less than 5.0\%, clearly reveals that meaningful measurements of the $\pi^+$-meson DVCS are attainable at the EicC by installing a ZDC to detect the neutron with polar angles within 15 mrad of the incident proton beam axis.

\subsection{GPDs and TDAs at J-PARC 
}
At the Hadron Experimental Facility in J-PARC, a high-momentum beamline that delivers a 30~GeV primary proton beam was recently constructed. In addition, this beamline will be upgraded to a secondary beamline that delivers 2-20~GeV/c positive and negative $\pi$/$K$/$p$ beams ($\pi$20 beam line). Using these high-momentum hadron beams, GPDs measurements are being considered~\cite{Tomida_MENU2023}.
Theoretical investigations were carried out for the $p+p\to p+\pi+B$~\cite{PhysRevD.80.074003},
$\pi^- + p \to \gamma + \gamma + n$~\cite{PhysRevD.109.074023} and 
$\pi^- + p \to \mu^+ + \mu^- + n$~\cite{BERGER2001265,GOLOSKOKOV2015323,DY_PRD2016,PhysRevD.109.074023} reactions.
The $p+p\to p+\pi+B$ and $\pi^- + p \to \gamma + \gamma + n$ reactions can access the $x$-dependence and the 
ERBL region of the GPDs.
Measurements at J-PARC can complement data from lepton machines.

Moreover, the backward peak of the cross sections for the reactions 
$\pi^- + p \to \gamma^* + n$ and $\pi^- + p \to J/\psi + n$, which are both observable as 
$\pi^- + p \to \mu^+ + \mu^- + n$, 
should also be analyzed at J-PARC to probe the factorization of $\pi \to N$ TDAs in these time-reversed channels~\cite{Pire:2016gut,Pire:2022kwu} with respect to those accessed in electroproduction experiments.

\section{Conclusions and outlook}
\label{sec:conclusion}

The study of 
{GPDs}
and related non-perturbative quantities has become an important task of modern hadronic physics. Over the past few decades, advancements in experimental techniques, theoretical frameworks, and lattice QCD have evolved the field from a collection of exploratory concepts into a coherent program for hadron tomography. 

Today, one of the central topics is the study of {GFFs} and their connection to the mechanical properties of hadrons. The first extractions of the pressure distribution inside the proton have stimulated both experimental and theoretical efforts, and forthcoming work will focus on controlling higher-order and higher-twist effects. The mesonic sector, accessible through GDAs, provides an essential complement and remains to be explored systematically.

A major theoretical challenge lies in the understanding of gluon GPDs. While next-to-leading order studies of DVCS and TCS have shown their visible impact, gluon contributions have often been underestimated in phenomenological analyses. The development of reliable theoretical tools for gluon degrees of freedom is therefore a key priority, especially in view of the future EIC, EicC, and related facilities.

Hard exclusive processes can also play an important role in resonance spectroscopy through studies of non-diagonal hard exclusive reactions. This may provide access not only to long-standing puzzles such as the nature of the Roper resonance  but also to bridging spectroscopy and the QCD factorization framework.

Alongside these developments, progress in lattice QCD is  essential. Beyond direct studies of GPDs, extending lattice efforts to GDAs and TDAs is crucial for connecting partonic correlation functions to observable quantities.  
Incorporating the model-independent non-analytic chiral structure provided by chiral EFT imposes the necessary constraints in the chiral limit and enables a substantial improvement in the reliability of lattice QCD calculations. 
Particularly, convoluting the nucleon-meson splitting functions computed by the chiral EFT with the meson structure information, such as GPD and GDA, provided by the light-front quark dynamics can advance the global analysis framework toward a deeper understanding of hadron structures. The convergence of the chiral expansion can also be improved by the nonlocal realization of the chiral EFT. The lattice QCD calculations constrained by the chiral EFT thus provide useful inputs for the global QCD analyses of hadron structure.
Functional approaches, such as the Dyson–Schwinger and Bethe–Salpeter equation methods, provide further complementary insights, particularly in the gluonic sector. Moreover, light-front holographic QCD (LFHQCD) has attracted growing interest and brings a novel perspective on the dynamics of strong interactions.
  
 A more general task concerns framing hadronic physics as high-energy nuclear physics. Extending the methods developed for the nucleon to nuclear systems offers a natural and promising direction, allowing the dynamics of QCD to be studied in increasingly complex environments.

{The developments summarized in this white paper should be viewed in the context of a field entering a new experimental era. The forthcoming and proposed measurements will require increasingly precise theoretical tools, flexible phenomenological frameworks, and closer integration with lattice-QCD results. }
{Looking ahead, the ongoing CLAS12 program, the
planned $\mu$CLAS12 upgrade and the
next generation of facilities, including the Jefferson Lab 22~GeV upgrade, COMPASS/AMBER, J-PARC, and particularly the EIC and EicC, will deliver data of unprecedented precision and coverage.} This will be supplemented by data from the LHC, which provides unique access to the lowest region in Bjorken-$x$.
The complementary strengths of the experiments, combined with progress in phenomenological methods, advanced analysis techniques, and machine learning,  pave the way for quantitative three-dimensional imaging of hadrons. Meanwhile, theoretical advancements -- from higher-order corrections and dynamic models to lattice extractions of GPDs -- will ensure that the resulting data are analyzed in a controlled and reliable manner. 
 Progress in detector capabilities will allow one to study lower--cross-section processes, such as DDVCS and TCS. The two workshops outlined the areas of major progress in developing the theory and the tools required for the study of those processes. 

The field is now ready to provide 
stronger constraints on {GFFs}, the $D$-term, and the pressure distribution within the proton, to investigate gluon imaging in nucleons and nuclei, and to establish strong links between QCD dynamics and emergent hadron properties, such as mass and spin. The Incheon and Trento workshops have highlighted the significance of ongoing international collaboration and the exchange of ideas among different communities. Establishing these kinds of meetings as a regular biannual series dedicated to hadron structure studies through hard exclusive reactions would foster the exchange of insights and help align experimental and theoretical priorities.

\section{Acknowledgments}

The workshop ``3D structure of the Nucleon via Generalized Parton Distributions'' held in Incheon, Republic of Korea, on June 25-28, 2024, was supported 
by the \href{http://apctp.org/}{Asia Pacific Center for Theoretical Physics} (APCTP), the \href{https://chep.knu.ac.kr/}{Center for High Energy Physics} (CHEP) and
\href{https://rsri.knu.ac.kr/}{Radiation Science Research Institute} of \href{https://en.knu.ac.kr/main/main.htm}{Kyungpook National University} (KNU), and the National Research Foundation of Korea (NRF) grant funded by the Korea government (MSIT) (RS-2023-00280845).  

The workshop ``Towards improved hadron tomography with hard exclusive reactions'' held in Trento, Italy on August 5-9, 2024 by the ECT*-Trento was supported by \href{https://web.infn.it/EURO-LABS/}{EUROpean Laboratories for Accelerator Based Sciences} (EURO-LABS),
\href{https://www.infn.it/en/}{Istituto Nazionale di Fisica Nucleare} (INFN) and
\href{https://jsallc.org/}{Jefferson Science Associates} (JSA).

A.~Afanasev acknowledges support from the U.S. National Science Foundation under grants 
PHY--2433872 and PHY--2514669.

M.~Bo\"er is supported by the  U.S. Department of Energy, Office of Nuclear Physics, under Grant No. DE-SC0025657.

H.-M. Choi was supported by the National Research Foundation of Korea (NRF) grant funded by the Korea government (MSIT) (RS-2023-NR076506).

K.~Cichy was supported by the National Science Centre (Poland) grant OPUS no.\ 2021/43/B/ST2/00497.

M.~Constantinou acknowledges financial support by the U.S. Department of Energy, Office of Nuclear Physics, under Grant No.\ DE-SC0025218.

N. Crnković was supported by the Croatian
Science Foundation project DOK-2020-01-9883.

G.M.~Huber is supported by the Natural Sciences and Engineering Research Council of Canada (NSERC), SAPIN-2021-00026.

C.-R.~Ji was supported by the U.S. Department of Energy (Grant No. DE-FG02-03ER41260). 

H.-S. Jo was supported by the National Research Foundation of Korea (NRF) grant funded by the Korea government (MSIT) (RS-2023-00280845) and by the Basic Science Research Program through the National Research Foundation of Korea (NRF) funded by the Ministry of Education (RS-2018-NR031074).

B.~Kriesten was supported at Argonne National Laboratory by the U.S.~Department of Energy under contract DE-AC02-06CH11357.

P.-J. Lin was supported by National Science and Technology Council (NSTC) through Grant No. NSTC 112-2112-M-008-041-MY3

V.~Mart\'inez-Fern\'andez was supported in part by l’Agence Nationale de la Recherche (ANR), project ANR-23-CE31-0019. 

Z.-E.~Meziani acknowledges financial support by the U.S. Department of Energy, Office of Nuclear Physics,  under contract No. DE-AC02-06CH11357.

M.~Nefedov was supported by the binational Science Foundation
grants \#2012124 and \#2021789, and by the ISF grant \#910/23.

{P.~Rossi acknowledges financial support by the U.S. Department of Energy, Office of Nuclear Physics,  under contract No. 89243126CSC000213.}

K.~Semenov-Tian-Shansky was supported by Basic Science Research Program through the National Research Foundation of Korea (NRF) funded by the Ministry of Education  RS-2023-00238703 and RS-2018-NR031074.

H.-D. Son was supported by the National Research Foundation of Korea (NRF) grant funded by the Korea government (MSIT) (RS-2023-00210298).

P.~Sznajder and J.~Wagner
were supported by the grant no.~2024/53/B/ST2/00968
of the National Science Centre, Poland.

N.~Tomida was supported in part by
Japan Society for the Promotion of Science (JSPS) 
Grant-in-Aid for Early-Career Scientists Grant No.~JP24K17068
and Grant-in-Aid for Scientific Research (A) Grant No.~JP22H00124.

A.~W.~Thomas was supported by the Australian Research Council through the Discovery Project DP230101791.

C. Van Hulse was supported by the program Atracci\'{o}n de Talento, Comunidad de Madrid (Spain), under the grant agreement No 2020-T1/TIC-20295.

{Z.W. Zhao was supported in part by the U.S. Department of Energy grant
number DE-FG02-03ER41231.}


\bibliographystyle{elsarticle-num}

\bibliography{refs}

\end{document}